\newif\ifdraft
\draftfalse

\ifdefined\DRAFT\drafttrue\fi

\ifdraft
   \documentclass[twocolumn,tighten,times,linenumbers,trackchanges,twocolappendix]{aastex631}
\else
   \documentclass[twocolumn,tighten,times,twocolappendix]{aastex631}
\fi

\usepackage{amsmath}
\let\tablenum\relax

\usepackage{afterpage}

\usepackage{siunitx}
\DeclareSIUnit{\parsec}{pc}
\DeclareSIUnit{\Mpc}{\mega\parsec}
\DeclareSIUnit{\year}{yr}

\usepackage{textgreek}

\usepackage[utf8]{inputenc}
\usepackage{xspace}
\usepackage{acronym}
\usepackage{booktabs}

\usepackage{color, colortbl}

\usepackage{graphicx}
\usepackage[caption=false]{subfig}

\usepackage{tabularx}
\usepackage{multirow}

\usepackage{makecell}

\newcolumntype{Y}{>{\centering\arraybackslash}X}
\newcolumntype{Z}{>{\raggedright\arraybackslash}X}
\newcolumntype{K}{>{\raggedleft\arraybackslash}X}
\newcolumntype{U}{>{\hsize=1.01\hsize}Y}
\newcolumntype{V}{>{\hsize=1.2\hsize}Y}
\newcolumntype{W}{>{\hsize=0.71\hsize}Y}

\usepackage{scrextend}
\usepackage{textgreek}

\usepackage{microtype}

\AtBeginDocument{}

\newcommand{\checkbar}{\checkmark\kern-1.1ex\raisebox{.7ex}{\rotatebox[origin=c]{135}{--}}}

\newcommand{\checkprevext}{+}
\newcommand{\checkprevlvk}{++}
\newcommand{\checknodata}{\!\!\nodata\!\!}

\newcommand{\tablefootnote}[1]{}

\usepackage{makerobust}

\renewcommand{\today}{\number\day\space\ifcase\month\or
  January\or February\or March\or April\or May\or June\or
  July\or August\or September\or October\or November\or December\fi
  \space\number\year}

\usepackage{float}

\definecolor{NOTECOLOR}{rgb}{0.4, 0.2, 0.1}
\definecolor{notecolor}{rgb}{0.4, 0.2, 0.1}
\definecolor{danger-red}{rgb}{0.8, 0.4, 0.0}
\definecolor{ok-green}{rgb}{0.0, 0.6, 0.5}

\newcommand{\reviewed}[1]{{#1}}

\newcommand{\macro}[1]{#1}

\definecolor{mgreen}{rgb}{0.1,0.7,0.1}

\newcommand{\subpapersection}[1]{\section{#1}}
\newcommand{\subpapersubsection}[1]{\subsection{#1}}
\newcommand{\subpapersubsubsection}[1]{\subsubsection{#1}}

\newcommand{\paperscommonname}{GWTC-4.0: Tests of General Relativity. }

\newtoggle{fullauthorlist}
\toggletrue{fullauthorlist}

\newtoggle{endauthorlist}
\toggletrue{endauthorlist}

\newcommand{\LVKCollabAuthors}{The LIGO Scientific Collaboration, the Virgo Collaboration, and the KAGRA Collaboration}

\newcommand{\LVKCorrespondence}{The full author list is given at the end.\\
 For correspondence: LSC P\&P Committee, for LVK Publications, via}

\newcommand\gwtc[1][?]{\mbox{GWTC\if#1?\else-#1\fi}}
\newcommand\thisgwtcversionmajor{4}
\newcommand\thisgwtcversionminor{0}
\newcommand\thisgwtcversionfull{\thisgwtcversionmajor.\thisgwtcversionminor}
\newcommand\thisgwtcversion\thisgwtcversionfull

\newcommand{\soft}[1]{\textsc{#1}}

\newcommand{\GSTLAL}{\soft{GstLAL}\xspace}

\newcommand{\CWB}{\soft{cWB}\xspace}

\newcommand{\PYCBC}{\soft{PyCBC}\xspace}
\newcommand{\MBTA}{\soft{MBTA}\xspace}

\newcommand{\BAYESWAVE}{\soft{BayesWave}\xspace}
\newcommand{\BILBY}{\soft{Bilby}\xspace}

\newcommand{\LALSUITE}{\soft{LALSuite}\xspace}

\newcommand{\PBILBY}{\soft{ParallelBilby}\xspace}
\newcommand{\ASIMOV}{\soft{Asimov}\xspace}
\newcommand{\PESUMMARY}{\soft{PESummary}\xspace}

\newcommand{\NUMPY}{\soft{NumPy}\xspace}
\newcommand{\SCIPY}{\soft{SciPy}\xspace}

\newcommand{\SEABORN}{\soft{seaborn}\xspace}
\newcommand{\GWPY}{\soft{GWpy}\xspace}
\newcommand{\DYNESTY}{\soft{Dynesty}\xspace}

\newcommand{\IMRPhenomXPHM}{\soft{IMRPhenomXPHM}\xspace}
\newcommand{\IMRPhenomXPHMST}{\soft{IMRPhenomXPHM\_SpinTaylor}\xspace}

\newcommand{\SEOBNRFIVEPHM}{\soft{SEOBNRv5PHM}\xspace}

\newcommand{\SURSEVENDQFOUR}{\soft{NRSur7dq4}\xspace}

\DeclareSIUnit\parsec{pc}
\DeclareSIUnit\Mpc{\mega\parsec}
\DeclareSIUnit\yr{yr}
\DeclareSIUnit\GpcCubedYear{\giga\parsec\cubed\yr}

\newcommand{\Msun}{\ensuremath{\mathit{M_\odot}}}

\newcommand{\Mtot}{\ensuremath{M}}
\newcommand{\Mf}{\ensuremath{M_\mathrm{f}}}

\newcommand\PEpdfp{\ensuremath{p}}

\newcommand{\PEparameter}{\ensuremath{\boldsymbol{\theta}}}%

\newcommand\PEpdf[2][?]{\ensuremath{\PEpdfp({#2}\ifx#1?\else | {#1}\fi)}}

\newcommand\PEpriorpdfpi{\ensuremath{\pi}}
\newcommand\PEpdfprior[1]{\ensuremath{\PEpriorpdfpi({#1})}}
\newcommand\PEprior[1][\PEparameter]{\PEpdfprior{#1}}
\newcommand\PEpriorpe[1][\PEparameter]{{\let\keepPEpriorpdfpi\PEpriorpdfpi\def\PEpriorpdfpi{\keepPEpriorpdfpi_{\text{PE}}}\PEprior[#1]\let\PEpriorpdfpi\keepPEpriorpdfpi}}

\newcommand{\BH}[0]{\ac{BH}\xspace}

\newcommand{\SNR}[0]{\ac{SNR}\xspace}

\newcommand{\GR}[0]{\ac{GR}\xspace}

\usepackage{amstext}

\makeatletter
\@ifpackageloaded{cleveref}{
  \AtBeginDocument{
    \def\ltx@label#1{\cref@label{#1}}%
    \def\label@in@display@noarg#1{\cref@old@label@in@display{#1}}%
    \def\label@in@mmeasure@noarg#1{%
      \begingroup%
        \measuring@false%
        \cref@old@label@in@display{#1}%
      \endgroup}%
  }
}{}
\makeatother

\protected\def\protectedacused{\acused}

\acrodef{LIGO}[LIGO]{Laser Interferometer Gravitational-Wave Observatory}
\acrodef{LHO}[LHO]{\ac{LIGO} Hanford Observatory}
\acrodef{LLO}[LLO]{\ac{LIGO} Livingston Observatory}
\acrodef{KAGRA}[KAGRA]{KAGRA}\acused{KAGRA}
\acrodef{iKAGRA}[iKAGRA]{initial-phase \ac{KAGRA}}
\acrodef{bKAGRA}[bKAGRA]{baseline-design \ac{KAGRA}}
\acrodef{GEO}[GEO]{GEO\,600 \ac{GW} detector}
\acrodef{aLIGO}{Advanced \ac{LIGO}}
\acrodef{A+}{Advanced+ \ac{LIGO}}
\acrodef{Asharp}[\ensuremath{\text{A}^\sharp}]{\ac{LIGO} \acs{Asharp}}
\acrodef{AdV}{Advanced \acl{Virgo}}
\acrodef{AdV+}{Advanced \acl{Virgo}+}
\acrodef{Virgo}{Virgo}\acused{Virgo}
\acrodef{VirgoNEXT}[Virgo\_nEXT]{Virgo\_nEXT}\acused{VirgoNEXT}

\acrodef{LSC}[LSC]{\acs{LIGO} Scientific Collaboration}
\acrodef{LV}[LV]{\acs{LIGO}--\acs{Virgo} Collaboration\protect\protectedacused{LVC}}
\acrodef{LVC}[LV]{\acs{LIGO}--\acs{Virgo} Collaboration\protect\protectedacused{LV}}
\acrodef{LVK}[LVK]{\acs{LIGO}--\ac{Virgo}--\ac{KAGRA} Collaboration}
\acrodef{IGWN}[IGWN]{International \ac{GWH} Observatory Network}

\acrodef{O1}[O1]{first observing run}
\acrodef{O2}[O2]{second observing run}
\acrodef{O3}[O3]{third observing run}
\acrodef{O3a}[O3a]{first half of the third observing run}
\acrodef{O3b}[O3b]{second half of the third observing run}
\acrodef{O3GK}[O3GK]{observing run}
\acrodef{O4}[O4]{fourth observing run}
\acrodef{O4a}[O4a]{first part of the fourth observing run}
\acrodef{O4b}[O4b]{second part of the fourth observing run}
\acrodef{O4c}[O4c]{third part of the fourth observing run}
\acrodef{O5}[O5]{fifth observing run}

\acrodef{BH}[BH]{black hole}
\acrodef{BBH}[BBH]{binary \ac{BH}}
\acrodef{BNS}[BNS]{binary \ac{NS}}
\acrodef{IMBH}[IMBH]{intermediate-mass \ac{BH}}
\acrodef{NS}[NS]{neutron star}
\acrodef{BHNS}[BHNS]{\ac{BH}--\ac{NS} binary}
\acrodefplural{BHNS}[BHNSs]{\ac{BH}--\ac{NS} binaries}
\acrodef{NSBH}[NSBH]{\ac{NS}--\ac{BH} binary}
\acrodefplural{NSBH}[NSBHs]{\ac{NS}--\ac{BH} binaries}
\acrodef{PBH}[PBH]{primordial \ac{BH}}
\acrodef{CBC}[CBC]{compact binary coalescence}

\acrodef{GW}[GW]{gravitational wave\protect\protectedacused{GWH}}
\acrodef{GWH}[GW]{gravitational-wave\protect\protectedacused{GW}}
\acrodef{IFO}[IFO]{interferometer}
\acrodef{SNR}[SNR]{signal-to-noise ratio}
\acrodef{FAR}[FAR]{false-alarm rate}
\acrodef{IFAR}[IFAR]{inverse false-alarm rate}
\acrodef{FAP}[FAP]{false alarm probability}
\acrodef{PSD}[PSD]{power spectral density}
\acrodef{GR}[GR]{general relativity}
\acrodef{NR}[NR]{numerical relativity}
\acrodef{PN}[PN]{post-Newtonian}
\acrodef{EOB}[EOB]{effective-one-body}
\acrodef{ROM}[ROM]{reduced-order model}
\acrodef{IMR}[IMR]{inspiral--merger--ringdown}
\acrodef{PDF}[pdf]{probability density function}
\acrodef{PE}[PE]{parameter estimation}
\acrodef{CI}[CI]{credible interval}
\acrodef{CL}[CL]{credible level}
\acrodef{EOS}[EoS]{equation of state}
\acrodef{KLD}[KLD]{Kullback--Leibler divergence}
\acrodef{JSD}[JSD]{Jensen--Shannon divergence}
\acrodef{GCN}[GCN]{General Coordinates Network}
\acrodef{GWTC}[GWTC]{Gravitational-Wave Transient Catalog}
\acrodef{GWOSC}[GWOSC]{Gravitational Wave Open Science Center}
\acrodef{WDM}[WDM]{Wilson--Debauchies--Meyer}

\acrodef{CWB}[cWB]{coherent WaveBurst}
\acrodef{LAL}[LAL]{\ac{LIGO} algorithm library}

\acrodef{CHRoCC}{central heating radius of curvature correction}
\acrodef{NonSENS}{non-stationary estimation and noise subtraction}

\acrodef{PTA}{Pulsar Timing Array}

\input{common__event_macros}

\newcommand{\BILBYTGR}{\soft{BilbyTGR}\xspace}

\newcommand{\IMRPhenomXPHMMSA}{\soft{IMRPhenomXPHM\_MSA}\xspace}

\newcommand{\pSEOBNRFIVEPHM}{\soft{pSEOBNRv5PHM}\xspace}
\newcommand{\BW}{\soft{BW}\xspace}

\newcommand{\TGRFARTHRESH}{\reviewed{$\le \qty{e-3}{\yr^{-1}}$}\xspace}
\newcommand{\TGRNUMEVENTS}{\reviewed{42}\xspace}

\newcommand{\TGRFigureWidth}{3.375in}
\newcommand{\TGRFigureWidthPage}{\textwidth}

\DeclareRobustCommand{\COMMONNAME}[1]{\IfEqCase{#1}{{GW230529}{\MINIMALNAME{GW230529_181500}{}}{GW230814other}{\MINIMALNAME{GW230814_061920}{}}{GW230814single}{\MINIMALNAME{GW230814_230901}{}}{GW231123}{\MINIMALNAME{GW231123_135430}{}}}[\textcolor{red}{???}]}

\DeclareRobustCommand{\TGRImrctGWTCFOURResults}[1]{\IfEqCase{#1}{{DMFGWTC4PHENOM}{\reviewed{\ensuremath{-0.00^{+0.05}_{-0.05}}}}{DCHIFGWTC4PHENOM}{\reviewed{\ensuremath{-0.04^{+0.07}_{-0.06}}}}{GRQUANTGWTC4}{\reviewed{\ensuremath{89.1}}}{GRQUANTGWTC3}{\ensuremath{79.6}}}}

\DeclareRobustCommand{\TGRImrctGWTCFOURResultsWOGWOneNineZeroEightOneFour}[1]{\IfEqCase{#1}{{DMFGWTC4PHENOM}{\reviewed{\ensuremath{0.01^{+0.05}_{-0.05}}}}{DCHIFGWTC4PHENOM}{\reviewed{\ensuremath{-0.03^{+0.06}_{-0.06}}}}{GRQUANTGWTC4}{\reviewed{\ensuremath{86.0}}}}}

\DeclareRobustCommand{\TGRImrctHierPopGWTCFOURResults}[1]{\IfEqCase{#1}{{DMFGWTC4HIERPOP}{\reviewed{\ensuremath{0.00^{+0.07}_{-0.06}}}}{DCHIFGWTC4HIERPOP}{\reviewed{\ensuremath{-0.05^{+0.11}_{-0.11}}}}{GRQUANTDevParamsTwoDGWTC4HIERPOP}{\reviewed{\ensuremath{51.7}}}{GRQUANTMeansTwoDGWTC4HIERPOP}{\reviewed{\ensuremath{94.2}}}{GRQUANTHyperParamsFourDGWTC4HIERPOP}{\reviewed{\ensuremath{73.1}}}}}

\DeclareRobustCommand{\TGRImrctHierPopGWTCFOURResultsWOGWOneNineZeroEightOneFour}[1]{\IfEqCase{#1}{{DMFGWTC4HIERPOP}{\reviewed{\ensuremath{0.02^{+0.06}_{-0.06}}}}{DCHIFGWTC4HIERPOP}{\reviewed{\ensuremath{-0.02^{+0.07}_{-0.07}}}}{GRQUANTDevParamsTwoDGWTC4HIERPOP}{\reviewed{\ensuremath{44.8}}}{GRQUANTMeansTwoDGWTC4HIERPOP}{\reviewed{\ensuremath{78.7}}}{GRQUANTHyperParamsFourDGWTC4HIERPOP}{\reviewed{\ensuremath{38.6}}}}}

\DeclareRobustCommand{\TGRPOLlogBImprovement}[1]{\IfEqCase{#1}{
{S}{\reviewed{10.48}}
{V}{\reviewed{3.63}}
{TS}{\reviewed{0.40}}
{TV}{\reviewed{0.21}}
{VS}{\reviewed{3.48}}
{TVS}{\reviewed{0.23}}
{min}{\reviewed{0.21}}
{max}{\reviewed{10.48}}
}}

\DeclareRobustCommand{\TGRPOLlogBResults}[1]{\IfEqCase{#1}{
{S}{\reviewed{-14.72}}
{Serr}{\reviewed{0.59}}
{V}{\reviewed{-5.33}}
{Verr}{\reviewed{0.58}}
{TS}{\reviewed{-0.20}}
{TSerr}{\reviewed{0.57}}
{TV}{\reviewed{0.10}}
{TVerr}{\reviewed{0.57}}
{VS}{\reviewed{-5.21}}
{VSerr}{\reviewed{0.58}}
{TVS}{\reviewed{-0.31}}
{TVSerr}{\reviewed{0.57}}
{min}{\reviewed{-14.72}}
{minerr}{\reviewed{0.59}}
{max}{\reviewed{0.10}}
{maxerr}{\reviewed{0.57}}
}}

\DeclareRobustCommand{\TGRTIGERBound}[1]{\IfEqCase{#1}{
{dchiminus2}{\reviewed{\num{5.3e-04}}}
{dchi0}{\reviewed{\num{8.6e-02}}}
{dchi1}{\reviewed{0.18}}
{dchi2}{\reviewed{0.10}}
{dchi3}{\reviewed{\num{6.5e-02}}}
{dchi4}{\reviewed{0.57}}
{dchi5l}{\reviewed{0.23}}
{dchi6}{\reviewed{0.38}}
{dchi6l}{\reviewed{1.18}}
{dchi7}{\reviewed{0.93}}
{min}{\reviewed{\num{5.3e-04}}}
{max}{\reviewed{1.2}}
}}

\DeclareRobustCommand{\TGRTIGERBoundPI}[1]{\IfEqCase{#1}{
{db1}{\reviewed{\num{2.8e-02}}}
{db2}{\reviewed{\num{7.7e-03}}}
{db3}{\reviewed{\num{7.9e-03}}}
{db4}{\reviewed{\num{1.7e-02}}}
{dc1}{\reviewed{0.13}}
{dc2}{\reviewed{\num{4.3e-02}}}
{dc4}{\reviewed{0.13}}
{dcl}{\reviewed{0.28}}
{min}{\reviewed{\num{7.7e-03}}}
{max}{\reviewed{0.28}}
}}

\DeclareRobustCommand{\TGRTIGERBoundImprovement}[1]{\IfEqCase{#1}{
{dchiminus2}{\reviewed{3.9}}
{dchi0}{\reviewed{1.6}}
{dchi1}{\reviewed{1.7}}
{dchi2}{\reviewed{2.0}}
{dchi3}{\reviewed{1.6}}
{dchi4}{\reviewed{1.3}}
{dchi5l}{\reviewed{1.4}}
{dchi6}{\reviewed{1.7}}
{dchi6l}{\reviewed{1.4}}
{dchi7}{\reviewed{1.7}}
{min}{\reviewed{1.3}}
{max}{\reviewed{3.9}}
}}

\DeclareRobustCommand{\TGRFTIBound}[1]{\IfEqCase{#1}{
{dchiMinus2}{\reviewed{\num{1.6e-03}}}
{dchi0}{\reviewed{\num{9.5e-02}}}
{dchi1}{\reviewed{0.26}}
{dchi2}{\reviewed{0.14}}
{dchi3NS}{\reviewed{\num{8.0e-02}}}
{dchi4NS}{\reviewed{0.46}}
{dchi5lNS}{\reviewed{0.14}}
{dchi6NS}{\reviewed{0.19}}
{dchi6l}{\reviewed{1.52}}
{dchi7NS}{\reviewed{0.34}}
{min}{\reviewed{\num{1.6e-03}}}
{max}{\reviewed{1.5}}
}}

\DeclareRobustCommand{\TGRFTIBoundImprovement}[1]{\IfEqCase{#1}{
{dchiMinus2}{\reviewed{1.2}}
{dchi0}{\reviewed{1.4}}
{dchi1}{\reviewed{1.2}}
{dchi2}{\reviewed{1.5}}
{dchi3NS}{\reviewed{1.8}}
{dchi4NS}{\reviewed{1.5}}
{dchi5lNS}{\reviewed{3.5}}
{dchi6NS}{\reviewed{2.8}}
{dchi6l}{\reviewed{1.4}}
{dchi7NS}{\reviewed{5.5}}
{min}{\reviewed{1.2}}
{max}{\reviewed{5.5}}
}}

\DeclareRobustCommand{\TGRSIMPhenomCombinedCI}[1]{\IfEqCase{#1}{{HIER_POP}{\reviewed{\ensuremath{-19^{+28}_{-34}}}}{RESTRICTED_POP}{\reviewed{\ensuremath{-14^{+12}_{-14}}}}{HIER_POP_NEG}{\reviewed{46}}{HIER_POP_POS}{\reviewed{26}}{RESTRICTED_POP_NEG}{\reviewed{24}}{RESTRICTED_POP_POS}{\reviewed{5.7}}{HIER_MU}{\reviewed{\ensuremath{-19^{+20}_{-22}}}}{HIER_SIGMA}{\reviewed{25}}}}
\DeclareRobustCommand{\TGRSIMEOBCombinedCI}[1]{\IfEqCase{#1}{{HIER_POP}{\reviewed{\ensuremath{-49^{+95}_{-176}}}}{RESTRICTED_POP}{\reviewed{\ensuremath{-32^{+29}_{-53}}}}{HIER_POP_NEG}{\reviewed{179}}{HIER_POP_POS}{\reviewed{127}}{RESTRICTED_POP_NEG}{\reviewed{68}}{RESTRICTED_POP_POS}{\reviewed{13}}{HIER_MU}{\reviewed{\ensuremath{-51^{+53}_{-128}}}}{HIER_SIGMA}{\reviewed{119}}}}

\DeclareRobustCommand{\TGRLOSAResults}[1]{\IfEqCase{#1}{
{GW170817}{\reviewed{0.42_{-1.87}^{+2.00}} \times 10^{-6}}
{GW190425}{\reviewed{{-0.01}_{-7.40}^{+6.29}} \times 10^{-6}}
}}

\DeclareRobustCommand{\TGRMDRGravitonBound}[1]{\IfEqCase{#1}{
{gwtc3}{\reviewed{\num{2.23e-23} $\mathrm{eV}/c^2$}}
{gwtc4}{\reviewed{\num{1.92e-23} $\mathrm{eV}/c^2$}}
}}
\DeclareRobustCommand{\TGRMDRGravitonBoundNoUnits}[1]{\IfEqCase{#1}{
{gwtc3}{\reviewed{\num{2.23e-23}}}
{gwtc4}{\reviewed{\num{1.92e-23}}}
}}
\DeclareRobustCommand{\TGRMDRBoundImprovement}[1]{\IfEqCase{#1}{
{mg}{\reviewed{1.16}}
{alpha_0p0}{\reviewed{1.48}}
{alpha_0p5}{\reviewed{1.64}}
{alpha_1p5}{\reviewed{1.68}}
{alpha_2p5}{\reviewed{1.75}}
{alpha_3p0}{\reviewed{2.10}}
{alpha_3p5}{\reviewed{2.60}}
{alpha_4p0}{\reviewed{2.88}}
{alpha_m1p0}{\reviewed{1.69}}
{alpha_m2p0}{\reviewed{1.81}}
{alpha_m3p0}{\reviewed{2.00}}
{amplitude_mean}{\reviewed{1.96}}
{amplitude_min}{\reviewed{1.48}}
{amplitude_max}{\reviewed{2.88}}
}}
\DeclareRobustCommand{\TGRMDRAmplitudeBoundPeV}[1]{\IfEqCase{#1}{
{amplitude_min}{\reviewed{0.01}}
{amplitude_max}{\reviewed{351}}
}}

\DeclareRobustCommand{\TGRSSBKFiveVZeroZero}[1]{\IfEqCase{#1}{
{gwtc4}{\reviewed{\num{1.52e-14} m}}
}}
\DeclareRobustCommand{\TGRSSBKFiveVZeroZeroNoUnits}[1]{\IfEqCase{#1}{
{gwtc4}{\reviewed{\num{1.52e-14}}}
}}

\DeclareRobustCommand{\pyRingHierarchicalDeltaFMedian}{\reviewed{0.10}}
\DeclareRobustCommand{\pyRingHierarchicalDeltaFPlus}{\reviewed{0.23}}
\DeclareRobustCommand{\pyRingHierarchicalDeltaFMinus}{\reviewed{0.18}}
\DeclareRobustCommand{\pyRingHierarchicalDeltaTauMedian}{\reviewed{0.18}}
\DeclareRobustCommand{\pyRingHierarchicalDeltaTauPlus}{\reviewed{0.27}}
\DeclareRobustCommand{\pyRingHierarchicalDeltaTauMinus}{\reviewed{0.26}}

\DeclareRobustCommand{\pyRingHierarchicalQuantileFourDValue}{\reviewed{94.7}}

\DeclareRobustCommand{\pyRingHierarchicalQuantileOneDMuDeltaTauValuelnBcut}{\reviewed{80}}

\DeclareRobustCommand{\pseobHierarchicalDeltaFMedian}{\reviewed{0.00}}
\DeclareRobustCommand{\pseobHierarchicalDeltaFPlus}{\reviewed{0.06}}
\DeclareRobustCommand{\pseobHierarchicalDeltaFMinus}{\reviewed{0.06}}
\DeclareRobustCommand{\pseobHierarchicalDeltaTauMedian}{\reviewed{0.16}}
\DeclareRobustCommand{\pseobHierarchicalDeltaTauPlus}{\reviewed{0.18}}
\DeclareRobustCommand{\pseobHierarchicalDeltaTauMinus}{\reviewed{0.16}}

\DeclareRobustCommand{\QuantileJointGWTCFour}{\reviewed{98.6}\%}
\DeclareRobustCommand{\QuantileHierGWTCFourTauOneD}{\reviewed{99.3}\%}

\DeclareRobustCommand{\QuantileJointGWTCFourPlusTwoFiveZeroOneOneFour}{\reviewed{92.2}\%}
\DeclareRobustCommand{\QuantileHierGWTCFourTauOneDPlusTwoFiveZeroOneOneFour}{\reviewed{96.2}\%}

\DeclareRobustCommand{\INSTRUMENTSOONEOTWO}[1]{\IfEqCase{#1}{{GW150914}{HL}{GW151012}{HL}{GW151226}{HL}{GW170104}{HL}{GW170608}{HL}{GW170729}{HLV}{GW170809}{HLV}{GW170814}{HLV}{GW170817}{HLV}{GW170818}{HLV}{GW170823}{HL}}[\textcolor{red}{???}]}

\DeclareRobustCommand{\INSTRUMENTSOTHREEA}[1]{\IfEqCase{#1}{{GW190408A}{HLV}{GW190412A}{HLV}{GW190413A}{HL}{GW190413B}{HLV}{GW190421A}{HL}{GW190424A}{L}{GW190425A}{LV}{GW190426A}{HLV}{GW190503A}{HLV}{GW190512A}{HLV}{GW190513A}{HLV}{GW190514A}{HL}{GW190517A}{HLV}{GW190519A}{HLV}{GW190521A}{HLV}{GW190521B}{HL}{GW190527A}{HL}{GW190602A}{HLV}{GW190620A}{LV}{GW190630A}{LV}{GW190701A}{HLV}{GW190706A}{HLV}{GW190707A}{HL}{GW190708A}{LV}{GW190719A}{HL}{GW190720A}{HLV}{GW190727A}{HLV}{GW190728A}{HLV}{GW190731A}{HL}{GW190803A}{HLV}{GW190814A}{HLV}{GW190828A}{HLV}{GW190828B}{HLV}{GW190909A}{HL}{GW190910A}{LV}{GW190915A}{HLV}{GW190924A}{HLV}{GW190929A}{HLV}{GW190930A}{HL}}[\textcolor{red}{???}]}

\DeclareRobustCommand{\INSTRUMENTSOTHREEB}[1]{\IfEqCase{#1}{{GW191103A}{HL}{GW191105C}{HLV}{GW191109A}{HL}{GW191113B}{HLV}{GW191118N}{LV}{GW191126C}{HL}{GW191127B}{HLV}{GW191129G}{HL}{GW191204A}{HL}{GW191204G}{HL}{GW191215G}{HLV}{GW191216G}{HV}{GW191219E}{HLV}{GW191222A}{HL}{GW191230H}{HLV}{GW200105F}{LV}{GW200112H}{LV}{GW200115A}{HLV}{GW200121A}{HV}{GW200128C}{HL}{GW200129D}{HLV}{GW200201F}{HLV}{GW200202F}{HLV}{GW200208G}{HLV}{GW200208K}{HLV}{GW200209E}{HLV}{GW200210B}{HLV}{GW200214K}{HLV}{GW200216G}{HLV}{GW200219D}{HLV}{GW200219K}{HLV}{GW200220E}{HLV}{GW200220H}{HL}{GW200224H}{HLV}{GW200225B}{HL}{GW200302A}{HV}{GW200306A}{HL}{GW200308G}{HLV}{GW200311H}{HL}{GW200311L}{HLV}{GW200316I}{HLV}{GW200322G}{HLV}}[\textcolor{red}{???}]}

\input{common__pe_macros}
\input{common__pe_macros_gwtc2point1}

\DeclareRobustCommand{\finalspingwtcthreeminus}[1]{\IfEqCase{#1}{{GW200224_222234}{0.07}{GW191129_134029}{0.05}{GW200311_115853}{0.08}{GW191230_180458}{0.12}{GW191222_033537}{0.11}{GW200225_060421}{0.13}{GW200302_015811}{0.15}{GW200128_022011}{0.10}{GW191204_171526}{0.03}{GW200112_155838}{0.06}{GW200105_162426}{0.03}{GW191105_143521}{0.05}{GW191109_010717}{0.19}{GW200209_085452}{0.11}{GW200115_042309}{0.06}{GW191127_050227}{0.29}{GW200216_220804}{0.24}{GW191215_223052}{0.07}{GW200208_130117}{0.13}{GW200219_094415}{0.13}{GW191103_012549}{0.05}{GW200316_215756}{0.04}{GW200202_154313}{0.04}{GW200129_065458}{0.05}{GW191216_213338}{0.04}}[{\red{???}}]}
\DeclareRobustCommand{\finalspingwtcthreemed}[1]{\IfEqCase{#1}{{GW200224_222234}{0.73}{GW191129_134029}{0.69}{GW200311_115853}{0.69}{GW191230_180458}{0.69}{GW191222_033537}{0.67}{GW200225_060421}{0.66}{GW200302_015811}{0.67}{GW200128_022011}{0.75}{GW191204_171526}{0.73}{GW200112_155838}{0.71}{GW200105_162426}{0.44}{GW191105_143521}{0.67}{GW191109_010717}{0.61}{GW200209_085452}{0.67}{GW200115_042309}{0.43}{GW191127_050227}{0.75}{GW200216_220804}{0.70}{GW191215_223052}{0.68}{GW200208_130117}{0.66}{GW200219_094415}{0.66}{GW191103_012549}{0.75}{GW200316_215756}{0.70}{GW200202_154313}{0.69}{GW200129_065458}{0.73}{GW191216_213338}{0.70}}[{\red{???}}]}
\DeclareRobustCommand{\finalspingwtcthreeplus}[1]{\IfEqCase{#1}{{GW200224_222234}{0.07}{GW191129_134029}{0.03}{GW200311_115853}{0.07}{GW191230_180458}{0.11}{GW191222_033537}{0.08}{GW200225_060421}{0.07}{GW200302_015811}{0.13}{GW200128_022011}{0.10}{GW191204_171526}{0.03}{GW200112_155838}{0.06}{GW200105_162426}{0.07}{GW191105_143521}{0.04}{GW191109_010717}{0.18}{GW200209_085452}{0.10}{GW200115_042309}{0.10}{GW191127_050227}{0.13}{GW200216_220804}{0.14}{GW191215_223052}{0.07}{GW200208_130117}{0.09}{GW200219_094415}{0.10}{GW191103_012549}{0.06}{GW200316_215756}{0.04}{GW200202_154313}{0.03}{GW200129_065458}{0.06}{GW191216_213338}{0.03}}[{\red{???}}]}
\DeclareRobustCommand{\finalspingwtcthreetenthpercentile}[1]{\IfEqCase{#1}{{GW200224_222234}{0.68}{GW191129_134029}{0.65}{GW200311_115853}{0.63}{GW191230_180458}{0.60}{GW191222_033537}{0.60}{GW200225_060421}{0.56}{GW200302_015811}{0.56}{GW200128_022011}{0.67}{GW191204_171526}{0.71}{GW200112_155838}{0.66}{GW200105_162426}{0.41}{GW191105_143521}{0.63}{GW191109_010717}{0.47}{GW200209_085452}{0.59}{GW200115_042309}{0.38}{GW191127_050227}{0.56}{GW200216_220804}{0.53}{GW191215_223052}{0.63}{GW200208_130117}{0.57}{GW200219_094415}{0.57}{GW191103_012549}{0.71}{GW200316_215756}{0.67}{GW200202_154313}{0.66}{GW200129_065458}{0.70}{GW191216_213338}{0.67}}[{\red{???}}]}
\DeclareRobustCommand{\finalspingwtcthreenintiethpercentile}[1]{\IfEqCase{#1}{{GW200224_222234}{0.78}{GW191129_134029}{0.71}{GW200311_115853}{0.74}{GW191230_180458}{0.77}{GW191222_033537}{0.74}{GW200225_060421}{0.71}{GW200302_015811}{0.78}{GW200128_022011}{0.83}{GW191204_171526}{0.75}{GW200112_155838}{0.75}{GW200105_162426}{0.49}{GW191105_143521}{0.70}{GW191109_010717}{0.74}{GW200209_085452}{0.74}{GW200115_042309}{0.51}{GW191127_050227}{0.86}{GW200216_220804}{0.82}{GW191215_223052}{0.73}{GW200208_130117}{0.72}{GW200219_094415}{0.74}{GW191103_012549}{0.79}{GW200316_215756}{0.73}{GW200202_154313}{0.71}{GW200129_065458}{0.78}{GW191216_213338}{0.72}}[{\red{???}}]}
\DeclareRobustCommand{\spintwoygwtcthreeminus}[1]{\IfEqCase{#1}{{GW200224_222234}{0.57}{GW191129_134029}{0.46}{GW200311_115853}{0.56}{GW191230_180458}{0.61}{GW191222_033537}{0.54}{GW200225_060421}{0.54}{GW200302_015811}{0.58}{GW200128_022011}{0.61}{GW191204_171526}{0.53}{GW200112_155838}{0.52}{GW200105_162426}{0.51}{GW191105_143521}{0.52}{GW191109_010717}{0.69}{GW200209_085452}{0.60}{GW200115_042309}{0.50}{GW191127_050227}{0.58}{GW200216_220804}{0.60}{GW191215_223052}{0.58}{GW200208_130117}{0.55}{GW200219_094415}{0.57}{GW191103_012549}{0.54}{GW200316_215756}{0.53}{GW200202_154313}{0.49}{GW200129_065458}{0.55}{GW191216_213338}{0.48}}[{\red{???}}]}
\DeclareRobustCommand{\spintwoygwtcthreemed}[1]{\IfEqCase{#1}{{GW200224_222234}{0.00}{GW191129_134029}{0.00}{GW200311_115853}{0.00}{GW191230_180458}{0.00}{GW191222_033537}{0.00}{GW200225_060421}{0.00}{GW200302_015811}{0.00}{GW200128_022011}{0.00}{GW191204_171526}{0.00}{GW200112_155838}{0.00}{GW200105_162426}{0.00}{GW191105_143521}{0.00}{GW191109_010717}{0.01}{GW200209_085452}{0.00}{GW200115_042309}{0.00}{GW191127_050227}{0.00}{GW200216_220804}{0.00}{GW191215_223052}{0.00}{GW200208_130117}{0.00}{GW200219_094415}{0.00}{GW191103_012549}{0.00}{GW200316_215756}{0.00}{GW200202_154313}{0.00}{GW200129_065458}{0.00}{GW191216_213338}{0.00}}[{\red{???}}]}
\DeclareRobustCommand{\spintwoygwtcthreeplus}[1]{\IfEqCase{#1}{{GW200224_222234}{0.55}{GW191129_134029}{0.47}{GW200311_115853}{0.58}{GW191230_180458}{0.60}{GW191222_033537}{0.56}{GW200225_060421}{0.54}{GW200302_015811}{0.57}{GW200128_022011}{0.62}{GW191204_171526}{0.50}{GW200112_155838}{0.53}{GW200105_162426}{0.51}{GW191105_143521}{0.52}{GW191109_010717}{0.69}{GW200209_085452}{0.60}{GW200115_042309}{0.51}{GW191127_050227}{0.58}{GW200216_220804}{0.57}{GW191215_223052}{0.60}{GW200208_130117}{0.55}{GW200219_094415}{0.59}{GW191103_012549}{0.55}{GW200316_215756}{0.54}{GW200202_154313}{0.50}{GW200129_065458}{0.57}{GW191216_213338}{0.47}}[{\red{???}}]}
\DeclareRobustCommand{\spintwoygwtcthreetenthpercentile}[1]{\IfEqCase{#1}{{GW200224_222234}{-0.43}{GW191129_134029}{-0.33}{GW200311_115853}{-0.40}{GW191230_180458}{-0.46}{GW191222_033537}{-0.39}{GW200225_060421}{-0.40}{GW200302_015811}{-0.42}{GW200128_022011}{-0.46}{GW191204_171526}{-0.40}{GW200112_155838}{-0.38}{GW200105_162426}{-0.35}{GW191105_143521}{-0.36}{GW191109_010717}{-0.54}{GW200209_085452}{-0.44}{GW200115_042309}{-0.35}{GW191127_050227}{-0.43}{GW200216_220804}{-0.44}{GW191215_223052}{-0.43}{GW200208_130117}{-0.41}{GW200219_094415}{-0.43}{GW191103_012549}{-0.40}{GW200316_215756}{-0.39}{GW200202_154313}{-0.35}{GW200129_065458}{-0.38}{GW191216_213338}{-0.34}}[{\red{???}}]}
\DeclareRobustCommand{\spintwoygwtcthreenintiethpercentile}[1]{\IfEqCase{#1}{{GW200224_222234}{0.40}{GW191129_134029}{0.34}{GW200311_115853}{0.43}{GW191230_180458}{0.44}{GW191222_033537}{0.41}{GW200225_060421}{0.40}{GW200302_015811}{0.42}{GW200128_022011}{0.46}{GW191204_171526}{0.37}{GW200112_155838}{0.39}{GW200105_162426}{0.36}{GW191105_143521}{0.36}{GW191109_010717}{0.56}{GW200209_085452}{0.45}{GW200115_042309}{0.37}{GW191127_050227}{0.44}{GW200216_220804}{0.43}{GW191215_223052}{0.43}{GW200208_130117}{0.40}{GW200219_094415}{0.43}{GW191103_012549}{0.40}{GW200316_215756}{0.39}{GW200202_154313}{0.35}{GW200129_065458}{0.42}{GW191216_213338}{0.33}}[{\red{???}}]}
\DeclareRobustCommand{\finalmasssourcegwtcthreeminus}[1]{\IfEqCase{#1}{{GW200224_222234}{4.5}{GW191129_134029}{1.1}{GW200311_115853}{3.9}{GW191230_180458}{10}{GW191222_033537}{9.9}{GW200225_060421}{2.8}{GW200302_015811}{6.5}{GW200128_022011}{9.9}{GW191204_171526}{0.92}{GW200112_155838}{4.3}{GW200105_162426}{1.7}{GW191105_143521}{1.2}{GW191109_010717}{15}{GW200209_085452}{8.2}{GW200115_042309}{1.6}{GW191127_050227}{21}{GW200216_220804}{13}{GW191215_223052}{3.8}{GW200208_130117}{6.4}{GW200219_094415}{7.8}{GW191103_012549}{1.7}{GW200316_215756}{1.9}{GW200202_154313}{0.66}{GW200129_065458}{3.3}{GW191216_213338}{0.93}}[{\red{???}}]}
\DeclareRobustCommand{\finalmasssourcegwtcthreemed}[1]{\IfEqCase{#1}{{GW200224_222234}{68.3}{GW191129_134029}{16.7}{GW200311_115853}{59.0}{GW191230_180458}{80}{GW191222_033537}{75.5}{GW200225_060421}{32.1}{GW200302_015811}{55.0}{GW200128_022011}{67.9}{GW191204_171526}{19.14}{GW200112_155838}{60.8}{GW200105_162426}{10.7}{GW191105_143521}{17.6}{GW191109_010717}{107}{GW200209_085452}{58.5}{GW200115_042309}{7.1}{GW191127_050227}{76}{GW200216_220804}{78}{GW191215_223052}{40.7}{GW200208_130117}{62.5}{GW200219_094415}{62.2}{GW191103_012549}{19.0}{GW200316_215756}{20.2}{GW200202_154313}{16.76}{GW200129_065458}{60.3}{GW191216_213338}{18.87}}[{\red{???}}]}
\DeclareRobustCommand{\finalmasssourcegwtcthreeplus}[1]{\IfEqCase{#1}{{GW200224_222234}{6.3}{GW191129_134029}{2.5}{GW200311_115853}{4.8}{GW191230_180458}{16}{GW191222_033537}{15.3}{GW200225_060421}{3.5}{GW200302_015811}{8.9}{GW200128_022011}{14.1}{GW191204_171526}{1.79}{GW200112_155838}{5.3}{GW200105_162426}{2.0}{GW191105_143521}{2.1}{GW191109_010717}{18}{GW200209_085452}{12.2}{GW200115_042309}{2.2}{GW191127_050227}{39}{GW200216_220804}{19}{GW191215_223052}{5.3}{GW200208_130117}{7.3}{GW200219_094415}{11.7}{GW191103_012549}{3.8}{GW200316_215756}{7.5}{GW200202_154313}{1.87}{GW200129_065458}{4.0}{GW191216_213338}{2.84}}[{\red{???}}]}
\DeclareRobustCommand{\finalmasssourcegwtcthreetenthpercentile}[1]{\IfEqCase{#1}{{GW200224_222234}{64.7}{GW191129_134029}{15.7}{GW200311_115853}{56.0}{GW191230_180458}{71}{GW191222_033537}{67.3}{GW200225_060421}{29.9}{GW200302_015811}{49.8}{GW200128_022011}{59.8}{GW191204_171526}{18.38}{GW200112_155838}{57.3}{GW200105_162426}{9.6}{GW191105_143521}{16.7}{GW191109_010717}{96}{GW200209_085452}{51.9}{GW200115_042309}{5.7}{GW191127_050227}{59}{GW200216_220804}{67}{GW191215_223052}{37.6}{GW200208_130117}{57.5}{GW200219_094415}{55.9}{GW191103_012549}{17.5}{GW200316_215756}{18.6}{GW200202_154313}{16.22}{GW200129_065458}{57.6}{GW191216_213338}{18.08}}[{\red{???}}]}
\DeclareRobustCommand{\finalmasssourcegwtcthreenintiethpercentile}[1]{\IfEqCase{#1}{{GW200224_222234}{73.1}{GW191129_134029}{18.5}{GW200311_115853}{62.7}{GW191230_180458}{91}{GW191222_033537}{87.3}{GW200225_060421}{34.7}{GW200302_015811}{61.5}{GW200128_022011}{78.6}{GW191204_171526}{20.40}{GW200112_155838}{64.6}{GW200105_162426}{12.0}{GW191105_143521}{19.1}{GW191109_010717}{121}{GW200209_085452}{67.5}{GW200115_042309}{8.7}{GW191127_050227}{105}{GW200216_220804}{92}{GW191215_223052}{44.8}{GW200208_130117}{68.2}{GW200219_094415}{71.3}{GW191103_012549}{21.4}{GW200316_215756}{24.8}{GW200202_154313}{17.99}{GW200129_065458}{63.4}{GW191216_213338}{20.65}}[{\red{???}}]}
\DeclareRobustCommand{\spinoneygwtcthreeminus}[1]{\IfEqCase{#1}{{GW200224_222234}{0.59}{GW191129_134029}{0.36}{GW200311_115853}{0.56}{GW191230_180458}{0.65}{GW191222_033537}{0.53}{GW200225_060421}{0.65}{GW200302_015811}{0.54}{GW200128_022011}{0.66}{GW191204_171526}{0.48}{GW200112_155838}{0.49}{GW200105_162426}{0.13}{GW191105_143521}{0.41}{GW191109_010717}{0.74}{GW200209_085452}{0.65}{GW200115_042309}{0.35}{GW191127_050227}{0.70}{GW200216_220804}{0.60}{GW191215_223052}{0.63}{GW200208_130117}{0.47}{GW200219_094415}{0.61}{GW191103_012549}{0.51}{GW200316_215756}{0.39}{GW200202_154313}{0.36}{GW200129_065458}{0.58}{GW191216_213338}{0.29}}[{\red{???}}]}
\DeclareRobustCommand{\spinoneygwtcthreemed}[1]{\IfEqCase{#1}{{GW200224_222234}{0.00}{GW191129_134029}{0.00}{GW200311_115853}{0.00}{GW191230_180458}{0.00}{GW191222_033537}{0.00}{GW200225_060421}{0.00}{GW200302_015811}{0.00}{GW200128_022011}{0.00}{GW191204_171526}{0.00}{GW200112_155838}{0.00}{GW200105_162426}{0.00}{GW191105_143521}{0.00}{GW191109_010717}{0.01}{GW200209_085452}{0.00}{GW200115_042309}{0.00}{GW191127_050227}{0.00}{GW200216_220804}{0.00}{GW191215_223052}{0.00}{GW200208_130117}{0.00}{GW200219_094415}{0.00}{GW191103_012549}{0.00}{GW200316_215756}{-0.01}{GW200202_154313}{0.00}{GW200129_065458}{0.00}{GW191216_213338}{0.00}}[{\red{???}}]}
\DeclareRobustCommand{\spinoneygwtcthreeplus}[1]{\IfEqCase{#1}{{GW200224_222234}{0.57}{GW191129_134029}{0.37}{GW200311_115853}{0.58}{GW191230_180458}{0.64}{GW191222_033537}{0.53}{GW200225_060421}{0.64}{GW200302_015811}{0.53}{GW200128_022011}{0.66}{GW191204_171526}{0.49}{GW200112_155838}{0.48}{GW200105_162426}{0.13}{GW191105_143521}{0.41}{GW191109_010717}{0.70}{GW200209_085452}{0.66}{GW200115_042309}{0.33}{GW191127_050227}{0.71}{GW200216_220804}{0.61}{GW191215_223052}{0.63}{GW200208_130117}{0.49}{GW200219_094415}{0.59}{GW191103_012549}{0.53}{GW200316_215756}{0.38}{GW200202_154313}{0.37}{GW200129_065458}{0.72}{GW191216_213338}{0.29}}[{\red{???}}]}
\DeclareRobustCommand{\spinoneygwtcthreetenthpercentile}[1]{\IfEqCase{#1}{{GW200224_222234}{-0.45}{GW191129_134029}{-0.25}{GW200311_115853}{-0.40}{GW191230_180458}{-0.50}{GW191222_033537}{-0.37}{GW200225_060421}{-0.52}{GW200302_015811}{-0.40}{GW200128_022011}{-0.52}{GW191204_171526}{-0.37}{GW200112_155838}{-0.35}{GW200105_162426}{-0.09}{GW191105_143521}{-0.27}{GW191109_010717}{-0.60}{GW200209_085452}{-0.49}{GW200115_042309}{-0.26}{GW191127_050227}{-0.54}{GW200216_220804}{-0.46}{GW191215_223052}{-0.47}{GW200208_130117}{-0.33}{GW200219_094415}{-0.46}{GW191103_012549}{-0.38}{GW200316_215756}{-0.29}{GW200202_154313}{-0.25}{GW200129_065458}{-0.42}{GW191216_213338}{-0.19}}[{\red{???}}]}
\DeclareRobustCommand{\spinoneygwtcthreenintiethpercentile}[1]{\IfEqCase{#1}{{GW200224_222234}{0.43}{GW191129_134029}{0.26}{GW200311_115853}{0.43}{GW191230_180458}{0.48}{GW191222_033537}{0.37}{GW200225_060421}{0.51}{GW200302_015811}{0.39}{GW200128_022011}{0.53}{GW191204_171526}{0.37}{GW200112_155838}{0.35}{GW200105_162426}{0.08}{GW191105_143521}{0.28}{GW191109_010717}{0.59}{GW200209_085452}{0.50}{GW200115_042309}{0.23}{GW191127_050227}{0.55}{GW200216_220804}{0.46}{GW191215_223052}{0.49}{GW200208_130117}{0.36}{GW200219_094415}{0.43}{GW191103_012549}{0.38}{GW200316_215756}{0.26}{GW200202_154313}{0.25}{GW200129_065458}{0.56}{GW191216_213338}{0.21}}[{\red{???}}]}
\DeclareRobustCommand{\costilttwogwtcthreeminus}[1]{\IfEqCase{#1}{{GW200224_222234}{1.01}{GW191129_134029}{1.07}{GW200311_115853}{0.84}{GW191230_180458}{0.79}{GW191222_033537}{0.81}{GW200225_060421}{0.71}{GW200302_015811}{0.98}{GW200128_022011}{0.99}{GW191204_171526}{1.06}{GW200112_155838}{1.02}{GW200105_162426}{0.89}{GW191105_143521}{0.83}{GW191109_010717}{0.66}{GW200209_085452}{0.67}{GW200115_042309}{0.64}{GW191127_050227}{1.06}{GW200216_220804}{1.03}{GW191215_223052}{0.79}{GW200208_130117}{0.76}{GW200219_094415}{0.75}{GW191103_012549}{1.19}{GW200316_215756}{1.05}{GW200202_154313}{0.97}{GW200129_065458}{1.21}{GW191216_213338}{1.16}}[{\red{???}}]}
\DeclareRobustCommand{\costilttwogwtcthreemed}[1]{\IfEqCase{#1}{{GW200224_222234}{0.19}{GW191129_134029}{0.30}{GW200311_115853}{-0.02}{GW191230_180458}{-0.12}{GW191222_033537}{-0.09}{GW200225_060421}{-0.22}{GW200302_015811}{0.12}{GW200128_022011}{0.15}{GW191204_171526}{0.45}{GW200112_155838}{0.25}{GW200105_162426}{0.02}{GW191105_143521}{-0.03}{GW191109_010717}{-0.26}{GW200209_085452}{-0.26}{GW200115_042309}{-0.31}{GW191127_050227}{0.22}{GW200216_220804}{0.18}{GW191215_223052}{-0.09}{GW200208_130117}{-0.15}{GW200219_094415}{-0.17}{GW191103_012549}{0.48}{GW200316_215756}{0.37}{GW200202_154313}{0.22}{GW200129_065458}{0.41}{GW191216_213338}{0.40}}[{\red{???}}]}
\DeclareRobustCommand{\costilttwogwtcthreeplus}[1]{\IfEqCase{#1}{{GW200224_222234}{0.71}{GW191129_134029}{0.63}{GW200311_115853}{0.87}{GW191230_180458}{0.96}{GW191222_033537}{0.95}{GW200225_060421}{1.04}{GW200302_015811}{0.78}{GW200128_022011}{0.74}{GW191204_171526}{0.50}{GW200112_155838}{0.67}{GW200105_162426}{0.84}{GW191105_143521}{0.89}{GW191109_010717}{1.04}{GW200209_085452}{1.05}{GW200115_042309}{1.13}{GW191127_050227}{0.71}{GW200216_220804}{0.75}{GW191215_223052}{0.93}{GW200208_130117}{1.00}{GW200219_094415}{0.99}{GW191103_012549}{0.48}{GW200316_215756}{0.57}{GW200202_154313}{0.68}{GW200129_065458}{0.54}{GW191216_213338}{0.54}}[{\red{???}}]}
\DeclareRobustCommand{\costilttwogwtcthreetenthpercentile}[1]{\IfEqCase{#1}{{GW200224_222234}{-0.67}{GW191129_134029}{-0.56}{GW200311_115853}{-0.75}{GW191230_180458}{-0.81}{GW191222_033537}{-0.80}{GW200225_060421}{-0.85}{GW200302_015811}{-0.72}{GW200128_022011}{-0.69}{GW191204_171526}{-0.38}{GW200112_155838}{-0.61}{GW200105_162426}{-0.74}{GW191105_143521}{-0.73}{GW191109_010717}{-0.85}{GW200209_085452}{-0.86}{GW200115_042309}{-0.90}{GW191127_050227}{-0.69}{GW200216_220804}{-0.71}{GW191215_223052}{-0.78}{GW200208_130117}{-0.82}{GW200219_094415}{-0.84}{GW191103_012549}{-0.50}{GW200316_215756}{-0.46}{GW200202_154313}{-0.55}{GW200129_065458}{-0.61}{GW191216_213338}{-0.54}}[{\red{???}}]}
\DeclareRobustCommand{\costilttwogwtcthreenintiethpercentile}[1]{\IfEqCase{#1}{{GW200224_222234}{0.81}{GW191129_134029}{0.87}{GW200311_115853}{0.72}{GW191230_180458}{0.70}{GW191222_033537}{0.71}{GW200225_060421}{0.65}{GW200302_015811}{0.82}{GW200128_022011}{0.79}{GW191204_171526}{0.89}{GW200112_155838}{0.82}{GW200105_162426}{0.74}{GW191105_143521}{0.74}{GW191109_010717}{0.62}{GW200209_085452}{0.61}{GW200115_042309}{0.65}{GW191127_050227}{0.86}{GW200216_220804}{0.85}{GW191215_223052}{0.70}{GW200208_130117}{0.70}{GW200219_094415}{0.69}{GW191103_012549}{0.91}{GW200316_215756}{0.87}{GW200202_154313}{0.82}{GW200129_065458}{0.90}{GW191216_213338}{0.89}}[{\red{???}}]}
\DeclareRobustCommand{\massonesourcegwtcthreeminus}[1]{\IfEqCase{#1}{{GW200224_222234}{4.4}{GW191129_134029}{2.1}{GW200311_115853}{3.8}{GW191230_180458}{8.6}{GW191222_033537}{8.0}{GW200225_060421}{3.0}{GW200302_015811}{8.4}{GW200128_022011}{7.6}{GW191204_171526}{1.8}{GW200112_155838}{4.5}{GW200105_162426}{1.7}{GW191105_143521}{1.6}{GW191109_010717}{11}{GW200209_085452}{6.3}{GW200115_042309}{2.5}{GW191127_050227}{20}{GW200216_220804}{13}{GW191215_223052}{4.1}{GW200208_130117}{6.2}{GW200219_094415}{6.9}{GW191103_012549}{2.2}{GW200316_215756}{2.9}{GW200202_154313}{1.4}{GW200129_065458}{3.2}{GW191216_213338}{2.2}}[{\red{???}}]}
\DeclareRobustCommand{\massonesourcegwtcthreemed}[1]{\IfEqCase{#1}{{GW200224_222234}{39.8}{GW191129_134029}{10.6}{GW200311_115853}{34.2}{GW191230_180458}{47.7}{GW191222_033537}{45.1}{GW200225_060421}{19.3}{GW200302_015811}{37.0}{GW200128_022011}{40.4}{GW191204_171526}{11.9}{GW200112_155838}{35.6}{GW200105_162426}{9.0}{GW191105_143521}{10.7}{GW191109_010717}{65}{GW200209_085452}{34.6}{GW200115_042309}{5.9}{GW191127_050227}{53}{GW200216_220804}{51}{GW191215_223052}{24.6}{GW200208_130117}{37.8}{GW200219_094415}{37.5}{GW191103_012549}{11.8}{GW200316_215756}{13.1}{GW200202_154313}{10.1}{GW200129_065458}{34.5}{GW191216_213338}{12.1}}[{\red{???}}]}
\DeclareRobustCommand{\massonesourcegwtcthreeplus}[1]{\IfEqCase{#1}{{GW200224_222234}{6.9}{GW191129_134029}{4.1}{GW200311_115853}{6.4}{GW191230_180458}{13.4}{GW191222_033537}{10.9}{GW200225_060421}{5.0}{GW200302_015811}{8.9}{GW200128_022011}{11.1}{GW191204_171526}{3.3}{GW200112_155838}{6.7}{GW200105_162426}{1.7}{GW191105_143521}{3.7}{GW191109_010717}{11}{GW200209_085452}{10.0}{GW200115_042309}{2.0}{GW191127_050227}{47}{GW200216_220804}{22}{GW191215_223052}{7.0}{GW200208_130117}{9.2}{GW200219_094415}{10.1}{GW191103_012549}{6.2}{GW200316_215756}{10.2}{GW200202_154313}{3.5}{GW200129_065458}{9.9}{GW191216_213338}{4.6}}[{\red{???}}]}
\DeclareRobustCommand{\massonesourcegwtcthreetenthpercentile}[1]{\IfEqCase{#1}{{GW200224_222234}{36.1}{GW191129_134029}{8.8}{GW200311_115853}{31.1}{GW191230_180458}{40.6}{GW191222_033537}{38.5}{GW200225_060421}{16.8}{GW200302_015811}{30.3}{GW200128_022011}{34.3}{GW191204_171526}{10.3}{GW200112_155838}{31.9}{GW200105_162426}{8.0}{GW191105_143521}{9.3}{GW191109_010717}{57}{GW200209_085452}{29.5}{GW200115_042309}{3.7}{GW191127_050227}{36}{GW200216_220804}{40}{GW191215_223052}{21.2}{GW200208_130117}{32.7}{GW200219_094415}{31.8}{GW191103_012549}{9.9}{GW200316_215756}{10.6}{GW200202_154313}{8.9}{GW200129_065458}{31.9}{GW191216_213338}{10.1}}[{\red{???}}]}
\DeclareRobustCommand{\massonesourcegwtcthreenintiethpercentile}[1]{\IfEqCase{#1}{{GW200224_222234}{45.0}{GW191129_134029}{13.7}{GW200311_115853}{38.9}{GW191230_180458}{57.6}{GW191222_033537}{53.2}{GW200225_060421}{23.0}{GW200302_015811}{43.9}{GW200128_022011}{48.7}{GW191204_171526}{14.3}{GW200112_155838}{40.8}{GW200105_162426}{10.0}{GW191105_143521}{13.3}{GW191109_010717}{73}{GW200209_085452}{41.9}{GW200115_042309}{7.4}{GW191127_050227}{88}{GW200216_220804}{68}{GW191215_223052}{29.8}{GW200208_130117}{44.9}{GW200219_094415}{45.1}{GW191103_012549}{16.1}{GW200316_215756}{20.0}{GW200202_154313}{12.7}{GW200129_065458}{42.3}{GW191216_213338}{15.2}}[{\red{???}}]}
\DeclareRobustCommand{\spintwoxgwtcthreeminus}[1]{\IfEqCase{#1}{{GW200224_222234}{0.58}{GW191129_134029}{0.47}{GW200311_115853}{0.55}{GW191230_180458}{0.60}{GW191222_033537}{0.55}{GW200225_060421}{0.55}{GW200302_015811}{0.58}{GW200128_022011}{0.61}{GW191204_171526}{0.53}{GW200112_155838}{0.51}{GW200105_162426}{0.52}{GW191105_143521}{0.51}{GW191109_010717}{0.68}{GW200209_085452}{0.58}{GW200115_042309}{0.51}{GW191127_050227}{0.60}{GW200216_220804}{0.60}{GW191215_223052}{0.58}{GW200208_130117}{0.54}{GW200219_094415}{0.60}{GW191103_012549}{0.53}{GW200316_215756}{0.55}{GW200202_154313}{0.50}{GW200129_065458}{0.57}{GW191216_213338}{0.44}}[{\red{???}}]}
\DeclareRobustCommand{\spintwoxgwtcthreemed}[1]{\IfEqCase{#1}{{GW200224_222234}{0.01}{GW191129_134029}{0.00}{GW200311_115853}{0.00}{GW191230_180458}{0.00}{GW191222_033537}{0.00}{GW200225_060421}{0.00}{GW200302_015811}{0.00}{GW200128_022011}{0.00}{GW191204_171526}{0.00}{GW200112_155838}{0.00}{GW200105_162426}{0.00}{GW191105_143521}{0.00}{GW191109_010717}{0.00}{GW200209_085452}{0.00}{GW200115_042309}{0.00}{GW191127_050227}{0.00}{GW200216_220804}{0.00}{GW191215_223052}{0.00}{GW200208_130117}{0.00}{GW200219_094415}{0.00}{GW191103_012549}{0.00}{GW200316_215756}{0.00}{GW200202_154313}{0.00}{GW200129_065458}{0.00}{GW191216_213338}{-0.01}}[{\red{???}}]}
\DeclareRobustCommand{\spintwoxgwtcthreeplus}[1]{\IfEqCase{#1}{{GW200224_222234}{0.58}{GW191129_134029}{0.46}{GW200311_115853}{0.56}{GW191230_180458}{0.60}{GW191222_033537}{0.55}{GW200225_060421}{0.55}{GW200302_015811}{0.58}{GW200128_022011}{0.61}{GW191204_171526}{0.50}{GW200112_155838}{0.53}{GW200105_162426}{0.50}{GW191105_143521}{0.51}{GW191109_010717}{0.65}{GW200209_085452}{0.60}{GW200115_042309}{0.51}{GW191127_050227}{0.60}{GW200216_220804}{0.60}{GW191215_223052}{0.58}{GW200208_130117}{0.57}{GW200219_094415}{0.58}{GW191103_012549}{0.55}{GW200316_215756}{0.53}{GW200202_154313}{0.50}{GW200129_065458}{0.55}{GW191216_213338}{0.47}}[{\red{???}}]}
\DeclareRobustCommand{\spintwoxgwtcthreetenthpercentile}[1]{\IfEqCase{#1}{{GW200224_222234}{-0.41}{GW191129_134029}{-0.33}{GW200311_115853}{-0.41}{GW191230_180458}{-0.45}{GW191222_033537}{-0.41}{GW200225_060421}{-0.40}{GW200302_015811}{-0.42}{GW200128_022011}{-0.46}{GW191204_171526}{-0.39}{GW200112_155838}{-0.37}{GW200105_162426}{-0.36}{GW191105_143521}{-0.35}{GW191109_010717}{-0.53}{GW200209_085452}{-0.44}{GW200115_042309}{-0.38}{GW191127_050227}{-0.46}{GW200216_220804}{-0.45}{GW191215_223052}{-0.42}{GW200208_130117}{-0.40}{GW200219_094415}{-0.47}{GW191103_012549}{-0.40}{GW200316_215756}{-0.41}{GW200202_154313}{-0.36}{GW200129_065458}{-0.41}{GW191216_213338}{-0.33}}[{\red{???}}]}
\DeclareRobustCommand{\spintwoxgwtcthreenintiethpercentile}[1]{\IfEqCase{#1}{{GW200224_222234}{0.44}{GW191129_134029}{0.33}{GW200311_115853}{0.40}{GW191230_180458}{0.45}{GW191222_033537}{0.39}{GW200225_060421}{0.40}{GW200302_015811}{0.43}{GW200128_022011}{0.47}{GW191204_171526}{0.38}{GW200112_155838}{0.39}{GW200105_162426}{0.35}{GW191105_143521}{0.36}{GW191109_010717}{0.51}{GW200209_085452}{0.45}{GW200115_042309}{0.36}{GW191127_050227}{0.45}{GW200216_220804}{0.45}{GW191215_223052}{0.43}{GW200208_130117}{0.42}{GW200219_094415}{0.43}{GW191103_012549}{0.40}{GW200316_215756}{0.39}{GW200202_154313}{0.35}{GW200129_065458}{0.41}{GW191216_213338}{0.31}}[{\red{???}}]}
\DeclareRobustCommand{\phitwogwtcthreeminus}[1]{\IfEqCase{#1}{{GW200224_222234}{2.9}{GW191129_134029}{2.8}{GW200311_115853}{2.8}{GW191230_180458}{2.8}{GW191222_033537}{2.9}{GW200225_060421}{2.8}{GW200302_015811}{2.9}{GW200128_022011}{2.9}{GW191204_171526}{2.8}{GW200112_155838}{2.9}{GW200105_162426}{2.8}{GW191105_143521}{2.8}{GW191109_010717}{2.7}{GW200209_085452}{2.8}{GW200115_042309}{2.8}{GW191127_050227}{2.8}{GW200216_220804}{2.9}{GW191215_223052}{2.8}{GW200208_130117}{2.8}{GW200219_094415}{2.8}{GW191103_012549}{2.8}{GW200316_215756}{2.8}{GW200202_154313}{2.8}{GW200129_065458}{2.7}{GW191216_213338}{2.8}}[{\red{???}}]}
\DeclareRobustCommand{\phitwogwtcthreemed}[1]{\IfEqCase{#1}{{GW200224_222234}{3.2}{GW191129_134029}{3.1}{GW200311_115853}{3.1}{GW191230_180458}{3.1}{GW191222_033537}{3.2}{GW200225_060421}{3.1}{GW200302_015811}{3.1}{GW200128_022011}{3.2}{GW191204_171526}{3.2}{GW200112_155838}{3.2}{GW200105_162426}{3.1}{GW191105_143521}{3.2}{GW191109_010717}{3.0}{GW200209_085452}{3.1}{GW200115_042309}{3.1}{GW191127_050227}{3.1}{GW200216_220804}{3.2}{GW191215_223052}{3.1}{GW200208_130117}{3.1}{GW200219_094415}{3.2}{GW191103_012549}{3.1}{GW200316_215756}{3.1}{GW200202_154313}{3.2}{GW200129_065458}{3.1}{GW191216_213338}{3.2}}[{\red{???}}]}
\DeclareRobustCommand{\phitwogwtcthreeplus}[1]{\IfEqCase{#1}{{GW200224_222234}{2.8}{GW191129_134029}{2.9}{GW200311_115853}{2.9}{GW191230_180458}{2.8}{GW191222_033537}{2.8}{GW200225_060421}{2.9}{GW200302_015811}{2.8}{GW200128_022011}{2.8}{GW191204_171526}{2.8}{GW200112_155838}{2.8}{GW200105_162426}{2.8}{GW191105_143521}{2.8}{GW191109_010717}{2.9}{GW200209_085452}{2.8}{GW200115_042309}{2.8}{GW191127_050227}{2.9}{GW200216_220804}{2.8}{GW191215_223052}{2.8}{GW200208_130117}{2.8}{GW200219_094415}{2.8}{GW191103_012549}{2.8}{GW200316_215756}{2.8}{GW200202_154313}{2.8}{GW200129_065458}{2.9}{GW191216_213338}{2.8}}[{\red{???}}]}
\DeclareRobustCommand{\phitwogwtcthreetenthpercentile}[1]{\IfEqCase{#1}{{GW200224_222234}{0.6}{GW191129_134029}{0.6}{GW200311_115853}{0.6}{GW191230_180458}{0.6}{GW191222_033537}{0.6}{GW200225_060421}{0.7}{GW200302_015811}{0.6}{GW200128_022011}{0.6}{GW191204_171526}{0.6}{GW200112_155838}{0.6}{GW200105_162426}{0.6}{GW191105_143521}{0.6}{GW191109_010717}{0.6}{GW200209_085452}{0.6}{GW200115_042309}{0.6}{GW191127_050227}{0.6}{GW200216_220804}{0.6}{GW191215_223052}{0.6}{GW200208_130117}{0.6}{GW200219_094415}{0.6}{GW191103_012549}{0.7}{GW200316_215756}{0.6}{GW200202_154313}{0.6}{GW200129_065458}{0.6}{GW191216_213338}{0.7}}[{\red{???}}]}
\DeclareRobustCommand{\phitwogwtcthreenintiethpercentile}[1]{\IfEqCase{#1}{{GW200224_222234}{5.7}{GW191129_134029}{5.7}{GW200311_115853}{5.6}{GW191230_180458}{5.6}{GW191222_033537}{5.7}{GW200225_060421}{5.6}{GW200302_015811}{5.7}{GW200128_022011}{5.7}{GW191204_171526}{5.6}{GW200112_155838}{5.6}{GW200105_162426}{5.6}{GW191105_143521}{5.7}{GW191109_010717}{5.6}{GW200209_085452}{5.7}{GW200115_042309}{5.6}{GW191127_050227}{5.6}{GW200216_220804}{5.7}{GW191215_223052}{5.6}{GW200208_130117}{5.7}{GW200219_094415}{5.7}{GW191103_012549}{5.7}{GW200316_215756}{5.6}{GW200202_154313}{5.7}{GW200129_065458}{5.7}{GW191216_213338}{5.6}}[{\red{???}}]}
\DeclareRobustCommand{\chipgwtcthreeminus}[1]{\IfEqCase{#1}{{GW200224_222234}{0.36}{GW191129_134029}{0.19}{GW200311_115853}{0.35}{GW191230_180458}{0.39}{GW191222_033537}{0.32}{GW200225_060421}{0.38}{GW200302_015811}{0.29}{GW200128_022011}{0.40}{GW191204_171526}{0.26}{GW200112_155838}{0.30}{GW200105_162426}{0.07}{GW191105_143521}{0.24}{GW191109_010717}{0.37}{GW200209_085452}{0.38}{GW200115_042309}{0.16}{GW191127_050227}{0.41}{GW200216_220804}{0.35}{GW191215_223052}{0.38}{GW200208_130117}{0.29}{GW200219_094415}{0.35}{GW191103_012549}{0.26}{GW200316_215756}{0.20}{GW200202_154313}{0.22}{GW200129_065458}{0.37}{GW191216_213338}{0.15}}[{\red{???}}]}
\DeclareRobustCommand{\chipgwtcthreemed}[1]{\IfEqCase{#1}{{GW200224_222234}{0.50}{GW191129_134029}{0.26}{GW200311_115853}{0.45}{GW191230_180458}{0.52}{GW191222_033537}{0.41}{GW200225_060421}{0.53}{GW200302_015811}{0.38}{GW200128_022011}{0.57}{GW191204_171526}{0.39}{GW200112_155838}{0.39}{GW200105_162426}{0.09}{GW191105_143521}{0.30}{GW191109_010717}{0.63}{GW200209_085452}{0.52}{GW200115_042309}{0.20}{GW191127_050227}{0.52}{GW200216_220804}{0.45}{GW191215_223052}{0.50}{GW200208_130117}{0.38}{GW200219_094415}{0.48}{GW191103_012549}{0.40}{GW200316_215756}{0.29}{GW200202_154313}{0.28}{GW200129_065458}{0.52}{GW191216_213338}{0.23}}[{\red{???}}]}
\DeclareRobustCommand{\chipgwtcthreeplus}[1]{\IfEqCase{#1}{{GW200224_222234}{0.37}{GW191129_134029}{0.36}{GW200311_115853}{0.40}{GW191230_180458}{0.38}{GW191222_033537}{0.41}{GW200225_060421}{0.34}{GW200302_015811}{0.44}{GW200128_022011}{0.33}{GW191204_171526}{0.35}{GW200112_155838}{0.39}{GW200105_162426}{0.17}{GW191105_143521}{0.45}{GW191109_010717}{0.29}{GW200209_085452}{0.38}{GW200115_042309}{0.34}{GW191127_050227}{0.41}{GW200216_220804}{0.42}{GW191215_223052}{0.38}{GW200208_130117}{0.41}{GW200219_094415}{0.40}{GW191103_012549}{0.41}{GW200316_215756}{0.38}{GW200202_154313}{0.40}{GW200129_065458}{0.42}{GW191216_213338}{0.35}}[{\red{???}}]}
\DeclareRobustCommand{\chipgwtcthreetenthpercentile}[1]{\IfEqCase{#1}{{GW200224_222234}{0.20}{GW191129_134029}{0.10}{GW200311_115853}{0.16}{GW191230_180458}{0.20}{GW191222_033537}{0.15}{GW200225_060421}{0.22}{GW200302_015811}{0.13}{GW200128_022011}{0.24}{GW191204_171526}{0.18}{GW200112_155838}{0.13}{GW200105_162426}{0.03}{GW191105_143521}{0.09}{GW191109_010717}{0.33}{GW200209_085452}{0.20}{GW200115_042309}{0.06}{GW191127_050227}{0.17}{GW200216_220804}{0.15}{GW191215_223052}{0.19}{GW200208_130117}{0.13}{GW200219_094415}{0.18}{GW191103_012549}{0.18}{GW200316_215756}{0.12}{GW200202_154313}{0.09}{GW200129_065458}{0.21}{GW191216_213338}{0.10}}[{\red{???}}]}
\DeclareRobustCommand{\chipgwtcthreenintiethpercentile}[1]{\IfEqCase{#1}{{GW200224_222234}{0.80}{GW191129_134029}{0.54}{GW200311_115853}{0.77}{GW191230_180458}{0.84}{GW191222_033537}{0.75}{GW200225_060421}{0.82}{GW200302_015811}{0.74}{GW200128_022011}{0.85}{GW191204_171526}{0.67}{GW200112_155838}{0.70}{GW200105_162426}{0.19}{GW191105_143521}{0.65}{GW191109_010717}{0.87}{GW200209_085452}{0.84}{GW200115_042309}{0.46}{GW191127_050227}{0.88}{GW200216_220804}{0.80}{GW191215_223052}{0.82}{GW200208_130117}{0.71}{GW200219_094415}{0.80}{GW191103_012549}{0.72}{GW200316_215756}{0.58}{GW200202_154313}{0.59}{GW200129_065458}{0.91}{GW191216_213338}{0.48}}[{\red{???}}]}
\DeclareRobustCommand{\chirpmassdetgwtcthreeminus}[1]{\IfEqCase{#1}{{GW200224_222234}{3.8}{GW191129_134029}{0.05}{GW200311_115853}{2.8}{GW191230_180458}{9.5}{GW191222_033537}{6.5}{GW200225_060421}{1.97}{GW200302_015811}{4.3}{GW200128_022011}{6.6}{GW191204_171526}{0.05}{GW200112_155838}{2.4}{GW200105_162426}{0.01}{GW191105_143521}{0.14}{GW191109_010717}{9.3}{GW200209_085452}{7.3}{GW200115_042309}{0.01}{GW191127_050227}{19}{GW200216_220804}{20}{GW191215_223052}{1.4}{GW200208_130117}{4.8}{GW200219_094415}{6.2}{GW191103_012549}{0.12}{GW200316_215756}{0.12}{GW200202_154313}{0.05}{GW200129_065458}{2.6}{GW191216_213338}{0.05}}[{\red{???}}]}
\DeclareRobustCommand{\chirpmassdetgwtcthreemed}[1]{\IfEqCase{#1}{{GW200224_222234}{41.1}{GW191129_134029}{8.49}{GW200311_115853}{32.7}{GW191230_180458}{62.8}{GW191222_033537}{51.0}{GW200225_060421}{17.65}{GW200302_015811}{30.4}{GW200128_022011}{50.5}{GW191204_171526}{9.69}{GW200112_155838}{33.9}{GW200105_162426}{3.62}{GW191105_143521}{9.58}{GW191109_010717}{60.1}{GW200209_085452}{42.9}{GW200115_042309}{2.58}{GW191127_050227}{48}{GW200216_220804}{56}{GW191215_223052}{24.9}{GW200208_130117}{38.8}{GW200219_094415}{43.7}{GW191103_012549}{9.98}{GW200316_215756}{10.68}{GW200202_154313}{8.15}{GW200129_065458}{32.1}{GW191216_213338}{8.94}}[{\red{???}}]}
\DeclareRobustCommand{\chirpmassdetgwtcthreeplus}[1]{\IfEqCase{#1}{{GW200224_222234}{3.6}{GW191129_134029}{0.06}{GW200311_115853}{2.7}{GW191230_180458}{9.4}{GW191222_033537}{7.2}{GW200225_060421}{0.98}{GW200302_015811}{7.7}{GW200128_022011}{7.5}{GW191204_171526}{0.05}{GW200112_155838}{2.9}{GW200105_162426}{0.01}{GW191105_143521}{0.12}{GW191109_010717}{9.8}{GW200209_085452}{8.7}{GW200115_042309}{0.01}{GW191127_050227}{21}{GW200216_220804}{14}{GW191215_223052}{1.5}{GW200208_130117}{5.2}{GW200219_094415}{6.3}{GW191103_012549}{0.13}{GW200316_215756}{0.12}{GW200202_154313}{0.05}{GW200129_065458}{1.8}{GW191216_213338}{0.05}}[{\red{???}}]}
\DeclareRobustCommand{\chirpmassdetgwtcthreetenthpercentile}[1]{\IfEqCase{#1}{{GW200224_222234}{38.2}{GW191129_134029}{8.45}{GW200311_115853}{30.6}{GW191230_180458}{55.5}{GW191222_033537}{46.1}{GW200225_060421}{15.96}{GW200302_015811}{27.1}{GW200128_022011}{45.4}{GW191204_171526}{9.65}{GW200112_155838}{32.1}{GW200105_162426}{3.61}{GW191105_143521}{9.48}{GW191109_010717}{52.7}{GW200209_085452}{37.3}{GW200115_042309}{2.57}{GW191127_050227}{33}{GW200216_220804}{39}{GW191215_223052}{23.8}{GW200208_130117}{35.0}{GW200219_094415}{39.0}{GW191103_012549}{9.88}{GW200316_215756}{10.59}{GW200202_154313}{8.11}{GW200129_065458}{30.2}{GW191216_213338}{8.90}}[{\red{???}}]}
\DeclareRobustCommand{\chirpmassdetgwtcthreenintiethpercentile}[1]{\IfEqCase{#1}{{GW200224_222234}{43.8}{GW191129_134029}{8.53}{GW200311_115853}{34.6}{GW191230_180458}{69.9}{GW191222_033537}{56.3}{GW200225_060421}{18.43}{GW200302_015811}{36.4}{GW200128_022011}{56.3}{GW191204_171526}{9.74}{GW200112_155838}{36.2}{GW200105_162426}{3.62}{GW191105_143521}{9.67}{GW191109_010717}{67.5}{GW200209_085452}{49.4}{GW200115_042309}{2.59}{GW191127_050227}{65}{GW200216_220804}{67}{GW191215_223052}{26.1}{GW200208_130117}{42.8}{GW200219_094415}{48.5}{GW191103_012549}{10.07}{GW200316_215756}{10.76}{GW200202_154313}{8.19}{GW200129_065458}{33.4}{GW191216_213338}{8.98}}[{\red{???}}]}
\DeclareRobustCommand{\chirpmasssourcegwtcthreeminus}[1]{\IfEqCase{#1}{{GW200224_222234}{2.5}{GW191129_134029}{0.27}{GW200311_115853}{2.0}{GW191230_180458}{4.9}{GW191222_033537}{5.0}{GW200225_060421}{1.4}{GW200302_015811}{2.9}{GW200128_022011}{4.6}{GW191204_171526}{0.25}{GW200112_155838}{2.1}{GW200105_162426}{0.08}{GW191105_143521}{0.45}{GW191109_010717}{7.5}{GW200209_085452}{3.8}{GW200115_042309}{0.07}{GW191127_050227}{9.1}{GW200216_220804}{8.5}{GW191215_223052}{1.5}{GW200208_130117}{3.1}{GW200219_094415}{3.8}{GW191103_012549}{0.57}{GW200316_215756}{0.55}{GW200202_154313}{0.20}{GW200129_065458}{2.3}{GW191216_213338}{0.19}}[{\red{???}}]}
\DeclareRobustCommand{\chirpmasssourcegwtcthreemed}[1]{\IfEqCase{#1}{{GW200224_222234}{31.0}{GW191129_134029}{7.28}{GW200311_115853}{26.6}{GW191230_180458}{35.5}{GW191222_033537}{33.8}{GW200225_060421}{14.2}{GW200302_015811}{23.3}{GW200128_022011}{30.6}{GW191204_171526}{8.53}{GW200112_155838}{27.4}{GW200105_162426}{3.42}{GW191105_143521}{7.82}{GW191109_010717}{47.5}{GW200209_085452}{26.1}{GW200115_042309}{2.43}{GW191127_050227}{29.9}{GW200216_220804}{32.9}{GW191215_223052}{18.1}{GW200208_130117}{27.7}{GW200219_094415}{27.6}{GW191103_012549}{8.34}{GW200316_215756}{8.75}{GW200202_154313}{7.49}{GW200129_065458}{27.2}{GW191216_213338}{8.33}}[{\red{???}}]}
\DeclareRobustCommand{\chirpmasssourcegwtcthreeplus}[1]{\IfEqCase{#1}{{GW200224_222234}{3.1}{GW191129_134029}{0.42}{GW200311_115853}{2.4}{GW191230_180458}{7.5}{GW191222_033537}{7.1}{GW200225_060421}{1.5}{GW200302_015811}{4.6}{GW200128_022011}{6.7}{GW191204_171526}{0.38}{GW200112_155838}{2.6}{GW200105_162426}{0.08}{GW191105_143521}{0.61}{GW191109_010717}{9.6}{GW200209_085452}{5.6}{GW200115_042309}{0.05}{GW191127_050227}{11.7}{GW200216_220804}{9.3}{GW191215_223052}{2.2}{GW200208_130117}{3.6}{GW200219_094415}{5.6}{GW191103_012549}{0.66}{GW200316_215756}{0.62}{GW200202_154313}{0.24}{GW200129_065458}{2.1}{GW191216_213338}{0.22}}[{\red{???}}]}
\DeclareRobustCommand{\chirpmasssourcegwtcthreetenthpercentile}[1]{\IfEqCase{#1}{{GW200224_222234}{29.0}{GW191129_134029}{7.07}{GW200311_115853}{25.0}{GW191230_180458}{31.6}{GW191222_033537}{29.7}{GW200225_060421}{13.1}{GW200302_015811}{21.0}{GW200128_022011}{26.8}{GW191204_171526}{8.32}{GW200112_155838}{25.7}{GW200105_162426}{3.36}{GW191105_143521}{7.46}{GW191109_010717}{41.5}{GW200209_085452}{23.1}{GW200115_042309}{2.38}{GW191127_050227}{22.8}{GW200216_220804}{26.2}{GW191215_223052}{16.9}{GW200208_130117}{25.3}{GW200219_094415}{24.6}{GW191103_012549}{7.88}{GW200316_215756}{8.31}{GW200202_154313}{7.33}{GW200129_065458}{25.4}{GW191216_213338}{8.18}}[{\red{???}}]}
\DeclareRobustCommand{\chirpmasssourcegwtcthreenintiethpercentile}[1]{\IfEqCase{#1}{{GW200224_222234}{33.3}{GW191129_134029}{7.62}{GW200311_115853}{28.4}{GW191230_180458}{41.0}{GW191222_033537}{39.4}{GW200225_060421}{15.3}{GW200302_015811}{26.7}{GW200128_022011}{35.6}{GW191204_171526}{8.83}{GW200112_155838}{29.3}{GW200105_162426}{3.48}{GW191105_143521}{8.30}{GW191109_010717}{54.5}{GW200209_085452}{30.2}{GW200115_042309}{2.47}{GW191127_050227}{38.9}{GW200216_220804}{40.0}{GW191215_223052}{19.7}{GW200208_130117}{30.5}{GW200219_094415}{32.0}{GW191103_012549}{8.88}{GW200316_215756}{9.26}{GW200202_154313}{7.68}{GW200129_065458}{28.8}{GW191216_213338}{8.51}}[{\red{???}}]}
\DeclareRobustCommand{\totalmassdetgwtcthreeminus}[1]{\IfEqCase{#1}{{GW200224_222234}{7.2}{GW191129_134029}{0.65}{GW200311_115853}{5.7}{GW191230_180458}{19}{GW191222_033537}{13}{GW200225_060421}{4.0}{GW200302_015811}{8.1}{GW200128_022011}{14}{GW191204_171526}{0.48}{GW200112_155838}{5.1}{GW200105_162426}{1.5}{GW191105_143521}{0.50}{GW191109_010717}{17}{GW200209_085452}{16}{GW200115_042309}{1.8}{GW191127_050227}{45}{GW200216_220804}{32}{GW191215_223052}{3.7}{GW200208_130117}{10}{GW200219_094415}{12}{GW191103_012549}{0.68}{GW200316_215756}{1.1}{GW200202_154313}{0.34}{GW200129_065458}{3.8}{GW191216_213338}{0.66}}[{\red{???}}]}
\DeclareRobustCommand{\totalmassdetgwtcthreemed}[1]{\IfEqCase{#1}{{GW200224_222234}{95.3}{GW191129_134029}{20.11}{GW200311_115853}{75.9}{GW191230_180458}{147}{GW191222_033537}{119}{GW200225_060421}{41.2}{GW200302_015811}{74.5}{GW200128_022011}{118}{GW191204_171526}{22.73}{GW200112_155838}{79.1}{GW200105_162426}{11.6}{GW191105_143521}{22.38}{GW191109_010717}{140}{GW200209_085452}{100}{GW200115_042309}{7.8}{GW191127_050227}{130}{GW200216_220804}{135}{GW191215_223052}{58.6}{GW200208_130117}{91}{GW200219_094415}{103}{GW191103_012549}{23.47}{GW200316_215756}{25.5}{GW200202_154313}{19.01}{GW200129_065458}{74.6}{GW191216_213338}{21.17}}[{\red{???}}]}
\DeclareRobustCommand{\totalmassdetgwtcthreeplus}[1]{\IfEqCase{#1}{{GW200224_222234}{8.4}{GW191129_134029}{2.94}{GW200311_115853}{6.2}{GW191230_180458}{21}{GW191222_033537}{16}{GW200225_060421}{3.0}{GW200302_015811}{15.7}{GW200128_022011}{18}{GW191204_171526}{1.93}{GW200112_155838}{6.5}{GW200105_162426}{1.6}{GW191105_143521}{2.35}{GW191109_010717}{21}{GW200209_085452}{20}{GW200115_042309}{1.8}{GW191127_050227}{53}{GW200216_220804}{30}{GW191215_223052}{5.1}{GW200208_130117}{11}{GW200219_094415}{14}{GW191103_012549}{4.58}{GW200316_215756}{8.8}{GW200202_154313}{1.99}{GW200129_065458}{4.5}{GW191216_213338}{2.96}}[{\red{???}}]}
\DeclareRobustCommand{\totalmassdetgwtcthreetenthpercentile}[1]{\IfEqCase{#1}{{GW200224_222234}{89.6}{GW191129_134029}{19.51}{GW200311_115853}{71.5}{GW191230_180458}{132}{GW191222_033537}{109}{GW200225_060421}{37.9}{GW200302_015811}{68.1}{GW200128_022011}{107}{GW191204_171526}{22.30}{GW200112_155838}{75.2}{GW200105_162426}{10.7}{GW191105_143521}{21.96}{GW191109_010717}{127}{GW200209_085452}{88}{GW200115_042309}{6.2}{GW191127_050227}{91}{GW200216_220804}{109}{GW191215_223052}{55.7}{GW200208_130117}{84}{GW200219_094415}{93}{GW191103_012549}{22.87}{GW200316_215756}{24.5}{GW200202_154313}{18.71}{GW200129_065458}{71.7}{GW191216_213338}{20.56}}[{\red{???}}]}
\DeclareRobustCommand{\totalmassdetgwtcthreenintiethpercentile}[1]{\IfEqCase{#1}{{GW200224_222234}{101.5}{GW191129_134029}{22.18}{GW200311_115853}{80.4}{GW191230_180458}{163}{GW191222_033537}{131}{GW200225_060421}{43.5}{GW200302_015811}{86.5}{GW200128_022011}{131}{GW191204_171526}{24.06}{GW200112_155838}{84.0}{GW200105_162426}{12.5}{GW191105_143521}{23.91}{GW191109_010717}{156}{GW200209_085452}{115}{GW200115_042309}{9.2}{GW191127_050227}{171}{GW200216_220804}{158}{GW191215_223052}{62.1}{GW200208_130117}{100}{GW200219_094415}{113}{GW191103_012549}{26.45}{GW200316_215756}{31.0}{GW200202_154313}{20.33}{GW200129_065458}{78.0}{GW191216_213338}{23.01}}[{\red{???}}]}
\DeclareRobustCommand{\redshiftgwtcthreeminus}[1]{\IfEqCase{#1}{{GW200224_222234}{0.10}{GW191129_134029}{0.06}{GW200311_115853}{0.07}{GW191230_180458}{0.27}{GW191222_033537}{0.26}{GW200225_060421}{0.10}{GW200302_015811}{0.13}{GW200128_022011}{0.29}{GW191204_171526}{0.05}{GW200112_155838}{0.08}{GW200105_162426}{0.02}{GW191105_143521}{0.09}{GW191109_010717}{0.12}{GW200209_085452}{0.25}{GW200115_042309}{0.02}{GW191127_050227}{0.29}{GW200216_220804}{0.29}{GW191215_223052}{0.15}{GW200208_130117}{0.14}{GW200219_094415}{0.22}{GW191103_012549}{0.09}{GW200316_215756}{0.08}{GW200202_154313}{0.03}{GW200129_065458}{0.07}{GW191216_213338}{0.03}}[{\red{???}}]}
\DeclareRobustCommand{\redshiftgwtcthreemed}[1]{\IfEqCase{#1}{{GW200224_222234}{0.33}{GW191129_134029}{0.17}{GW200311_115853}{0.23}{GW191230_180458}{0.76}{GW191222_033537}{0.51}{GW200225_060421}{0.22}{GW200302_015811}{0.31}{GW200128_022011}{0.66}{GW191204_171526}{0.14}{GW200112_155838}{0.24}{GW200105_162426}{0.06}{GW191105_143521}{0.23}{GW191109_010717}{0.25}{GW200209_085452}{0.64}{GW200115_042309}{0.06}{GW191127_050227}{0.57}{GW200216_220804}{0.63}{GW191215_223052}{0.38}{GW200208_130117}{0.40}{GW200219_094415}{0.57}{GW191103_012549}{0.20}{GW200316_215756}{0.22}{GW200202_154313}{0.09}{GW200129_065458}{0.18}{GW191216_213338}{0.07}}[{\red{???}}]}
\DeclareRobustCommand{\redshiftgwtcthreeplus}[1]{\IfEqCase{#1}{{GW200224_222234}{0.07}{GW191129_134029}{0.04}{GW200311_115853}{0.05}{GW191230_180458}{0.26}{GW191222_033537}{0.23}{GW200225_060421}{0.09}{GW200302_015811}{0.17}{GW200128_022011}{0.26}{GW191204_171526}{0.04}{GW200112_155838}{0.07}{GW200105_162426}{0.02}{GW191105_143521}{0.07}{GW191109_010717}{0.18}{GW200209_085452}{0.25}{GW200115_042309}{0.03}{GW191127_050227}{0.40}{GW200216_220804}{0.37}{GW191215_223052}{0.13}{GW200208_130117}{0.15}{GW200219_094415}{0.22}{GW191103_012549}{0.09}{GW200316_215756}{0.08}{GW200202_154313}{0.03}{GW200129_065458}{0.05}{GW191216_213338}{0.02}}[{\red{???}}]}
\DeclareRobustCommand{\redshiftgwtcthreetenthpercentile}[1]{\IfEqCase{#1}{{GW200224_222234}{0.25}{GW191129_134029}{0.11}{GW200311_115853}{0.17}{GW191230_180458}{0.54}{GW191222_033537}{0.31}{GW200225_060421}{0.15}{GW200302_015811}{0.20}{GW200128_022011}{0.43}{GW191204_171526}{0.10}{GW200112_155838}{0.18}{GW200105_162426}{0.04}{GW191105_143521}{0.15}{GW191109_010717}{0.15}{GW200209_085452}{0.44}{GW200115_042309}{0.05}{GW191127_050227}{0.34}{GW200216_220804}{0.39}{GW191215_223052}{0.26}{GW200208_130117}{0.29}{GW200219_094415}{0.39}{GW191103_012549}{0.12}{GW200316_215756}{0.15}{GW200202_154313}{0.06}{GW200129_065458}{0.12}{GW191216_213338}{0.05}}[{\red{???}}]}
\DeclareRobustCommand{\redshiftgwtcthreenintiethpercentile}[1]{\IfEqCase{#1}{{GW200224_222234}{0.39}{GW191129_134029}{0.20}{GW200311_115853}{0.27}{GW191230_180458}{0.96}{GW191222_033537}{0.69}{GW200225_060421}{0.29}{GW200302_015811}{0.44}{GW200128_022011}{0.86}{GW191204_171526}{0.17}{GW200112_155838}{0.30}{GW200105_162426}{0.08}{GW191105_143521}{0.28}{GW191109_010717}{0.38}{GW200209_085452}{0.83}{GW200115_042309}{0.08}{GW191127_050227}{0.87}{GW200216_220804}{0.91}{GW191215_223052}{0.48}{GW200208_130117}{0.51}{GW200219_094415}{0.75}{GW191103_012549}{0.26}{GW200316_215756}{0.28}{GW200202_154313}{0.11}{GW200129_065458}{0.22}{GW191216_213338}{0.09}}[{\red{???}}]}
\DeclareRobustCommand{\geocenttimegwtcthreeminus}[1]{\IfEqCase{#1}{{GW200224_222234}{0.0179}{GW191129_134029}{0.019}{GW200311_115853}{0.0}{GW191230_180458}{0.0}{GW191222_033537}{0.023}{GW200225_060421}{0.0049}{GW200302_015811}{0.0350}{GW200128_022011}{0.0313}{GW191204_171526}{0.0}{GW200112_155838}{0.0358}{GW200105_162426}{0.0017}{GW191105_143521}{0.023}{GW191109_010717}{0.0311}{GW200209_085452}{0.0304}{GW200115_042309}{0.0400}{GW191127_050227}{0.012}{GW200216_220804}{0.0}{GW191215_223052}{0.0439}{GW200208_130117}{0.0032}{GW200219_094415}{0.026}{GW191103_012549}{0.019}{GW200316_215756}{0.0}{GW200202_154313}{0.1}{GW200129_065458}{0.0}{GW191216_213338}{0.0045}}[{\red{???}}]}
\DeclareRobustCommand{\geocenttimegwtcthreemed}[1]{\IfEqCase{#1}{{GW200224_222234}{1266618172.3978}{GW191129_134029}{1259070047.197}{GW200311_115853}{1267963151.4}{GW191230_180458}{1261764316.4}{GW191222_033537}{1261020955.117}{GW200225_060421}{1266645879.4018}{GW200302_015811}{1267149509.5273}{GW200128_022011}{1264213229.9033}{GW191204_171526}{1259514944.1}{GW200112_155838}{1262879936.1035}{GW200105_162426}{1262276684.0349}{GW191105_143521}{1256999739.933}{GW191109_010717}{1257296855.2191}{GW200209_085452}{1265273710.1831}{GW200115_042309}{1263097407.7623}{GW191127_050227}{1258866165.541}{GW200216_220804}{1265926102.9}{GW191215_223052}{1260484270.3546}{GW200208_130117}{1265202095.9383}{GW200219_094415}{1266140673.195}{GW191103_012549}{1256779567.535}{GW200316_215756}{1268431094.2}{GW200202_154313}{1264693411.6}{GW200129_065458}{1264316116.4}{GW191216_213338}{1260567236.4871}}[{\red{???}}]}
\DeclareRobustCommand{\geocenttimegwtcthreeplus}[1]{\IfEqCase{#1}{{GW200224_222234}{0.0080}{GW191129_134029}{0.020}{GW200311_115853}{0.0}{GW191230_180458}{0.0}{GW191222_033537}{0.017}{GW200225_060421}{0.0124}{GW200302_015811}{0.0070}{GW200128_022011}{0.0100}{GW191204_171526}{0.0}{GW200112_155838}{0.0035}{GW200105_162426}{0.0448}{GW191105_143521}{0.015}{GW191109_010717}{0.0016}{GW200209_085452}{0.0067}{GW200115_042309}{0.0043}{GW191127_050227}{0.024}{GW200216_220804}{0.0}{GW191215_223052}{0.0023}{GW200208_130117}{0.0114}{GW200219_094415}{0.015}{GW191103_012549}{0.013}{GW200316_215756}{0.0}{GW200202_154313}{0.0}{GW200129_065458}{0.0}{GW191216_213338}{0.0044}}[{\red{???}}]}
\DeclareRobustCommand{\geocenttimegwtcthreetenthpercentile}[1]{\IfEqCase{#1}{{GW200224_222234}{1266618172.3803}{GW191129_134029}{1259070047.178}{GW200311_115853}{1267963151.4}{GW191230_180458}{1261764316.4}{GW191222_033537}{1261020955.095}{GW200225_060421}{1266645879.3970}{GW200302_015811}{1267149509.4933}{GW200128_022011}{1264213229.8733}{GW191204_171526}{1259514944.1}{GW200112_155838}{1262879936.0709}{GW200105_162426}{1262276684.0332}{GW191105_143521}{1256999739.910}{GW191109_010717}{1257296855.1896}{GW200209_085452}{1265273710.1818}{GW200115_042309}{1263097407.7501}{GW191127_050227}{1258866165.531}{GW200216_220804}{1265926102.9}{GW191215_223052}{1260484270.3114}{GW200208_130117}{1265202095.9356}{GW200219_094415}{1266140673.170}{GW191103_012549}{1256779567.517}{GW200316_215756}{1268431094.2}{GW200202_154313}{1264693411.6}{GW200129_065458}{1264316116.4}{GW191216_213338}{1260567236.4851}}[{\red{???}}]}
\DeclareRobustCommand{\geocenttimegwtcthreenintiethpercentile}[1]{\IfEqCase{#1}{{GW200224_222234}{1266618172.4058}{GW191129_134029}{1259070047.217}{GW200311_115853}{1267963151.4}{GW191230_180458}{1261764316.4}{GW191222_033537}{1261020955.129}{GW200225_060421}{1266645879.4139}{GW200302_015811}{1267149509.5335}{GW200128_022011}{1264213229.9121}{GW191204_171526}{1259514944.1}{GW200112_155838}{1262879936.1064}{GW200105_162426}{1262276684.0789}{GW191105_143521}{1256999739.945}{GW191109_010717}{1257296855.2207}{GW200209_085452}{1265273710.1888}{GW200115_042309}{1263097407.7666}{GW191127_050227}{1258866165.562}{GW200216_220804}{1265926102.9}{GW191215_223052}{1260484270.3569}{GW200208_130117}{1265202095.9497}{GW200219_094415}{1266140673.209}{GW191103_012549}{1256779567.546}{GW200316_215756}{1268431094.2}{GW200202_154313}{1264693411.6}{GW200129_065458}{1264316116.4}{GW191216_213338}{1260567236.4909}}[{\red{???}}]}
\DeclareRobustCommand{\luminositydistancegwtcthreeminus}[1]{\IfEqCase{#1}{{GW200224_222234}{0.63}{GW191129_134029}{0.33}{GW200311_115853}{0.40}{GW191230_180458}{2.0}{GW191222_033537}{1.7}{GW200225_060421}{0.53}{GW200302_015811}{0.77}{GW200128_022011}{2.0}{GW191204_171526}{0.25}{GW200112_155838}{0.46}{GW200105_162426}{0.11}{GW191105_143521}{0.48}{GW191109_010717}{0.65}{GW200209_085452}{1.8}{GW200115_042309}{0.10}{GW191127_050227}{1.9}{GW200216_220804}{2.0}{GW191215_223052}{0.91}{GW200208_130117}{0.85}{GW200219_094415}{1.5}{GW191103_012549}{0.47}{GW200316_215756}{0.44}{GW200202_154313}{0.16}{GW200129_065458}{0.38}{GW191216_213338}{0.13}}[{\red{???}}]}
\DeclareRobustCommand{\luminositydistancegwtcthreemed}[1]{\IfEqCase{#1}{{GW200224_222234}{1.77}{GW191129_134029}{0.82}{GW200311_115853}{1.17}{GW191230_180458}{4.8}{GW191222_033537}{3.0}{GW200225_060421}{1.15}{GW200302_015811}{1.67}{GW200128_022011}{4.0}{GW191204_171526}{0.66}{GW200112_155838}{1.25}{GW200105_162426}{0.27}{GW191105_143521}{1.15}{GW191109_010717}{1.29}{GW200209_085452}{3.9}{GW200115_042309}{0.29}{GW191127_050227}{3.4}{GW200216_220804}{3.8}{GW191215_223052}{2.10}{GW200208_130117}{2.23}{GW200219_094415}{3.4}{GW191103_012549}{0.99}{GW200316_215756}{1.12}{GW200202_154313}{0.41}{GW200129_065458}{0.90}{GW191216_213338}{0.34}}[{\red{???}}]}
\DeclareRobustCommand{\luminositydistancegwtcthreeplus}[1]{\IfEqCase{#1}{{GW200224_222234}{0.47}{GW191129_134029}{0.25}{GW200311_115853}{0.28}{GW191230_180458}{2.1}{GW191222_033537}{1.7}{GW200225_060421}{0.51}{GW200302_015811}{1.09}{GW200128_022011}{2.1}{GW191204_171526}{0.19}{GW200112_155838}{0.43}{GW200105_162426}{0.12}{GW191105_143521}{0.43}{GW191109_010717}{1.13}{GW200209_085452}{2.0}{GW200115_042309}{0.15}{GW191127_050227}{3.1}{GW200216_220804}{3.0}{GW191215_223052}{0.86}{GW200208_130117}{1.00}{GW200219_094415}{1.7}{GW191103_012549}{0.50}{GW200316_215756}{0.47}{GW200202_154313}{0.15}{GW200129_065458}{0.29}{GW191216_213338}{0.12}}[{\red{???}}]}
\DeclareRobustCommand{\luminositydistancegwtcthreetenthpercentile}[1]{\IfEqCase{#1}{{GW200224_222234}{1.29}{GW191129_134029}{0.55}{GW200311_115853}{0.87}{GW191230_180458}{3.2}{GW191222_033537}{1.6}{GW200225_060421}{0.74}{GW200302_015811}{1.03}{GW200128_022011}{2.4}{GW191204_171526}{0.47}{GW200112_155838}{0.89}{GW200105_162426}{0.18}{GW191105_143521}{0.76}{GW191109_010717}{0.76}{GW200209_085452}{2.5}{GW200115_042309}{0.21}{GW191127_050227}{1.8}{GW200216_220804}{2.2}{GW191215_223052}{1.36}{GW200208_130117}{1.54}{GW200219_094415}{2.2}{GW191103_012549}{0.60}{GW200316_215756}{0.76}{GW200202_154313}{0.28}{GW200129_065458}{0.60}{GW191216_213338}{0.23}}[{\red{???}}]}
\DeclareRobustCommand{\luminositydistancegwtcthreenintiethpercentile}[1]{\IfEqCase{#1}{{GW200224_222234}{2.15}{GW191129_134029}{1.02}{GW200311_115853}{1.39}{GW191230_180458}{6.4}{GW191222_033537}{4.3}{GW200225_060421}{1.55}{GW200302_015811}{2.49}{GW200128_022011}{5.7}{GW191204_171526}{0.82}{GW200112_155838}{1.60}{GW200105_162426}{0.36}{GW191105_143521}{1.49}{GW191109_010717}{2.13}{GW200209_085452}{5.4}{GW200115_042309}{0.40}{GW191127_050227}{5.7}{GW200216_220804}{6.0}{GW191215_223052}{2.78}{GW200208_130117}{2.98}{GW200219_094415}{4.7}{GW191103_012549}{1.38}{GW200316_215756}{1.49}{GW200202_154313}{0.53}{GW200129_065458}{1.13}{GW191216_213338}{0.44}}[{\red{???}}]}
\DeclareRobustCommand{\thetajngwtcthreeminus}[1]{\IfEqCase{#1}{{GW200224_222234}{0.42}{GW191129_134029}{1.5}{GW200311_115853}{0.40}{GW191230_180458}{1.85}{GW191222_033537}{1.3}{GW200225_060421}{1.00}{GW200302_015811}{0.98}{GW200128_022011}{1.1}{GW191204_171526}{2.04}{GW200112_155838}{0.68}{GW200105_162426}{1.2}{GW191105_143521}{0.85}{GW191109_010717}{1.18}{GW200209_085452}{1.5}{GW200115_042309}{0.44}{GW191127_050227}{1.2}{GW200216_220804}{0.69}{GW191215_223052}{0.82}{GW200208_130117}{0.58}{GW200219_094415}{0.92}{GW191103_012549}{1.1}{GW200316_215756}{1.84}{GW200202_154313}{0.59}{GW200129_065458}{0.41}{GW191216_213338}{0.81}}[{\red{???}}]}
\DeclareRobustCommand{\thetajngwtcthreemed}[1]{\IfEqCase{#1}{{GW200224_222234}{0.57}{GW191129_134029}{1.8}{GW200311_115853}{0.55}{GW191230_180458}{2.13}{GW191222_033537}{1.6}{GW200225_060421}{1.31}{GW200302_015811}{1.26}{GW200128_022011}{1.4}{GW191204_171526}{2.29}{GW200112_155838}{0.88}{GW200105_162426}{1.5}{GW191105_143521}{1.07}{GW191109_010717}{1.91}{GW200209_085452}{1.8}{GW200115_042309}{0.62}{GW191127_050227}{1.5}{GW200216_220804}{0.89}{GW191215_223052}{1.12}{GW200208_130117}{2.53}{GW200219_094415}{1.19}{GW191103_012549}{1.4}{GW200316_215756}{2.32}{GW200202_154313}{2.57}{GW200129_065458}{0.66}{GW191216_213338}{2.50}}[{\red{???}}]}
\DeclareRobustCommand{\thetajngwtcthreeplus}[1]{\IfEqCase{#1}{{GW200224_222234}{0.54}{GW191129_134029}{1.1}{GW200311_115853}{0.52}{GW191230_180458}{0.79}{GW191222_033537}{1.2}{GW200225_060421}{1.47}{GW200302_015811}{1.55}{GW200128_022011}{1.5}{GW191204_171526}{0.64}{GW200112_155838}{2.04}{GW200105_162426}{1.3}{GW191105_143521}{1.82}{GW191109_010717}{0.87}{GW200209_085452}{1.1}{GW200115_042309}{1.94}{GW191127_050227}{1.4}{GW200216_220804}{1.87}{GW191215_223052}{1.65}{GW200208_130117}{0.44}{GW200219_094415}{1.59}{GW191103_012549}{1.5}{GW200316_215756}{0.58}{GW200202_154313}{0.42}{GW200129_065458}{0.59}{GW191216_213338}{0.45}}[{\red{???}}]}
\DeclareRobustCommand{\thetajngwtcthreetenthpercentile}[1]{\IfEqCase{#1}{{GW200224_222234}{0.22}{GW191129_134029}{0.4}{GW200311_115853}{0.21}{GW191230_180458}{0.40}{GW191222_033537}{0.4}{GW200225_060421}{0.45}{GW200302_015811}{0.40}{GW200128_022011}{0.3}{GW191204_171526}{0.36}{GW200112_155838}{0.28}{GW200105_162426}{0.5}{GW191105_143521}{0.32}{GW191109_010717}{1.02}{GW200209_085452}{0.4}{GW200115_042309}{0.26}{GW191127_050227}{0.4}{GW200216_220804}{0.30}{GW191215_223052}{0.42}{GW200208_130117}{2.09}{GW200219_094415}{0.40}{GW191103_012549}{0.3}{GW200316_215756}{0.71}{GW200202_154313}{2.10}{GW200129_065458}{0.31}{GW191216_213338}{2.00}}[{\red{???}}]}
\DeclareRobustCommand{\thetajngwtcthreenintiethpercentile}[1]{\IfEqCase{#1}{{GW200224_222234}{1.00}{GW191129_134029}{2.8}{GW200311_115853}{0.97}{GW191230_180458}{2.81}{GW191222_033537}{2.7}{GW200225_060421}{2.64}{GW200302_015811}{2.68}{GW200128_022011}{2.8}{GW191204_171526}{2.84}{GW200112_155838}{2.82}{GW200105_162426}{2.7}{GW191105_143521}{2.77}{GW191109_010717}{2.63}{GW200209_085452}{2.8}{GW200115_042309}{2.12}{GW191127_050227}{2.7}{GW200216_220804}{2.57}{GW191215_223052}{2.61}{GW200208_130117}{2.90}{GW200219_094415}{2.60}{GW191103_012549}{2.8}{GW200316_215756}{2.80}{GW200202_154313}{2.92}{GW200129_065458}{1.11}{GW191216_213338}{2.88}}[{\red{???}}]}
\DeclareRobustCommand{\chieffgwtcthreeminus}[1]{\IfEqCase{#1}{{GW200224_222234}{0.15}{GW191129_134029}{0.08}{GW200311_115853}{0.20}{GW191230_180458}{0.30}{GW191222_033537}{0.25}{GW200225_060421}{0.28}{GW200302_015811}{0.25}{GW200128_022011}{0.25}{GW191204_171526}{0.05}{GW200112_155838}{0.15}{GW200105_162426}{0.18}{GW191105_143521}{0.09}{GW191109_010717}{0.31}{GW200209_085452}{0.30}{GW200115_042309}{0.41}{GW191127_050227}{0.36}{GW200216_220804}{0.36}{GW191215_223052}{0.21}{GW200208_130117}{0.27}{GW200219_094415}{0.29}{GW191103_012549}{0.10}{GW200316_215756}{0.10}{GW200202_154313}{0.06}{GW200129_065458}{0.16}{GW191216_213338}{0.06}}[{\red{???}}]}
\DeclareRobustCommand{\chieffgwtcthreemed}[1]{\IfEqCase{#1}{{GW200224_222234}{0.11}{GW191129_134029}{0.06}{GW200311_115853}{-0.02}{GW191230_180458}{-0.03}{GW191222_033537}{-0.04}{GW200225_060421}{-0.12}{GW200302_015811}{0.03}{GW200128_022011}{0.14}{GW191204_171526}{0.16}{GW200112_155838}{0.06}{GW200105_162426}{0.00}{GW191105_143521}{-0.02}{GW191109_010717}{-0.29}{GW200209_085452}{-0.10}{GW200115_042309}{-0.15}{GW191127_050227}{0.18}{GW200216_220804}{0.10}{GW191215_223052}{-0.03}{GW200208_130117}{-0.07}{GW200219_094415}{-0.08}{GW191103_012549}{0.21}{GW200316_215756}{0.13}{GW200202_154313}{0.04}{GW200129_065458}{0.11}{GW191216_213338}{0.11}}[{\red{???}}]}
\DeclareRobustCommand{\chieffgwtcthreeplus}[1]{\IfEqCase{#1}{{GW200224_222234}{0.15}{GW191129_134029}{0.16}{GW200311_115853}{0.16}{GW191230_180458}{0.26}{GW191222_033537}{0.20}{GW200225_060421}{0.17}{GW200302_015811}{0.26}{GW200128_022011}{0.24}{GW191204_171526}{0.08}{GW200112_155838}{0.15}{GW200105_162426}{0.13}{GW191105_143521}{0.13}{GW191109_010717}{0.42}{GW200209_085452}{0.24}{GW200115_042309}{0.24}{GW191127_050227}{0.34}{GW200216_220804}{0.34}{GW191215_223052}{0.17}{GW200208_130117}{0.22}{GW200219_094415}{0.23}{GW191103_012549}{0.16}{GW200316_215756}{0.27}{GW200202_154313}{0.13}{GW200129_065458}{0.11}{GW191216_213338}{0.13}}[{\red{???}}]}
\DeclareRobustCommand{\chieffgwtcthreetenthpercentile}[1]{\IfEqCase{#1}{{GW200224_222234}{-0.01}{GW191129_134029}{0.00}{GW200311_115853}{-0.17}{GW191230_180458}{-0.26}{GW191222_033537}{-0.23}{GW200225_060421}{-0.34}{GW200302_015811}{-0.15}{GW200128_022011}{-0.05}{GW191204_171526}{0.12}{GW200112_155838}{-0.05}{GW200105_162426}{-0.10}{GW191105_143521}{-0.09}{GW191109_010717}{-0.54}{GW200209_085452}{-0.33}{GW200115_042309}{-0.51}{GW191127_050227}{-0.10}{GW200216_220804}{-0.17}{GW191215_223052}{-0.19}{GW200208_130117}{-0.27}{GW200219_094415}{-0.30}{GW191103_012549}{0.13}{GW200316_215756}{0.04}{GW200202_154313}{-0.01}{GW200129_065458}{0.00}{GW191216_213338}{0.06}}[{\red{???}}]}
\DeclareRobustCommand{\chieffgwtcthreenintiethpercentile}[1]{\IfEqCase{#1}{{GW200224_222234}{0.22}{GW191129_134029}{0.19}{GW200311_115853}{0.10}{GW191230_180458}{0.18}{GW191222_033537}{0.11}{GW200225_060421}{0.02}{GW200302_015811}{0.23}{GW200128_022011}{0.33}{GW191204_171526}{0.22}{GW200112_155838}{0.18}{GW200105_162426}{0.08}{GW191105_143521}{0.07}{GW191109_010717}{0.00}{GW200209_085452}{0.09}{GW200115_042309}{0.04}{GW191127_050227}{0.45}{GW200216_220804}{0.36}{GW191215_223052}{0.10}{GW200208_130117}{0.09}{GW200219_094415}{0.10}{GW191103_012549}{0.33}{GW200316_215756}{0.32}{GW200202_154313}{0.13}{GW200129_065458}{0.20}{GW191216_213338}{0.20}}[{\red{???}}]}
\DeclareRobustCommand{\spinonegwtcthreeminus}[1]{\IfEqCase{#1}{{GW200224_222234}{0.42}{GW191129_134029}{0.22}{GW200311_115853}{0.36}{GW191230_180458}{0.46}{GW191222_033537}{0.34}{GW200225_060421}{0.51}{GW200302_015811}{0.35}{GW200128_022011}{0.53}{GW191204_171526}{0.35}{GW200112_155838}{0.31}{GW200105_162426}{0.07}{GW191105_143521}{0.21}{GW191109_010717}{0.58}{GW200209_085452}{0.46}{GW200115_042309}{0.29}{GW191127_050227}{0.58}{GW200216_220804}{0.43}{GW191215_223052}{0.42}{GW200208_130117}{0.32}{GW200219_094415}{0.43}{GW191103_012549}{0.40}{GW200316_215756}{0.28}{GW200202_154313}{0.20}{GW200129_065458}{0.47}{GW191216_213338}{0.22}}[{\red{???}}]}
\DeclareRobustCommand{\spinonegwtcthreemed}[1]{\IfEqCase{#1}{{GW200224_222234}{0.47}{GW191129_134029}{0.25}{GW200311_115853}{0.39}{GW191230_180458}{0.51}{GW191222_033537}{0.38}{GW200225_060421}{0.59}{GW200302_015811}{0.38}{GW200128_022011}{0.60}{GW191204_171526}{0.40}{GW200112_155838}{0.34}{GW200105_162426}{0.08}{GW191105_143521}{0.23}{GW191109_010717}{0.83}{GW200209_085452}{0.52}{GW200115_042309}{0.32}{GW191127_050227}{0.66}{GW200216_220804}{0.48}{GW191215_223052}{0.47}{GW200208_130117}{0.36}{GW200219_094415}{0.47}{GW191103_012549}{0.46}{GW200316_215756}{0.32}{GW200202_154313}{0.22}{GW200129_065458}{0.53}{GW191216_213338}{0.24}}[{\red{???}}]}
\DeclareRobustCommand{\spinonegwtcthreeplus}[1]{\IfEqCase{#1}{{GW200224_222234}{0.44}{GW191129_134029}{0.37}{GW200311_115853}{0.48}{GW191230_180458}{0.44}{GW191222_033537}{0.50}{GW200225_060421}{0.35}{GW200302_015811}{0.51}{GW200128_022011}{0.36}{GW191204_171526}{0.38}{GW200112_155838}{0.46}{GW200105_162426}{0.30}{GW191105_143521}{0.53}{GW191109_010717}{0.15}{GW200209_085452}{0.43}{GW200115_042309}{0.50}{GW191127_050227}{0.31}{GW200216_220804}{0.46}{GW191215_223052}{0.45}{GW200208_130117}{0.51}{GW200219_094415}{0.46}{GW191103_012549}{0.40}{GW200316_215756}{0.37}{GW200202_154313}{0.45}{GW200129_065458}{0.42}{GW191216_213338}{0.36}}[{\red{???}}]}
\DeclareRobustCommand{\spinonegwtcthreetenthpercentile}[1]{\IfEqCase{#1}{{GW200224_222234}{0.10}{GW191129_134029}{0.05}{GW200311_115853}{0.07}{GW191230_180458}{0.10}{GW191222_033537}{0.07}{GW200225_060421}{0.14}{GW200302_015811}{0.07}{GW200128_022011}{0.14}{GW191204_171526}{0.10}{GW200112_155838}{0.07}{GW200105_162426}{0.01}{GW191105_143521}{0.04}{GW191109_010717}{0.42}{GW200209_085452}{0.10}{GW200115_042309}{0.05}{GW191127_050227}{0.16}{GW200216_220804}{0.10}{GW191215_223052}{0.09}{GW200208_130117}{0.07}{GW200219_094415}{0.09}{GW191103_012549}{0.13}{GW200316_215756}{0.08}{GW200202_154313}{0.04}{GW200129_065458}{0.12}{GW191216_213338}{0.05}}[{\red{???}}]}
\DeclareRobustCommand{\spinonegwtcthreenintiethpercentile}[1]{\IfEqCase{#1}{{GW200224_222234}{0.84}{GW191129_134029}{0.54}{GW200311_115853}{0.79}{GW191230_180458}{0.90}{GW191222_033537}{0.80}{GW200225_060421}{0.89}{GW200302_015811}{0.81}{GW200128_022011}{0.92}{GW191204_171526}{0.70}{GW200112_155838}{0.71}{GW200105_162426}{0.27}{GW191105_143521}{0.64}{GW191109_010717}{0.97}{GW200209_085452}{0.90}{GW200115_042309}{0.73}{GW191127_050227}{0.94}{GW200216_220804}{0.89}{GW191215_223052}{0.85}{GW200208_130117}{0.78}{GW200219_094415}{0.88}{GW191103_012549}{0.78}{GW200316_215756}{0.61}{GW200202_154313}{0.55}{GW200129_065458}{0.93}{GW191216_213338}{0.50}}[{\red{???}}]}
\DeclareRobustCommand{\cosiotagwtcthreeminus}[1]{\IfEqCase{#1}{{GW200224_222234}{0.43}{GW191129_134029}{0.78}{GW200311_115853}{0.37}{GW191230_180458}{0.41}{GW191222_033537}{0.91}{GW200225_060421}{1.15}{GW200302_015811}{1.24}{GW200128_022011}{1.13}{GW191204_171526}{0.33}{GW200112_155838}{1.59}{GW200105_162426}{0.97}{GW191105_143521}{1.44}{GW191109_010717}{0.55}{GW200209_085452}{0.83}{GW200115_042309}{1.64}{GW191127_050227}{1.00}{GW200216_220804}{1.50}{GW191215_223052}{1.28}{GW200208_130117}{0.17}{GW200219_094415}{1.26}{GW191103_012549}{1.17}{GW200316_215756}{0.29}{GW200202_154313}{0.15}{GW200129_065458}{0.50}{GW191216_213338}{0.18}}[{\red{???}}]}
\DeclareRobustCommand{\cosiotagwtcthreemed}[1]{\IfEqCase{#1}{{GW200224_222234}{0.84}{GW191129_134029}{-0.19}{GW200311_115853}{0.86}{GW191230_180458}{-0.57}{GW191222_033537}{-0.05}{GW200225_060421}{0.21}{GW200302_015811}{0.30}{GW200128_022011}{0.16}{GW191204_171526}{-0.65}{GW200112_155838}{0.62}{GW200105_162426}{0.03}{GW191105_143521}{0.47}{GW191109_010717}{-0.40}{GW200209_085452}{-0.14}{GW200115_042309}{0.81}{GW191127_050227}{0.06}{GW200216_220804}{0.57}{GW191215_223052}{0.36}{GW200208_130117}{-0.81}{GW200219_094415}{0.33}{GW191103_012549}{0.20}{GW200316_215756}{-0.68}{GW200202_154313}{-0.84}{GW200129_065458}{0.80}{GW191216_213338}{-0.80}}[{\red{???}}]}
\DeclareRobustCommand{\cosiotagwtcthreeplus}[1]{\IfEqCase{#1}{{GW200224_222234}{0.15}{GW191129_134029}{1.16}{GW200311_115853}{0.13}{GW191230_180458}{1.53}{GW191222_033537}{1.00}{GW200225_060421}{0.74}{GW200302_015811}{0.66}{GW200128_022011}{0.80}{GW191204_171526}{1.62}{GW200112_155838}{0.36}{GW200105_162426}{0.91}{GW191105_143521}{0.50}{GW191109_010717}{1.24}{GW200209_085452}{1.11}{GW200115_042309}{0.18}{GW191127_050227}{0.88}{GW200216_220804}{0.40}{GW191215_223052}{0.60}{GW200208_130117}{0.49}{GW200219_094415}{0.63}{GW191103_012549}{0.78}{GW200316_215756}{1.57}{GW200202_154313}{0.45}{GW200129_065458}{0.19}{GW191216_213338}{0.68}}[{\red{???}}]}
\DeclareRobustCommand{\cosiotagwtcthreetenthpercentile}[1]{\IfEqCase{#1}{{GW200224_222234}{0.52}{GW191129_134029}{-0.94}{GW200311_115853}{0.57}{GW191230_180458}{-0.95}{GW191222_033537}{-0.91}{GW200225_060421}{-0.87}{GW200302_015811}{-0.89}{GW200128_022011}{-0.92}{GW191204_171526}{-0.96}{GW200112_155838}{-0.94}{GW200105_162426}{-0.89}{GW191105_143521}{-0.93}{GW191109_010717}{-0.90}{GW200209_085452}{-0.93}{GW200115_042309}{-0.52}{GW191127_050227}{-0.87}{GW200216_220804}{-0.82}{GW191215_223052}{-0.85}{GW200208_130117}{-0.97}{GW200219_094415}{-0.85}{GW191103_012549}{-0.94}{GW200316_215756}{-0.95}{GW200202_154313}{-0.98}{GW200129_065458}{0.43}{GW191216_213338}{-0.97}}[{\red{???}}]}
\DeclareRobustCommand{\cosiotagwtcthreenintiethpercentile}[1]{\IfEqCase{#1}{{GW200224_222234}{0.98}{GW191129_134029}{0.94}{GW200311_115853}{0.98}{GW191230_180458}{0.93}{GW191222_033537}{0.90}{GW200225_060421}{0.90}{GW200302_015811}{0.92}{GW200128_022011}{0.92}{GW191204_171526}{0.94}{GW200112_155838}{0.96}{GW200105_162426}{0.89}{GW191105_143521}{0.95}{GW191109_010717}{0.68}{GW200209_085452}{0.93}{GW200115_042309}{0.97}{GW191127_050227}{0.88}{GW200216_220804}{0.94}{GW191215_223052}{0.91}{GW200208_130117}{-0.47}{GW200219_094415}{0.92}{GW191103_012549}{0.94}{GW200316_215756}{0.75}{GW200202_154313}{-0.50}{GW200129_065458}{0.97}{GW191216_213338}{-0.41}}[{\red{???}}]}
\DeclareRobustCommand{\radiatedenergygwtcthreeminus}[1]{\IfEqCase{#1}{{GW200224_222234}{0.67}{GW191129_134029}{0.116}{GW200311_115853}{0.57}{GW191230_180458}{1.0}{GW191222_033537}{1.00}{GW200225_060421}{0.39}{GW200302_015811}{0.79}{GW200128_022011}{0.96}{GW191204_171526}{0.106}{GW200112_155838}{0.53}{GW200105_162426}{0.026}{GW191105_143521}{0.118}{GW191109_010717}{1.3}{GW200209_085452}{0.70}{GW200115_042309}{0.029}{GW191127_050227}{2.1}{GW200216_220804}{2.0}{GW191215_223052}{0.32}{GW200208_130117}{0.77}{GW200219_094415}{0.81}{GW191103_012549}{0.17}{GW200316_215756}{0.22}{GW200202_154313}{0.094}{GW200129_065458}{0.86}{GW191216_213338}{0.130}}[{\red{???}}]}
\DeclareRobustCommand{\radiatedenergygwtcthreemed}[1]{\IfEqCase{#1}{{GW200224_222234}{3.61}{GW191129_134029}{0.778}{GW200311_115853}{2.87}{GW191230_180458}{3.8}{GW191222_033537}{3.57}{GW200225_060421}{1.43}{GW200302_015811}{2.35}{GW200128_022011}{3.61}{GW191204_171526}{0.991}{GW200112_155838}{3.08}{GW200105_162426}{0.204}{GW191105_143521}{0.818}{GW191109_010717}{4.3}{GW200209_085452}{2.68}{GW200115_042309}{0.146}{GW191127_050227}{3.0}{GW200216_220804}{3.5}{GW191215_223052}{1.88}{GW200208_130117}{2.83}{GW200219_094415}{2.84}{GW191103_012549}{0.98}{GW200316_215756}{0.94}{GW200202_154313}{0.814}{GW200129_065458}{3.18}{GW191216_213338}{0.921}}[{\red{???}}]}
\DeclareRobustCommand{\radiatedenergygwtcthreeplus}[1]{\IfEqCase{#1}{{GW200224_222234}{0.67}{GW191129_134029}{0.072}{GW200311_115853}{0.52}{GW191230_180458}{1.2}{GW191222_033537}{1.04}{GW200225_060421}{0.27}{GW200302_015811}{1.10}{GW200128_022011}{1.26}{GW191204_171526}{0.069}{GW200112_155838}{0.59}{GW200105_162426}{0.032}{GW191105_143521}{0.088}{GW191109_010717}{2.3}{GW200209_085452}{0.83}{GW200115_042309}{0.045}{GW191127_050227}{2.6}{GW200216_220804}{2.0}{GW191215_223052}{0.37}{GW200208_130117}{0.77}{GW200219_094415}{0.91}{GW191103_012549}{0.13}{GW200316_215756}{0.11}{GW200202_154313}{0.047}{GW200129_065458}{0.42}{GW191216_213338}{0.057}}[{\red{???}}]}
\DeclareRobustCommand{\radiatedenergygwtcthreetenthpercentile}[1]{\IfEqCase{#1}{{GW200224_222234}{3.09}{GW191129_134029}{0.690}{GW200311_115853}{2.44}{GW191230_180458}{3.0}{GW191222_033537}{2.84}{GW200225_060421}{1.12}{GW200302_015811}{1.72}{GW200128_022011}{2.85}{GW191204_171526}{0.911}{GW200112_155838}{2.67}{GW200105_162426}{0.187}{GW191105_143521}{0.731}{GW191109_010717}{3.3}{GW200209_085452}{2.14}{GW200115_042309}{0.124}{GW191127_050227}{1.3}{GW200216_220804}{1.9}{GW191215_223052}{1.63}{GW200208_130117}{2.24}{GW200219_094415}{2.23}{GW191103_012549}{0.85}{GW200316_215756}{0.79}{GW200202_154313}{0.748}{GW200129_065458}{2.58}{GW191216_213338}{0.829}}[{\red{???}}]}
\DeclareRobustCommand{\radiatedenergygwtcthreenintiethpercentile}[1]{\IfEqCase{#1}{{GW200224_222234}{4.11}{GW191129_134029}{0.835}{GW200311_115853}{3.27}{GW191230_180458}{4.7}{GW191222_033537}{4.39}{GW200225_060421}{1.64}{GW200302_015811}{3.18}{GW200128_022011}{4.56}{GW191204_171526}{1.045}{GW200112_155838}{3.52}{GW200105_162426}{0.225}{GW191105_143521}{0.887}{GW191109_010717}{5.8}{GW200209_085452}{3.30}{GW200115_042309}{0.183}{GW191127_050227}{5.0}{GW200216_220804}{5.0}{GW191215_223052}{2.15}{GW200208_130117}{3.42}{GW200219_094415}{3.53}{GW191103_012549}{1.08}{GW200316_215756}{1.03}{GW200202_154313}{0.851}{GW200129_065458}{3.53}{GW191216_213338}{0.967}}[{\red{???}}]}
\DeclareRobustCommand{\costhetajngwtcthreeminus}[1]{\IfEqCase{#1}{{GW200224_222234}{0.40}{GW191129_134029}{0.78}{GW200311_115853}{0.37}{GW191230_180458}{0.45}{GW191222_033537}{0.91}{GW200225_060421}{1.19}{GW200302_015811}{1.25}{GW200128_022011}{1.19}{GW191204_171526}{0.32}{GW200112_155838}{1.61}{GW200105_162426}{0.98}{GW191105_143521}{1.45}{GW191109_010717}{0.60}{GW200209_085452}{0.75}{GW200115_042309}{1.65}{GW191127_050227}{1.07}{GW200216_220804}{1.56}{GW191215_223052}{1.37}{GW200208_130117}{0.17}{GW200219_094415}{1.30}{GW191103_012549}{1.16}{GW200316_215756}{0.29}{GW200202_154313}{0.15}{GW200129_065458}{0.47}{GW191216_213338}{0.18}}[{\red{???}}]}
\DeclareRobustCommand{\costhetajngwtcthreemed}[1]{\IfEqCase{#1}{{GW200224_222234}{0.84}{GW191129_134029}{-0.19}{GW200311_115853}{0.85}{GW191230_180458}{-0.53}{GW191222_033537}{-0.05}{GW200225_060421}{0.25}{GW200302_015811}{0.30}{GW200128_022011}{0.22}{GW191204_171526}{-0.66}{GW200112_155838}{0.64}{GW200105_162426}{0.03}{GW191105_143521}{0.48}{GW191109_010717}{-0.33}{GW200209_085452}{-0.22}{GW200115_042309}{0.81}{GW191127_050227}{0.11}{GW200216_220804}{0.63}{GW191215_223052}{0.44}{GW200208_130117}{-0.82}{GW200219_094415}{0.37}{GW191103_012549}{0.19}{GW200316_215756}{-0.68}{GW200202_154313}{-0.84}{GW200129_065458}{0.79}{GW191216_213338}{-0.80}}[{\red{???}}]}
\DeclareRobustCommand{\costhetajngwtcthreeplus}[1]{\IfEqCase{#1}{{GW200224_222234}{0.15}{GW191129_134029}{1.16}{GW200311_115853}{0.14}{GW191230_180458}{1.49}{GW191222_033537}{1.01}{GW200225_060421}{0.70}{GW200302_015811}{0.66}{GW200128_022011}{0.75}{GW191204_171526}{1.63}{GW200112_155838}{0.34}{GW200105_162426}{0.92}{GW191105_143521}{0.50}{GW191109_010717}{1.07}{GW200209_085452}{1.19}{GW200115_042309}{0.17}{GW191127_050227}{0.85}{GW200216_220804}{0.35}{GW191215_223052}{0.52}{GW200208_130117}{0.44}{GW200219_094415}{0.59}{GW191103_012549}{0.78}{GW200316_215756}{1.57}{GW200202_154313}{0.45}{GW200129_065458}{0.18}{GW191216_213338}{0.68}}[{\red{???}}]}
\DeclareRobustCommand{\costhetajngwtcthreetenthpercentile}[1]{\IfEqCase{#1}{{GW200224_222234}{0.54}{GW191129_134029}{-0.94}{GW200311_115853}{0.57}{GW191230_180458}{-0.95}{GW191222_033537}{-0.92}{GW200225_060421}{-0.88}{GW200302_015811}{-0.89}{GW200128_022011}{-0.94}{GW191204_171526}{-0.96}{GW200112_155838}{-0.95}{GW200105_162426}{-0.90}{GW191105_143521}{-0.93}{GW191109_010717}{-0.87}{GW200209_085452}{-0.94}{GW200115_042309}{-0.52}{GW191127_050227}{-0.92}{GW200216_220804}{-0.84}{GW191215_223052}{-0.86}{GW200208_130117}{-0.97}{GW200219_094415}{-0.85}{GW191103_012549}{-0.94}{GW200316_215756}{-0.94}{GW200202_154313}{-0.98}{GW200129_065458}{0.45}{GW191216_213338}{-0.97}}[{\red{???}}]}
\DeclareRobustCommand{\costhetajngwtcthreenintiethpercentile}[1]{\IfEqCase{#1}{{GW200224_222234}{0.98}{GW191129_134029}{0.94}{GW200311_115853}{0.98}{GW191230_180458}{0.92}{GW191222_033537}{0.91}{GW200225_060421}{0.90}{GW200302_015811}{0.92}{GW200128_022011}{0.94}{GW191204_171526}{0.94}{GW200112_155838}{0.96}{GW200105_162426}{0.90}{GW191105_143521}{0.95}{GW191109_010717}{0.52}{GW200209_085452}{0.93}{GW200115_042309}{0.97}{GW191127_050227}{0.91}{GW200216_220804}{0.95}{GW191215_223052}{0.91}{GW200208_130117}{-0.50}{GW200219_094415}{0.92}{GW191103_012549}{0.94}{GW200316_215756}{0.76}{GW200202_154313}{-0.50}{GW200129_065458}{0.95}{GW191216_213338}{-0.41}}[{\red{???}}]}
\DeclareRobustCommand{\totalmasssourcegwtcthreeminus}[1]{\IfEqCase{#1}{{GW200224_222234}{5.0}{GW191129_134029}{1.1}{GW200311_115853}{4.2}{GW191230_180458}{11}{GW191222_033537}{11}{GW200225_060421}{3.0}{GW200302_015811}{6.7}{GW200128_022011}{11}{GW191204_171526}{0.93}{GW200112_155838}{4.6}{GW200105_162426}{1.4}{GW191105_143521}{1.3}{GW191109_010717}{16}{GW200209_085452}{8.6}{GW200115_042309}{1.6}{GW191127_050227}{22}{GW200216_220804}{14}{GW191215_223052}{4.0}{GW200208_130117}{6.8}{GW200219_094415}{8.2}{GW191103_012549}{1.8}{GW200316_215756}{2.0}{GW200202_154313}{0.67}{GW200129_065458}{3.6}{GW191216_213338}{0.93}}[{\red{???}}]}
\DeclareRobustCommand{\totalmasssourcegwtcthreemed}[1]{\IfEqCase{#1}{{GW200224_222234}{71.9}{GW191129_134029}{17.5}{GW200311_115853}{61.9}{GW191230_180458}{83}{GW191222_033537}{79}{GW200225_060421}{33.5}{GW200302_015811}{57.3}{GW200128_022011}{72}{GW191204_171526}{20.14}{GW200112_155838}{63.9}{GW200105_162426}{11.0}{GW191105_143521}{18.5}{GW191109_010717}{112}{GW200209_085452}{61.1}{GW200115_042309}{7.3}{GW191127_050227}{80}{GW200216_220804}{81}{GW191215_223052}{42.6}{GW200208_130117}{65.4}{GW200219_094415}{65.0}{GW191103_012549}{20.0}{GW200316_215756}{21.2}{GW200202_154313}{17.58}{GW200129_065458}{63.4}{GW191216_213338}{19.80}}[{\red{???}}]}
\DeclareRobustCommand{\totalmasssourcegwtcthreeplus}[1]{\IfEqCase{#1}{{GW200224_222234}{6.8}{GW191129_134029}{2.4}{GW200311_115853}{5.3}{GW191230_180458}{17}{GW191222_033537}{16}{GW200225_060421}{3.6}{GW200302_015811}{9.6}{GW200128_022011}{15}{GW191204_171526}{1.70}{GW200112_155838}{5.7}{GW200105_162426}{1.5}{GW191105_143521}{2.1}{GW191109_010717}{20}{GW200209_085452}{12.8}{GW200115_042309}{1.7}{GW191127_050227}{39}{GW200216_220804}{20}{GW191215_223052}{5.4}{GW200208_130117}{7.8}{GW200219_094415}{12.6}{GW191103_012549}{3.7}{GW200316_215756}{7.2}{GW200202_154313}{1.78}{GW200129_065458}{4.3}{GW191216_213338}{2.72}}[{\red{???}}]}
\DeclareRobustCommand{\totalmasssourcegwtcthreetenthpercentile}[1]{\IfEqCase{#1}{{GW200224_222234}{67.9}{GW191129_134029}{16.5}{GW200311_115853}{58.5}{GW191230_180458}{75}{GW191222_033537}{70}{GW200225_060421}{31.1}{GW200302_015811}{51.9}{GW200128_022011}{63}{GW191204_171526}{19.37}{GW200112_155838}{60.1}{GW200105_162426}{10.1}{GW191105_143521}{17.4}{GW191109_010717}{99}{GW200209_085452}{54.2}{GW200115_042309}{5.8}{GW191127_050227}{61}{GW200216_220804}{70}{GW191215_223052}{39.4}{GW200208_130117}{60.0}{GW200219_094415}{58.3}{GW191103_012549}{18.5}{GW200316_215756}{19.5}{GW200202_154313}{17.04}{GW200129_065458}{60.5}{GW191216_213338}{19.02}}[{\red{???}}]}
\DeclareRobustCommand{\totalmasssourcegwtcthreenintiethpercentile}[1]{\IfEqCase{#1}{{GW200224_222234}{77.1}{GW191129_134029}{19.2}{GW200311_115853}{65.9}{GW191230_180458}{96}{GW191222_033537}{92}{GW200225_060421}{36.2}{GW200302_015811}{64.3}{GW200128_022011}{83}{GW191204_171526}{21.36}{GW200112_155838}{68.1}{GW200105_162426}{11.8}{GW191105_143521}{19.9}{GW191109_010717}{126}{GW200209_085452}{70.6}{GW200115_042309}{8.6}{GW191127_050227}{108}{GW200216_220804}{96}{GW191215_223052}{46.8}{GW200208_130117}{71.4}{GW200219_094415}{74.7}{GW191103_012549}{22.4}{GW200316_215756}{25.6}{GW200202_154313}{18.75}{GW200129_065458}{66.7}{GW191216_213338}{21.49}}[{\red{???}}]}
\DeclareRobustCommand{\phijlgwtcthreeminus}[1]{\IfEqCase{#1}{{GW200224_222234}{2.6}{GW191129_134029}{2.8}{GW200311_115853}{2.3}{GW191230_180458}{3.0}{GW191222_033537}{2.8}{GW200225_060421}{2.8}{GW200302_015811}{2.8}{GW200128_022011}{2.9}{GW191204_171526}{2.8}{GW200112_155838}{2.8}{GW200105_162426}{2.8}{GW191105_143521}{2.8}{GW191109_010717}{3.1}{GW200209_085452}{2.8}{GW200115_042309}{2.7}{GW191127_050227}{2.7}{GW200216_220804}{3.0}{GW191215_223052}{2.9}{GW200208_130117}{2.8}{GW200219_094415}{2.8}{GW191103_012549}{2.8}{GW200316_215756}{3.0}{GW200202_154313}{2.8}{GW200129_065458}{2.0}{GW191216_213338}{3.0}}[{\red{???}}]}
\DeclareRobustCommand{\phijlgwtcthreemed}[1]{\IfEqCase{#1}{{GW200224_222234}{3.0}{GW191129_134029}{3.1}{GW200311_115853}{2.7}{GW191230_180458}{3.4}{GW191222_033537}{3.1}{GW200225_060421}{3.1}{GW200302_015811}{3.1}{GW200128_022011}{3.2}{GW191204_171526}{3.2}{GW200112_155838}{3.1}{GW200105_162426}{3.2}{GW191105_143521}{3.1}{GW191109_010717}{3.5}{GW200209_085452}{3.1}{GW200115_042309}{3.0}{GW191127_050227}{3.1}{GW200216_220804}{3.4}{GW191215_223052}{3.2}{GW200208_130117}{3.2}{GW200219_094415}{3.1}{GW191103_012549}{3.1}{GW200316_215756}{3.3}{GW200202_154313}{3.1}{GW200129_065458}{2.7}{GW191216_213338}{3.3}}[{\red{???}}]}
\DeclareRobustCommand{\phijlgwtcthreeplus}[1]{\IfEqCase{#1}{{GW200224_222234}{2.8}{GW191129_134029}{2.8}{GW200311_115853}{3.3}{GW191230_180458}{2.6}{GW191222_033537}{2.8}{GW200225_060421}{2.9}{GW200302_015811}{2.8}{GW200128_022011}{2.8}{GW191204_171526}{2.8}{GW200112_155838}{2.9}{GW200105_162426}{2.8}{GW191105_143521}{2.8}{GW191109_010717}{2.4}{GW200209_085452}{2.8}{GW200115_042309}{3.0}{GW191127_050227}{2.8}{GW200216_220804}{2.5}{GW191215_223052}{2.8}{GW200208_130117}{2.7}{GW200219_094415}{2.8}{GW191103_012549}{2.9}{GW200316_215756}{2.6}{GW200202_154313}{2.9}{GW200129_065458}{3.1}{GW191216_213338}{2.7}}[{\red{???}}]}
\DeclareRobustCommand{\phijlgwtcthreetenthpercentile}[1]{\IfEqCase{#1}{{GW200224_222234}{0.8}{GW191129_134029}{0.6}{GW200311_115853}{0.6}{GW191230_180458}{0.7}{GW191222_033537}{0.7}{GW200225_060421}{0.6}{GW200302_015811}{0.6}{GW200128_022011}{0.6}{GW191204_171526}{0.6}{GW200112_155838}{0.6}{GW200105_162426}{0.7}{GW191105_143521}{0.6}{GW191109_010717}{0.8}{GW200209_085452}{0.7}{GW200115_042309}{0.7}{GW191127_050227}{0.7}{GW200216_220804}{0.8}{GW191215_223052}{0.6}{GW200208_130117}{0.7}{GW200219_094415}{0.8}{GW191103_012549}{0.7}{GW200316_215756}{0.7}{GW200202_154313}{0.6}{GW200129_065458}{1.0}{GW191216_213338}{0.6}}[{\red{???}}]}
\DeclareRobustCommand{\phijlgwtcthreenintiethpercentile}[1]{\IfEqCase{#1}{{GW200224_222234}{5.1}{GW191129_134029}{5.6}{GW200311_115853}{5.6}{GW191230_180458}{5.7}{GW191222_033537}{5.7}{GW200225_060421}{5.7}{GW200302_015811}{5.6}{GW200128_022011}{5.7}{GW191204_171526}{5.7}{GW200112_155838}{5.7}{GW200105_162426}{5.7}{GW191105_143521}{5.6}{GW191109_010717}{5.5}{GW200209_085452}{5.6}{GW200115_042309}{5.6}{GW191127_050227}{5.6}{GW200216_220804}{5.7}{GW191215_223052}{5.7}{GW200208_130117}{5.6}{GW200219_094415}{5.6}{GW191103_012549}{5.7}{GW200316_215756}{5.6}{GW200202_154313}{5.7}{GW200129_065458}{5.4}{GW191216_213338}{5.7}}[{\red{???}}]}
\DeclareRobustCommand{\masstwodetgwtcthreeminus}[1]{\IfEqCase{#1}{{GW200224_222234}{9.7}{GW191129_134029}{1.9}{GW200311_115853}{7.3}{GW191230_180458}{20}{GW191222_033537}{15}{GW200225_060421}{4.6}{GW200302_015811}{8.1}{GW200128_022011}{13}{GW191204_171526}{1.8}{GW200112_155838}{7.4}{GW200105_162426}{0.25}{GW191105_143521}{2.2}{GW191109_010717}{17}{GW200209_085452}{13}{GW200115_042309}{0.30}{GW191127_050227}{25}{GW200216_220804}{31}{GW191215_223052}{5.1}{GW200208_130117}{11.2}{GW200219_094415}{14.1}{GW191103_012549}{2.9}{GW200316_215756}{3.5}{GW200202_154313}{1.9}{GW200129_065458}{10.8}{GW191216_213338}{2.0}}[{\red{???}}]}
\DeclareRobustCommand{\masstwodetgwtcthreemed}[1]{\IfEqCase{#1}{{GW200224_222234}{43.0}{GW191129_134029}{7.8}{GW200311_115853}{34.0}{GW191230_180458}{64}{GW191222_033537}{52}{GW200225_060421}{17.2}{GW200302_015811}{26.5}{GW200128_022011}{52}{GW191204_171526}{9.3}{GW200112_155838}{35.2}{GW200105_162426}{2.02}{GW191105_143521}{9.4}{GW191109_010717}{60}{GW200209_085452}{44}{GW200115_042309}{1.53}{GW191127_050227}{38}{GW200216_220804}{51}{GW191215_223052}{24.5}{GW200208_130117}{38.5}{GW200219_094415}{44.4}{GW191103_012549}{9.4}{GW200316_215756}{9.5}{GW200202_154313}{8.0}{GW200129_065458}{34.1}{GW191216_213338}{8.2}}[{\red{???}}]}
\DeclareRobustCommand{\masstwodetgwtcthreeplus}[1]{\IfEqCase{#1}{{GW200224_222234}{5.8}{GW191129_134029}{1.7}{GW200311_115853}{4.7}{GW191230_180458}{14}{GW191222_033537}{11}{GW200225_060421}{3.0}{GW200302_015811}{12.5}{GW200128_022011}{10}{GW191204_171526}{1.5}{GW200112_155838}{5.1}{GW200105_162426}{0.35}{GW191105_143521}{1.4}{GW191109_010717}{16}{GW200209_085452}{11}{GW200115_042309}{0.91}{GW191127_050227}{31}{GW200216_220804}{23}{GW191215_223052}{4.1}{GW200208_130117}{8.6}{GW200219_094415}{9.3}{GW191103_012549}{1.8}{GW200316_215756}{2.3}{GW200202_154313}{1.2}{GW200129_065458}{3.3}{GW191216_213338}{1.7}}[{\red{???}}]}
\DeclareRobustCommand{\masstwodetgwtcthreetenthpercentile}[1]{\IfEqCase{#1}{{GW200224_222234}{35.7}{GW191129_134029}{6.2}{GW200311_115853}{28.6}{GW191230_180458}{48}{GW191222_033537}{41}{GW200225_060421}{13.7}{GW200302_015811}{19.8}{GW200128_022011}{41}{GW191204_171526}{7.8}{GW200112_155838}{29.5}{GW200105_162426}{1.87}{GW191105_143521}{7.6}{GW191109_010717}{46}{GW200209_085452}{34}{GW200115_042309}{1.29}{GW191127_050227}{17}{GW200216_220804}{24}{GW191215_223052}{20.5}{GW200208_130117}{29.6}{GW200219_094415}{33.7}{GW191103_012549}{7.2}{GW200316_215756}{6.8}{GW200202_154313}{6.5}{GW200129_065458}{25.3}{GW191216_213338}{6.7}}[{\red{???}}]}
\DeclareRobustCommand{\masstwodetgwtcthreenintiethpercentile}[1]{\IfEqCase{#1}{{GW200224_222234}{47.8}{GW191129_134029}{9.3}{GW200311_115853}{37.7}{GW191230_180458}{75}{GW191222_033537}{61}{GW200225_060421}{19.8}{GW200302_015811}{36.3}{GW200128_022011}{60}{GW191204_171526}{10.6}{GW200112_155838}{39.3}{GW200105_162426}{2.20}{GW191105_143521}{10.6}{GW191109_010717}{72}{GW200209_085452}{52}{GW200115_042309}{2.25}{GW191127_050227}{63}{GW200216_220804}{70}{GW191215_223052}{28.0}{GW200208_130117}{45.5}{GW200219_094415}{51.8}{GW191103_012549}{11.0}{GW200316_215756}{11.5}{GW200202_154313}{9.1}{GW200129_065458}{36.8}{GW191216_213338}{9.8}}[{\red{???}}]}
\DeclareRobustCommand{\ragwtcthreeminus}[1]{\IfEqCase{#1}{{GW200224_222234}{0.048}{GW191129_134029}{2.77}{GW200311_115853}{0.029}{GW191230_180458}{0.29}{GW191222_033537}{3.0}{GW200225_060421}{0.30}{GW200302_015811}{3.13}{GW200128_022011}{3.0}{GW191204_171526}{0.53}{GW200112_155838}{2.9}{GW200105_162426}{1.3}{GW191105_143521}{0.36}{GW191109_010717}{1.4}{GW200209_085452}{0.92}{GW200115_042309}{0.10}{GW191127_050227}{1.1}{GW200216_220804}{1.49}{GW191215_223052}{0.76}{GW200208_130117}{0.039}{GW200219_094415}{0.12}{GW191103_012549}{1.85}{GW200316_215756}{0.35}{GW200202_154313}{0.117}{GW200129_065458}{0.15}{GW191216_213338}{3.518}}[{\red{???}}]}
\DeclareRobustCommand{\ragwtcthreemed}[1]{\IfEqCase{#1}{{GW200224_222234}{3.050}{GW191129_134029}{5.59}{GW200311_115853}{0.038}{GW191230_180458}{1.07}{GW191222_033537}{3.6}{GW200225_060421}{1.92}{GW200302_015811}{3.83}{GW200128_022011}{3.8}{GW191204_171526}{1.27}{GW200112_155838}{3.5}{GW200105_162426}{2.0}{GW191105_143521}{0.44}{GW191109_010717}{3.7}{GW200209_085452}{2.51}{GW200115_042309}{0.74}{GW191127_050227}{1.2}{GW200216_220804}{5.31}{GW191215_223052}{2.63}{GW200208_130117}{2.438}{GW200219_094415}{0.39}{GW191103_012549}{4.35}{GW200316_215756}{1.51}{GW200202_154313}{2.523}{GW200129_065458}{5.56}{GW191216_213338}{5.557}}[{\red{???}}]}
\DeclareRobustCommand{\ragwtcthreeplus}[1]{\IfEqCase{#1}{{GW200224_222234}{0.041}{GW191129_134029}{0.42}{GW200311_115853}{0.041}{GW191230_180458}{3.73}{GW191222_033537}{1.8}{GW200225_060421}{3.25}{GW200302_015811}{0.96}{GW200128_022011}{1.1}{GW191204_171526}{1.91}{GW200112_155838}{1.7}{GW200105_162426}{3.0}{GW191105_143521}{5.68}{GW191109_010717}{1.2}{GW200209_085452}{0.63}{GW200115_042309}{4.06}{GW191127_050227}{4.9}{GW200216_220804}{0.28}{GW191215_223052}{3.28}{GW200208_130117}{0.040}{GW200219_094415}{2.85}{GW191103_012549}{0.39}{GW200316_215756}{2.01}{GW200202_154313}{0.058}{GW200129_065458}{0.47}{GW191216_213338}{0.066}}[{\red{???}}]}
\DeclareRobustCommand{\ragwtcthreetenthpercentile}[1]{\IfEqCase{#1}{{GW200224_222234}{3.015}{GW191129_134029}{3.01}{GW200311_115853}{0.014}{GW191230_180458}{0.84}{GW191222_033537}{1.1}{GW200225_060421}{1.71}{GW200302_015811}{0.89}{GW200128_022011}{0.9}{GW191204_171526}{0.82}{GW200112_155838}{0.6}{GW200105_162426}{0.8}{GW191105_143521}{0.12}{GW191109_010717}{2.4}{GW200209_085452}{2.22}{GW200115_042309}{0.66}{GW191127_050227}{0.1}{GW200216_220804}{3.96}{GW191215_223052}{1.93}{GW200208_130117}{2.409}{GW200219_094415}{0.30}{GW191103_012549}{2.56}{GW200316_215756}{1.21}{GW200202_154313}{2.425}{GW200129_065458}{5.45}{GW191216_213338}{5.260}}[{\red{???}}]}
\DeclareRobustCommand{\ragwtcthreenintiethpercentile}[1]{\IfEqCase{#1}{{GW200224_222234}{3.082}{GW191129_134029}{5.92}{GW200311_115853}{0.068}{GW191230_180458}{4.65}{GW191222_033537}{4.7}{GW200225_060421}{2.42}{GW200302_015811}{4.59}{GW200128_022011}{4.5}{GW191204_171526}{2.05}{GW200112_155838}{5.1}{GW200105_162426}{4.9}{GW191105_143521}{6.07}{GW191109_010717}{4.6}{GW200209_085452}{2.91}{GW200115_042309}{4.65}{GW191127_050227}{5.0}{GW200216_220804}{5.54}{GW191215_223052}{5.77}{GW200208_130117}{2.468}{GW200219_094415}{3.06}{GW191103_012549}{4.71}{GW200316_215756}{3.42}{GW200202_154313}{2.569}{GW200129_065458}{5.59}{GW191216_213338}{5.615}}[{\red{???}}]}
\DeclareRobustCommand{\finalmassdetgwtcthreeminus}[1]{\IfEqCase{#1}{{GW200224_222234}{6.4}{GW191129_134029}{0.67}{GW200311_115853}{5.1}{GW191230_180458}{17}{GW191222_033537}{12}{GW200225_060421}{3.6}{GW200302_015811}{7.6}{GW200128_022011}{12}{GW191204_171526}{0.50}{GW200112_155838}{4.6}{GW200105_162426}{1.8}{GW191105_143521}{0.47}{GW191109_010717}{15}{GW200209_085452}{14}{GW200115_042309}{1.7}{GW191127_050227}{43}{GW200216_220804}{30}{GW191215_223052}{3.3}{GW200208_130117}{9.1}{GW200219_094415}{11}{GW191103_012549}{0.66}{GW200316_215756}{1.1}{GW200202_154313}{0.35}{GW200129_065458}{3.4}{GW191216_213338}{0.70}}[{\red{???}}]}
\DeclareRobustCommand{\finalmassdetgwtcthreemed}[1]{\IfEqCase{#1}{{GW200224_222234}{90.5}{GW191129_134029}{19.20}{GW200311_115853}{72.4}{GW191230_180458}{140}{GW191222_033537}{114}{GW200225_060421}{39.4}{GW200302_015811}{71.6}{GW200128_022011}{112}{GW191204_171526}{21.60}{GW200112_155838}{75.3}{GW200105_162426}{11.4}{GW191105_143521}{21.36}{GW191109_010717}{135}{GW200209_085452}{96}{GW200115_042309}{7.6}{GW191127_050227}{124}{GW200216_220804}{129}{GW191215_223052}{55.9}{GW200208_130117}{87.5}{GW200219_094415}{98}{GW191103_012549}{22.27}{GW200316_215756}{24.4}{GW200202_154313}{18.12}{GW200129_065458}{70.9}{GW191216_213338}{20.18}}[{\red{???}}]}
\DeclareRobustCommand{\finalmassdetgwtcthreeplus}[1]{\IfEqCase{#1}{{GW200224_222234}{7.6}{GW191129_134029}{3.08}{GW200311_115853}{5.6}{GW191230_180458}{20}{GW191222_033537}{14}{GW200225_060421}{2.9}{GW200302_015811}{14.1}{GW200128_022011}{16}{GW191204_171526}{2.04}{GW200112_155838}{5.8}{GW200105_162426}{2.1}{GW191105_143521}{2.48}{GW191109_010717}{19}{GW200209_085452}{19}{GW200115_042309}{2.3}{GW191127_050227}{52}{GW200216_220804}{27}{GW191215_223052}{5.0}{GW200208_130117}{10.3}{GW200219_094415}{13}{GW191103_012549}{4.79}{GW200316_215756}{9.0}{GW200202_154313}{2.09}{GW200129_065458}{4.2}{GW191216_213338}{3.10}}[{\red{???}}]}
\DeclareRobustCommand{\finalmassdetgwtcthreetenthpercentile}[1]{\IfEqCase{#1}{{GW200224_222234}{85.4}{GW191129_134029}{18.57}{GW200311_115853}{68.4}{GW191230_180458}{127}{GW191222_033537}{105}{GW200225_060421}{36.5}{GW200302_015811}{65.5}{GW200128_022011}{102}{GW191204_171526}{21.14}{GW200112_155838}{71.8}{GW200105_162426}{10.2}{GW191105_143521}{20.95}{GW191109_010717}{123}{GW200209_085452}{84}{GW200115_042309}{6.1}{GW191127_050227}{88}{GW200216_220804}{106}{GW191215_223052}{53.3}{GW200208_130117}{80.3}{GW200219_094415}{89}{GW191103_012549}{21.68}{GW200316_215756}{23.3}{GW200202_154313}{17.80}{GW200129_065458}{68.3}{GW191216_213338}{19.53}}[{\red{???}}]}
\DeclareRobustCommand{\finalmassdetgwtcthreenintiethpercentile}[1]{\IfEqCase{#1}{{GW200224_222234}{96.2}{GW191129_134029}{21.37}{GW200311_115853}{76.5}{GW191230_180458}{155}{GW191222_033537}{125}{GW200225_060421}{41.6}{GW200302_015811}{82.3}{GW200128_022011}{124}{GW191204_171526}{23.02}{GW200112_155838}{79.7}{GW200105_162426}{12.7}{GW191105_143521}{22.99}{GW191109_010717}{149}{GW200209_085452}{110}{GW200115_042309}{9.3}{GW191127_050227}{164}{GW200216_220804}{150}{GW191215_223052}{59.4}{GW200208_130117}{95.3}{GW200219_094415}{108}{GW191103_012549}{25.37}{GW200316_215756}{30.0}{GW200202_154313}{19.51}{GW200129_065458}{74.2}{GW191216_213338}{22.11}}[{\red{???}}]}
\DeclareRobustCommand{\spinonexgwtcthreeminus}[1]{\IfEqCase{#1}{{GW200224_222234}{0.59}{GW191129_134029}{0.36}{GW200311_115853}{0.55}{GW191230_180458}{0.64}{GW191222_033537}{0.53}{GW200225_060421}{0.65}{GW200302_015811}{0.55}{GW200128_022011}{0.68}{GW191204_171526}{0.50}{GW200112_155838}{0.51}{GW200105_162426}{0.14}{GW191105_143521}{0.41}{GW191109_010717}{0.70}{GW200209_085452}{0.64}{GW200115_042309}{0.33}{GW191127_050227}{0.70}{GW200216_220804}{0.60}{GW191215_223052}{0.64}{GW200208_130117}{0.47}{GW200219_094415}{0.61}{GW191103_012549}{0.53}{GW200316_215756}{0.40}{GW200202_154313}{0.37}{GW200129_065458}{0.81}{GW191216_213338}{0.29}}[{\red{???}}]}
\DeclareRobustCommand{\spinonexgwtcthreemed}[1]{\IfEqCase{#1}{{GW200224_222234}{0.01}{GW191129_134029}{0.00}{GW200311_115853}{0.00}{GW191230_180458}{0.00}{GW191222_033537}{0.00}{GW200225_060421}{0.00}{GW200302_015811}{0.00}{GW200128_022011}{0.00}{GW191204_171526}{0.00}{GW200112_155838}{0.00}{GW200105_162426}{0.00}{GW191105_143521}{0.00}{GW191109_010717}{0.00}{GW200209_085452}{0.00}{GW200115_042309}{0.00}{GW191127_050227}{0.00}{GW200216_220804}{0.00}{GW191215_223052}{0.00}{GW200208_130117}{0.00}{GW200219_094415}{0.00}{GW191103_012549}{0.00}{GW200316_215756}{0.00}{GW200202_154313}{0.00}{GW200129_065458}{-0.02}{GW191216_213338}{0.00}}[{\red{???}}]}
\DeclareRobustCommand{\spinonexgwtcthreeplus}[1]{\IfEqCase{#1}{{GW200224_222234}{0.61}{GW191129_134029}{0.37}{GW200311_115853}{0.54}{GW191230_180458}{0.65}{GW191222_033537}{0.54}{GW200225_060421}{0.65}{GW200302_015811}{0.55}{GW200128_022011}{0.67}{GW191204_171526}{0.48}{GW200112_155838}{0.48}{GW200105_162426}{0.13}{GW191105_143521}{0.42}{GW191109_010717}{0.68}{GW200209_085452}{0.65}{GW200115_042309}{0.33}{GW191127_050227}{0.67}{GW200216_220804}{0.59}{GW191215_223052}{0.63}{GW200208_130117}{0.52}{GW200219_094415}{0.61}{GW191103_012549}{0.51}{GW200316_215756}{0.38}{GW200202_154313}{0.37}{GW200129_065458}{0.65}{GW191216_213338}{0.29}}[{\red{???}}]}
\DeclareRobustCommand{\spinonexgwtcthreetenthpercentile}[1]{\IfEqCase{#1}{{GW200224_222234}{-0.43}{GW191129_134029}{-0.25}{GW200311_115853}{-0.42}{GW191230_180458}{-0.49}{GW191222_033537}{-0.39}{GW200225_060421}{-0.52}{GW200302_015811}{-0.41}{GW200128_022011}{-0.53}{GW191204_171526}{-0.39}{GW200112_155838}{-0.37}{GW200105_162426}{-0.09}{GW191105_143521}{-0.27}{GW191109_010717}{-0.56}{GW200209_085452}{-0.48}{GW200115_042309}{-0.24}{GW191127_050227}{-0.54}{GW200216_220804}{-0.46}{GW191215_223052}{-0.50}{GW200208_130117}{-0.34}{GW200219_094415}{-0.45}{GW191103_012549}{-0.39}{GW200316_215756}{-0.27}{GW200202_154313}{-0.25}{GW200129_065458}{-0.72}{GW191216_213338}{-0.21}}[{\red{???}}]}
\DeclareRobustCommand{\spinonexgwtcthreenintiethpercentile}[1]{\IfEqCase{#1}{{GW200224_222234}{0.49}{GW191129_134029}{0.26}{GW200311_115853}{0.38}{GW191230_180458}{0.49}{GW191222_033537}{0.38}{GW200225_060421}{0.52}{GW200302_015811}{0.40}{GW200128_022011}{0.52}{GW191204_171526}{0.36}{GW200112_155838}{0.35}{GW200105_162426}{0.08}{GW191105_143521}{0.29}{GW191109_010717}{0.57}{GW200209_085452}{0.49}{GW200115_042309}{0.25}{GW191127_050227}{0.52}{GW200216_220804}{0.44}{GW191215_223052}{0.48}{GW200208_130117}{0.37}{GW200219_094415}{0.45}{GW191103_012549}{0.38}{GW200316_215756}{0.27}{GW200202_154313}{0.24}{GW200129_065458}{0.46}{GW191216_213338}{0.20}}[{\red{???}}]}
\DeclareRobustCommand{\loglikelihoodgwtcthreeminus}[1]{\IfEqCase{#1}{{GW200224_222234}{5.0}{GW191129_134029}{5.3}{GW200311_115853}{4.7}{GW191230_180458}{4.4}{GW191222_033537}{4.3}{GW200225_060421}{5.0}{GW200302_015811}{4.6}{GW200128_022011}{4.5}{GW191204_171526}{5.1}{GW200112_155838}{4.9}{GW200105_162426}{5.5}{GW191105_143521}{5.9}{GW191109_010717}{6.7}{GW200209_085452}{4.7}{GW200115_042309}{6.4}{GW191127_050227}{5.3}{GW200216_220804}{4.2}{GW191215_223052}{5.0}{GW200208_130117}{5.2}{GW200219_094415}{5.1}{GW191103_012549}{5.1}{GW200316_215756}{6.6}{GW200202_154313}{5.7}{GW200129_065458}{21.5}{GW191216_213338}{6.7}}[{\red{???}}]}
\DeclareRobustCommand{\loglikelihoodgwtcthreemed}[1]{\IfEqCase{#1}{{GW200224_222234}{188.2}{GW191129_134029}{73.6}{GW200311_115853}{148.0}{GW191230_180458}{46.2}{GW191222_033537}{68.9}{GW200225_060421}{66.4}{GW200302_015811}{47.0}{GW200128_022011}{48.7}{GW191204_171526}{138.2}{GW200112_155838}{183.8}{GW200105_162426}{82.7}{GW191105_143521}{34.7}{GW191109_010717}{138.1}{GW200209_085452}{37.1}{GW200115_042309}{47.8}{GW191127_050227}{32.1}{GW200216_220804}{25.0}{GW191215_223052}{51.9}{GW200208_130117}{49.4}{GW200219_094415}{48.6}{GW191103_012549}{27.5}{GW200316_215756}{40.3}{GW200202_154313}{44.9}{GW200129_065458}{343.3}{GW191216_213338}{156.6}}[{\red{???}}]}
\DeclareRobustCommand{\loglikelihoodgwtcthreeplus}[1]{\IfEqCase{#1}{{GW200224_222234}{3.3}{GW191129_134029}{5.1}{GW200311_115853}{4.6}{GW191230_180458}{2.8}{GW191222_033537}{2.8}{GW200225_060421}{3.8}{GW200302_015811}{4.0}{GW200128_022011}{3.5}{GW191204_171526}{4.4}{GW200112_155838}{3.4}{GW200105_162426}{4.0}{GW191105_143521}{5.4}{GW191109_010717}{7.4}{GW200209_085452}{3.2}{GW200115_042309}{8.1}{GW191127_050227}{4.9}{GW200216_220804}{3.1}{GW191215_223052}{4.2}{GW200208_130117}{2.9}{GW200219_094415}{3.4}{GW191103_012549}{4.8}{GW200316_215756}{4.8}{GW200202_154313}{6.9}{GW200129_065458}{6.3}{GW191216_213338}{8.0}}[{\red{???}}]}
\DeclareRobustCommand{\loglikelihoodgwtcthreetenthpercentile}[1]{\IfEqCase{#1}{{GW200224_222234}{184.6}{GW191129_134029}{69.5}{GW200311_115853}{144.4}{GW191230_180458}{43.0}{GW191222_033537}{65.7}{GW200225_060421}{62.7}{GW200302_015811}{43.7}{GW200128_022011}{45.3}{GW191204_171526}{134.4}{GW200112_155838}{180.1}{GW200105_162426}{78.6}{GW191105_143521}{30.2}{GW191109_010717}{132.9}{GW200209_085452}{33.6}{GW200115_042309}{42.9}{GW191127_050227}{28.0}{GW200216_220804}{21.9}{GW191215_223052}{48.2}{GW200208_130117}{45.7}{GW200219_094415}{44.8}{GW191103_012549}{23.6}{GW200316_215756}{35.4}{GW200202_154313}{40.4}{GW200129_065458}{330.2}{GW191216_213338}{151.5}}[{\red{???}}]}
\DeclareRobustCommand{\loglikelihoodgwtcthreenintiethpercentile}[1]{\IfEqCase{#1}{{GW200224_222234}{190.9}{GW191129_134029}{77.9}{GW200311_115853}{151.8}{GW191230_180458}{48.5}{GW191222_033537}{71.1}{GW200225_060421}{69.4}{GW200302_015811}{50.1}{GW200128_022011}{51.4}{GW191204_171526}{141.8}{GW200112_155838}{186.6}{GW200105_162426}{85.8}{GW191105_143521}{39.1}{GW191109_010717}{144.1}{GW200209_085452}{39.7}{GW200115_042309}{54.8}{GW191127_050227}{35.9}{GW200216_220804}{27.5}{GW191215_223052}{55.2}{GW200208_130117}{51.8}{GW200219_094415}{51.4}{GW191103_012549}{31.6}{GW200316_215756}{44.2}{GW200202_154313}{51.1}{GW200129_065458}{348.4}{GW191216_213338}{163.5}}[{\red{???}}]}
\DeclareRobustCommand{\tilttwogwtcthreeminus}[1]{\IfEqCase{#1}{{GW200224_222234}{0.92}{GW191129_134029}{0.89}{GW200311_115853}{1.0}{GW191230_180458}{1.1}{GW191222_033537}{1.1}{GW200225_060421}{1.18}{GW200302_015811}{1.0}{GW200128_022011}{0.95}{GW191204_171526}{0.77}{GW200112_155838}{0.91}{GW200105_162426}{1.0}{GW191105_143521}{1.07}{GW191109_010717}{1.16}{GW200209_085452}{1.17}{GW200115_042309}{1.27}{GW191127_050227}{0.98}{GW200216_220804}{1.0}{GW191215_223052}{1.09}{GW200208_130117}{1.17}{GW200219_094415}{1.1}{GW191103_012549}{0.77}{GW200316_215756}{0.83}{GW200202_154313}{0.90}{GW200129_065458}{0.84}{GW191216_213338}{0.83}}[{\red{???}}]}
\DeclareRobustCommand{\tilttwogwtcthreemed}[1]{\IfEqCase{#1}{{GW200224_222234}{1.38}{GW191129_134029}{1.27}{GW200311_115853}{1.6}{GW191230_180458}{1.7}{GW191222_033537}{1.7}{GW200225_060421}{1.79}{GW200302_015811}{1.4}{GW200128_022011}{1.42}{GW191204_171526}{1.11}{GW200112_155838}{1.32}{GW200105_162426}{1.5}{GW191105_143521}{1.60}{GW191109_010717}{1.84}{GW200209_085452}{1.84}{GW200115_042309}{1.89}{GW191127_050227}{1.35}{GW200216_220804}{1.4}{GW191215_223052}{1.66}{GW200208_130117}{1.72}{GW200219_094415}{1.7}{GW191103_012549}{1.07}{GW200316_215756}{1.19}{GW200202_154313}{1.34}{GW200129_065458}{1.15}{GW191216_213338}{1.16}}[{\red{???}}]}
\DeclareRobustCommand{\tilttwogwtcthreeplus}[1]{\IfEqCase{#1}{{GW200224_222234}{1.16}{GW191129_134029}{1.19}{GW200311_115853}{1.0}{GW191230_180458}{1.0}{GW191222_033537}{1.0}{GW200225_060421}{0.97}{GW200302_015811}{1.1}{GW200128_022011}{1.15}{GW191204_171526}{1.12}{GW200112_155838}{1.14}{GW200105_162426}{1.1}{GW191105_143521}{0.99}{GW191109_010717}{0.92}{GW200209_085452}{0.93}{GW200115_042309}{0.93}{GW191127_050227}{1.22}{GW200216_220804}{1.2}{GW191215_223052}{0.99}{GW200208_130117}{0.99}{GW200219_094415}{1.0}{GW191103_012549}{1.29}{GW200316_215756}{1.13}{GW200202_154313}{1.07}{GW200129_065458}{1.36}{GW191216_213338}{1.28}}[{\red{???}}]}
\DeclareRobustCommand{\tilttwogwtcthreetenthpercentile}[1]{\IfEqCase{#1}{{GW200224_222234}{0.63}{GW191129_134029}{0.52}{GW200311_115853}{0.8}{GW191230_180458}{0.8}{GW191222_033537}{0.8}{GW200225_060421}{0.87}{GW200302_015811}{0.6}{GW200128_022011}{0.66}{GW191204_171526}{0.47}{GW200112_155838}{0.60}{GW200105_162426}{0.7}{GW191105_143521}{0.74}{GW191109_010717}{0.91}{GW200209_085452}{0.92}{GW200115_042309}{0.86}{GW191127_050227}{0.53}{GW200216_220804}{0.6}{GW191215_223052}{0.79}{GW200208_130117}{0.79}{GW200219_094415}{0.8}{GW191103_012549}{0.43}{GW200316_215756}{0.51}{GW200202_154313}{0.62}{GW200129_065458}{0.45}{GW191216_213338}{0.47}}[{\red{???}}]}
\DeclareRobustCommand{\tilttwogwtcthreenintiethpercentile}[1]{\IfEqCase{#1}{{GW200224_222234}{2.30}{GW191129_134029}{2.17}{GW200311_115853}{2.4}{GW191230_180458}{2.5}{GW191222_033537}{2.5}{GW200225_060421}{2.59}{GW200302_015811}{2.4}{GW200128_022011}{2.34}{GW191204_171526}{1.96}{GW200112_155838}{2.23}{GW200105_162426}{2.4}{GW191105_143521}{2.39}{GW191109_010717}{2.59}{GW200209_085452}{2.60}{GW200115_042309}{2.69}{GW191127_050227}{2.34}{GW200216_220804}{2.4}{GW191215_223052}{2.46}{GW200208_130117}{2.54}{GW200219_094415}{2.6}{GW191103_012549}{2.09}{GW200316_215756}{2.05}{GW200202_154313}{2.16}{GW200129_065458}{2.23}{GW191216_213338}{2.15}}[{\red{???}}]}
\DeclareRobustCommand{\tiltonegwtcthreeminus}[1]{\IfEqCase{#1}{{GW200224_222234}{0.84}{GW191129_134029}{0.87}{GW200311_115853}{1.02}{GW191230_180458}{1.0}{GW191222_033537}{1.10}{GW200225_060421}{0.90}{GW200302_015811}{0.97}{GW200128_022011}{0.79}{GW191204_171526}{0.72}{GW200112_155838}{0.95}{GW200105_162426}{1.1}{GW191105_143521}{1.05}{GW191109_010717}{0.90}{GW200209_085452}{1.01}{GW200115_042309}{1.37}{GW191127_050227}{0.84}{GW200216_220804}{0.92}{GW191215_223052}{0.96}{GW200208_130117}{1.15}{GW200219_094415}{1.06}{GW191103_012549}{0.68}{GW200316_215756}{0.80}{GW200202_154313}{0.94}{GW200129_065458}{0.92}{GW191216_213338}{0.73}}[{\red{???}}]}
\DeclareRobustCommand{\tiltonegwtcthreemed}[1]{\IfEqCase{#1}{{GW200224_222234}{1.27}{GW191129_134029}{1.29}{GW200311_115853}{1.66}{GW191230_180458}{1.6}{GW191222_033537}{1.73}{GW200225_060421}{1.88}{GW200302_015811}{1.51}{GW200128_022011}{1.20}{GW191204_171526}{1.13}{GW200112_155838}{1.43}{GW200105_162426}{1.6}{GW191105_143521}{1.64}{GW191109_010717}{2.22}{GW200209_085452}{1.76}{GW200115_042309}{2.24}{GW191127_050227}{1.19}{GW200216_220804}{1.33}{GW191215_223052}{1.63}{GW200208_130117}{1.82}{GW200219_094415}{1.80}{GW191103_012549}{1.00}{GW200316_215756}{1.08}{GW200202_154313}{1.39}{GW200129_065458}{1.45}{GW191216_213338}{1.01}}[{\red{???}}]}
\DeclareRobustCommand{\tiltonegwtcthreeplus}[1]{\IfEqCase{#1}{{GW200224_222234}{1.03}{GW191129_134029}{1.01}{GW200311_115853}{0.97}{GW191230_180458}{1.0}{GW191222_033537}{0.94}{GW200225_060421}{0.80}{GW200302_015811}{1.06}{GW200128_022011}{1.03}{GW191204_171526}{0.91}{GW200112_155838}{1.02}{GW200105_162426}{1.2}{GW191105_143521}{0.95}{GW191109_010717}{0.67}{GW200209_085452}{0.91}{GW200115_042309}{0.66}{GW191127_050227}{1.06}{GW200216_220804}{1.14}{GW191215_223052}{0.83}{GW200208_130117}{0.93}{GW200219_094415}{0.92}{GW191103_012549}{0.98}{GW200316_215756}{1.07}{GW200202_154313}{1.06}{GW200129_065458}{1.04}{GW191216_213338}{1.20}}[{\red{???}}]}
\DeclareRobustCommand{\tiltonegwtcthreetenthpercentile}[1]{\IfEqCase{#1}{{GW200224_222234}{0.60}{GW191129_134029}{0.59}{GW200311_115853}{0.89}{GW191230_180458}{0.8}{GW191222_033537}{0.88}{GW200225_060421}{1.24}{GW200302_015811}{0.75}{GW200128_022011}{0.59}{GW191204_171526}{0.56}{GW200112_155838}{0.67}{GW200105_162426}{0.6}{GW191105_143521}{0.82}{GW191109_010717}{1.55}{GW200209_085452}{0.98}{GW200115_042309}{1.15}{GW191127_050227}{0.50}{GW200216_220804}{0.57}{GW191215_223052}{0.92}{GW200208_130117}{0.91}{GW200219_094415}{0.98}{GW191103_012549}{0.44}{GW200316_215756}{0.40}{GW200202_154313}{0.64}{GW200129_065458}{0.70}{GW191216_213338}{0.39}}[{\red{???}}]}
\DeclareRobustCommand{\tiltonegwtcthreenintiethpercentile}[1]{\IfEqCase{#1}{{GW200224_222234}{2.03}{GW191129_134029}{2.03}{GW200311_115853}{2.42}{GW191230_180458}{2.4}{GW191222_033537}{2.49}{GW200225_060421}{2.51}{GW200302_015811}{2.35}{GW200128_022011}{1.96}{GW191204_171526}{1.76}{GW200112_155838}{2.22}{GW200105_162426}{2.5}{GW191105_143521}{2.38}{GW191109_010717}{2.77}{GW200209_085452}{2.49}{GW200115_042309}{2.79}{GW191127_050227}{1.96}{GW200216_220804}{2.23}{GW191215_223052}{2.27}{GW200208_130117}{2.58}{GW200219_094415}{2.56}{GW191103_012549}{1.69}{GW200316_215756}{1.86}{GW200202_154313}{2.20}{GW200129_065458}{2.18}{GW191216_213338}{1.89}}[{\red{???}}]}
\DeclareRobustCommand{\psigwtcthreeminus}[1]{\IfEqCase{#1}{{GW200224_222234}{1.6}{GW191129_134029}{1.9}{GW200311_115853}{1.5}{GW191230_180458}{1.7}{GW191222_033537}{0.99}{GW200225_060421}{1.5}{GW200302_015811}{1.3}{GW200128_022011}{1.4}{GW191204_171526}{1.5}{GW200112_155838}{1.5}{GW200105_162426}{2.3}{GW191105_143521}{1.4}{GW191109_010717}{1.9}{GW200209_085452}{1.1}{GW200115_042309}{2.1}{GW191127_050227}{1.4}{GW200216_220804}{1.4}{GW191215_223052}{1.4}{GW200208_130117}{1.1}{GW200219_094415}{1.4}{GW191103_012549}{1.5}{GW200316_215756}{1.5}{GW200202_154313}{1.3}{GW200129_065458}{0.99}{GW191216_213338}{1.2}}[{\red{???}}]}
\DeclareRobustCommand{\psigwtcthreemed}[1]{\IfEqCase{#1}{{GW200224_222234}{1.7}{GW191129_134029}{2.1}{GW200311_115853}{1.8}{GW191230_180458}{1.9}{GW191222_033537}{1.11}{GW200225_060421}{1.7}{GW200302_015811}{1.4}{GW200128_022011}{1.6}{GW191204_171526}{1.6}{GW200112_155838}{1.6}{GW200105_162426}{2.4}{GW191105_143521}{1.5}{GW191109_010717}{2.0}{GW200209_085452}{1.3}{GW200115_042309}{2.3}{GW191127_050227}{1.6}{GW200216_220804}{1.6}{GW191215_223052}{1.6}{GW200208_130117}{1.3}{GW200219_094415}{1.7}{GW191103_012549}{1.6}{GW200316_215756}{1.6}{GW200202_154313}{1.5}{GW200129_065458}{1.18}{GW191216_213338}{1.7}}[{\red{???}}]}
\DeclareRobustCommand{\psigwtcthreeplus}[1]{\IfEqCase{#1}{{GW200224_222234}{1.4}{GW191129_134029}{3.6}{GW200311_115853}{1.1}{GW191230_180458}{3.6}{GW191222_033537}{1.88}{GW200225_060421}{1.3}{GW200302_015811}{1.5}{GW200128_022011}{1.4}{GW191204_171526}{1.3}{GW200112_155838}{1.4}{GW200105_162426}{3.5}{GW191105_143521}{1.5}{GW191109_010717}{1.1}{GW200209_085452}{1.6}{GW200115_042309}{3.5}{GW191127_050227}{1.4}{GW200216_220804}{1.4}{GW191215_223052}{1.4}{GW200208_130117}{1.6}{GW200219_094415}{1.2}{GW191103_012549}{1.4}{GW200316_215756}{1.4}{GW200202_154313}{1.5}{GW200129_065458}{1.65}{GW191216_213338}{3.7}}[{\red{???}}]}
\DeclareRobustCommand{\psigwtcthreetenthpercentile}[1]{\IfEqCase{#1}{{GW200224_222234}{0.2}{GW191129_134029}{0.4}{GW200311_115853}{0.5}{GW191230_180458}{0.4}{GW191222_033537}{0.24}{GW200225_060421}{0.3}{GW200302_015811}{0.3}{GW200128_022011}{0.3}{GW191204_171526}{0.4}{GW200112_155838}{0.3}{GW200105_162426}{0.3}{GW191105_143521}{0.2}{GW191109_010717}{0.2}{GW200209_085452}{0.3}{GW200115_042309}{0.4}{GW191127_050227}{0.3}{GW200216_220804}{0.3}{GW191215_223052}{0.3}{GW200208_130117}{0.4}{GW200219_094415}{0.5}{GW191103_012549}{0.3}{GW200316_215756}{0.2}{GW200202_154313}{0.3}{GW200129_065458}{0.40}{GW191216_213338}{0.7}}[{\red{???}}]}
\DeclareRobustCommand{\psigwtcthreenintiethpercentile}[1]{\IfEqCase{#1}{{GW200224_222234}{3.0}{GW191129_134029}{5.0}{GW200311_115853}{2.7}{GW191230_180458}{4.9}{GW191222_033537}{2.83}{GW200225_060421}{2.8}{GW200302_015811}{2.8}{GW200128_022011}{2.8}{GW191204_171526}{2.8}{GW200112_155838}{2.8}{GW200105_162426}{5.2}{GW191105_143521}{2.9}{GW191109_010717}{3.0}{GW200209_085452}{2.8}{GW200115_042309}{5.2}{GW191127_050227}{2.8}{GW200216_220804}{2.8}{GW191215_223052}{2.9}{GW200208_130117}{2.7}{GW200219_094415}{2.7}{GW191103_012549}{2.8}{GW200316_215756}{3.0}{GW200202_154313}{2.8}{GW200129_065458}{2.46}{GW191216_213338}{4.8}}[{\red{???}}]}
\DeclareRobustCommand{\decgwtcthreeminus}[1]{\IfEqCase{#1}{{GW200224_222234}{0.076}{GW191129_134029}{0.62}{GW200311_115853}{0.093}{GW191230_180458}{0.52}{GW191222_033537}{0.54}{GW200225_060421}{0.73}{GW200302_015811}{0.55}{GW200128_022011}{0.73}{GW191204_171526}{0.21}{GW200112_155838}{0.57}{GW200105_162426}{0.82}{GW191105_143521}{0.14}{GW191109_010717}{0.20}{GW200209_085452}{1.22}{GW200115_042309}{0.62}{GW191127_050227}{1.86}{GW200216_220804}{0.59}{GW191215_223052}{0.87}{GW200208_130117}{0.063}{GW200219_094415}{0.19}{GW191103_012549}{1.06}{GW200316_215756}{1.433}{GW200202_154313}{0.11}{GW200129_065458}{0.53}{GW191216_213338}{1.09}}[{\red{???}}]}
\DeclareRobustCommand{\decgwtcthreemed}[1]{\IfEqCase{#1}{{GW200224_222234}{-0.168}{GW191129_134029}{-0.58}{GW200311_115853}{-0.133}{GW191230_180458}{-0.62}{GW191222_033537}{-0.68}{GW200225_060421}{0.94}{GW200302_015811}{-0.51}{GW200128_022011}{-0.38}{GW191204_171526}{-0.54}{GW200112_155838}{-0.21}{GW200105_162426}{-0.05}{GW191105_143521}{-0.61}{GW191109_010717}{-0.59}{GW200209_085452}{0.88}{GW200115_042309}{-0.03}{GW191127_050227}{1.00}{GW200216_220804}{0.75}{GW191215_223052}{-0.24}{GW200208_130117}{-0.597}{GW200219_094415}{-0.46}{GW191103_012549}{0.65}{GW200316_215756}{0.819}{GW200202_154313}{0.38}{GW200129_065458}{0.09}{GW191216_213338}{0.44}}[{\red{???}}]}
\DeclareRobustCommand{\decgwtcthreeplus}[1]{\IfEqCase{#1}{{GW200224_222234}{0.095}{GW191129_134029}{1.30}{GW200311_115853}{0.079}{GW191230_180458}{0.98}{GW191222_033537}{1.56}{GW200225_060421}{0.53}{GW200302_015811}{1.65}{GW200128_022011}{1.35}{GW191204_171526}{0.90}{GW200112_155838}{1.12}{GW200105_162426}{0.86}{GW191105_143521}{1.86}{GW191109_010717}{0.99}{GW200209_085452}{0.48}{GW200115_042309}{0.52}{GW191127_050227}{0.50}{GW200216_220804}{0.37}{GW191215_223052}{0.93}{GW200208_130117}{0.074}{GW200219_094415}{1.39}{GW191103_012549}{0.73}{GW200316_215756}{0.060}{GW200202_154313}{0.13}{GW200129_065458}{0.39}{GW191216_213338}{0.47}}[{\red{???}}]}
\DeclareRobustCommand{\decgwtcthreetenthpercentile}[1]{\IfEqCase{#1}{{GW200224_222234}{-0.227}{GW191129_134029}{-1.06}{GW200311_115853}{-0.205}{GW191230_180458}{-1.07}{GW191222_033537}{-1.15}{GW200225_060421}{0.40}{GW200302_015811}{-1.03}{GW200128_022011}{-1.01}{GW191204_171526}{-0.74}{GW200112_155838}{-0.66}{GW200105_162426}{-0.73}{GW191105_143521}{-0.73}{GW191109_010717}{-0.77}{GW200209_085452}{-0.19}{GW200115_042309}{-0.53}{GW191127_050227}{-0.79}{GW200216_220804}{0.25}{GW191215_223052}{-0.96}{GW200208_130117}{-0.645}{GW200219_094415}{-0.60}{GW191103_012549}{-0.29}{GW200316_215756}{-0.432}{GW200202_154313}{0.30}{GW200129_065458}{0.00}{GW191216_213338}{-0.20}}[{\red{???}}]}
\DeclareRobustCommand{\decgwtcthreenintiethpercentile}[1]{\IfEqCase{#1}{{GW200224_222234}{-0.099}{GW191129_134029}{0.63}{GW200311_115853}{-0.071}{GW191230_180458}{0.07}{GW191222_033537}{0.72}{GW200225_060421}{1.39}{GW200302_015811}{0.96}{GW200128_022011}{0.86}{GW191204_171526}{0.08}{GW200112_155838}{0.80}{GW200105_162426}{0.71}{GW191105_143521}{0.92}{GW191109_010717}{0.14}{GW200209_085452}{1.33}{GW200115_042309}{0.15}{GW191127_050227}{1.46}{GW200216_220804}{1.05}{GW191215_223052}{0.49}{GW200208_130117}{-0.544}{GW200219_094415}{0.72}{GW191103_012549}{1.32}{GW200316_215756}{0.866}{GW200202_154313}{0.48}{GW200129_065458}{0.40}{GW191216_213338}{0.79}}[{\red{???}}]}
\DeclareRobustCommand{\symmetricmassratiogwtcthreeminus}[1]{\IfEqCase{#1}{{GW200224_222234}{0.017}{GW191129_134029}{0.048}{GW200311_115853}{0.019}{GW191230_180458}{0.030}{GW191222_033537}{0.028}{GW200225_060421}{0.030}{GW200302_015811}{0.041}{GW200128_022011}{0.024}{GW191204_171526}{0.031}{GW200112_155838}{0.019}{GW200105_162426}{0.028}{GW191105_143521}{0.037}{GW191109_010717}{0.023}{GW200209_085452}{0.025}{GW200115_042309}{0.046}{GW191127_050227}{0.126}{GW200216_220804}{0.093}{GW191215_223052}{0.029}{GW200208_130117}{0.032}{GW200219_094415}{0.032}{GW191103_012549}{0.061}{GW200316_215756}{0.089}{GW200202_154313}{0.037}{GW200129_065458}{0.035}{GW191216_213338}{0.046}}[{\red{???}}]}
\DeclareRobustCommand{\symmetricmassratiogwtcthreemed}[1]{\IfEqCase{#1}{{GW200224_222234}{0.248}{GW191129_134029}{0.237}{GW200311_115853}{0.248}{GW191230_180458}{0.246}{GW191222_033537}{0.247}{GW200225_060421}{0.244}{GW200302_015811}{0.229}{GW200128_022011}{0.247}{GW191204_171526}{0.242}{GW200112_155838}{0.247}{GW200105_162426}{0.144}{GW191105_143521}{0.243}{GW191109_010717}{0.244}{GW200209_085452}{0.247}{GW200115_042309}{0.157}{GW191127_050227}{0.217}{GW200216_220804}{0.235}{GW191215_223052}{0.244}{GW200208_130117}{0.244}{GW200219_094415}{0.246}{GW191103_012549}{0.240}{GW200316_215756}{0.234}{GW200202_154313}{0.244}{GW200129_065458}{0.248}{GW191216_213338}{0.238}}[{\red{???}}]}
\DeclareRobustCommand{\symmetricmassratiogwtcthreeplus}[1]{\IfEqCase{#1}{{GW200224_222234}{0.002}{GW191129_134029}{0.012}{GW200311_115853}{0.002}{GW191230_180458}{0.004}{GW191222_033537}{0.003}{GW200225_060421}{0.006}{GW200302_015811}{0.021}{GW200128_022011}{0.003}{GW191204_171526}{0.008}{GW200112_155838}{0.003}{GW200105_162426}{0.036}{GW191105_143521}{0.006}{GW191109_010717}{0.006}{GW200209_085452}{0.003}{GW200115_042309}{0.083}{GW191127_050227}{0.033}{GW200216_220804}{0.015}{GW191215_223052}{0.006}{GW200208_130117}{0.006}{GW200219_094415}{0.004}{GW191103_012549}{0.009}{GW200316_215756}{0.016}{GW200202_154313}{0.006}{GW200129_065458}{0.002}{GW191216_213338}{0.012}}[{\red{???}}]}
\DeclareRobustCommand{\symmetricmassratiogwtcthreetenthpercentile}[1]{\IfEqCase{#1}{{GW200224_222234}{0.236}{GW191129_134029}{0.202}{GW200311_115853}{0.235}{GW191230_180458}{0.226}{GW191222_033537}{0.228}{GW200225_060421}{0.223}{GW200302_015811}{0.197}{GW200128_022011}{0.230}{GW191204_171526}{0.220}{GW200112_155838}{0.234}{GW200105_162426}{0.128}{GW191105_143521}{0.217}{GW191109_010717}{0.227}{GW200209_085452}{0.230}{GW200115_042309}{0.121}{GW191127_050227}{0.115}{GW200216_220804}{0.161}{GW191215_223052}{0.224}{GW200208_130117}{0.221}{GW200219_094415}{0.224}{GW191103_012549}{0.198}{GW200316_215756}{0.171}{GW200202_154313}{0.218}{GW200129_065458}{0.222}{GW191216_213338}{0.207}}[{\red{???}}]}
\DeclareRobustCommand{\symmetricmassratiogwtcthreenintiethpercentile}[1]{\IfEqCase{#1}{{GW200224_222234}{0.250}{GW191129_134029}{0.249}{GW200311_115853}{0.250}{GW191230_180458}{0.250}{GW191222_033537}{0.250}{GW200225_060421}{0.250}{GW200302_015811}{0.248}{GW200128_022011}{0.250}{GW191204_171526}{0.249}{GW200112_155838}{0.250}{GW200105_162426}{0.163}{GW191105_143521}{0.250}{GW191109_010717}{0.249}{GW200209_085452}{0.250}{GW200115_042309}{0.231}{GW191127_050227}{0.249}{GW200216_220804}{0.250}{GW191215_223052}{0.250}{GW200208_130117}{0.250}{GW200219_094415}{0.250}{GW191103_012549}{0.250}{GW200316_215756}{0.249}{GW200202_154313}{0.250}{GW200129_065458}{0.250}{GW191216_213338}{0.249}}[{\red{???}}]}
\DeclareRobustCommand{\masstwosourcegwtcthreeminus}[1]{\IfEqCase{#1}{{GW200224_222234}{7.0}{GW191129_134029}{1.7}{GW200311_115853}{5.9}{GW191230_180458}{10.2}{GW191222_033537}{10.5}{GW200225_060421}{3.5}{GW200302_015811}{5.8}{GW200128_022011}{8.0}{GW191204_171526}{1.6}{GW200112_155838}{5.9}{GW200105_162426}{0.24}{GW191105_143521}{1.9}{GW191109_010717}{13}{GW200209_085452}{7.0}{GW200115_042309}{0.28}{GW191127_050227}{14}{GW200216_220804}{16}{GW191215_223052}{3.8}{GW200208_130117}{7.4}{GW200219_094415}{8.4}{GW191103_012549}{2.4}{GW200316_215756}{2.9}{GW200202_154313}{1.7}{GW200129_065458}{9.3}{GW191216_213338}{1.9}}[{\red{???}}]}
\DeclareRobustCommand{\masstwosourcegwtcthreemed}[1]{\IfEqCase{#1}{{GW200224_222234}{32.4}{GW191129_134029}{6.7}{GW200311_115853}{27.7}{GW191230_180458}{36.0}{GW191222_033537}{34.7}{GW200225_060421}{14.0}{GW200302_015811}{20.3}{GW200128_022011}{31.1}{GW191204_171526}{8.2}{GW200112_155838}{28.3}{GW200105_162426}{1.91}{GW191105_143521}{7.7}{GW191109_010717}{47}{GW200209_085452}{26.6}{GW200115_042309}{1.44}{GW191127_050227}{24}{GW200216_220804}{30}{GW191215_223052}{17.8}{GW200208_130117}{27.4}{GW200219_094415}{27.9}{GW191103_012549}{7.9}{GW200316_215756}{7.8}{GW200202_154313}{7.3}{GW200129_065458}{28.9}{GW191216_213338}{7.7}}[{\red{???}}]}
\DeclareRobustCommand{\masstwosourcegwtcthreeplus}[1]{\IfEqCase{#1}{{GW200224_222234}{4.7}{GW191129_134029}{1.5}{GW200311_115853}{4.1}{GW191230_180458}{9.7}{GW191222_033537}{9.3}{GW200225_060421}{2.8}{GW200302_015811}{8.0}{GW200128_022011}{8.7}{GW191204_171526}{1.4}{GW200112_155838}{4.4}{GW200105_162426}{0.33}{GW191105_143521}{1.4}{GW191109_010717}{15}{GW200209_085452}{7.0}{GW200115_042309}{0.85}{GW191127_050227}{17}{GW200216_220804}{14}{GW191215_223052}{3.7}{GW200208_130117}{6.1}{GW200219_094415}{7.4}{GW191103_012549}{1.7}{GW200316_215756}{1.9}{GW200202_154313}{1.1}{GW200129_065458}{3.4}{GW191216_213338}{1.6}}[{\red{???}}]}
\DeclareRobustCommand{\masstwosourcegwtcthreetenthpercentile}[1]{\IfEqCase{#1}{{GW200224_222234}{27.1}{GW191129_134029}{5.3}{GW200311_115853}{23.2}{GW191230_180458}{28.0}{GW191222_033537}{26.6}{GW200225_060421}{11.2}{GW200302_015811}{15.5}{GW200128_022011}{24.8}{GW191204_171526}{6.9}{GW200112_155838}{23.7}{GW200105_162426}{1.77}{GW191105_143521}{6.2}{GW191109_010717}{36}{GW200209_085452}{21.3}{GW200115_042309}{1.21}{GW191127_050227}{12}{GW200216_220804}{16}{GW191215_223052}{14.7}{GW200208_130117}{21.5}{GW200219_094415}{21.6}{GW191103_012549}{6.0}{GW200316_215756}{5.6}{GW200202_154313}{6.0}{GW200129_065458}{21.3}{GW191216_213338}{6.3}}[{\red{???}}]}
\DeclareRobustCommand{\masstwosourcegwtcthreenintiethpercentile}[1]{\IfEqCase{#1}{{GW200224_222234}{36.1}{GW191129_134029}{8.0}{GW200311_115853}{31.0}{GW191230_180458}{43.4}{GW191222_033537}{42.1}{GW200225_060421}{16.3}{GW200302_015811}{26.6}{GW200128_022011}{37.8}{GW191204_171526}{9.4}{GW200112_155838}{31.8}{GW200105_162426}{2.08}{GW191105_143521}{8.8}{GW191109_010717}{58}{GW200209_085452}{32.0}{GW200115_042309}{2.12}{GW191127_050227}{37}{GW200216_220804}{41}{GW191215_223052}{20.7}{GW200208_130117}{32.3}{GW200219_094415}{33.8}{GW191103_012549}{9.3}{GW200316_215756}{9.4}{GW200202_154313}{8.3}{GW200129_065458}{31.7}{GW191216_213338}{9.1}}[{\red{???}}]}
\DeclareRobustCommand{\iotagwtcthreeminus}[1]{\IfEqCase{#1}{{GW200224_222234}{0.42}{GW191129_134029}{1.5}{GW200311_115853}{0.39}{GW191230_180458}{1.92}{GW191222_033537}{1.3}{GW200225_060421}{1.0}{GW200302_015811}{0.99}{GW200128_022011}{1.1}{GW191204_171526}{2.03}{GW200112_155838}{0.69}{GW200105_162426}{1.2}{GW191105_143521}{0.86}{GW191109_010717}{1.41}{GW200209_085452}{1.5}{GW200115_042309}{0.46}{GW191127_050227}{1.2}{GW200216_220804}{0.73}{GW191215_223052}{0.91}{GW200208_130117}{0.62}{GW200219_094415}{0.96}{GW191103_012549}{1.1}{GW200316_215756}{1.84}{GW200202_154313}{0.59}{GW200129_065458}{0.50}{GW191216_213338}{0.80}}[{\red{???}}]}
\DeclareRobustCommand{\iotagwtcthreemed}[1]{\IfEqCase{#1}{{GW200224_222234}{0.58}{GW191129_134029}{1.8}{GW200311_115853}{0.54}{GW191230_180458}{2.17}{GW191222_033537}{1.6}{GW200225_060421}{1.4}{GW200302_015811}{1.27}{GW200128_022011}{1.4}{GW191204_171526}{2.28}{GW200112_155838}{0.90}{GW200105_162426}{1.5}{GW191105_143521}{1.08}{GW191109_010717}{1.98}{GW200209_085452}{1.7}{GW200115_042309}{0.63}{GW191127_050227}{1.5}{GW200216_220804}{0.96}{GW191215_223052}{1.21}{GW200208_130117}{2.52}{GW200219_094415}{1.23}{GW191103_012549}{1.4}{GW200316_215756}{2.32}{GW200202_154313}{2.57}{GW200129_065458}{0.65}{GW191216_213338}{2.50}}[{\red{???}}]}
\DeclareRobustCommand{\iotagwtcthreeplus}[1]{\IfEqCase{#1}{{GW200224_222234}{0.57}{GW191129_134029}{1.1}{GW200311_115853}{0.51}{GW191230_180458}{0.76}{GW191222_033537}{1.2}{GW200225_060421}{1.4}{GW200302_015811}{1.55}{GW200128_022011}{1.5}{GW191204_171526}{0.66}{GW200112_155838}{2.00}{GW200105_162426}{1.3}{GW191105_143521}{1.80}{GW191109_010717}{0.85}{GW200209_085452}{1.2}{GW200115_042309}{1.92}{GW191127_050227}{1.3}{GW200216_220804}{1.79}{GW191215_223052}{1.55}{GW200208_130117}{0.45}{GW200219_094415}{1.53}{GW191103_012549}{1.5}{GW200316_215756}{0.59}{GW200202_154313}{0.42}{GW200129_065458}{0.62}{GW191216_213338}{0.46}}[{\red{???}}]}
\DeclareRobustCommand{\iotagwtcthreetenthpercentile}[1]{\IfEqCase{#1}{{GW200224_222234}{0.22}{GW191129_134029}{0.4}{GW200311_115853}{0.21}{GW191230_180458}{0.38}{GW191222_033537}{0.4}{GW200225_060421}{0.5}{GW200302_015811}{0.40}{GW200128_022011}{0.4}{GW191204_171526}{0.36}{GW200112_155838}{0.29}{GW200105_162426}{0.5}{GW191105_143521}{0.32}{GW191109_010717}{0.83}{GW200209_085452}{0.4}{GW200115_042309}{0.25}{GW191127_050227}{0.5}{GW200216_220804}{0.34}{GW191215_223052}{0.43}{GW200208_130117}{2.05}{GW200219_094415}{0.40}{GW191103_012549}{0.3}{GW200316_215756}{0.72}{GW200202_154313}{2.09}{GW200129_065458}{0.23}{GW191216_213338}{2.00}}[{\red{???}}]}
\DeclareRobustCommand{\iotagwtcthreenintiethpercentile}[1]{\IfEqCase{#1}{{GW200224_222234}{1.03}{GW191129_134029}{2.8}{GW200311_115853}{0.96}{GW191230_180458}{2.83}{GW191222_033537}{2.7}{GW200225_060421}{2.6}{GW200302_015811}{2.67}{GW200128_022011}{2.7}{GW191204_171526}{2.85}{GW200112_155838}{2.81}{GW200105_162426}{2.7}{GW191105_143521}{2.76}{GW191109_010717}{2.69}{GW200209_085452}{2.8}{GW200115_042309}{2.12}{GW191127_050227}{2.6}{GW200216_220804}{2.53}{GW191215_223052}{2.59}{GW200208_130117}{2.89}{GW200219_094415}{2.58}{GW191103_012549}{2.8}{GW200316_215756}{2.81}{GW200202_154313}{2.92}{GW200129_065458}{1.12}{GW191216_213338}{2.88}}[{\red{???}}]}
\DeclareRobustCommand{\costiltonegwtcthreeminus}[1]{\IfEqCase{#1}{{GW200224_222234}{0.96}{GW191129_134029}{0.95}{GW200311_115853}{0.78}{GW191230_180458}{0.83}{GW191222_033537}{0.73}{GW200225_060421}{0.59}{GW200302_015811}{0.90}{GW200128_022011}{0.98}{GW191204_171526}{0.88}{GW200112_155838}{0.91}{GW200105_162426}{0.93}{GW191105_143521}{0.78}{GW191109_010717}{0.36}{GW200209_085452}{0.70}{GW200115_042309}{0.35}{GW191127_050227}{1.00}{GW200216_220804}{1.02}{GW191215_223052}{0.72}{GW200208_130117}{0.68}{GW200219_094415}{0.69}{GW191103_012549}{0.94}{GW200316_215756}{1.02}{GW200202_154313}{0.95}{GW200129_065458}{0.92}{GW191216_213338}{1.13}}[{\red{???}}]}
\DeclareRobustCommand{\costiltonegwtcthreemed}[1]{\IfEqCase{#1}{{GW200224_222234}{0.30}{GW191129_134029}{0.27}{GW200311_115853}{-0.09}{GW191230_180458}{-0.04}{GW191222_033537}{-0.16}{GW200225_060421}{-0.30}{GW200302_015811}{0.06}{GW200128_022011}{0.36}{GW191204_171526}{0.43}{GW200112_155838}{0.14}{GW200105_162426}{0.01}{GW191105_143521}{-0.07}{GW191109_010717}{-0.61}{GW200209_085452}{-0.19}{GW200115_042309}{-0.62}{GW191127_050227}{0.37}{GW200216_220804}{0.24}{GW191215_223052}{-0.06}{GW200208_130117}{-0.24}{GW200219_094415}{-0.23}{GW191103_012549}{0.54}{GW200316_215756}{0.47}{GW200202_154313}{0.18}{GW200129_065458}{0.13}{GW191216_213338}{0.54}}[{\red{???}}]}
\DeclareRobustCommand{\costiltonegwtcthreeplus}[1]{\IfEqCase{#1}{{GW200224_222234}{0.61}{GW191129_134029}{0.64}{GW200311_115853}{0.89}{GW191230_180458}{0.86}{GW191222_033537}{0.97}{GW200225_060421}{0.86}{GW200302_015811}{0.80}{GW200128_022011}{0.56}{GW191204_171526}{0.49}{GW200112_155838}{0.75}{GW200105_162426}{0.90}{GW191105_143521}{0.90}{GW191109_010717}{0.85}{GW200209_085452}{0.92}{GW200115_042309}{1.26}{GW191127_050227}{0.56}{GW200216_220804}{0.68}{GW191215_223052}{0.85}{GW200208_130117}{1.03}{GW200219_094415}{0.96}{GW191103_012549}{0.41}{GW200316_215756}{0.49}{GW200202_154313}{0.72}{GW200129_065458}{0.74}{GW191216_213338}{0.43}}[{\red{???}}]}
\DeclareRobustCommand{\costiltonegwtcthreetenthpercentile}[1]{\IfEqCase{#1}{{GW200224_222234}{-0.44}{GW191129_134029}{-0.45}{GW200311_115853}{-0.75}{GW191230_180458}{-0.73}{GW191222_033537}{-0.80}{GW200225_060421}{-0.81}{GW200302_015811}{-0.70}{GW200128_022011}{-0.38}{GW191204_171526}{-0.19}{GW200112_155838}{-0.60}{GW200105_162426}{-0.82}{GW191105_143521}{-0.72}{GW191109_010717}{-0.93}{GW200209_085452}{-0.79}{GW200115_042309}{-0.94}{GW191127_050227}{-0.38}{GW200216_220804}{-0.62}{GW191215_223052}{-0.64}{GW200208_130117}{-0.85}{GW200219_094415}{-0.83}{GW191103_012549}{-0.12}{GW200316_215756}{-0.29}{GW200202_154313}{-0.59}{GW200129_065458}{-0.57}{GW191216_213338}{-0.32}}[{\red{???}}]}
\DeclareRobustCommand{\costiltonegwtcthreenintiethpercentile}[1]{\IfEqCase{#1}{{GW200224_222234}{0.82}{GW191129_134029}{0.83}{GW200311_115853}{0.63}{GW191230_180458}{0.69}{GW191222_033537}{0.64}{GW200225_060421}{0.32}{GW200302_015811}{0.73}{GW200128_022011}{0.83}{GW191204_171526}{0.85}{GW200112_155838}{0.78}{GW200105_162426}{0.82}{GW191105_143521}{0.68}{GW191109_010717}{0.02}{GW200209_085452}{0.55}{GW200115_042309}{0.41}{GW191127_050227}{0.88}{GW200216_220804}{0.84}{GW191215_223052}{0.61}{GW200208_130117}{0.61}{GW200219_094415}{0.56}{GW191103_012549}{0.90}{GW200316_215756}{0.92}{GW200202_154313}{0.80}{GW200129_065458}{0.76}{GW191216_213338}{0.92}}[{\red{???}}]}
\DeclareRobustCommand{\phionetwogwtcthreeminus}[1]{\IfEqCase{#1}{{GW200224_222234}{2.7}{GW191129_134029}{2.8}{GW200311_115853}{2.8}{GW191230_180458}{2.9}{GW191222_033537}{2.8}{GW200225_060421}{2.8}{GW200302_015811}{2.8}{GW200128_022011}{3.0}{GW191204_171526}{2.8}{GW200112_155838}{2.7}{GW200105_162426}{2.8}{GW191105_143521}{2.8}{GW191109_010717}{3.5}{GW200209_085452}{2.8}{GW200115_042309}{2.8}{GW191127_050227}{2.8}{GW200216_220804}{2.9}{GW191215_223052}{2.9}{GW200208_130117}{2.7}{GW200219_094415}{2.9}{GW191103_012549}{2.8}{GW200316_215756}{2.8}{GW200202_154313}{2.9}{GW200129_065458}{2.6}{GW191216_213338}{2.8}}[{\red{???}}]}
\DeclareRobustCommand{\phionetwogwtcthreemed}[1]{\IfEqCase{#1}{{GW200224_222234}{3.1}{GW191129_134029}{3.1}{GW200311_115853}{3.1}{GW191230_180458}{3.1}{GW191222_033537}{3.2}{GW200225_060421}{3.2}{GW200302_015811}{3.1}{GW200128_022011}{3.3}{GW191204_171526}{3.1}{GW200112_155838}{3.1}{GW200105_162426}{3.2}{GW191105_143521}{3.2}{GW191109_010717}{3.7}{GW200209_085452}{3.1}{GW200115_042309}{3.2}{GW191127_050227}{3.2}{GW200216_220804}{3.2}{GW191215_223052}{3.2}{GW200208_130117}{3.0}{GW200219_094415}{3.2}{GW191103_012549}{3.1}{GW200316_215756}{3.1}{GW200202_154313}{3.2}{GW200129_065458}{3.0}{GW191216_213338}{3.2}}[{\red{???}}]}
\DeclareRobustCommand{\phionetwogwtcthreeplus}[1]{\IfEqCase{#1}{{GW200224_222234}{2.8}{GW191129_134029}{2.8}{GW200311_115853}{2.8}{GW191230_180458}{2.9}{GW191222_033537}{2.8}{GW200225_060421}{2.8}{GW200302_015811}{2.8}{GW200128_022011}{2.7}{GW191204_171526}{2.8}{GW200112_155838}{2.9}{GW200105_162426}{2.8}{GW191105_143521}{2.8}{GW191109_010717}{2.3}{GW200209_085452}{2.9}{GW200115_042309}{2.8}{GW191127_050227}{2.8}{GW200216_220804}{2.8}{GW191215_223052}{2.8}{GW200208_130117}{2.9}{GW200219_094415}{2.8}{GW191103_012549}{2.9}{GW200316_215756}{2.8}{GW200202_154313}{2.8}{GW200129_065458}{2.9}{GW191216_213338}{2.8}}[{\red{???}}]}
\DeclareRobustCommand{\phionetwogwtcthreetenthpercentile}[1]{\IfEqCase{#1}{{GW200224_222234}{0.7}{GW191129_134029}{0.7}{GW200311_115853}{0.7}{GW191230_180458}{0.6}{GW191222_033537}{0.7}{GW200225_060421}{0.7}{GW200302_015811}{0.7}{GW200128_022011}{0.6}{GW191204_171526}{0.7}{GW200112_155838}{0.7}{GW200105_162426}{0.7}{GW191105_143521}{0.7}{GW191109_010717}{0.5}{GW200209_085452}{0.5}{GW200115_042309}{0.7}{GW191127_050227}{0.6}{GW200216_220804}{0.6}{GW191215_223052}{0.6}{GW200208_130117}{0.6}{GW200219_094415}{0.6}{GW191103_012549}{0.6}{GW200316_215756}{0.6}{GW200202_154313}{0.7}{GW200129_065458}{0.6}{GW191216_213338}{0.7}}[{\red{???}}]}
\DeclareRobustCommand{\phionetwogwtcthreenintiethpercentile}[1]{\IfEqCase{#1}{{GW200224_222234}{5.6}{GW191129_134029}{5.6}{GW200311_115853}{5.6}{GW191230_180458}{5.7}{GW191222_033537}{5.6}{GW200225_060421}{5.6}{GW200302_015811}{5.6}{GW200128_022011}{5.7}{GW191204_171526}{5.6}{GW200112_155838}{5.6}{GW200105_162426}{5.6}{GW191105_143521}{5.6}{GW191109_010717}{5.8}{GW200209_085452}{5.7}{GW200115_042309}{5.6}{GW191127_050227}{5.7}{GW200216_220804}{5.6}{GW191215_223052}{5.6}{GW200208_130117}{5.6}{GW200219_094415}{5.7}{GW191103_012549}{5.6}{GW200316_215756}{5.6}{GW200202_154313}{5.6}{GW200129_065458}{5.6}{GW191216_213338}{5.6}}[{\red{???}}]}
\DeclareRobustCommand{\massratiogwtcthreeminus}[1]{\IfEqCase{#1}{{GW200224_222234}{0.26}{GW191129_134029}{0.29}{GW200311_115853}{0.27}{GW191230_180458}{0.32}{GW191222_033537}{0.32}{GW200225_060421}{0.28}{GW200302_015811}{0.21}{GW200128_022011}{0.29}{GW191204_171526}{0.26}{GW200112_155838}{0.26}{GW200105_162426}{0.056}{GW191105_143521}{0.31}{GW191109_010717}{0.24}{GW200209_085452}{0.30}{GW200115_042309}{0.097}{GW191127_050227}{0.35}{GW200216_220804}{0.40}{GW191215_223052}{0.27}{GW200208_130117}{0.29}{GW200219_094415}{0.32}{GW191103_012549}{0.37}{GW200316_215756}{0.38}{GW200202_154313}{0.31}{GW200129_065458}{0.41}{GW191216_213338}{0.29}}[{\red{???}}]}
\DeclareRobustCommand{\massratiogwtcthreemed}[1]{\IfEqCase{#1}{{GW200224_222234}{0.82}{GW191129_134029}{0.63}{GW200311_115853}{0.82}{GW191230_180458}{0.78}{GW191222_033537}{0.79}{GW200225_060421}{0.73}{GW200302_015811}{0.55}{GW200128_022011}{0.79}{GW191204_171526}{0.69}{GW200112_155838}{0.81}{GW200105_162426}{0.211}{GW191105_143521}{0.72}{GW191109_010717}{0.73}{GW200209_085452}{0.79}{GW200115_042309}{0.243}{GW191127_050227}{0.47}{GW200216_220804}{0.61}{GW191215_223052}{0.73}{GW200208_130117}{0.73}{GW200219_094415}{0.77}{GW191103_012549}{0.67}{GW200316_215756}{0.59}{GW200202_154313}{0.72}{GW200129_065458}{0.85}{GW191216_213338}{0.63}}[{\red{???}}]}
\DeclareRobustCommand{\massratiogwtcthreeplus}[1]{\IfEqCase{#1}{{GW200224_222234}{0.16}{GW191129_134029}{0.31}{GW200311_115853}{0.16}{GW191230_180458}{0.20}{GW191222_033537}{0.18}{GW200225_060421}{0.23}{GW200302_015811}{0.36}{GW200128_022011}{0.18}{GW191204_171526}{0.25}{GW200112_155838}{0.17}{GW200105_162426}{0.095}{GW191105_143521}{0.24}{GW191109_010717}{0.21}{GW200209_085452}{0.19}{GW200115_042309}{0.432}{GW191127_050227}{0.47}{GW200216_220804}{0.35}{GW191215_223052}{0.24}{GW200208_130117}{0.23}{GW200219_094415}{0.21}{GW191103_012549}{0.29}{GW200316_215756}{0.34}{GW200202_154313}{0.24}{GW200129_065458}{0.12}{GW191216_213338}{0.31}}[{\red{???}}]}
\DeclareRobustCommand{\massratiogwtcthreetenthpercentile}[1]{\IfEqCase{#1}{{GW200224_222234}{0.62}{GW191129_134029}{0.39}{GW200311_115853}{0.61}{GW191230_180458}{0.52}{GW191222_033537}{0.54}{GW200225_060421}{0.50}{GW200302_015811}{0.37}{GW200128_022011}{0.56}{GW191204_171526}{0.48}{GW200112_155838}{0.60}{GW200105_162426}{0.177}{GW191105_143521}{0.47}{GW191109_010717}{0.54}{GW200209_085452}{0.56}{GW200115_042309}{0.164}{GW191127_050227}{0.15}{GW200216_220804}{0.25}{GW191215_223052}{0.51}{GW200208_130117}{0.49}{GW200219_094415}{0.51}{GW191103_012549}{0.37}{GW200316_215756}{0.28}{GW200202_154313}{0.48}{GW200129_065458}{0.50}{GW191216_213338}{0.41}}[{\red{???}}]}
\DeclareRobustCommand{\massratiogwtcthreenintiethpercentile}[1]{\IfEqCase{#1}{{GW200224_222234}{0.96}{GW191129_134029}{0.91}{GW200311_115853}{0.96}{GW191230_180458}{0.95}{GW191222_033537}{0.96}{GW200225_060421}{0.93}{GW200302_015811}{0.84}{GW200128_022011}{0.96}{GW191204_171526}{0.91}{GW200112_155838}{0.96}{GW200105_162426}{0.259}{GW191105_143521}{0.93}{GW191109_010717}{0.91}{GW200209_085452}{0.96}{GW200115_042309}{0.571}{GW191127_050227}{0.88}{GW200216_220804}{0.92}{GW191215_223052}{0.94}{GW200208_130117}{0.94}{GW200219_094415}{0.95}{GW191103_012549}{0.92}{GW200316_215756}{0.89}{GW200202_154313}{0.94}{GW200129_065458}{0.96}{GW191216_213338}{0.90}}[{\red{???}}]}
\DeclareRobustCommand{\comovingdistgwtcthreeminus}[1]{\IfEqCase{#1}{{GW200224_222234}{400}{GW191129_134029}{260}{GW200311_115853}{280}{GW191230_180458}{850}{GW191222_033537}{920}{GW200225_060421}{390}{GW200302_015811}{510}{GW200128_022011}{970}{GW191204_171526}{200}{GW200112_155838}{330}{GW200105_162426}{100}{GW191105_143521}{350}{GW191109_010717}{470}{GW200209_085452}{840}{GW200115_042309}{92}{GW191127_050227}{1000}{GW200216_220804}{1000}{GW191215_223052}{560}{GW200208_130117}{500}{GW200219_094415}{750}{GW191103_012549}{360}{GW200316_215756}{320}{GW200202_154313}{140}{GW200129_065458}{290}{GW191216_213338}{120}}[{\red{???}}]}
\DeclareRobustCommand{\comovingdistgwtcthreemed}[1]{\IfEqCase{#1}{{GW200224_222234}{1340}{GW191129_134029}{700}{GW200311_115853}{950}{GW191230_180458}{2740}{GW191222_033537}{1980}{GW200225_060421}{940}{GW200302_015811}{1270}{GW200128_022011}{2440}{GW191204_171526}{580}{GW200112_155838}{1010}{GW200105_162426}{260}{GW191105_143521}{940}{GW191109_010717}{1040}{GW200209_085452}{2390}{GW200115_042309}{274}{GW191127_050227}{2200}{GW200216_220804}{2350}{GW191215_223052}{1520}{GW200208_130117}{1590}{GW200219_094415}{2180}{GW191103_012549}{830}{GW200316_215756}{920}{GW200202_154313}{380}{GW200129_065458}{760}{GW191216_213338}{320}}[{\red{???}}]}
\DeclareRobustCommand{\comovingdistgwtcthreeplus}[1]{\IfEqCase{#1}{{GW200224_222234}{270}{GW191129_134029}{180}{GW200311_115853}{180}{GW191230_180458}{690}{GW191222_033537}{710}{GW200225_060421}{330}{GW200302_015811}{590}{GW200128_022011}{750}{GW191204_171526}{140}{GW200112_155838}{270}{GW200105_162426}{100}{GW191105_143521}{280}{GW191109_010717}{660}{GW200209_085452}{720}{GW200115_042309}{126}{GW191127_050227}{1100}{GW200216_220804}{1040}{GW191215_223052}{440}{GW200208_130117}{500}{GW200219_094415}{680}{GW191103_012549}{340}{GW200316_215756}{310}{GW200202_154313}{120}{GW200129_065458}{200}{GW191216_213338}{100}}[{\red{???}}]}
\DeclareRobustCommand{\comovingdistgwtcthreetenthpercentile}[1]{\IfEqCase{#1}{{GW200224_222234}{1030}{GW191129_134029}{490}{GW200311_115853}{740}{GW191230_180458}{2080}{GW191222_033537}{1260}{GW200225_060421}{640}{GW200302_015811}{860}{GW200128_022011}{1690}{GW191204_171526}{420}{GW200112_155838}{750}{GW200105_162426}{170}{GW191105_143521}{660}{GW191109_010717}{660}{GW200209_085452}{1750}{GW200115_042309}{202}{GW191127_050227}{1400}{GW200216_220804}{1560}{GW191215_223052}{1080}{GW200208_130117}{1190}{GW200219_094415}{1550}{GW191103_012549}{530}{GW200316_215756}{660}{GW200202_154313}{270}{GW200129_065458}{540}{GW191216_213338}{220}}[{\red{???}}]}
\DeclareRobustCommand{\comovingdistgwtcthreenintiethpercentile}[1]{\IfEqCase{#1}{{GW200224_222234}{1550}{GW191129_134029}{850}{GW200311_115853}{1100}{GW191230_180458}{3280}{GW191222_033537}{2550}{GW200225_060421}{1200}{GW200302_015811}{1730}{GW200128_022011}{3040}{GW191204_171526}{700}{GW200112_155838}{1230}{GW200105_162426}{340}{GW191105_143521}{1160}{GW191109_010717}{1540}{GW200209_085452}{2950}{GW200115_042309}{365}{GW191127_050227}{3100}{GW200216_220804}{3160}{GW191215_223052}{1880}{GW200208_130117}{1970}{GW200219_094415}{2710}{GW191103_012549}{1090}{GW200316_215756}{1160}{GW200202_154313}{480}{GW200129_065458}{930}{GW191216_213338}{400}}[{\red{???}}]}
\DeclareRobustCommand{\phasegwtcthreeminus}[1]{\IfEqCase{#1}{{GW200224_222234}{2.8}{GW191129_134029}{2.9}{GW200311_115853}{2.7}{GW191230_180458}{2.8}{GW191222_033537}{2.8}{GW200225_060421}{2.4}{GW200302_015811}{2.2}{GW200128_022011}{3.1}{GW191204_171526}{2.7}{GW200112_155838}{2.9}{GW200105_162426}{3.1}{GW191105_143521}{2.8}{GW191109_010717}{1.4}{GW200209_085452}{2.9}{GW200115_042309}{2.8}{GW191127_050227}{2.8}{GW200216_220804}{2.6}{GW191215_223052}{2.4}{GW200208_130117}{3.5}{GW200219_094415}{3.5}{GW191103_012549}{2.6}{GW200316_215756}{2.4}{GW200202_154313}{2.6}{GW200129_065458}{3.0}{GW191216_213338}{2.1}}[{\red{???}}]}
\DeclareRobustCommand{\phasegwtcthreemed}[1]{\IfEqCase{#1}{{GW200224_222234}{3.0}{GW191129_134029}{3.4}{GW200311_115853}{3.1}{GW191230_180458}{3.1}{GW191222_033537}{3.1}{GW200225_060421}{2.8}{GW200302_015811}{2.5}{GW200128_022011}{3.4}{GW191204_171526}{3.1}{GW200112_155838}{3.4}{GW200105_162426}{4.2}{GW191105_143521}{3.1}{GW191109_010717}{1.6}{GW200209_085452}{3.2}{GW200115_042309}{3.2}{GW191127_050227}{3.0}{GW200216_220804}{3.0}{GW191215_223052}{2.6}{GW200208_130117}{3.8}{GW200219_094415}{3.8}{GW191103_012549}{3.0}{GW200316_215756}{2.9}{GW200202_154313}{3.0}{GW200129_065458}{3.5}{GW191216_213338}{2.2}}[{\red{???}}]}
\DeclareRobustCommand{\phasegwtcthreeplus}[1]{\IfEqCase{#1}{{GW200224_222234}{3.0}{GW191129_134029}{2.5}{GW200311_115853}{2.8}{GW191230_180458}{2.9}{GW191222_033537}{3.0}{GW200225_060421}{3.2}{GW200302_015811}{3.4}{GW200128_022011}{2.7}{GW191204_171526}{2.8}{GW200112_155838}{2.5}{GW200105_162426}{1.4}{GW191105_143521}{2.9}{GW191109_010717}{4.3}{GW200209_085452}{2.8}{GW200115_042309}{2.8}{GW191127_050227}{2.9}{GW200216_220804}{3.0}{GW191215_223052}{3.4}{GW200208_130117}{2.2}{GW200219_094415}{2.2}{GW191103_012549}{3.0}{GW200316_215756}{2.8}{GW200202_154313}{2.9}{GW200129_065458}{2.4}{GW191216_213338}{3.9}}[{\red{???}}]}
\DeclareRobustCommand{\phasegwtcthreetenthpercentile}[1]{\IfEqCase{#1}{{GW200224_222234}{0.4}{GW191129_134029}{0.9}{GW200311_115853}{0.8}{GW191230_180458}{0.6}{GW191222_033537}{0.5}{GW200225_060421}{0.7}{GW200302_015811}{0.7}{GW200128_022011}{0.6}{GW191204_171526}{0.8}{GW200112_155838}{1.0}{GW200105_162426}{1.8}{GW191105_143521}{0.6}{GW191109_010717}{0.3}{GW200209_085452}{0.5}{GW200115_042309}{0.6}{GW191127_050227}{0.5}{GW200216_220804}{0.7}{GW191215_223052}{0.4}{GW200208_130117}{0.7}{GW200219_094415}{0.8}{GW191103_012549}{0.7}{GW200316_215756}{1.0}{GW200202_154313}{0.7}{GW200129_065458}{0.9}{GW191216_213338}{0.3}}[{\red{???}}]}
\DeclareRobustCommand{\phasegwtcthreenintiethpercentile}[1]{\IfEqCase{#1}{{GW200224_222234}{5.8}{GW191129_134029}{5.5}{GW200311_115853}{5.5}{GW191230_180458}{5.6}{GW191222_033537}{5.8}{GW200225_060421}{5.5}{GW200302_015811}{5.5}{GW200128_022011}{5.8}{GW191204_171526}{5.4}{GW200112_155838}{5.5}{GW200105_162426}{5.3}{GW191105_143521}{5.7}{GW191109_010717}{5.0}{GW200209_085452}{5.7}{GW200115_042309}{5.7}{GW191127_050227}{5.7}{GW200216_220804}{5.6}{GW191215_223052}{5.7}{GW200208_130117}{5.7}{GW200219_094415}{5.7}{GW191103_012549}{5.6}{GW200316_215756}{5.2}{GW200202_154313}{5.5}{GW200129_065458}{5.6}{GW191216_213338}{6.0}}[{\red{???}}]}
\DeclareRobustCommand{\phionegwtcthreeminus}[1]{\IfEqCase{#1}{{GW200224_222234}{2.9}{GW191129_134029}{2.9}{GW200311_115853}{2.7}{GW191230_180458}{2.8}{GW191222_033537}{2.8}{GW200225_060421}{2.9}{GW200302_015811}{2.8}{GW200128_022011}{2.8}{GW191204_171526}{2.8}{GW200112_155838}{2.8}{GW200105_162426}{2.9}{GW191105_143521}{2.8}{GW191109_010717}{2.7}{GW200209_085452}{2.8}{GW200115_042309}{3.0}{GW191127_050227}{2.8}{GW200216_220804}{2.9}{GW191215_223052}{2.8}{GW200208_130117}{2.8}{GW200219_094415}{2.9}{GW191103_012549}{2.8}{GW200316_215756}{3.0}{GW200202_154313}{2.9}{GW200129_065458}{2.8}{GW191216_213338}{2.7}}[{\red{???}}]}
\DeclareRobustCommand{\phionegwtcthreemed}[1]{\IfEqCase{#1}{{GW200224_222234}{3.1}{GW191129_134029}{3.2}{GW200311_115853}{3.1}{GW191230_180458}{3.2}{GW191222_033537}{3.1}{GW200225_060421}{3.2}{GW200302_015811}{3.2}{GW200128_022011}{3.1}{GW191204_171526}{3.1}{GW200112_155838}{3.1}{GW200105_162426}{3.2}{GW191105_143521}{3.1}{GW191109_010717}{3.1}{GW200209_085452}{3.2}{GW200115_042309}{3.3}{GW191127_050227}{3.1}{GW200216_220804}{3.2}{GW191215_223052}{3.1}{GW200208_130117}{3.0}{GW200219_094415}{3.2}{GW191103_012549}{3.1}{GW200316_215756}{3.3}{GW200202_154313}{3.2}{GW200129_065458}{3.1}{GW191216_213338}{3.0}}[{\red{???}}]}
\DeclareRobustCommand{\phionegwtcthreeplus}[1]{\IfEqCase{#1}{{GW200224_222234}{2.9}{GW191129_134029}{2.8}{GW200311_115853}{2.9}{GW191230_180458}{2.8}{GW191222_033537}{2.9}{GW200225_060421}{2.8}{GW200302_015811}{2.8}{GW200128_022011}{2.8}{GW191204_171526}{2.9}{GW200112_155838}{2.9}{GW200105_162426}{2.8}{GW191105_143521}{2.8}{GW191109_010717}{2.9}{GW200209_085452}{2.8}{GW200115_042309}{2.7}{GW191127_050227}{2.9}{GW200216_220804}{2.8}{GW191215_223052}{2.9}{GW200208_130117}{2.9}{GW200219_094415}{2.8}{GW191103_012549}{2.9}{GW200316_215756}{2.7}{GW200202_154313}{2.8}{GW200129_065458}{2.9}{GW191216_213338}{3.0}}[{\red{???}}]}
\DeclareRobustCommand{\phionegwtcthreetenthpercentile}[1]{\IfEqCase{#1}{{GW200224_222234}{0.6}{GW191129_134029}{0.6}{GW200311_115853}{0.7}{GW191230_180458}{0.6}{GW191222_033537}{0.6}{GW200225_060421}{0.6}{GW200302_015811}{0.6}{GW200128_022011}{0.7}{GW191204_171526}{0.7}{GW200112_155838}{0.6}{GW200105_162426}{0.6}{GW191105_143521}{0.6}{GW191109_010717}{0.6}{GW200209_085452}{0.6}{GW200115_042309}{0.6}{GW191127_050227}{0.7}{GW200216_220804}{0.7}{GW191215_223052}{0.6}{GW200208_130117}{0.6}{GW200219_094415}{0.6}{GW191103_012549}{0.7}{GW200316_215756}{0.6}{GW200202_154313}{0.6}{GW200129_065458}{0.6}{GW191216_213338}{0.6}}[{\red{???}}]}
\DeclareRobustCommand{\phionegwtcthreenintiethpercentile}[1]{\IfEqCase{#1}{{GW200224_222234}{5.7}{GW191129_134029}{5.7}{GW200311_115853}{5.6}{GW191230_180458}{5.6}{GW191222_033537}{5.6}{GW200225_060421}{5.7}{GW200302_015811}{5.7}{GW200128_022011}{5.6}{GW191204_171526}{5.6}{GW200112_155838}{5.6}{GW200105_162426}{5.6}{GW191105_143521}{5.7}{GW191109_010717}{5.6}{GW200209_085452}{5.7}{GW200115_042309}{5.7}{GW191127_050227}{5.7}{GW200216_220804}{5.6}{GW191215_223052}{5.7}{GW200208_130117}{5.7}{GW200219_094415}{5.7}{GW191103_012549}{5.6}{GW200316_215756}{5.7}{GW200202_154313}{5.7}{GW200129_065458}{5.7}{GW191216_213338}{5.6}}[{\red{???}}]}
\DeclareRobustCommand{\spintwogwtcthreeminus}[1]{\IfEqCase{#1}{{GW200224_222234}{0.39}{GW191129_134029}{0.31}{GW200311_115853}{0.37}{GW191230_180458}{0.44}{GW191222_033537}{0.38}{GW200225_060421}{0.39}{GW200302_015811}{0.41}{GW200128_022011}{0.45}{GW191204_171526}{0.40}{GW200112_155838}{0.34}{GW200105_162426}{0.30}{GW191105_143521}{0.31}{GW191109_010717}{0.58}{GW200209_085452}{0.44}{GW200115_042309}{0.39}{GW191127_050227}{0.49}{GW200216_220804}{0.46}{GW191215_223052}{0.40}{GW200208_130117}{0.39}{GW200219_094415}{0.43}{GW191103_012549}{0.44}{GW200316_215756}{0.39}{GW200202_154313}{0.30}{GW200129_065458}{0.42}{GW191216_213338}{0.32}}[{\red{???}}]}
\DeclareRobustCommand{\spintwogwtcthreemed}[1]{\IfEqCase{#1}{{GW200224_222234}{0.44}{GW191129_134029}{0.35}{GW200311_115853}{0.41}{GW191230_180458}{0.49}{GW191222_033537}{0.41}{GW200225_060421}{0.42}{GW200302_015811}{0.45}{GW200128_022011}{0.50}{GW191204_171526}{0.46}{GW200112_155838}{0.39}{GW200105_162426}{0.33}{GW191105_143521}{0.34}{GW191109_010717}{0.65}{GW200209_085452}{0.49}{GW200115_042309}{0.44}{GW191127_050227}{0.54}{GW200216_220804}{0.51}{GW191215_223052}{0.44}{GW200208_130117}{0.43}{GW200219_094415}{0.48}{GW191103_012549}{0.50}{GW200316_215756}{0.44}{GW200202_154313}{0.33}{GW200129_065458}{0.49}{GW191216_213338}{0.36}}[{\red{???}}]}
\DeclareRobustCommand{\spintwogwtcthreeplus}[1]{\IfEqCase{#1}{{GW200224_222234}{0.48}{GW191129_134029}{0.51}{GW200311_115853}{0.51}{GW191230_180458}{0.45}{GW191222_033537}{0.50}{GW200225_060421}{0.49}{GW200302_015811}{0.48}{GW200128_022011}{0.44}{GW191204_171526}{0.41}{GW200112_155838}{0.50}{GW200105_162426}{0.56}{GW191105_143521}{0.54}{GW191109_010717}{0.32}{GW200209_085452}{0.45}{GW200115_042309}{0.48}{GW191127_050227}{0.41}{GW200216_220804}{0.44}{GW191215_223052}{0.49}{GW200208_130117}{0.49}{GW200219_094415}{0.46}{GW191103_012549}{0.44}{GW200316_215756}{0.47}{GW200202_154313}{0.53}{GW200129_065458}{0.44}{GW191216_213338}{0.50}}[{\red{???}}]}
\DeclareRobustCommand{\spintwogwtcthreetenthpercentile}[1]{\IfEqCase{#1}{{GW200224_222234}{0.09}{GW191129_134029}{0.07}{GW200311_115853}{0.08}{GW191230_180458}{0.10}{GW191222_033537}{0.07}{GW200225_060421}{0.08}{GW200302_015811}{0.09}{GW200128_022011}{0.10}{GW191204_171526}{0.11}{GW200112_155838}{0.08}{GW200105_162426}{0.06}{GW191105_143521}{0.06}{GW191109_010717}{0.14}{GW200209_085452}{0.10}{GW200115_042309}{0.09}{GW191127_050227}{0.11}{GW200216_220804}{0.10}{GW191215_223052}{0.09}{GW200208_130117}{0.08}{GW200219_094415}{0.10}{GW191103_012549}{0.10}{GW200316_215756}{0.10}{GW200202_154313}{0.06}{GW200129_065458}{0.13}{GW191216_213338}{0.08}}[{\red{???}}]}
\DeclareRobustCommand{\spintwogwtcthreenintiethpercentile}[1]{\IfEqCase{#1}{{GW200224_222234}{0.85}{GW191129_134029}{0.76}{GW200311_115853}{0.84}{GW191230_180458}{0.89}{GW191222_033537}{0.84}{GW200225_060421}{0.85}{GW200302_015811}{0.87}{GW200128_022011}{0.89}{GW191204_171526}{0.79}{GW200112_155838}{0.81}{GW200105_162426}{0.79}{GW191105_143521}{0.78}{GW191109_010717}{0.94}{GW200209_085452}{0.90}{GW200115_042309}{0.84}{GW191127_050227}{0.91}{GW200216_220804}{0.90}{GW191215_223052}{0.86}{GW200208_130117}{0.85}{GW200219_094415}{0.88}{GW191103_012549}{0.88}{GW200316_215756}{0.83}{GW200202_154313}{0.77}{GW200129_065458}{0.86}{GW191216_213338}{0.76}}[{\red{???}}]}
\DeclareRobustCommand{\spinonezgwtcthreeminus}[1]{\IfEqCase{#1}{{GW200224_222234}{0.31}{GW191129_134029}{0.20}{GW200311_115853}{0.41}{GW191230_180458}{0.49}{GW191222_033537}{0.45}{GW200225_060421}{0.48}{GW200302_015811}{0.36}{GW200128_022011}{0.38}{GW191204_171526}{0.27}{GW200112_155838}{0.31}{GW200105_162426}{0.21}{GW191105_143521}{0.30}{GW191109_010717}{0.41}{GW200209_085452}{0.52}{GW200115_042309}{0.55}{GW191127_050227}{0.45}{GW200216_220804}{0.44}{GW191215_223052}{0.39}{GW200208_130117}{0.48}{GW200219_094415}{0.50}{GW191103_012549}{0.32}{GW200316_215756}{0.24}{GW200202_154313}{0.23}{GW200129_065458}{0.35}{GW191216_213338}{0.20}}[{\red{???}}]}
\DeclareRobustCommand{\spinonezgwtcthreemed}[1]{\IfEqCase{#1}{{GW200224_222234}{0.11}{GW191129_134029}{0.05}{GW200311_115853}{-0.02}{GW191230_180458}{-0.01}{GW191222_033537}{-0.03}{GW200225_060421}{-0.13}{GW200302_015811}{0.01}{GW200128_022011}{0.17}{GW191204_171526}{0.16}{GW200112_155838}{0.03}{GW200105_162426}{0.00}{GW191105_143521}{-0.01}{GW191109_010717}{-0.44}{GW200209_085452}{-0.06}{GW200115_042309}{-0.15}{GW191127_050227}{0.18}{GW200216_220804}{0.07}{GW191215_223052}{-0.02}{GW200208_130117}{-0.05}{GW200219_094415}{-0.07}{GW191103_012549}{0.24}{GW200316_215756}{0.12}{GW200202_154313}{0.02}{GW200129_065458}{0.05}{GW191216_213338}{0.11}}[{\red{???}}]}
\DeclareRobustCommand{\spinonezgwtcthreeplus}[1]{\IfEqCase{#1}{{GW200224_222234}{0.42}{GW191129_134029}{0.24}{GW200311_115853}{0.31}{GW191230_180458}{0.44}{GW191222_033537}{0.33}{GW200225_060421}{0.32}{GW200302_015811}{0.43}{GW200128_022011}{0.47}{GW191204_171526}{0.21}{GW200112_155838}{0.36}{GW200105_162426}{0.16}{GW191105_143521}{0.21}{GW191109_010717}{0.60}{GW200209_085452}{0.37}{GW200115_042309}{0.25}{GW191127_050227}{0.51}{GW200216_220804}{0.53}{GW191215_223052}{0.29}{GW200208_130117}{0.31}{GW200219_094415}{0.38}{GW191103_012549}{0.29}{GW200316_215756}{0.34}{GW200202_154313}{0.23}{GW200129_065458}{0.41}{GW191216_213338}{0.26}}[{\red{???}}]}
\DeclareRobustCommand{\spinonezgwtcthreetenthpercentile}[1]{\IfEqCase{#1}{{GW200224_222234}{-0.12}{GW191129_134029}{-0.08}{GW200311_115853}{-0.32}{GW191230_180458}{-0.37}{GW191222_033537}{-0.37}{GW200225_060421}{-0.52}{GW200302_015811}{-0.25}{GW200128_022011}{-0.12}{GW191204_171526}{-0.04}{GW200112_155838}{-0.18}{GW200105_162426}{-0.11}{GW191105_143521}{-0.22}{GW191109_010717}{-0.79}{GW200209_085452}{-0.46}{GW200115_042309}{-0.62}{GW191127_050227}{-0.15}{GW200216_220804}{-0.24}{GW191215_223052}{-0.30}{GW200208_130117}{-0.42}{GW200219_094415}{-0.45}{GW191103_012549}{-0.02}{GW200316_215756}{-0.05}{GW200202_154313}{-0.12}{GW200129_065458}{-0.22}{GW191216_213338}{-0.03}}[{\red{???}}]}
\DeclareRobustCommand{\spinonezgwtcthreenintiethpercentile}[1]{\IfEqCase{#1}{{GW200224_222234}{0.43}{GW191129_134029}{0.23}{GW200311_115853}{0.20}{GW191230_180458}{0.31}{GW191222_033537}{0.20}{GW200225_060421}{0.11}{GW200302_015811}{0.33}{GW200128_022011}{0.54}{GW191204_171526}{0.33}{GW200112_155838}{0.29}{GW200105_162426}{0.09}{GW191105_143521}{0.14}{GW191109_010717}{0.01}{GW200209_085452}{0.21}{GW200115_042309}{0.05}{GW191127_050227}{0.60}{GW200216_220804}{0.49}{GW191215_223052}{0.20}{GW200208_130117}{0.16}{GW200219_094415}{0.20}{GW191103_012549}{0.46}{GW200316_215756}{0.39}{GW200202_154313}{0.20}{GW200129_065458}{0.36}{GW191216_213338}{0.31}}[{\red{???}}]}
\DeclareRobustCommand{\spintwozgwtcthreeminus}[1]{\IfEqCase{#1}{{GW200224_222234}{0.44}{GW191129_134029}{0.30}{GW200311_115853}{0.45}{GW191230_180458}{0.56}{GW191222_033537}{0.50}{GW200225_060421}{0.55}{GW200302_015811}{0.47}{GW200128_022011}{0.47}{GW191204_171526}{0.35}{GW200112_155838}{0.35}{GW200105_162426}{0.42}{GW191105_143521}{0.34}{GW191109_010717}{0.58}{GW200209_085452}{0.58}{GW200115_042309}{0.59}{GW191127_050227}{0.52}{GW200216_220804}{0.52}{GW191215_223052}{0.50}{GW200208_130117}{0.52}{GW200219_094415}{0.57}{GW191103_012549}{0.42}{GW200316_215756}{0.35}{GW200202_154313}{0.28}{GW200129_065458}{0.50}{GW191216_213338}{0.37}}[{\red{???}}]}
\DeclareRobustCommand{\spintwozgwtcthreemed}[1]{\IfEqCase{#1}{{GW200224_222234}{0.05}{GW191129_134029}{0.07}{GW200311_115853}{0.00}{GW191230_180458}{-0.03}{GW191222_033537}{-0.02}{GW200225_060421}{-0.05}{GW200302_015811}{0.03}{GW200128_022011}{0.04}{GW191204_171526}{0.17}{GW200112_155838}{0.06}{GW200105_162426}{0.00}{GW191105_143521}{0.00}{GW191109_010717}{-0.10}{GW200209_085452}{-0.08}{GW200115_042309}{-0.08}{GW191127_050227}{0.06}{GW200216_220804}{0.05}{GW191215_223052}{-0.02}{GW200208_130117}{-0.03}{GW200219_094415}{-0.04}{GW191103_012549}{0.18}{GW200316_215756}{0.12}{GW200202_154313}{0.05}{GW200129_065458}{0.15}{GW191216_213338}{0.11}}[{\red{???}}]}
\DeclareRobustCommand{\spintwozgwtcthreeplus}[1]{\IfEqCase{#1}{{GW200224_222234}{0.45}{GW191129_134029}{0.50}{GW200311_115853}{0.41}{GW191230_180458}{0.46}{GW191222_033537}{0.42}{GW200225_060421}{0.40}{GW200302_015811}{0.54}{GW200128_022011}{0.52}{GW191204_171526}{0.43}{GW200112_155838}{0.46}{GW200105_162426}{0.38}{GW191105_143521}{0.42}{GW191109_010717}{0.62}{GW200209_085452}{0.41}{GW200115_042309}{0.42}{GW191127_050227}{0.64}{GW200216_220804}{0.61}{GW191215_223052}{0.41}{GW200208_130117}{0.42}{GW200219_094415}{0.47}{GW191103_012549}{0.52}{GW200316_215756}{0.49}{GW200202_154313}{0.42}{GW200129_065458}{0.49}{GW191216_213338}{0.41}}[{\red{???}}]}
\DeclareRobustCommand{\spintwozgwtcthreetenthpercentile}[1]{\IfEqCase{#1}{{GW200224_222234}{-0.25}{GW191129_134029}{-0.13}{GW200311_115853}{-0.33}{GW191230_180458}{-0.46}{GW191222_033537}{-0.39}{GW200225_060421}{-0.47}{GW200302_015811}{-0.31}{GW200128_022011}{-0.30}{GW191204_171526}{-0.09}{GW200112_155838}{-0.19}{GW200105_162426}{-0.26}{GW191105_143521}{-0.24}{GW191109_010717}{-0.56}{GW200209_085452}{-0.53}{GW200115_042309}{-0.57}{GW191127_050227}{-0.31}{GW200216_220804}{-0.32}{GW191215_223052}{-0.39}{GW200208_130117}{-0.44}{GW200219_094415}{-0.48}{GW191103_012549}{-0.13}{GW200316_215756}{-0.12}{GW200202_154313}{-0.13}{GW200129_065458}{-0.21}{GW191216_213338}{-0.14}}[{\red{???}}]}
\DeclareRobustCommand{\spintwozgwtcthreenintiethpercentile}[1]{\IfEqCase{#1}{{GW200224_222234}{0.40}{GW191129_134029}{0.44}{GW200311_115853}{0.29}{GW191230_180458}{0.31}{GW191222_033537}{0.28}{GW200225_060421}{0.22}{GW200302_015811}{0.44}{GW200128_022011}{0.45}{GW191204_171526}{0.51}{GW200112_155838}{0.40}{GW200105_162426}{0.26}{GW191105_143521}{0.29}{GW191109_010717}{0.34}{GW200209_085452}{0.21}{GW200115_042309}{0.22}{GW191127_050227}{0.58}{GW200216_220804}{0.53}{GW191215_223052}{0.27}{GW200208_130117}{0.27}{GW200219_094415}{0.29}{GW191103_012549}{0.60}{GW200316_215756}{0.50}{GW200202_154313}{0.35}{GW200129_065458}{0.53}{GW191216_213338}{0.43}}[{\red{???}}]}
\DeclareRobustCommand{\massonedetgwtcthreeminus}[1]{\IfEqCase{#1}{{GW200224_222234}{5.4}{GW191129_134029}{2.3}{GW200311_115853}{4.6}{GW191230_180458}{13}{GW191222_033537}{9.5}{GW200225_060421}{3.2}{GW200302_015811}{9.7}{GW200128_022011}{9.9}{GW191204_171526}{2.0}{GW200112_155838}{5.2}{GW200105_162426}{1.8}{GW191105_143521}{1.8}{GW191109_010717}{8.9}{GW200209_085452}{9.9}{GW200115_042309}{2.7}{GW191127_050227}{37}{GW200216_220804}{21}{GW191215_223052}{4.8}{GW200208_130117}{8.4}{GW200219_094415}{8.9}{GW191103_012549}{2.3}{GW200316_215756}{3.4}{GW200202_154313}{1.5}{GW200129_065458}{3.3}{GW191216_213338}{2.4}}[{\red{???}}]}
\DeclareRobustCommand{\massonedetgwtcthreemed}[1]{\IfEqCase{#1}{{GW200224_222234}{52.5}{GW191129_134029}{12.3}{GW200311_115853}{41.8}{GW191230_180458}{83}{GW191222_033537}{67.2}{GW200225_060421}{23.6}{GW200302_015811}{48.5}{GW200128_022011}{66.1}{GW191204_171526}{13.4}{GW200112_155838}{44.0}{GW200105_162426}{9.6}{GW191105_143521}{13.0}{GW191109_010717}{81.2}{GW200209_085452}{56.5}{GW200115_042309}{6.3}{GW191127_050227}{86}{GW200216_220804}{84}{GW191215_223052}{33.7}{GW200208_130117}{53.0}{GW200219_094415}{58.6}{GW191103_012549}{14.0}{GW200316_215756}{16.0}{GW200202_154313}{11.0}{GW200129_065458}{40.2}{GW191216_213338}{13.0}}[{\red{???}}]}
\DeclareRobustCommand{\massonedetgwtcthreeplus}[1]{\IfEqCase{#1}{{GW200224_222234}{9.1}{GW191129_134029}{4.9}{GW200311_115853}{8.2}{GW191230_180458}{19}{GW191222_033537}{14.7}{GW200225_060421}{5.6}{GW200302_015811}{10.8}{GW200128_022011}{16.6}{GW191204_171526}{3.8}{GW200112_155838}{8.2}{GW200105_162426}{1.9}{GW191105_143521}{4.5}{GW191109_010717}{12.9}{GW200209_085452}{15.4}{GW200115_042309}{2.1}{GW191127_050227}{60}{GW200216_220804}{28}{GW191215_223052}{9.4}{GW200208_130117}{12.2}{GW200219_094415}{13.5}{GW191103_012549}{7.4}{GW200316_215756}{12.3}{GW200202_154313}{3.8}{GW200129_065458}{12.2}{GW191216_213338}{5.0}}[{\red{???}}]}
\DeclareRobustCommand{\massonedetgwtcthreetenthpercentile}[1]{\IfEqCase{#1}{{GW200224_222234}{48.1}{GW191129_134029}{10.2}{GW200311_115853}{38.1}{GW191230_180458}{73}{GW191222_033537}{59.4}{GW200225_060421}{20.8}{GW200302_015811}{40.7}{GW200128_022011}{58.0}{GW191204_171526}{11.7}{GW200112_155838}{39.8}{GW200105_162426}{8.5}{GW191105_143521}{11.4}{GW191109_010717}{74.0}{GW200209_085452}{48.5}{GW200115_042309}{3.9}{GW191127_050227}{55}{GW200216_220804}{68}{GW191215_223052}{29.6}{GW200208_130117}{46.1}{GW200219_094415}{51.1}{GW191103_012549}{11.9}{GW200316_215756}{13.0}{GW200202_154313}{9.7}{GW200129_065458}{37.5}{GW191216_213338}{10.8}}[{\red{???}}]}
\DeclareRobustCommand{\massonedetgwtcthreenintiethpercentile}[1]{\IfEqCase{#1}{{GW200224_222234}{59.3}{GW191129_134029}{15.9}{GW200311_115853}{47.8}{GW191230_180458}{98}{GW191222_033537}{77.9}{GW200225_060421}{27.9}{GW200302_015811}{56.6}{GW200128_022011}{78.3}{GW191204_171526}{16.2}{GW200112_155838}{50.3}{GW200105_162426}{10.6}{GW191105_143521}{16.3}{GW191109_010717}{90.3}{GW200209_085452}{67.7}{GW200115_042309}{7.9}{GW191127_050227}{133}{GW200216_220804}{104}{GW191215_223052}{40.5}{GW200208_130117}{62.2}{GW200219_094415}{69.0}{GW191103_012549}{19.3}{GW200316_215756}{24.2}{GW200202_154313}{13.8}{GW200129_065458}{50.4}{GW191216_213338}{16.3}}[{\red{???}}]}
\DeclareRobustCommand{\chieffinfinityonlyprecavggwtcthreeminus}[1]{\IfEqCase{#1}{{GW200224_222234}{0.15}{GW191129_134029}{0.08}{GW200311_115853}{0.20}{GW191230_180458}{0.30}{GW191222_033537}{0.25}{GW200225_060421}{0.28}{GW200302_015811}{0.25}{GW200128_022011}{0.25}{GW191204_171526}{0.05}{GW200112_155838}{0.15}{GW200105_162426}{0.18}{GW191105_143521}{0.09}{GW191109_010717}{0.31}{GW200209_085452}{0.30}{GW200115_042309}{0.41}{GW191127_050227}{0.36}{GW200216_220804}{0.36}{GW191215_223052}{0.21}{GW200208_130117}{0.27}{GW200219_094415}{0.29}{GW191103_012549}{0.10}{GW200316_215756}{0.10}{GW200202_154313}{0.06}{GW200129_065458}{0.16}{GW191216_213338}{0.06}}[{\red{???}}]}
\DeclareRobustCommand{\chieffinfinityonlyprecavggwtcthreemed}[1]{\IfEqCase{#1}{{GW200224_222234}{0.11}{GW191129_134029}{0.06}{GW200311_115853}{-0.02}{GW191230_180458}{-0.03}{GW191222_033537}{-0.04}{GW200225_060421}{-0.12}{GW200302_015811}{0.03}{GW200128_022011}{0.14}{GW191204_171526}{0.16}{GW200112_155838}{0.06}{GW200105_162426}{0.00}{GW191105_143521}{-0.02}{GW191109_010717}{-0.29}{GW200209_085452}{-0.10}{GW200115_042309}{-0.15}{GW191127_050227}{0.18}{GW200216_220804}{0.10}{GW191215_223052}{-0.03}{GW200208_130117}{-0.07}{GW200219_094415}{-0.08}{GW191103_012549}{0.21}{GW200316_215756}{0.13}{GW200202_154313}{0.04}{GW200129_065458}{0.11}{GW191216_213338}{0.11}}[{\red{???}}]}
\DeclareRobustCommand{\chieffinfinityonlyprecavggwtcthreeplus}[1]{\IfEqCase{#1}{{GW200224_222234}{0.15}{GW191129_134029}{0.16}{GW200311_115853}{0.16}{GW191230_180458}{0.26}{GW191222_033537}{0.20}{GW200225_060421}{0.17}{GW200302_015811}{0.26}{GW200128_022011}{0.24}{GW191204_171526}{0.08}{GW200112_155838}{0.15}{GW200105_162426}{0.13}{GW191105_143521}{0.13}{GW191109_010717}{0.42}{GW200209_085452}{0.24}{GW200115_042309}{0.24}{GW191127_050227}{0.34}{GW200216_220804}{0.34}{GW191215_223052}{0.17}{GW200208_130117}{0.22}{GW200219_094415}{0.23}{GW191103_012549}{0.16}{GW200316_215756}{0.27}{GW200202_154313}{0.13}{GW200129_065458}{0.11}{GW191216_213338}{0.13}}[{\red{???}}]}
\DeclareRobustCommand{\chieffinfinityonlyprecavggwtcthreetenthpercentile}[1]{\IfEqCase{#1}{{GW200224_222234}{-0.01}{GW191129_134029}{0.00}{GW200311_115853}{-0.17}{GW191230_180458}{-0.26}{GW191222_033537}{-0.23}{GW200225_060421}{-0.34}{GW200302_015811}{-0.15}{GW200128_022011}{-0.05}{GW191204_171526}{0.12}{GW200112_155838}{-0.05}{GW200105_162426}{-0.10}{GW191105_143521}{-0.09}{GW191109_010717}{-0.54}{GW200209_085452}{-0.33}{GW200115_042309}{-0.51}{GW191127_050227}{-0.10}{GW200216_220804}{-0.17}{GW191215_223052}{-0.19}{GW200208_130117}{-0.27}{GW200219_094415}{-0.30}{GW191103_012549}{0.13}{GW200316_215756}{0.04}{GW200202_154313}{-0.01}{GW200129_065458}{0.00}{GW191216_213338}{0.06}}[{\red{???}}]}
\DeclareRobustCommand{\chieffinfinityonlyprecavggwtcthreenintiethpercentile}[1]{\IfEqCase{#1}{{GW200224_222234}{0.22}{GW191129_134029}{0.19}{GW200311_115853}{0.10}{GW191230_180458}{0.18}{GW191222_033537}{0.11}{GW200225_060421}{0.02}{GW200302_015811}{0.23}{GW200128_022011}{0.33}{GW191204_171526}{0.22}{GW200112_155838}{0.18}{GW200105_162426}{0.08}{GW191105_143521}{0.07}{GW191109_010717}{0.00}{GW200209_085452}{0.09}{GW200115_042309}{0.04}{GW191127_050227}{0.45}{GW200216_220804}{0.36}{GW191215_223052}{0.10}{GW200208_130117}{0.09}{GW200219_094415}{0.10}{GW191103_012549}{0.33}{GW200316_215756}{0.32}{GW200202_154313}{0.13}{GW200129_065458}{0.20}{GW191216_213338}{0.20}}[{\red{???}}]}
\DeclareRobustCommand{\chipinfinityonlyprecavggwtcthreeminus}[1]{\IfEqCase{#1}{{GW200224_222234}{0.37}{GW191129_134029}{0.19}{GW200311_115853}{0.35}{GW191230_180458}{0.39}{GW191222_033537}{0.32}{GW200225_060421}{0.38}{GW200302_015811}{0.29}{GW200128_022011}{0.40}{GW191204_171526}{0.26}{GW200112_155838}{0.30}{GW200105_162426}{0.07}{GW191105_143521}{0.24}{GW191109_010717}{0.38}{GW200209_085452}{0.38}{GW200115_042309}{0.16}{GW191127_050227}{0.41}{GW200216_220804}{0.35}{GW191215_223052}{0.38}{GW200208_130117}{0.29}{GW200219_094415}{0.35}{GW191103_012549}{0.27}{GW200316_215756}{0.20}{GW200202_154313}{0.22}{GW200129_065458}{0.39}{GW191216_213338}{0.15}}[{\red{???}}]}
\DeclareRobustCommand{\chipinfinityonlyprecavggwtcthreemed}[1]{\IfEqCase{#1}{{GW200224_222234}{0.50}{GW191129_134029}{0.26}{GW200311_115853}{0.45}{GW191230_180458}{0.52}{GW191222_033537}{0.41}{GW200225_060421}{0.53}{GW200302_015811}{0.38}{GW200128_022011}{0.56}{GW191204_171526}{0.39}{GW200112_155838}{0.40}{GW200105_162426}{0.09}{GW191105_143521}{0.30}{GW191109_010717}{0.63}{GW200209_085452}{0.52}{GW200115_042309}{0.20}{GW191127_050227}{0.52}{GW200216_220804}{0.45}{GW191215_223052}{0.50}{GW200208_130117}{0.38}{GW200219_094415}{0.47}{GW191103_012549}{0.40}{GW200316_215756}{0.29}{GW200202_154313}{0.28}{GW200129_065458}{0.54}{GW191216_213338}{0.23}}[{\red{???}}]}
\DeclareRobustCommand{\chipinfinityonlyprecavggwtcthreeplus}[1]{\IfEqCase{#1}{{GW200224_222234}{0.37}{GW191129_134029}{0.36}{GW200311_115853}{0.39}{GW191230_180458}{0.38}{GW191222_033537}{0.42}{GW200225_060421}{0.35}{GW200302_015811}{0.44}{GW200128_022011}{0.34}{GW191204_171526}{0.35}{GW200112_155838}{0.38}{GW200105_162426}{0.17}{GW191105_143521}{0.45}{GW191109_010717}{0.28}{GW200209_085452}{0.38}{GW200115_042309}{0.34}{GW191127_050227}{0.40}{GW200216_220804}{0.42}{GW191215_223052}{0.38}{GW200208_130117}{0.42}{GW200219_094415}{0.40}{GW191103_012549}{0.41}{GW200316_215756}{0.38}{GW200202_154313}{0.40}{GW200129_065458}{0.39}{GW191216_213338}{0.35}}[{\red{???}}]}
\DeclareRobustCommand{\chipinfinityonlyprecavggwtcthreetenthpercentile}[1]{\IfEqCase{#1}{{GW200224_222234}{0.20}{GW191129_134029}{0.10}{GW200311_115853}{0.16}{GW191230_180458}{0.20}{GW191222_033537}{0.14}{GW200225_060421}{0.22}{GW200302_015811}{0.13}{GW200128_022011}{0.23}{GW191204_171526}{0.18}{GW200112_155838}{0.14}{GW200105_162426}{0.03}{GW191105_143521}{0.09}{GW191109_010717}{0.33}{GW200209_085452}{0.20}{GW200115_042309}{0.06}{GW191127_050227}{0.17}{GW200216_220804}{0.15}{GW191215_223052}{0.19}{GW200208_130117}{0.13}{GW200219_094415}{0.18}{GW191103_012549}{0.18}{GW200316_215756}{0.12}{GW200202_154313}{0.09}{GW200129_065458}{0.21}{GW191216_213338}{0.10}}[{\red{???}}]}
\DeclareRobustCommand{\chipinfinityonlyprecavggwtcthreenintiethpercentile}[1]{\IfEqCase{#1}{{GW200224_222234}{0.80}{GW191129_134029}{0.54}{GW200311_115853}{0.77}{GW191230_180458}{0.84}{GW191222_033537}{0.75}{GW200225_060421}{0.82}{GW200302_015811}{0.74}{GW200128_022011}{0.84}{GW191204_171526}{0.67}{GW200112_155838}{0.70}{GW200105_162426}{0.19}{GW191105_143521}{0.65}{GW191109_010717}{0.87}{GW200209_085452}{0.84}{GW200115_042309}{0.46}{GW191127_050227}{0.88}{GW200216_220804}{0.80}{GW191215_223052}{0.82}{GW200208_130117}{0.71}{GW200219_094415}{0.80}{GW191103_012549}{0.72}{GW200316_215756}{0.58}{GW200202_154313}{0.59}{GW200129_065458}{0.91}{GW191216_213338}{0.48}}[{\red{???}}]}
\DeclareRobustCommand{\logpriorgwtcthreeminus}[1]{\IfEqCase{#1}{{GW200224_222234}{10.9}{GW200311_115853}{10.8}{GW191230_180458}{10.8}{GW191222_033537}{9.0}{GW200225_060421}{9.3}{GW200302_015811}{9.0}{GW200128_022011}{9.2}{GW200112_155838}{9.0}{GW191105_143521}{11.1}{GW191109_010717}{9.4}{GW200209_085452}{11.0}{GW191127_050227}{11.0}{GW200216_220804}{10.5}{GW191215_223052}{10.9}{GW200208_130117}{10.9}{GW200219_094415}{10.8}{GW191103_012549}{9.0}{GW200202_154313}{10.9}{GW200129_065458}{11.1}}[{\red{???}}]}
\DeclareRobustCommand{\logpriorgwtcthreemed}[1]{\IfEqCase{#1}{{GW200224_222234}{127.6}{GW200311_115853}{132.7}{GW191230_180458}{129.7}{GW191222_033537}{90.7}{GW200225_060421}{100.3}{GW200302_015811}{82.6}{GW200128_022011}{93.6}{GW200112_155838}{72.1}{GW191105_143521}{129.4}{GW191109_010717}{90.9}{GW200209_085452}{132.4}{GW191127_050227}{129.5}{GW200216_220804}{131.3}{GW191215_223052}{131.5}{GW200208_130117}{128.4}{GW200219_094415}{132.2}{GW191103_012549}{100.7}{GW200202_154313}{135.3}{GW200129_065458}{132.8}}[{\red{???}}]}
\DeclareRobustCommand{\logpriorgwtcthreeplus}[1]{\IfEqCase{#1}{{GW200224_222234}{8.7}{GW200311_115853}{8.6}{GW191230_180458}{8.8}{GW191222_033537}{6.9}{GW200225_060421}{6.9}{GW200302_015811}{6.9}{GW200128_022011}{6.9}{GW200112_155838}{6.8}{GW191105_143521}{8.8}{GW191109_010717}{7.2}{GW200209_085452}{8.7}{GW191127_050227}{8.9}{GW200216_220804}{8.7}{GW191215_223052}{8.8}{GW200208_130117}{8.7}{GW200219_094415}{8.8}{GW191103_012549}{6.9}{GW200202_154313}{8.7}{GW200129_065458}{8.8}}[{\red{???}}]}
\DeclareRobustCommand{\logpriorgwtcthreetenthpercentile}[1]{\IfEqCase{#1}{{GW200224_222234}{119.3}{GW200311_115853}{124.6}{GW191230_180458}{121.4}{GW191222_033537}{83.9}{GW200225_060421}{93.4}{GW200302_015811}{75.8}{GW200128_022011}{86.7}{GW200112_155838}{65.4}{GW191105_143521}{121.1}{GW191109_010717}{83.8}{GW200209_085452}{124.1}{GW191127_050227}{121.2}{GW200216_220804}{123.3}{GW191215_223052}{123.1}{GW200208_130117}{120.2}{GW200219_094415}{124.0}{GW191103_012549}{93.9}{GW200202_154313}{127.1}{GW200129_065458}{124.4}}[{\red{???}}]}
\DeclareRobustCommand{\logpriorgwtcthreenintiethpercentile}[1]{\IfEqCase{#1}{{GW200224_222234}{134.5}{GW200311_115853}{139.6}{GW191230_180458}{136.7}{GW191222_033537}{96.3}{GW200225_060421}{105.9}{GW200302_015811}{88.2}{GW200128_022011}{99.2}{GW200112_155838}{77.6}{GW191105_143521}{136.5}{GW191109_010717}{96.6}{GW200209_085452}{139.4}{GW191127_050227}{136.7}{GW200216_220804}{138.2}{GW191215_223052}{138.6}{GW200208_130117}{135.4}{GW200219_094415}{139.2}{GW191103_012549}{106.3}{GW200202_154313}{142.2}{GW200129_065458}{139.9}}[{\red{???}}]}
\DeclareRobustCommand{\networkmatchedfiltersnrgwtcthreeminus}[1]{\IfEqCase{#1}{{GW200224_222234}{0.2}{GW191129_134029}{0.3}{GW200311_115853}{0.2}{GW191230_180458}{0.4}{GW191222_033537}{0.3}{GW200225_060421}{0.4}{GW200302_015811}{0.4}{GW200128_022011}{0.4}{GW191204_171526}{0.2}{GW200112_155838}{0.2}{GW200105_162426}{0.4}{GW191105_143521}{0.5}{GW191109_010717}{0.5}{GW200209_085452}{0.5}{GW200115_042309}{0.5}{GW191127_050227}{0.6}{GW200216_220804}{0.6}{GW191215_223052}{0.4}{GW200208_130117}{0.5}{GW200219_094415}{0.5}{GW191103_012549}{0.5}{GW200316_215756}{0.7}{GW200202_154313}{0.4}{GW200129_065458}{0.2}{GW191216_213338}{0.2}}[{\red{???}}]}
\DeclareRobustCommand{\networkmatchedfiltersnrgwtcthreemed}[1]{\IfEqCase{#1}{{GW200224_222234}{20.0}{GW191129_134029}{13.1}{GW200311_115853}{17.8}{GW191230_180458}{10.4}{GW191222_033537}{12.5}{GW200225_060421}{12.5}{GW200302_015811}{10.8}{GW200128_022011}{10.6}{GW191204_171526}{17.5}{GW200112_155838}{19.8}{GW200105_162426}{13.7}{GW191105_143521}{9.7}{GW191109_010717}{17.2}{GW200209_085452}{9.6}{GW200115_042309}{11.3}{GW191127_050227}{9.1}{GW200216_220804}{8.1}{GW191215_223052}{11.2}{GW200208_130117}{10.8}{GW200219_094415}{10.7}{GW191103_012549}{8.9}{GW200316_215756}{10.3}{GW200202_154313}{10.8}{GW200129_065458}{26.8}{GW191216_213338}{18.6}}[{\red{???}}]}
\DeclareRobustCommand{\networkmatchedfiltersnrgwtcthreeplus}[1]{\IfEqCase{#1}{{GW200224_222234}{0.2}{GW191129_134029}{0.2}{GW200311_115853}{0.2}{GW191230_180458}{0.3}{GW191222_033537}{0.2}{GW200225_060421}{0.3}{GW200302_015811}{0.3}{GW200128_022011}{0.3}{GW191204_171526}{0.2}{GW200112_155838}{0.1}{GW200105_162426}{0.2}{GW191105_143521}{0.3}{GW191109_010717}{0.5}{GW200209_085452}{0.4}{GW200115_042309}{0.3}{GW191127_050227}{0.5}{GW200216_220804}{0.3}{GW191215_223052}{0.3}{GW200208_130117}{0.3}{GW200219_094415}{0.3}{GW191103_012549}{0.3}{GW200316_215756}{0.4}{GW200202_154313}{0.2}{GW200129_065458}{0.2}{GW191216_213338}{0.2}}[{\red{???}}]}
\DeclareRobustCommand{\networkmatchedfiltersnrgwtcthreetenthpercentile}[1]{\IfEqCase{#1}{{GW200224_222234}{19.8}{GW191129_134029}{12.9}{GW200311_115853}{17.7}{GW191230_180458}{10.1}{GW191222_033537}{12.2}{GW200225_060421}{12.2}{GW200302_015811}{10.5}{GW200128_022011}{10.3}{GW191204_171526}{17.3}{GW200112_155838}{19.6}{GW200105_162426}{13.4}{GW191105_143521}{9.3}{GW191109_010717}{16.9}{GW200209_085452}{9.2}{GW200115_042309}{10.9}{GW191127_050227}{8.7}{GW200216_220804}{7.7}{GW191215_223052}{10.9}{GW200208_130117}{10.5}{GW200219_094415}{10.3}{GW191103_012549}{8.5}{GW200316_215756}{9.8}{GW200202_154313}{10.5}{GW200129_065458}{26.6}{GW191216_213338}{18.4}}[{\red{???}}]}
\DeclareRobustCommand{\networkmatchedfiltersnrgwtcthreenintiethpercentile}[1]{\IfEqCase{#1}{{GW200224_222234}{20.1}{GW191129_134029}{13.3}{GW200311_115853}{18.0}{GW191230_180458}{10.7}{GW191222_033537}{12.6}{GW200225_060421}{12.7}{GW200302_015811}{11.1}{GW200128_022011}{10.9}{GW191204_171526}{17.6}{GW200112_155838}{19.9}{GW200105_162426}{13.9}{GW191105_143521}{9.9}{GW191109_010717}{17.7}{GW200209_085452}{9.9}{GW200115_042309}{11.5}{GW191127_050227}{9.5}{GW200216_220804}{8.4}{GW191215_223052}{11.5}{GW200208_130117}{11.0}{GW200219_094415}{10.9}{GW191103_012549}{9.1}{GW200316_215756}{10.6}{GW200202_154313}{11.0}{GW200129_065458}{26.9}{GW191216_213338}{18.7}}[{\red{???}}]}
\DeclareRobustCommand{\networkoptimalsnrgwtcthreeminus}[1]{\IfEqCase{#1}{{GW200224_222234}{1.7}{GW191129_134029}{1.7}{GW200311_115853}{1.7}{GW191230_180458}{1.7}{GW191222_033537}{1.7}{GW200225_060421}{1.7}{GW200302_015811}{1.7}{GW200128_022011}{1.7}{GW191204_171526}{1.7}{GW200112_155838}{1.7}{GW200105_162426}{1.7}{GW191105_143521}{1.8}{GW191109_010717}{1.7}{GW200209_085452}{1.7}{GW200115_042309}{1.7}{GW191127_050227}{1.8}{GW200216_220804}{1.7}{GW191215_223052}{1.7}{GW200208_130117}{1.7}{GW200219_094415}{1.8}{GW191103_012549}{1.8}{GW200316_215756}{1.8}{GW200202_154313}{1.7}{GW200129_065458}{1.6}{GW191216_213338}{1.7}}[{\red{???}}]}
\DeclareRobustCommand{\networkoptimalsnrgwtcthreemed}[1]{\IfEqCase{#1}{{GW200224_222234}{19.8}{GW191129_134029}{12.9}{GW200311_115853}{17.6}{GW191230_180458}{10.2}{GW191222_033537}{12.1}{GW200225_060421}{12.2}{GW200302_015811}{10.5}{GW200128_022011}{10.4}{GW191204_171526}{17.2}{GW200112_155838}{19.6}{GW200105_162426}{13.3}{GW191105_143521}{9.2}{GW191109_010717}{17.0}{GW200209_085452}{9.3}{GW200115_042309}{10.9}{GW191127_050227}{8.6}{GW200216_220804}{7.6}{GW191215_223052}{10.9}{GW200208_130117}{10.4}{GW200219_094415}{10.3}{GW191103_012549}{8.4}{GW200316_215756}{9.8}{GW200202_154313}{10.4}{GW200129_065458}{26.6}{GW191216_213338}{18.3}}[{\red{???}}]}
\DeclareRobustCommand{\networkoptimalsnrgwtcthreeplus}[1]{\IfEqCase{#1}{{GW200224_222234}{1.7}{GW191129_134029}{1.7}{GW200311_115853}{1.7}{GW191230_180458}{1.7}{GW191222_033537}{1.7}{GW200225_060421}{1.7}{GW200302_015811}{1.7}{GW200128_022011}{1.6}{GW191204_171526}{1.7}{GW200112_155838}{1.7}{GW200105_162426}{1.7}{GW191105_143521}{1.7}{GW191109_010717}{1.7}{GW200209_085452}{1.7}{GW200115_042309}{1.7}{GW191127_050227}{1.8}{GW200216_220804}{1.8}{GW191215_223052}{1.7}{GW200208_130117}{1.7}{GW200219_094415}{1.7}{GW191103_012549}{1.7}{GW200316_215756}{1.7}{GW200202_154313}{1.7}{GW200129_065458}{1.7}{GW191216_213338}{1.7}}[{\red{???}}]}
\DeclareRobustCommand{\networkoptimalsnrgwtcthreetenthpercentile}[1]{\IfEqCase{#1}{{GW200224_222234}{18.5}{GW191129_134029}{11.6}{GW200311_115853}{16.3}{GW191230_180458}{8.8}{GW191222_033537}{10.8}{GW200225_060421}{10.8}{GW200302_015811}{9.2}{GW200128_022011}{9.0}{GW191204_171526}{15.9}{GW200112_155838}{18.3}{GW200105_162426}{12.0}{GW191105_143521}{7.8}{GW191109_010717}{15.7}{GW200209_085452}{8.0}{GW200115_042309}{9.5}{GW191127_050227}{7.2}{GW200216_220804}{6.2}{GW191215_223052}{9.6}{GW200208_130117}{9.0}{GW200219_094415}{8.9}{GW191103_012549}{7.0}{GW200316_215756}{8.5}{GW200202_154313}{9.1}{GW200129_065458}{25.3}{GW191216_213338}{17.0}}[{\red{???}}]}
\DeclareRobustCommand{\networkoptimalsnrgwtcthreenintiethpercentile}[1]{\IfEqCase{#1}{{GW200224_222234}{21.1}{GW191129_134029}{14.2}{GW200311_115853}{18.9}{GW191230_180458}{11.5}{GW191222_033537}{13.4}{GW200225_060421}{13.5}{GW200302_015811}{11.8}{GW200128_022011}{11.7}{GW191204_171526}{18.5}{GW200112_155838}{20.9}{GW200105_162426}{14.6}{GW191105_143521}{10.5}{GW191109_010717}{18.4}{GW200209_085452}{10.7}{GW200115_042309}{12.2}{GW191127_050227}{10.0}{GW200216_220804}{8.9}{GW191215_223052}{12.2}{GW200208_130117}{11.7}{GW200219_094415}{11.6}{GW191103_012549}{9.7}{GW200316_215756}{11.2}{GW200202_154313}{11.8}{GW200129_065458}{27.9}{GW191216_213338}{19.6}}[{\red{???}}]}
\DeclareRobustCommand{\PEpercentBNSgwtcthree}[1]{\IfEqCase{#1}{{GW200224_222234}{0}{GW191129_134029}{0}{GW200311_115853}{0}{GW191230_180458}{0}{GW191222_033537}{0}{GW200225_060421}{0}{GW200302_015811}{0}{GW200128_022011}{0}{GW191204_171526}{0}{GW200112_155838}{0}{GW200105_162426}{0}{GW191105_143521}{0}{GW191109_010717}{0}{GW200209_085452}{0}{GW200115_042309}{1}{GW191127_050227}{0}{GW200216_220804}{0}{GW191215_223052}{0}{GW200208_130117}{0}{GW200219_094415}{0}{GW191103_012549}{0}{GW200316_215756}{0}{GW200202_154313}{0}{GW200129_065458}{0}{GW191216_213338}{0}}[{\red{???}}]}
\DeclareRobustCommand{\PEpercentNSBHgwtcthree}[1]{\IfEqCase{#1}{{GW200224_222234}{0}{GW191129_134029}{0}{GW200311_115853}{0}{GW191230_180458}{0}{GW191222_033537}{0}{GW200225_060421}{0}{GW200302_015811}{0}{GW200128_022011}{0}{GW191204_171526}{0}{GW200112_155838}{0}{GW200105_162426}{99}{GW191105_143521}{0}{GW191109_010717}{0}{GW200209_085452}{0}{GW200115_042309}{99}{GW191127_050227}{0}{GW200216_220804}{0}{GW191215_223052}{0}{GW200208_130117}{0}{GW200219_094415}{0}{GW191103_012549}{0}{GW200316_215756}{0}{GW200202_154313}{0}{GW200129_065458}{0}{GW191216_213338}{0}}[{\red{???}}]}
\DeclareRobustCommand{\PEpercentBBHgwtcthree}[1]{\IfEqCase{#1}{{GW200224_222234}{100}{GW191129_134029}{100}{GW200311_115853}{100}{GW191230_180458}{100}{GW191222_033537}{100}{GW200225_060421}{100}{GW200302_015811}{100}{GW200128_022011}{100}{GW191204_171526}{100}{GW200112_155838}{100}{GW200105_162426}{1}{GW191105_143521}{100}{GW191109_010717}{100}{GW200209_085452}{100}{GW200115_042309}{0}{GW191127_050227}{100}{GW200216_220804}{100}{GW191215_223052}{100}{GW200208_130117}{100}{GW200219_094415}{100}{GW191103_012549}{100}{GW200316_215756}{100}{GW200202_154313}{100}{GW200129_065458}{100}{GW191216_213338}{100}}[{\red{???}}]}
\DeclareRobustCommand{\PEpercentMassGapgwtcthree}[1]{\IfEqCase{#1}{{GW200224_222234}{0}{GW191129_134029}{0}{GW200311_115853}{0}{GW191230_180458}{0}{GW191222_033537}{0}{GW200225_060421}{0}{GW200302_015811}{0}{GW200128_022011}{0}{GW191204_171526}{0}{GW200112_155838}{0}{GW200105_162426}{0}{GW191105_143521}{0}{GW191109_010717}{0}{GW200209_085452}{0}{GW200115_042309}{0}{GW191127_050227}{0}{GW200216_220804}{0}{GW191215_223052}{0}{GW200208_130117}{0}{GW200219_094415}{0}{GW191103_012549}{0}{GW200316_215756}{0}{GW200202_154313}{0}{GW200129_065458}{0}{GW191216_213338}{0}}[{\red{???}}]}
\DeclareRobustCommand{\percentmassonelessthanthreegwtcthree}[1]{\IfEqCase{#1}{{GW200224_222234}{0}{GW191129_134029}{0}{GW200311_115853}{0}{GW191230_180458}{0}{GW191222_033537}{0}{GW200225_060421}{0}{GW200302_015811}{0}{GW200128_022011}{0}{GW191204_171526}{0}{GW200112_155838}{0}{GW200105_162426}{0}{GW191105_143521}{0}{GW191109_010717}{0}{GW200209_085452}{0}{GW200115_042309}{1}{GW191127_050227}{0}{GW200216_220804}{0}{GW191215_223052}{0}{GW200208_130117}{0}{GW200219_094415}{0}{GW191103_012549}{0}{GW200316_215756}{0}{GW200202_154313}{0}{GW200129_065458}{0}{GW191216_213338}{0}}[{\red{???}}]}
\DeclareRobustCommand{\percentmasstwolessthanthreegwtcthree}[1]{\IfEqCase{#1}{{GW200224_222234}{0}{GW191129_134029}{0}{GW200311_115853}{0}{GW191230_180458}{0}{GW191222_033537}{0}{GW200225_060421}{0}{GW200302_015811}{0}{GW200128_022011}{0}{GW191204_171526}{0}{GW200112_155838}{0}{GW200105_162426}{99}{GW191105_143521}{0}{GW191109_010717}{0}{GW200209_085452}{0}{GW200115_042309}{100}{GW191127_050227}{0}{GW200216_220804}{0}{GW191215_223052}{0}{GW200208_130117}{0}{GW200219_094415}{0}{GW191103_012549}{0}{GW200316_215756}{0}{GW200202_154313}{0}{GW200129_065458}{0}{GW191216_213338}{0}}[{\red{???}}]}
\DeclareRobustCommand{\percentmassonelessthanfivegwtcthree}[1]{\IfEqCase{#1}{{GW200224_222234}{0}{GW191129_134029}{0}{GW200311_115853}{0}{GW191230_180458}{0}{GW191222_033537}{0}{GW200225_060421}{0}{GW200302_015811}{0}{GW200128_022011}{0}{GW191204_171526}{0}{GW200112_155838}{0}{GW200105_162426}{1}{GW191105_143521}{0}{GW191109_010717}{0}{GW200209_085452}{0}{GW200115_042309}{29}{GW191127_050227}{0}{GW200216_220804}{0}{GW191215_223052}{0}{GW200208_130117}{0}{GW200219_094415}{0}{GW191103_012549}{0}{GW200316_215756}{0}{GW200202_154313}{0}{GW200129_065458}{0}{GW191216_213338}{0}}[{\red{???}}]}
\DeclareRobustCommand{\percentmasstwolessthanfivegwtcthree}[1]{\IfEqCase{#1}{{GW200224_222234}{0}{GW191129_134029}{5}{GW200311_115853}{0}{GW191230_180458}{0}{GW191222_033537}{0}{GW200225_060421}{0}{GW200302_015811}{0}{GW200128_022011}{0}{GW191204_171526}{0}{GW200112_155838}{0}{GW200105_162426}{100}{GW191105_143521}{1}{GW191109_010717}{0}{GW200209_085452}{0}{GW200115_042309}{100}{GW191127_050227}{0}{GW200216_220804}{0}{GW191215_223052}{0}{GW200208_130117}{0}{GW200219_094415}{0}{GW191103_012549}{2}{GW200316_215756}{5}{GW200202_154313}{1}{GW200129_065458}{0}{GW191216_213338}{2}}[{\red{???}}]}
\DeclareRobustCommand{\percentmassonemorethansixtyfivegwtcthree}[1]{\IfEqCase{#1}{{GW200224_222234}{0}{GW191129_134029}{0}{GW200311_115853}{0}{GW191230_180458}{2}{GW191222_033537}{0}{GW200225_060421}{0}{GW200302_015811}{0}{GW200128_022011}{0}{GW191204_171526}{0}{GW200112_155838}{0}{GW200105_162426}{0}{GW191105_143521}{0}{GW191109_010717}{51}{GW200209_085452}{0}{GW200115_042309}{0}{GW191127_050227}{30}{GW200216_220804}{13}{GW191215_223052}{0}{GW200208_130117}{0}{GW200219_094415}{0}{GW191103_012549}{0}{GW200316_215756}{0}{GW200202_154313}{0}{GW200129_065458}{0}{GW191216_213338}{0}}[{\red{???}}]}
\DeclareRobustCommand{\percentmasstwomorethansixtyfivegwtcthree}[1]{\IfEqCase{#1}{{GW200224_222234}{0}{GW191129_134029}{0}{GW200311_115853}{0}{GW191230_180458}{0}{GW191222_033537}{0}{GW200225_060421}{0}{GW200302_015811}{0}{GW200128_022011}{0}{GW191204_171526}{0}{GW200112_155838}{0}{GW200105_162426}{0}{GW191105_143521}{0}{GW191109_010717}{2}{GW200209_085452}{0}{GW200115_042309}{0}{GW191127_050227}{0}{GW200216_220804}{0}{GW191215_223052}{0}{GW200208_130117}{0}{GW200219_094415}{0}{GW191103_012549}{0}{GW200316_215756}{0}{GW200202_154313}{0}{GW200129_065458}{0}{GW191216_213338}{0}}[{\red{???}}]}
\DeclareRobustCommand{\percentmassonemorethanonetwentygwtcthree}[1]{\IfEqCase{#1}{{GW200224_222234}{0}{GW191129_134029}{0}{GW200311_115853}{0}{GW191230_180458}{0}{GW191222_033537}{0}{GW200225_060421}{0}{GW200302_015811}{0}{GW200128_022011}{0}{GW191204_171526}{0}{GW200112_155838}{0}{GW200105_162426}{0}{GW191105_143521}{0}{GW191109_010717}{0}{GW200209_085452}{0}{GW200115_042309}{0}{GW191127_050227}{1}{GW200216_220804}{0}{GW191215_223052}{0}{GW200208_130117}{0}{GW200219_094415}{0}{GW191103_012549}{0}{GW200316_215756}{0}{GW200202_154313}{0}{GW200129_065458}{0}{GW191216_213338}{0}}[{\red{???}}]}
\DeclareRobustCommand{\percentmasstwomorethanonetwentygwtcthree}[1]{\IfEqCase{#1}{{GW200224_222234}{0}{GW191129_134029}{0}{GW200311_115853}{0}{GW191230_180458}{0}{GW191222_033537}{0}{GW200225_060421}{0}{GW200302_015811}{0}{GW200128_022011}{0}{GW191204_171526}{0}{GW200112_155838}{0}{GW200105_162426}{0}{GW191105_143521}{0}{GW191109_010717}{0}{GW200209_085452}{0}{GW200115_042309}{0}{GW191127_050227}{0}{GW200216_220804}{0}{GW191215_223052}{0}{GW200208_130117}{0}{GW200219_094415}{0}{GW191103_012549}{0}{GW200316_215756}{0}{GW200202_154313}{0}{GW200129_065458}{0}{GW191216_213338}{0}}[{\red{???}}]}
\DeclareRobustCommand{\percentmassfinalmorethanonehundredgwtcthree}[1]{\IfEqCase{#1}{{GW200224_222234}{0}{GW191129_134029}{0}{GW200311_115853}{0}{GW191230_180458}{2}{GW191222_033537}{0}{GW200225_060421}{0}{GW200302_015811}{0}{GW200128_022011}{0}{GW191204_171526}{0}{GW200112_155838}{0}{GW200105_162426}{0}{GW191105_143521}{0}{GW191109_010717}{78}{GW200209_085452}{0}{GW200115_042309}{0}{GW191127_050227}{14}{GW200216_220804}{4}{GW191215_223052}{0}{GW200208_130117}{0}{GW200219_094415}{0}{GW191103_012549}{0}{GW200316_215756}{0}{GW200202_154313}{0}{GW200129_065458}{0}{GW191216_213338}{0}}[{\red{???}}]}
\DeclareRobustCommand{\percentchieffmorethanzerogwtcthree}[1]{\IfEqCase{#1}{{GW200224_222234}{87}{GW191129_134029}{91}{GW200311_115853}{42}{GW191230_180458}{43}{GW191222_033537}{37}{GW200225_060421}{15}{GW200302_015811}{60}{GW200128_022011}{83}{GW191204_171526}{100}{GW200112_155838}{77}{GW200105_162426}{53}{GW191105_143521}{37}{GW191109_010717}{10}{GW200209_085452}{26}{GW200115_042309}{18}{GW191127_050227}{79}{GW200216_220804}{69}{GW191215_223052}{38}{GW200208_130117}{30}{GW200219_094415}{29}{GW191103_012549}{100}{GW200316_215756}{98}{GW200202_154313}{86}{GW200129_065458}{89}{GW191216_213338}{100}}[{\red{???}}]}
\DeclareRobustCommand{\percentchiefflessthanzerogwtcthree}[1]{\IfEqCase{#1}{{GW200224_222234}{13}{GW191129_134029}{9}{GW200311_115853}{58}{GW191230_180458}{57}{GW191222_033537}{63}{GW200225_060421}{85}{GW200302_015811}{40}{GW200128_022011}{17}{GW191204_171526}{0}{GW200112_155838}{23}{GW200105_162426}{47}{GW191105_143521}{63}{GW191109_010717}{90}{GW200209_085452}{74}{GW200115_042309}{82}{GW191127_050227}{21}{GW200216_220804}{31}{GW191215_223052}{62}{GW200208_130117}{70}{GW200219_094415}{71}{GW191103_012549}{0}{GW200316_215756}{2}{GW200202_154313}{14}{GW200129_065458}{11}{GW191216_213338}{0}}[{\red{???}}]}
\DeclareRobustCommand{\percentchionemorethanpointeightgwtcthree}[1]{\IfEqCase{#1}{{GW200224_222234}{14}{GW191129_134029}{1}{GW200311_115853}{10}{GW191230_180458}{21}{GW191222_033537}{10}{GW200225_060421}{21}{GW200302_015811}{11}{GW200128_022011}{26}{GW191204_171526}{4}{GW200112_155838}{5}{GW200105_162426}{0}{GW191105_143521}{4}{GW191109_010717}{55}{GW200209_085452}{21}{GW200115_042309}{6}{GW191127_050227}{33}{GW200216_220804}{19}{GW191215_223052}{15}{GW200208_130117}{9}{GW200219_094415}{17}{GW191103_012549}{9}{GW200316_215756}{2}{GW200202_154313}{2}{GW200129_065458}{26}{GW191216_213338}{1}}[{\red{???}}]}
\DeclareRobustCommand{\percentchitwomorethanpointeightgwtcthree}[1]{\IfEqCase{#1}{{GW200224_222234}{14}{GW191129_134029}{8}{GW200311_115853}{13}{GW191230_180458}{19}{GW191222_033537}{13}{GW200225_060421}{14}{GW200302_015811}{16}{GW200128_022011}{19}{GW191204_171526}{9}{GW200112_155838}{11}{GW200105_162426}{10}{GW191105_143521}{9}{GW191109_010717}{33}{GW200209_085452}{19}{GW200115_042309}{13}{GW191127_050227}{22}{GW200216_220804}{21}{GW191215_223052}{16}{GW200208_130117}{14}{GW200219_094415}{18}{GW191103_012549}{17}{GW200316_215756}{13}{GW200202_154313}{8}{GW200129_065458}{15}{GW191216_213338}{8}}[{\red{???}}]}
\DeclareRobustCommand{\percentanychimorethanpointeightgwtcthree}[1]{\IfEqCase{#1}{{GW200224_222234}{26}{GW191129_134029}{9}{GW200311_115853}{21}{GW191230_180458}{36}{GW191222_033537}{21}{GW200225_060421}{33}{GW200302_015811}{25}{GW200128_022011}{41}{GW191204_171526}{13}{GW200112_155838}{15}{GW200105_162426}{10}{GW191105_143521}{12}{GW191109_010717}{72}{GW200209_085452}{37}{GW200115_042309}{18}{GW191127_050227}{49}{GW200216_220804}{36}{GW191215_223052}{29}{GW200208_130117}{22}{GW200219_094415}{32}{GW191103_012549}{24}{GW200316_215756}{14}{GW200202_154313}{10}{GW200129_065458}{36}{GW191216_213338}{8}}[{\red{???}}]}

\newcommand{\massratiogwtcthreeuncert}[1]{\ensuremath{ \massratiogwtcthreemed{#1}_{-\massratiogwtcthreeminus{#1}}^{+\massratiogwtcthreeplus{#1}}  } }

\newcommand{\finalmassdetgwtcthreeuncert}[1]{\ensuremath{ \finalmassdetgwtcthreemed{#1}_{-\finalmassdetgwtcthreeminus{#1}}^{+\finalmassdetgwtcthreeplus{#1}}  } }

\newcommand{\networkmatchedfiltersnrgwtcthreeuncert}[1]{\ensuremath{ \networkmatchedfiltersnrgwtcthreemed{#1}_{-\networkmatchedfiltersnrgwtcthreeminus{#1}}^{+\networkmatchedfiltersnrgwtcthreeplus{#1}}  } }

\newcommand{\finalspingwtcthreeuncert}[1]{\ensuremath{ \finalspingwtcthreemed{#1}_{-\finalspingwtcthreeminus{#1}}^{+\finalspingwtcthreeplus{#1}}  } }

\newcommand{\imrppnrtSPINFINALCOMPACTOneSevenZeroEightOneSevenHighSpinCat}{\macro{\ensuremath{0.68_{-0.03}^{+0.01}}}}

\newcommand{\imrppnrtMFINALobsavgCOMPACTOneSevenZeroEightOneSevenHighSpinCat}{\macro{\ensuremath{2.67_{-0.05}^{+0.16}}}}

\newcommand{\imrppnrtMASSRATIOCOMPACTOneSevenZeroEightOneSevenHighSpinCat}{\macro{\ensuremath{0.72_{-0.21}^{+0.24}}}}

\newcommand{\imrppnrtPEMATCHSNRCOMPACTOneSevenZeroEightOneSevenHighSpinCat}{\macro{\ensuremath{32.7_{-0.1}^{+0.1}}}}

\newcommand{\loglikelihoodfourtwofiveminus}[1]{\IfEqCase{#1}{{AlignedSpinInspiralTidalHS}{5.5}{AlignedSpinInspiralTidalLS}{5.4}{AlignedSpinTidalHS}{6.5}{AlignedSpinTidalLS}{6.3}{IMRPhenomDNRTidal-HS}{5.4}{IMRPhenomDNRTidal-LS}{5.5}{IMRPhenomPv2NRTidal-HS}{5.7}{IMRPhenomPv2NRTidal-LS}{5.5}{SEOBNRv4TsurrogateHS}{5.2}{SEOBNRv4TsurrogateLS}{5.4}{SEOBNRv4TsurrogatehighspinRIFT}{18.6}{SEOBNRv4TsurrogatelowspinRIFT}{5.0}{TEOBResumS-HS}{15.8}{TEOBResumS-LS}{5.1}{TaylorF2-HS}{5.5}{TaylorF2-LS}{5.4}{PrecessingSpinIMRTidalHS}{5.7}{PrecessingSpinIMRTidalLS}{5.5}{PublicationSamples}{5.6}}}
\newcommand{\loglikelihoodfourtwofivemed}[1]{\IfEqCase{#1}{{AlignedSpinInspiralTidalHS}{-240965.7}{AlignedSpinInspiralTidalLS}{-240964.4}{AlignedSpinTidalHS}{-1045826.3}{AlignedSpinTidalLS}{-1045825.4}{IMRPhenomDNRTidal-HS}{-1045828.5}{IMRPhenomDNRTidal-LS}{-1045827.1}{IMRPhenomPv2NRTidal-HS}{-500483.9}{IMRPhenomPv2NRTidal-LS}{-500482.8}{SEOBNRv4TsurrogateHS}{-1045828.0}{SEOBNRv4TsurrogateLS}{-1045827.1}{SEOBNRv4TsurrogatehighspinRIFT}{66.9}{SEOBNRv4TsurrogatelowspinRIFT}{68.5}{TEOBResumS-HS}{67.1}{TEOBResumS-LS}{68.8}{TaylorF2-HS}{-240965.7}{TaylorF2-LS}{-240964.4}{PrecessingSpinIMRTidalHS}{-500483.9}{PrecessingSpinIMRTidalLS}{-500482.8}{PublicationSamples}{-500483.9}}}
\newcommand{\loglikelihoodfourtwofiveplus}[1]{\IfEqCase{#1}{{AlignedSpinInspiralTidalHS}{3.8}{AlignedSpinInspiralTidalLS}{3.2}{AlignedSpinTidalHS}{1045896.4}{AlignedSpinTidalLS}{1045896.5}{IMRPhenomDNRTidal-HS}{4.5}{IMRPhenomDNRTidal-LS}{3.7}{IMRPhenomPv2NRTidal-HS}{4.5}{IMRPhenomPv2NRTidal-LS}{3.8}{SEOBNRv4TsurrogateHS}{3.9}{SEOBNRv4TsurrogateLS}{3.5}{SEOBNRv4TsurrogatehighspinRIFT}{4.3}{SEOBNRv4TsurrogatelowspinRIFT}{3.4}{TEOBResumS-HS}{4.4}{TEOBResumS-LS}{3.3}{TaylorF2-HS}{3.8}{TaylorF2-LS}{3.2}{PrecessingSpinIMRTidalHS}{4.5}{PrecessingSpinIMRTidalLS}{3.8}{PublicationSamples}{4.5}}}
\newcommand{\chiefffourtwofiveminus}[1]{\IfEqCase{#1}{{AlignedSpinInspiralTidalHS}{0.04}{AlignedSpinInspiralTidalLS}{0.01}{AlignedSpinTidalHS}{0.03}{AlignedSpinTidalLS}{0.01}{IMRPhenomDNRTidal-HS}{0.03}{IMRPhenomDNRTidal-LS}{0.01}{IMRPhenomPv2NRTidal-HS}{0.05}{IMRPhenomPv2NRTidal-LS}{0.01}{SEOBNRv4TsurrogateHS}{0.03}{SEOBNRv4TsurrogateLS}{0.01}{SEOBNRv4TsurrogatehighspinRIFT}{0.03}{SEOBNRv4TsurrogatelowspinRIFT}{0.01}{TEOBResumS-HS}{0.03}{TEOBResumS-LS}{0.01}{TaylorF2-HS}{0.04}{TaylorF2-LS}{0.01}{PrecessingSpinIMRTidalHS}{0.05}{PrecessingSpinIMRTidalLS}{0.01}{PublicationSamples}{0.05}}}
\newcommand{\chiefffourtwofivemed}[1]{\IfEqCase{#1}{{AlignedSpinInspiralTidalHS}{0.05}{AlignedSpinInspiralTidalLS}{0.01}{AlignedSpinTidalHS}{0.04}{AlignedSpinTidalLS}{0.01}{IMRPhenomDNRTidal-HS}{0.04}{IMRPhenomDNRTidal-LS}{0.01}{IMRPhenomPv2NRTidal-HS}{0.06}{IMRPhenomPv2NRTidal-LS}{0.01}{SEOBNRv4TsurrogateHS}{0.04}{SEOBNRv4TsurrogateLS}{0.01}{SEOBNRv4TsurrogatehighspinRIFT}{0.03}{SEOBNRv4TsurrogatelowspinRIFT}{0.01}{TEOBResumS-HS}{0.04}{TEOBResumS-LS}{0.01}{TaylorF2-HS}{0.05}{TaylorF2-LS}{0.01}{PrecessingSpinIMRTidalHS}{0.06}{PrecessingSpinIMRTidalLS}{0.01}{PublicationSamples}{0.06}}}
\newcommand{\chiefffourtwofiveplus}[1]{\IfEqCase{#1}{{AlignedSpinInspiralTidalHS}{0.09}{AlignedSpinInspiralTidalLS}{0.01}{AlignedSpinTidalHS}{0.08}{AlignedSpinTidalLS}{0.01}{IMRPhenomDNRTidal-HS}{0.10}{IMRPhenomDNRTidal-LS}{0.01}{IMRPhenomPv2NRTidal-HS}{0.11}{IMRPhenomPv2NRTidal-LS}{0.01}{SEOBNRv4TsurrogateHS}{0.07}{SEOBNRv4TsurrogateLS}{0.01}{SEOBNRv4TsurrogatehighspinRIFT}{0.07}{SEOBNRv4TsurrogatelowspinRIFT}{0.02}{TEOBResumS-HS}{0.06}{TEOBResumS-LS}{0.02}{TaylorF2-HS}{0.09}{TaylorF2-LS}{0.01}{PrecessingSpinIMRTidalHS}{0.11}{PrecessingSpinIMRTidalLS}{0.01}{PublicationSamples}{0.11}}}
\newcommand{\totalmasssourcefourtwofiveminus}[1]{\IfEqCase{#1}{{AlignedSpinInspiralTidalHS}{0.09}{AlignedSpinInspiralTidalLS}{0.05}{AlignedSpinTidalHS}{0.08}{AlignedSpinTidalLS}{0.05}{IMRPhenomDNRTidal-HS}{0.09}{IMRPhenomDNRTidal-LS}{0.05}{IMRPhenomPv2NRTidal-HS}{0.1}{IMRPhenomPv2NRTidal-LS}{0.05}{SEOBNRv4TsurrogateHS}{0.07}{SEOBNRv4TsurrogateLS}{0.05}{SEOBNRv4TsurrogatehighspinRIFT}{0.07}{SEOBNRv4TsurrogatelowspinRIFT}{0.05}{TEOBResumS-HS}{0.08}{TEOBResumS-LS}{0.05}{TaylorF2-HS}{0.09}{TaylorF2-LS}{0.05}{PrecessingSpinIMRTidalHS}{0.1}{PrecessingSpinIMRTidalLS}{0.05}{PublicationSamples}{0.1}}}
\newcommand{\totalmasssourcefourtwofivemed}[1]{\IfEqCase{#1}{{AlignedSpinInspiralTidalHS}{3.37}{AlignedSpinInspiralTidalLS}{3.31}{AlignedSpinTidalHS}{3.35}{AlignedSpinTidalLS}{3.31}{IMRPhenomDNRTidal-HS}{3.36}{IMRPhenomDNRTidal-LS}{3.31}{IMRPhenomPv2NRTidal-HS}{3.4}{IMRPhenomPv2NRTidal-LS}{3.31}{SEOBNRv4TsurrogateHS}{3.35}{SEOBNRv4TsurrogateLS}{3.31}{SEOBNRv4TsurrogatehighspinRIFT}{3.34}{SEOBNRv4TsurrogatelowspinRIFT}{3.31}{TEOBResumS-HS}{3.35}{TEOBResumS-LS}{3.31}{TaylorF2-HS}{3.37}{TaylorF2-LS}{3.31}{PrecessingSpinIMRTidalHS}{3.4}{PrecessingSpinIMRTidalLS}{3.31}{PublicationSamples}{3.4}}}
\newcommand{\totalmasssourcefourtwofiveplus}[1]{\IfEqCase{#1}{{AlignedSpinInspiralTidalHS}{0.2}{AlignedSpinInspiralTidalLS}{0.06}{AlignedSpinTidalHS}{0.3}{AlignedSpinTidalLS}{0.06}{IMRPhenomDNRTidal-HS}{0.4}{IMRPhenomDNRTidal-LS}{0.06}{IMRPhenomPv2NRTidal-HS}{0.3}{IMRPhenomPv2NRTidal-LS}{0.06}{SEOBNRv4TsurrogateHS}{0.2}{SEOBNRv4TsurrogateLS}{0.06}{SEOBNRv4TsurrogatehighspinRIFT}{0.2}{SEOBNRv4TsurrogatelowspinRIFT}{0.06}{TEOBResumS-HS}{0.2}{TEOBResumS-LS}{0.06}{TaylorF2-HS}{0.2}{TaylorF2-LS}{0.06}{PrecessingSpinIMRTidalHS}{0.3}{PrecessingSpinIMRTidalLS}{0.06}{PublicationSamples}{0.3}}}
\newcommand{\chipfourtwofiveminus}[1]{\IfEqCase{#1}{{AlignedSpinInspiralTidalHS}{0.00}{AlignedSpinInspiralTidalLS}{0.00}{AlignedSpinTidalHS}{0.00}{AlignedSpinTidalLS}{0.00}{IMRPhenomDNRTidal-HS}{0.00}{IMRPhenomDNRTidal-LS}{0.00}{IMRPhenomPv2NRTidal-HS}{0.27}{IMRPhenomPv2NRTidal-LS}{0.02}{SEOBNRv4TsurrogateHS}{0.00}{SEOBNRv4TsurrogateLS}{0.00}{SEOBNRv4TsurrogatehighspinRIFT}{0.00}{SEOBNRv4TsurrogatelowspinRIFT}{0.00}{TEOBResumS-HS}{0.00}{TEOBResumS-LS}{0.00}{TaylorF2-HS}{0.00}{TaylorF2-LS}{0.00}{PrecessingSpinIMRTidalHS}{0.27}{PrecessingSpinIMRTidalLS}{0.02}{PublicationSamples}{0.27}}}
\newcommand{\chipfourtwofivemed}[1]{\IfEqCase{#1}{{AlignedSpinInspiralTidalHS}{0.00}{AlignedSpinInspiralTidalLS}{0.00}{AlignedSpinTidalHS}{0.00}{AlignedSpinTidalLS}{0.00}{IMRPhenomDNRTidal-HS}{0.00}{IMRPhenomDNRTidal-LS}{0.00}{IMRPhenomPv2NRTidal-HS}{0.34}{IMRPhenomPv2NRTidal-LS}{0.03}{SEOBNRv4TsurrogateHS}{0.00}{SEOBNRv4TsurrogateLS}{0.00}{SEOBNRv4TsurrogatehighspinRIFT}{0.00}{SEOBNRv4TsurrogatelowspinRIFT}{0.00}{TEOBResumS-HS}{0.00}{TEOBResumS-LS}{0.00}{TaylorF2-HS}{0.00}{TaylorF2-LS}{0.00}{PrecessingSpinIMRTidalHS}{0.34}{PrecessingSpinIMRTidalLS}{0.03}{PublicationSamples}{0.34}}}
\newcommand{\chipfourtwofiveplus}[1]{\IfEqCase{#1}{{AlignedSpinInspiralTidalHS}{0.00}{AlignedSpinInspiralTidalLS}{0.00}{AlignedSpinTidalHS}{0.00}{AlignedSpinTidalLS}{0.00}{IMRPhenomDNRTidal-HS}{0.00}{IMRPhenomDNRTidal-LS}{0.00}{IMRPhenomPv2NRTidal-HS}{0.43}{IMRPhenomPv2NRTidal-LS}{0.02}{SEOBNRv4TsurrogateHS}{0.00}{SEOBNRv4TsurrogateLS}{0.00}{SEOBNRv4TsurrogatehighspinRIFT}{0.00}{SEOBNRv4TsurrogatelowspinRIFT}{0.00}{TEOBResumS-HS}{0.00}{TEOBResumS-LS}{0.00}{TaylorF2-HS}{0.00}{TaylorF2-LS}{0.00}{PrecessingSpinIMRTidalHS}{0.43}{PrecessingSpinIMRTidalLS}{0.02}{PublicationSamples}{0.43}}}
\newcommand{\spinoneyfourtwofiveminus}[1]{\IfEqCase{#1}{{AlignedSpinInspiralTidalHS}{0.00}{AlignedSpinInspiralTidalLS}{0.00}{AlignedSpinTidalHS}{0.00}{AlignedSpinTidalLS}{0.00}{IMRPhenomDNRTidal-HS}{0.00}{IMRPhenomDNRTidal-LS}{0.00}{IMRPhenomPv2NRTidal-HS}{0.49}{IMRPhenomPv2NRTidal-LS}{0.03}{SEOBNRv4TsurrogateHS}{0.00}{SEOBNRv4TsurrogateLS}{0.00}{SEOBNRv4TsurrogatehighspinRIFT}{0.00}{SEOBNRv4TsurrogatelowspinRIFT}{0.00}{TEOBResumS-HS}{0.00}{TEOBResumS-LS}{0.00}{TaylorF2-HS}{0.00}{TaylorF2-LS}{0.00}{PrecessingSpinIMRTidalHS}{0.48}{PrecessingSpinIMRTidalLS}{0.03}{PublicationSamples}{0.48}}}
\newcommand{\spinoneyfourtwofivemed}[1]{\IfEqCase{#1}{{AlignedSpinInspiralTidalHS}{0.00}{AlignedSpinInspiralTidalLS}{0.00}{AlignedSpinTidalHS}{0.00}{AlignedSpinTidalLS}{0.00}{IMRPhenomDNRTidal-HS}{0.00}{IMRPhenomDNRTidal-LS}{0.00}{IMRPhenomPv2NRTidal-HS}{0.003}{IMRPhenomPv2NRTidal-LS}{0.00}{SEOBNRv4TsurrogateHS}{0.00}{SEOBNRv4TsurrogateLS}{0.00}{SEOBNRv4TsurrogatehighspinRIFT}{0.00}{SEOBNRv4TsurrogatelowspinRIFT}{0.00}{TEOBResumS-HS}{0.00}{TEOBResumS-LS}{0.00}{TaylorF2-HS}{0.00}{TaylorF2-LS}{0.00}{PrecessingSpinIMRTidalHS}{0.003}{PrecessingSpinIMRTidalLS}{0.00}{PublicationSamples}{0.003}}}
\newcommand{\spinoneyfourtwofiveplus}[1]{\IfEqCase{#1}{{AlignedSpinInspiralTidalHS}{0.00}{AlignedSpinInspiralTidalLS}{0.00}{AlignedSpinTidalHS}{0.00}{AlignedSpinTidalLS}{0.00}{IMRPhenomDNRTidal-HS}{0.00}{IMRPhenomDNRTidal-LS}{0.00}{IMRPhenomPv2NRTidal-HS}{0.48}{IMRPhenomPv2NRTidal-LS}{0.03}{SEOBNRv4TsurrogateHS}{0.00}{SEOBNRv4TsurrogateLS}{0.00}{SEOBNRv4TsurrogatehighspinRIFT}{0.00}{SEOBNRv4TsurrogatelowspinRIFT}{0.00}{TEOBResumS-HS}{0.00}{TEOBResumS-LS}{0.00}{TaylorF2-HS}{0.00}{TaylorF2-LS}{0.00}{PrecessingSpinIMRTidalHS}{0.48}{PrecessingSpinIMRTidalLS}{0.03}{PublicationSamples}{0.48}}}
\newcommand{\phitwofourtwofiveminus}[1]{\IfEqCase{#1}{{AlignedSpinInspiralTidalHS}{0.00}{AlignedSpinInspiralTidalLS}{0.00}{AlignedSpinTidalHS}{0.00}{AlignedSpinTidalLS}{0.00}{IMRPhenomDNRTidal-HS}{0.00}{IMRPhenomDNRTidal-LS}{0.00}{IMRPhenomPv2NRTidal-HS}{2.83}{IMRPhenomPv2NRTidal-LS}{2.82}{SEOBNRv4TsurrogateHS}{0.00}{SEOBNRv4TsurrogateLS}{0.00}{SEOBNRv4TsurrogatehighspinRIFT}{0.00}{SEOBNRv4TsurrogatelowspinRIFT}{0.00}{TEOBResumS-HS}{0.00}{TEOBResumS-LS}{0.00}{TaylorF2-HS}{0.00}{TaylorF2-LS}{0.00}{PrecessingSpinIMRTidalHS}{2.84}{PrecessingSpinIMRTidalLS}{2.82}{PublicationSamples}{2.84}}}
\newcommand{\phitwofourtwofivemed}[1]{\IfEqCase{#1}{{AlignedSpinInspiralTidalHS}{0.00}{AlignedSpinInspiralTidalLS}{0.00}{AlignedSpinTidalHS}{0.00}{AlignedSpinTidalLS}{0.00}{IMRPhenomDNRTidal-HS}{0.00}{IMRPhenomDNRTidal-LS}{0.00}{IMRPhenomPv2NRTidal-HS}{3.14}{IMRPhenomPv2NRTidal-LS}{3.13}{SEOBNRv4TsurrogateHS}{0.00}{SEOBNRv4TsurrogateLS}{0.00}{SEOBNRv4TsurrogatehighspinRIFT}{0.00}{SEOBNRv4TsurrogatelowspinRIFT}{0.00}{TEOBResumS-HS}{0.00}{TEOBResumS-LS}{0.00}{TaylorF2-HS}{0.00}{TaylorF2-LS}{0.00}{PrecessingSpinIMRTidalHS}{3.15}{PrecessingSpinIMRTidalLS}{3.13}{PublicationSamples}{3.15}}}
\newcommand{\phitwofourtwofiveplus}[1]{\IfEqCase{#1}{{AlignedSpinInspiralTidalHS}{0.00}{AlignedSpinInspiralTidalLS}{0.00}{AlignedSpinTidalHS}{0.00}{AlignedSpinTidalLS}{0.00}{IMRPhenomDNRTidal-HS}{0.00}{IMRPhenomDNRTidal-LS}{0.00}{IMRPhenomPv2NRTidal-HS}{2.84}{IMRPhenomPv2NRTidal-LS}{2.83}{SEOBNRv4TsurrogateHS}{0.00}{SEOBNRv4TsurrogateLS}{0.00}{SEOBNRv4TsurrogatehighspinRIFT}{0.00}{SEOBNRv4TsurrogatelowspinRIFT}{0.00}{TEOBResumS-HS}{0.00}{TEOBResumS-LS}{0.00}{TaylorF2-HS}{0.00}{TaylorF2-LS}{0.00}{PrecessingSpinIMRTidalHS}{2.83}{PrecessingSpinIMRTidalLS}{2.83}{PublicationSamples}{2.83}}}
\newcommand{\phionetwofourtwofiveminus}[1]{\IfEqCase{#1}{{AlignedSpinInspiralTidalHS}{0.00}{AlignedSpinInspiralTidalLS}{0.00}{AlignedSpinTidalHS}{0.00}{AlignedSpinTidalLS}{0.00}{IMRPhenomDNRTidal-HS}{0.00}{IMRPhenomDNRTidal-LS}{0.00}{IMRPhenomPv2NRTidal-HS}{2.88}{IMRPhenomPv2NRTidal-LS}{2.74}{SEOBNRv4TsurrogateHS}{0.00}{SEOBNRv4TsurrogateLS}{0.00}{SEOBNRv4TsurrogatehighspinRIFT}{0.00}{SEOBNRv4TsurrogatelowspinRIFT}{0.00}{TEOBResumS-HS}{0.00}{TEOBResumS-LS}{0.00}{TaylorF2-HS}{0.00}{TaylorF2-LS}{0.00}{PrecessingSpinIMRTidalHS}{2.87}{PrecessingSpinIMRTidalLS}{2.75}{PublicationSamples}{2.88}}}
\newcommand{\phionetwofourtwofivemed}[1]{\IfEqCase{#1}{{AlignedSpinInspiralTidalHS}{0.00}{AlignedSpinInspiralTidalLS}{0.00}{AlignedSpinTidalHS}{0.00}{AlignedSpinTidalLS}{0.00}{IMRPhenomDNRTidal-HS}{0.00}{IMRPhenomDNRTidal-LS}{0.00}{IMRPhenomPv2NRTidal-HS}{3.19}{IMRPhenomPv2NRTidal-LS}{3.05}{SEOBNRv4TsurrogateHS}{0.00}{SEOBNRv4TsurrogateLS}{0.00}{SEOBNRv4TsurrogatehighspinRIFT}{0.00}{SEOBNRv4TsurrogatelowspinRIFT}{0.00}{TEOBResumS-HS}{0.00}{TEOBResumS-LS}{0.00}{TaylorF2-HS}{0.00}{TaylorF2-LS}{0.00}{PrecessingSpinIMRTidalHS}{3.18}{PrecessingSpinIMRTidalLS}{3.05}{PublicationSamples}{3.18}}}
\newcommand{\phionetwofourtwofiveplus}[1]{\IfEqCase{#1}{{AlignedSpinInspiralTidalHS}{0.00}{AlignedSpinInspiralTidalLS}{0.00}{AlignedSpinTidalHS}{0.00}{AlignedSpinTidalLS}{0.00}{IMRPhenomDNRTidal-HS}{0.00}{IMRPhenomDNRTidal-LS}{0.00}{IMRPhenomPv2NRTidal-HS}{2.75}{IMRPhenomPv2NRTidal-LS}{2.88}{SEOBNRv4TsurrogateHS}{0.00}{SEOBNRv4TsurrogateLS}{0.00}{SEOBNRv4TsurrogatehighspinRIFT}{0.00}{SEOBNRv4TsurrogatelowspinRIFT}{0.00}{TEOBResumS-HS}{0.00}{TEOBResumS-LS}{0.00}{TaylorF2-HS}{0.00}{TaylorF2-LS}{0.00}{PrecessingSpinIMRTidalHS}{2.76}{PrecessingSpinIMRTidalLS}{2.88}{PublicationSamples}{2.76}}}
\newcommand{\rafourtwofiveminus}[1]{\IfEqCase{#1}{{AlignedSpinInspiralTidalHS}{1.04004}{AlignedSpinInspiralTidalLS}{1.10804}{AlignedSpinTidalHS}{1.04466}{AlignedSpinTidalLS}{1.05437}{IMRPhenomDNRTidal-HS}{1.04415}{IMRPhenomDNRTidal-LS}{1.34502}{IMRPhenomPv2NRTidal-HS}{1.14841}{IMRPhenomPv2NRTidal-LS}{1.49247}{SEOBNRv4TsurrogateHS}{1.04813}{SEOBNRv4TsurrogateLS}{0.99881}{SEOBNRv4TsurrogatehighspinRIFT}{1.03845}{SEOBNRv4TsurrogatelowspinRIFT}{1.07158}{TEOBResumS-HS}{1.04161}{TEOBResumS-LS}{1.06268}{TaylorF2-HS}{1.03744}{TaylorF2-LS}{1.11687}{PrecessingSpinIMRTidalHS}{1.14713}{PrecessingSpinIMRTidalLS}{1.36500}{PublicationSamples}{1.14187}}}
\newcommand{\rafourtwofivemed}[1]{\IfEqCase{#1}{{AlignedSpinInspiralTidalHS}{1.47807}{AlignedSpinInspiralTidalLS}{1.57610}{AlignedSpinTidalHS}{1.52327}{AlignedSpinTidalLS}{1.53218}{IMRPhenomDNRTidal-HS}{1.52282}{IMRPhenomDNRTidal-LS}{1.86790}{IMRPhenomPv2NRTidal-HS}{1.62902}{IMRPhenomPv2NRTidal-LS}{1.99502}{SEOBNRv4TsurrogateHS}{1.52665}{SEOBNRv4TsurrogateLS}{1.46873}{SEOBNRv4TsurrogatehighspinRIFT}{1.50830}{SEOBNRv4TsurrogatelowspinRIFT}{1.53881}{TEOBResumS-HS}{1.51000}{TEOBResumS-LS}{1.50223}{TaylorF2-HS}{1.47697}{TaylorF2-LS}{1.57735}{PrecessingSpinIMRTidalHS}{1.62833}{PrecessingSpinIMRTidalLS}{1.86958}{PublicationSamples}{1.62336}}}
\newcommand{\rafourtwofiveplus}[1]{\IfEqCase{#1}{{AlignedSpinInspiralTidalHS}{3.24632}{AlignedSpinInspiralTidalLS}{3.13801}{AlignedSpinTidalHS}{3.25020}{AlignedSpinTidalLS}{3.18422}{IMRPhenomDNRTidal-HS}{3.29588}{IMRPhenomDNRTidal-LS}{2.86054}{IMRPhenomPv2NRTidal-HS}{3.13029}{IMRPhenomPv2NRTidal-LS}{2.74917}{SEOBNRv4TsurrogateHS}{3.18219}{SEOBNRv4TsurrogateLS}{3.22628}{SEOBNRv4TsurrogatehighspinRIFT}{3.24313}{SEOBNRv4TsurrogatelowspinRIFT}{3.18050}{TEOBResumS-HS}{3.24215}{TEOBResumS-LS}{3.23570}{TaylorF2-HS}{3.25577}{TaylorF2-LS}{3.13908}{PrecessingSpinIMRTidalHS}{3.12911}{PrecessingSpinIMRTidalLS}{2.87696}{PublicationSamples}{3.13407}}}
\newcommand{\phijlfourtwofiveminus}[1]{\IfEqCase{#1}{{AlignedSpinInspiralTidalHS}{0.97}{AlignedSpinInspiralTidalLS}{0.88}{AlignedSpinTidalHS}{0.78}{AlignedSpinTidalLS}{0.77}{IMRPhenomDNRTidal-HS}{0.74}{IMRPhenomDNRTidal-LS}{0.78}{IMRPhenomPv2NRTidal-HS}{2.89}{IMRPhenomPv2NRTidal-LS}{2.58}{SEOBNRv4TsurrogateHS}{0.89}{SEOBNRv4TsurrogateLS}{0.85}{SEOBNRv4TsurrogatehighspinRIFT}{0.00}{SEOBNRv4TsurrogatelowspinRIFT}{0.00}{TEOBResumS-HS}{0.00}{TEOBResumS-LS}{0.00}{TaylorF2-HS}{0.97}{TaylorF2-LS}{0.88}{PrecessingSpinIMRTidalHS}{2.88}{PrecessingSpinIMRTidalLS}{2.57}{PublicationSamples}{2.89}}}
\newcommand{\phijlfourtwofivemed}[1]{\IfEqCase{#1}{{AlignedSpinInspiralTidalHS}{1.22}{AlignedSpinInspiralTidalLS}{1.11}{AlignedSpinTidalHS}{0.78}{AlignedSpinTidalLS}{0.77}{IMRPhenomDNRTidal-HS}{0.96}{IMRPhenomDNRTidal-LS}{1.02}{IMRPhenomPv2NRTidal-HS}{3.23}{IMRPhenomPv2NRTidal-LS}{2.87}{SEOBNRv4TsurrogateHS}{1.13}{SEOBNRv4TsurrogateLS}{1.09}{SEOBNRv4TsurrogatehighspinRIFT}{0.00}{SEOBNRv4TsurrogatelowspinRIFT}{0.00}{TEOBResumS-HS}{0.00}{TEOBResumS-LS}{0.00}{TaylorF2-HS}{1.22}{TaylorF2-LS}{1.11}{PrecessingSpinIMRTidalHS}{3.23}{PrecessingSpinIMRTidalLS}{2.86}{PublicationSamples}{3.23}}}
\newcommand{\phijlfourtwofiveplus}[1]{\IfEqCase{#1}{{AlignedSpinInspiralTidalHS}{1.64}{AlignedSpinInspiralTidalLS}{1.75}{AlignedSpinTidalHS}{2.37}{AlignedSpinTidalLS}{2.37}{IMRPhenomDNRTidal-HS}{1.88}{IMRPhenomDNRTidal-LS}{1.83}{IMRPhenomPv2NRTidal-HS}{2.75}{IMRPhenomPv2NRTidal-LS}{3.05}{SEOBNRv4TsurrogateHS}{1.72}{SEOBNRv4TsurrogateLS}{1.76}{SEOBNRv4TsurrogatehighspinRIFT}{3.14}{SEOBNRv4TsurrogatelowspinRIFT}{3.14}{TEOBResumS-HS}{3.14}{TEOBResumS-LS}{3.14}{TaylorF2-HS}{1.63}{TaylorF2-LS}{1.75}{PrecessingSpinIMRTidalHS}{2.76}{PrecessingSpinIMRTidalLS}{3.07}{PublicationSamples}{2.75}}}
\newcommand{\tilttwofourtwofiveminus}[1]{\IfEqCase{#1}{{AlignedSpinInspiralTidalHS}{0.00}{AlignedSpinInspiralTidalLS}{0.00}{AlignedSpinTidalHS}{0.00}{AlignedSpinTidalLS}{0.00}{IMRPhenomDNRTidal-HS}{0.00}{IMRPhenomDNRTidal-LS}{0.00}{IMRPhenomPv2NRTidal-HS}{0.87}{IMRPhenomPv2NRTidal-LS}{0.78}{SEOBNRv4TsurrogateHS}{0.00}{SEOBNRv4TsurrogateLS}{0.00}{SEOBNRv4TsurrogatehighspinRIFT}{0.00}{SEOBNRv4TsurrogatelowspinRIFT}{0.00}{TEOBResumS-HS}{0.00}{TEOBResumS-LS}{0.00}{TaylorF2-HS}{0.00}{TaylorF2-LS}{0.00}{PrecessingSpinIMRTidalHS}{0.87}{PrecessingSpinIMRTidalLS}{0.79}{PublicationSamples}{0.87}}}
\newcommand{\tilttwofourtwofivemed}[1]{\IfEqCase{#1}{{AlignedSpinInspiralTidalHS}{0.00}{AlignedSpinInspiralTidalLS}{0.00}{AlignedSpinTidalHS}{0.00}{AlignedSpinTidalLS}{0.00}{IMRPhenomDNRTidal-HS}{0.00}{IMRPhenomDNRTidal-LS}{0.00}{IMRPhenomPv2NRTidal-HS}{1.41}{IMRPhenomPv2NRTidal-LS}{1.09}{SEOBNRv4TsurrogateHS}{0.00}{SEOBNRv4TsurrogateLS}{0.00}{SEOBNRv4TsurrogatehighspinRIFT}{0.00}{SEOBNRv4TsurrogatelowspinRIFT}{0.00}{TEOBResumS-HS}{0.00}{TEOBResumS-LS}{0.00}{TaylorF2-HS}{0.00}{TaylorF2-LS}{0.00}{PrecessingSpinIMRTidalHS}{1.41}{PrecessingSpinIMRTidalLS}{1.09}{PublicationSamples}{1.41}}}
\newcommand{\tilttwofourtwofiveplus}[1]{\IfEqCase{#1}{{AlignedSpinInspiralTidalHS}{3.14}{AlignedSpinInspiralTidalLS}{3.14}{AlignedSpinTidalHS}{3.14}{AlignedSpinTidalLS}{3.14}{IMRPhenomDNRTidal-HS}{3.14}{IMRPhenomDNRTidal-LS}{3.14}{IMRPhenomPv2NRTidal-HS}{0.94}{IMRPhenomPv2NRTidal-LS}{1.21}{SEOBNRv4TsurrogateHS}{3.14}{SEOBNRv4TsurrogateLS}{3.14}{SEOBNRv4TsurrogatehighspinRIFT}{3.14}{SEOBNRv4TsurrogatelowspinRIFT}{3.14}{TEOBResumS-HS}{3.14}{TEOBResumS-LS}{3.14}{TaylorF2-HS}{3.14}{TaylorF2-LS}{3.14}{PrecessingSpinIMRTidalHS}{0.94}{PrecessingSpinIMRTidalLS}{1.20}{PublicationSamples}{0.94}}}
\newcommand{\costhetajnfourtwofiveminus}[1]{\IfEqCase{#1}{{AlignedSpinInspiralTidalHS}{1.30}{AlignedSpinInspiralTidalLS}{1.40}{AlignedSpinTidalHS}{1.44}{AlignedSpinTidalLS}{1.44}{IMRPhenomDNRTidal-HS}{1.53}{IMRPhenomDNRTidal-LS}{1.48}{IMRPhenomPv2NRTidal-HS}{1.43}{IMRPhenomPv2NRTidal-LS}{1.44}{SEOBNRv4TsurrogateHS}{1.38}{SEOBNRv4TsurrogateLS}{1.42}{SEOBNRv4TsurrogatehighspinRIFT}{1.41}{SEOBNRv4TsurrogatelowspinRIFT}{1.42}{TEOBResumS-HS}{1.42}{TEOBResumS-LS}{1.40}{TaylorF2-HS}{1.30}{TaylorF2-LS}{1.40}{PrecessingSpinIMRTidalHS}{1.43}{PrecessingSpinIMRTidalLS}{1.44}{PublicationSamples}{1.43}}}
\newcommand{\costhetajnfourtwofivemed}[1]{\IfEqCase{#1}{{AlignedSpinInspiralTidalHS}{0.34}{AlignedSpinInspiralTidalLS}{0.44}{AlignedSpinTidalHS}{0.49}{AlignedSpinTidalLS}{0.49}{IMRPhenomDNRTidal-HS}{0.58}{IMRPhenomDNRTidal-LS}{0.53}{IMRPhenomPv2NRTidal-HS}{0.47}{IMRPhenomPv2NRTidal-LS}{0.48}{SEOBNRv4TsurrogateHS}{0.43}{SEOBNRv4TsurrogateLS}{0.46}{SEOBNRv4TsurrogatehighspinRIFT}{0.45}{SEOBNRv4TsurrogatelowspinRIFT}{0.46}{TEOBResumS-HS}{0.46}{TEOBResumS-LS}{0.44}{TaylorF2-HS}{0.34}{TaylorF2-LS}{0.44}{PrecessingSpinIMRTidalHS}{0.47}{PrecessingSpinIMRTidalLS}{0.48}{PublicationSamples}{0.47}}}
\newcommand{\costhetajnfourtwofiveplus}[1]{\IfEqCase{#1}{{AlignedSpinInspiralTidalHS}{0.62}{AlignedSpinInspiralTidalLS}{0.53}{AlignedSpinTidalHS}{0.49}{AlignedSpinTidalLS}{0.49}{IMRPhenomDNRTidal-HS}{0.40}{IMRPhenomDNRTidal-LS}{0.45}{IMRPhenomPv2NRTidal-HS}{0.50}{IMRPhenomPv2NRTidal-LS}{0.49}{SEOBNRv4TsurrogateHS}{0.54}{SEOBNRv4TsurrogateLS}{0.51}{SEOBNRv4TsurrogatehighspinRIFT}{0.52}{SEOBNRv4TsurrogatelowspinRIFT}{0.51}{TEOBResumS-HS}{0.51}{TEOBResumS-LS}{0.54}{TaylorF2-HS}{0.63}{TaylorF2-LS}{0.53}{PrecessingSpinIMRTidalHS}{0.50}{PrecessingSpinIMRTidalLS}{0.49}{PublicationSamples}{0.50}}}
\newcommand{\spintwofourtwofiveminus}[1]{\IfEqCase{#1}{{AlignedSpinInspiralTidalHS}{0.07}{AlignedSpinInspiralTidalLS}{0.01}{AlignedSpinTidalHS}{0.07}{AlignedSpinTidalLS}{0.01}{IMRPhenomDNRTidal-HS}{0.10}{IMRPhenomDNRTidal-LS}{0.01}{IMRPhenomPv2NRTidal-HS}{0.25}{IMRPhenomPv2NRTidal-LS}{0.02}{SEOBNRv4TsurrogateHS}{0.06}{SEOBNRv4TsurrogateLS}{0.01}{SEOBNRv4TsurrogatehighspinRIFT}{0.06}{SEOBNRv4TsurrogatelowspinRIFT}{0.01}{TEOBResumS-HS}{0.06}{TEOBResumS-LS}{0.01}{TaylorF2-HS}{0.07}{TaylorF2-LS}{0.01}{PrecessingSpinIMRTidalHS}{0.25}{PrecessingSpinIMRTidalLS}{0.02}{PublicationSamples}{0.25}}}
\newcommand{\spintwofourtwofivemed}[1]{\IfEqCase{#1}{{AlignedSpinInspiralTidalHS}{0.08}{AlignedSpinInspiralTidalLS}{0.01}{AlignedSpinTidalHS}{0.07}{AlignedSpinTidalLS}{0.01}{IMRPhenomDNRTidal-HS}{0.11}{IMRPhenomDNRTidal-LS}{0.01}{IMRPhenomPv2NRTidal-HS}{0.28}{IMRPhenomPv2NRTidal-LS}{0.03}{SEOBNRv4TsurrogateHS}{0.06}{SEOBNRv4TsurrogateLS}{0.01}{SEOBNRv4TsurrogatehighspinRIFT}{0.06}{SEOBNRv4TsurrogatelowspinRIFT}{0.01}{TEOBResumS-HS}{0.06}{TEOBResumS-LS}{0.01}{TaylorF2-HS}{0.08}{TaylorF2-LS}{0.01}{PrecessingSpinIMRTidalHS}{0.28}{PrecessingSpinIMRTidalLS}{0.03}{PublicationSamples}{0.28}}}
\newcommand{\spintwofourtwofiveplus}[1]{\IfEqCase{#1}{{AlignedSpinInspiralTidalHS}{0.26}{AlignedSpinInspiralTidalLS}{0.03}{AlignedSpinTidalHS}{0.27}{AlignedSpinTidalLS}{0.03}{IMRPhenomDNRTidal-HS}{0.38}{IMRPhenomDNRTidal-LS}{0.03}{IMRPhenomPv2NRTidal-HS}{0.51}{IMRPhenomPv2NRTidal-LS}{0.02}{SEOBNRv4TsurrogateHS}{0.19}{SEOBNRv4TsurrogateLS}{0.03}{SEOBNRv4TsurrogatehighspinRIFT}{0.19}{SEOBNRv4TsurrogatelowspinRIFT}{0.03}{TEOBResumS-HS}{0.19}{TEOBResumS-LS}{0.03}{TaylorF2-HS}{0.26}{TaylorF2-LS}{0.03}{PrecessingSpinIMRTidalHS}{0.51}{PrecessingSpinIMRTidalLS}{0.02}{PublicationSamples}{0.51}}}
\newcommand{\massonedetfourtwofiveminus}[1]{\IfEqCase{#1}{{AlignedSpinInspiralTidalHS}{0.3}{AlignedSpinInspiralTidalLS}{0.09}{AlignedSpinTidalHS}{0.2}{AlignedSpinTidalLS}{0.09}{IMRPhenomDNRTidal-HS}{0.3}{IMRPhenomDNRTidal-LS}{0.09}{IMRPhenomPv2NRTidal-HS}{0.4}{IMRPhenomPv2NRTidal-LS}{0.09}{SEOBNRv4TsurrogateHS}{0.2}{SEOBNRv4TsurrogateLS}{0.09}{SEOBNRv4TsurrogatehighspinRIFT}{0.2}{SEOBNRv4TsurrogatelowspinRIFT}{0.09}{TEOBResumS-HS}{0.2}{TEOBResumS-LS}{0.09}{TaylorF2-HS}{0.3}{TaylorF2-LS}{0.09}{PrecessingSpinIMRTidalHS}{0.4}{PrecessingSpinIMRTidalLS}{0.09}{PublicationSamples}{0.4}}}
\newcommand{\massonedetfourtwofivemed}[1]{\IfEqCase{#1}{{AlignedSpinInspiralTidalHS}{2.0}{AlignedSpinInspiralTidalLS}{1.81}{AlignedSpinTidalHS}{2.0}{AlignedSpinTidalLS}{1.81}{IMRPhenomDNRTidal-HS}{2.0}{IMRPhenomDNRTidal-LS}{1.81}{IMRPhenomPv2NRTidal-HS}{2.1}{IMRPhenomPv2NRTidal-LS}{1.80}{SEOBNRv4TsurrogateHS}{2.0}{SEOBNRv4TsurrogateLS}{1.80}{SEOBNRv4TsurrogatehighspinRIFT}{1.9}{SEOBNRv4TsurrogatelowspinRIFT}{1.81}{TEOBResumS-HS}{2.0}{TEOBResumS-LS}{1.81}{TaylorF2-HS}{2.0}{TaylorF2-LS}{1.81}{PrecessingSpinIMRTidalHS}{2.1}{PrecessingSpinIMRTidalLS}{1.80}{PublicationSamples}{2.1}}}
\newcommand{\massonedetfourtwofiveplus}[1]{\IfEqCase{#1}{{AlignedSpinInspiralTidalHS}{0.5}{AlignedSpinInspiralTidalLS}{0.2}{AlignedSpinTidalHS}{0.6}{AlignedSpinTidalLS}{0.2}{IMRPhenomDNRTidal-HS}{0.7}{IMRPhenomDNRTidal-LS}{0.2}{IMRPhenomPv2NRTidal-HS}{0.6}{IMRPhenomPv2NRTidal-LS}{0.2}{SEOBNRv4TsurrogateHS}{0.5}{SEOBNRv4TsurrogateLS}{0.2}{SEOBNRv4TsurrogatehighspinRIFT}{0.5}{SEOBNRv4TsurrogatelowspinRIFT}{0.2}{TEOBResumS-HS}{0.5}{TEOBResumS-LS}{0.2}{TaylorF2-HS}{0.5}{TaylorF2-LS}{0.2}{PrecessingSpinIMRTidalHS}{0.6}{PrecessingSpinIMRTidalLS}{0.2}{PublicationSamples}{0.6}}}
\newcommand{\spintwoxfourtwofiveminus}[1]{\IfEqCase{#1}{{AlignedSpinInspiralTidalHS}{0.00}{AlignedSpinInspiralTidalLS}{0.00}{AlignedSpinTidalHS}{0.00}{AlignedSpinTidalLS}{0.00}{IMRPhenomDNRTidal-HS}{0.00}{IMRPhenomDNRTidal-LS}{0.00}{IMRPhenomPv2NRTidal-HS}{0.47}{IMRPhenomPv2NRTidal-LS}{0.03}{SEOBNRv4TsurrogateHS}{0.00}{SEOBNRv4TsurrogateLS}{0.00}{SEOBNRv4TsurrogatehighspinRIFT}{0.00}{SEOBNRv4TsurrogatelowspinRIFT}{0.00}{TEOBResumS-HS}{0.00}{TEOBResumS-LS}{0.00}{TaylorF2-HS}{0.00}{TaylorF2-LS}{0.00}{PrecessingSpinIMRTidalHS}{0.47}{PrecessingSpinIMRTidalLS}{0.03}{PublicationSamples}{0.47}}}
\newcommand{\spintwoxfourtwofivemed}[1]{\IfEqCase{#1}{{AlignedSpinInspiralTidalHS}{0.00}{AlignedSpinInspiralTidalLS}{0.00}{AlignedSpinTidalHS}{0.00}{AlignedSpinTidalLS}{0.00}{IMRPhenomDNRTidal-HS}{0.00}{IMRPhenomDNRTidal-LS}{0.00}{IMRPhenomPv2NRTidal-HS}{0.0007}{IMRPhenomPv2NRTidal-LS}{0.00}{SEOBNRv4TsurrogateHS}{0.00}{SEOBNRv4TsurrogateLS}{0.00}{SEOBNRv4TsurrogatehighspinRIFT}{0.00}{SEOBNRv4TsurrogatelowspinRIFT}{0.00}{TEOBResumS-HS}{0.00}{TEOBResumS-LS}{0.00}{TaylorF2-HS}{0.00}{TaylorF2-LS}{0.00}{PrecessingSpinIMRTidalHS}{0.0006}{PrecessingSpinIMRTidalLS}{0.00}{PublicationSamples}{0.0007}}}
\newcommand{\spintwoxfourtwofiveplus}[1]{\IfEqCase{#1}{{AlignedSpinInspiralTidalHS}{0.00}{AlignedSpinInspiralTidalLS}{0.00}{AlignedSpinTidalHS}{0.00}{AlignedSpinTidalLS}{0.00}{IMRPhenomDNRTidal-HS}{0.00}{IMRPhenomDNRTidal-LS}{0.00}{IMRPhenomPv2NRTidal-HS}{0.48}{IMRPhenomPv2NRTidal-LS}{0.03}{SEOBNRv4TsurrogateHS}{0.00}{SEOBNRv4TsurrogateLS}{0.00}{SEOBNRv4TsurrogatehighspinRIFT}{0.00}{SEOBNRv4TsurrogatelowspinRIFT}{0.00}{TEOBResumS-HS}{0.00}{TEOBResumS-LS}{0.00}{TaylorF2-HS}{0.00}{TaylorF2-LS}{0.00}{PrecessingSpinIMRTidalHS}{0.47}{PrecessingSpinIMRTidalLS}{0.03}{PublicationSamples}{0.47}}}
\newcommand{\massratiofourtwofiveminus}[1]{\IfEqCase{#1}{{AlignedSpinInspiralTidalHS}{0.24}{AlignedSpinInspiralTidalLS}{0.15}{AlignedSpinTidalHS}{0.30}{AlignedSpinTidalLS}{0.15}{IMRPhenomDNRTidal-HS}{0.31}{IMRPhenomDNRTidal-LS}{0.15}{IMRPhenomPv2NRTidal-HS}{0.25}{IMRPhenomPv2NRTidal-LS}{0.15}{SEOBNRv4TsurrogateHS}{0.29}{SEOBNRv4TsurrogateLS}{0.15}{SEOBNRv4TsurrogatehighspinRIFT}{0.27}{SEOBNRv4TsurrogatelowspinRIFT}{0.15}{TEOBResumS-HS}{0.27}{TEOBResumS-LS}{0.15}{TaylorF2-HS}{0.24}{TaylorF2-LS}{0.15}{PrecessingSpinIMRTidalHS}{0.25}{PrecessingSpinIMRTidalLS}{0.15}{PublicationSamples}{0.25}}}
\newcommand{\massratiofourtwofivemed}[1]{\IfEqCase{#1}{{AlignedSpinInspiralTidalHS}{0.70}{AlignedSpinInspiralTidalLS}{0.89}{AlignedSpinTidalHS}{0.74}{AlignedSpinTidalLS}{0.89}{IMRPhenomDNRTidal-HS}{0.72}{IMRPhenomDNRTidal-LS}{0.89}{IMRPhenomPv2NRTidal-HS}{0.67}{IMRPhenomPv2NRTidal-LS}{0.90}{SEOBNRv4TsurrogateHS}{0.77}{SEOBNRv4TsurrogateLS}{0.90}{SEOBNRv4TsurrogatehighspinRIFT}{0.78}{SEOBNRv4TsurrogatelowspinRIFT}{0.89}{TEOBResumS-HS}{0.75}{TEOBResumS-LS}{0.89}{TaylorF2-HS}{0.70}{TaylorF2-LS}{0.89}{PrecessingSpinIMRTidalHS}{0.67}{PrecessingSpinIMRTidalLS}{0.90}{PublicationSamples}{0.67}}}
\newcommand{\massratiofourtwofiveplus}[1]{\IfEqCase{#1}{{AlignedSpinInspiralTidalHS}{0.26}{AlignedSpinInspiralTidalLS}{0.10}{AlignedSpinTidalHS}{0.22}{AlignedSpinTidalLS}{0.09}{IMRPhenomDNRTidal-HS}{0.25}{IMRPhenomDNRTidal-LS}{0.09}{IMRPhenomPv2NRTidal-HS}{0.29}{IMRPhenomPv2NRTidal-LS}{0.09}{SEOBNRv4TsurrogateHS}{0.20}{SEOBNRv4TsurrogateLS}{0.09}{SEOBNRv4TsurrogatehighspinRIFT}{0.19}{SEOBNRv4TsurrogatelowspinRIFT}{0.10}{TEOBResumS-HS}{0.22}{TEOBResumS-LS}{0.10}{TaylorF2-HS}{0.26}{TaylorF2-LS}{0.10}{PrecessingSpinIMRTidalHS}{0.29}{PrecessingSpinIMRTidalLS}{0.09}{PublicationSamples}{0.29}}}
\newcommand{\spinonefourtwofiveminus}[1]{\IfEqCase{#1}{{AlignedSpinInspiralTidalHS}{0.06}{AlignedSpinInspiralTidalLS}{0.01}{AlignedSpinTidalHS}{0.06}{AlignedSpinTidalLS}{0.01}{IMRPhenomDNRTidal-HS}{0.08}{IMRPhenomDNRTidal-LS}{0.01}{IMRPhenomPv2NRTidal-HS}{0.25}{IMRPhenomPv2NRTidal-LS}{0.03}{SEOBNRv4TsurrogateHS}{0.05}{SEOBNRv4TsurrogateLS}{0.01}{SEOBNRv4TsurrogatehighspinRIFT}{0.05}{SEOBNRv4TsurrogatelowspinRIFT}{0.01}{TEOBResumS-HS}{0.05}{TEOBResumS-LS}{0.01}{TaylorF2-HS}{0.06}{TaylorF2-LS}{0.01}{PrecessingSpinIMRTidalHS}{0.25}{PrecessingSpinIMRTidalLS}{0.03}{PublicationSamples}{0.25}}}
\newcommand{\spinonefourtwofivemed}[1]{\IfEqCase{#1}{{AlignedSpinInspiralTidalHS}{0.06}{AlignedSpinInspiralTidalLS}{0.01}{AlignedSpinTidalHS}{0.06}{AlignedSpinTidalLS}{0.01}{IMRPhenomDNRTidal-HS}{0.09}{IMRPhenomDNRTidal-LS}{0.01}{IMRPhenomPv2NRTidal-HS}{0.27}{IMRPhenomPv2NRTidal-LS}{0.03}{SEOBNRv4TsurrogateHS}{0.06}{SEOBNRv4TsurrogateLS}{0.01}{SEOBNRv4TsurrogatehighspinRIFT}{0.06}{SEOBNRv4TsurrogatelowspinRIFT}{0.01}{TEOBResumS-HS}{0.06}{TEOBResumS-LS}{0.01}{TaylorF2-HS}{0.06}{TaylorF2-LS}{0.01}{PrecessingSpinIMRTidalHS}{0.27}{PrecessingSpinIMRTidalLS}{0.03}{PublicationSamples}{0.27}}}
\newcommand{\spinonefourtwofiveplus}[1]{\IfEqCase{#1}{{AlignedSpinInspiralTidalHS}{0.17}{AlignedSpinInspiralTidalLS}{0.03}{AlignedSpinTidalHS}{0.19}{AlignedSpinTidalLS}{0.03}{IMRPhenomDNRTidal-HS}{0.25}{IMRPhenomDNRTidal-LS}{0.03}{IMRPhenomPv2NRTidal-HS}{0.51}{IMRPhenomPv2NRTidal-LS}{0.02}{SEOBNRv4TsurrogateHS}{0.15}{SEOBNRv4TsurrogateLS}{0.03}{SEOBNRv4TsurrogatehighspinRIFT}{0.15}{SEOBNRv4TsurrogatelowspinRIFT}{0.03}{TEOBResumS-HS}{0.14}{TEOBResumS-LS}{0.03}{TaylorF2-HS}{0.17}{TaylorF2-LS}{0.03}{PrecessingSpinIMRTidalHS}{0.51}{PrecessingSpinIMRTidalLS}{0.02}{PublicationSamples}{0.51}}}
\newcommand{\costiltonefourtwofiveminus}[1]{\IfEqCase{#1}{{AlignedSpinInspiralTidalHS}{2.00}{AlignedSpinInspiralTidalLS}{2.00}{AlignedSpinTidalHS}{2.00}{AlignedSpinTidalLS}{2.00}{IMRPhenomDNRTidal-HS}{2.00}{IMRPhenomDNRTidal-LS}{2.00}{IMRPhenomPv2NRTidal-HS}{0.65}{IMRPhenomPv2NRTidal-LS}{1.10}{SEOBNRv4TsurrogateHS}{2.00}{SEOBNRv4TsurrogateLS}{2.00}{SEOBNRv4TsurrogatehighspinRIFT}{2.00}{SEOBNRv4TsurrogatelowspinRIFT}{2.00}{TEOBResumS-HS}{2.00}{TEOBResumS-LS}{2.00}{TaylorF2-HS}{2.00}{TaylorF2-LS}{2.00}{PrecessingSpinIMRTidalHS}{0.65}{PrecessingSpinIMRTidalLS}{1.10}{PublicationSamples}{0.65}}}
\newcommand{\costiltonefourtwofivemed}[1]{\IfEqCase{#1}{{AlignedSpinInspiralTidalHS}{1.00}{AlignedSpinInspiralTidalLS}{1.00}{AlignedSpinTidalHS}{1.00}{AlignedSpinTidalLS}{1.00}{IMRPhenomDNRTidal-HS}{1.00}{IMRPhenomDNRTidal-LS}{1.00}{IMRPhenomPv2NRTidal-HS}{0.26}{IMRPhenomPv2NRTidal-LS}{0.51}{SEOBNRv4TsurrogateHS}{1.00}{SEOBNRv4TsurrogateLS}{1.00}{SEOBNRv4TsurrogatehighspinRIFT}{1.00}{SEOBNRv4TsurrogatelowspinRIFT}{1.00}{TEOBResumS-HS}{1.00}{TEOBResumS-LS}{1.00}{TaylorF2-HS}{1.00}{TaylorF2-LS}{1.00}{PrecessingSpinIMRTidalHS}{0.26}{PrecessingSpinIMRTidalLS}{0.51}{PublicationSamples}{0.26}}}
\newcommand{\costiltonefourtwofiveplus}[1]{\IfEqCase{#1}{{AlignedSpinInspiralTidalHS}{0.00}{AlignedSpinInspiralTidalLS}{0.00}{AlignedSpinTidalHS}{0.00}{AlignedSpinTidalLS}{0.00}{IMRPhenomDNRTidal-HS}{0.00}{IMRPhenomDNRTidal-LS}{0.00}{IMRPhenomPv2NRTidal-HS}{0.61}{IMRPhenomPv2NRTidal-LS}{0.45}{SEOBNRv4TsurrogateHS}{0.00}{SEOBNRv4TsurrogateLS}{0.00}{SEOBNRv4TsurrogatehighspinRIFT}{0.00}{SEOBNRv4TsurrogatelowspinRIFT}{0.00}{TEOBResumS-HS}{0.00}{TEOBResumS-LS}{0.00}{TaylorF2-HS}{0.00}{TaylorF2-LS}{0.00}{PrecessingSpinIMRTidalHS}{0.61}{PrecessingSpinIMRTidalLS}{0.44}{PublicationSamples}{0.61}}}
\newcommand{\phasefourtwofiveminus}[1]{\IfEqCase{#1}{{AlignedSpinInspiralTidalHS}{3.10}{AlignedSpinInspiralTidalLS}{2.74}{AlignedSpinTidalHS}{2.81}{AlignedSpinTidalLS}{2.76}{IMRPhenomDNRTidal-HS}{2.83}{IMRPhenomDNRTidal-LS}{2.62}{IMRPhenomPv2NRTidal-HS}{2.82}{IMRPhenomPv2NRTidal-LS}{2.85}{SEOBNRv4TsurrogateHS}{2.74}{SEOBNRv4TsurrogateLS}{2.81}{SEOBNRv4TsurrogatehighspinRIFT}{2.78}{SEOBNRv4TsurrogatelowspinRIFT}{2.79}{TEOBResumS-HS}{2.78}{TEOBResumS-LS}{2.83}{TaylorF2-HS}{3.13}{TaylorF2-LS}{2.74}{PrecessingSpinIMRTidalHS}{2.82}{PrecessingSpinIMRTidalLS}{2.86}{PublicationSamples}{2.82}}}
\newcommand{\phasefourtwofivemed}[1]{\IfEqCase{#1}{{AlignedSpinInspiralTidalHS}{3.44}{AlignedSpinInspiralTidalLS}{3.01}{AlignedSpinTidalHS}{3.13}{AlignedSpinTidalLS}{3.07}{IMRPhenomDNRTidal-HS}{3.13}{IMRPhenomDNRTidal-LS}{2.88}{IMRPhenomPv2NRTidal-HS}{3.13}{IMRPhenomPv2NRTidal-LS}{3.20}{SEOBNRv4TsurrogateHS}{3.11}{SEOBNRv4TsurrogateLS}{3.16}{SEOBNRv4TsurrogatehighspinRIFT}{3.12}{SEOBNRv4TsurrogatelowspinRIFT}{3.12}{TEOBResumS-HS}{3.09}{TEOBResumS-LS}{3.14}{TaylorF2-HS}{3.46}{TaylorF2-LS}{3.00}{PrecessingSpinIMRTidalHS}{3.12}{PrecessingSpinIMRTidalLS}{3.20}{PublicationSamples}{3.13}}}
\newcommand{\phasefourtwofiveplus}[1]{\IfEqCase{#1}{{AlignedSpinInspiralTidalHS}{2.49}{AlignedSpinInspiralTidalLS}{2.93}{AlignedSpinTidalHS}{2.82}{AlignedSpinTidalLS}{2.88}{IMRPhenomDNRTidal-HS}{2.82}{IMRPhenomDNRTidal-LS}{3.04}{IMRPhenomPv2NRTidal-HS}{2.86}{IMRPhenomPv2NRTidal-LS}{2.74}{SEOBNRv4TsurrogateHS}{2.81}{SEOBNRv4TsurrogateLS}{2.74}{SEOBNRv4TsurrogatehighspinRIFT}{2.81}{SEOBNRv4TsurrogatelowspinRIFT}{2.79}{TEOBResumS-HS}{2.87}{TEOBResumS-LS}{2.86}{TaylorF2-HS}{2.47}{TaylorF2-LS}{2.94}{PrecessingSpinIMRTidalHS}{2.87}{PrecessingSpinIMRTidalLS}{2.73}{PublicationSamples}{2.86}}}
\newcommand{\masstwodetfourtwofiveminus}[1]{\IfEqCase{#1}{{AlignedSpinInspiralTidalHS}{0.3}{AlignedSpinInspiralTidalLS}{0.1}{AlignedSpinTidalHS}{0.3}{AlignedSpinTidalLS}{0.1}{IMRPhenomDNRTidal-HS}{0.3}{IMRPhenomDNRTidal-LS}{0.1}{IMRPhenomPv2NRTidal-HS}{0.3}{IMRPhenomPv2NRTidal-LS}{0.1}{SEOBNRv4TsurrogateHS}{0.3}{SEOBNRv4TsurrogateLS}{0.1}{SEOBNRv4TsurrogatehighspinRIFT}{0.3}{SEOBNRv4TsurrogatelowspinRIFT}{0.1}{TEOBResumS-HS}{0.3}{TEOBResumS-LS}{0.1}{TaylorF2-HS}{0.3}{TaylorF2-LS}{0.1}{PrecessingSpinIMRTidalHS}{0.3}{PrecessingSpinIMRTidalLS}{0.1}{PublicationSamples}{0.3}}}
\newcommand{\masstwodetfourtwofivemed}[1]{\IfEqCase{#1}{{AlignedSpinInspiralTidalHS}{1.4}{AlignedSpinInspiralTidalLS}{1.61}{AlignedSpinTidalHS}{1.5}{AlignedSpinTidalLS}{1.61}{IMRPhenomDNRTidal-HS}{1.5}{IMRPhenomDNRTidal-LS}{1.62}{IMRPhenomPv2NRTidal-HS}{1.4}{IMRPhenomPv2NRTidal-LS}{1.62}{SEOBNRv4TsurrogateHS}{1.5}{SEOBNRv4TsurrogateLS}{1.62}{SEOBNRv4TsurrogatehighspinRIFT}{1.5}{SEOBNRv4TsurrogatelowspinRIFT}{1.61}{TEOBResumS-HS}{1.5}{TEOBResumS-LS}{1.61}{TaylorF2-HS}{1.4}{TaylorF2-LS}{1.61}{PrecessingSpinIMRTidalHS}{1.4}{PrecessingSpinIMRTidalLS}{1.62}{PublicationSamples}{1.4}}}
\newcommand{\masstwodetfourtwofiveplus}[1]{\IfEqCase{#1}{{AlignedSpinInspiralTidalHS}{0.2}{AlignedSpinInspiralTidalLS}{0.09}{AlignedSpinTidalHS}{0.2}{AlignedSpinTidalLS}{0.08}{IMRPhenomDNRTidal-HS}{0.2}{IMRPhenomDNRTidal-LS}{0.08}{IMRPhenomPv2NRTidal-HS}{0.3}{IMRPhenomPv2NRTidal-LS}{0.08}{SEOBNRv4TsurrogateHS}{0.2}{SEOBNRv4TsurrogateLS}{0.08}{SEOBNRv4TsurrogatehighspinRIFT}{0.2}{SEOBNRv4TsurrogatelowspinRIFT}{0.08}{TEOBResumS-HS}{0.2}{TEOBResumS-LS}{0.09}{TaylorF2-HS}{0.2}{TaylorF2-LS}{0.09}{PrecessingSpinIMRTidalHS}{0.3}{PrecessingSpinIMRTidalLS}{0.08}{PublicationSamples}{0.3}}}
\newcommand{\masstwosourcefourtwofiveminus}[1]{\IfEqCase{#1}{{AlignedSpinInspiralTidalHS}{0.3}{AlignedSpinInspiralTidalLS}{0.1}{AlignedSpinTidalHS}{0.3}{AlignedSpinTidalLS}{0.1}{IMRPhenomDNRTidal-HS}{0.3}{IMRPhenomDNRTidal-LS}{0.1}{IMRPhenomPv2NRTidal-HS}{0.3}{IMRPhenomPv2NRTidal-LS}{0.1}{SEOBNRv4TsurrogateHS}{0.3}{SEOBNRv4TsurrogateLS}{0.1}{SEOBNRv4TsurrogatehighspinRIFT}{0.3}{SEOBNRv4TsurrogatelowspinRIFT}{0.1}{TEOBResumS-HS}{0.3}{TEOBResumS-LS}{0.1}{TaylorF2-HS}{0.3}{TaylorF2-LS}{0.1}{PrecessingSpinIMRTidalHS}{0.3}{PrecessingSpinIMRTidalLS}{0.1}{PublicationSamples}{0.3}}}
\newcommand{\masstwosourcefourtwofivemed}[1]{\IfEqCase{#1}{{AlignedSpinInspiralTidalHS}{1.4}{AlignedSpinInspiralTidalLS}{1.56}{AlignedSpinTidalHS}{1.4}{AlignedSpinTidalLS}{1.56}{IMRPhenomDNRTidal-HS}{1.4}{IMRPhenomDNRTidal-LS}{1.56}{IMRPhenomPv2NRTidal-HS}{1.4}{IMRPhenomPv2NRTidal-LS}{1.57}{SEOBNRv4TsurrogateHS}{1.4}{SEOBNRv4TsurrogateLS}{1.56}{SEOBNRv4TsurrogatehighspinRIFT}{1.5}{SEOBNRv4TsurrogatelowspinRIFT}{1.56}{TEOBResumS-HS}{1.4}{TEOBResumS-LS}{1.56}{TaylorF2-HS}{1.4}{TaylorF2-LS}{1.56}{PrecessingSpinIMRTidalHS}{1.4}{PrecessingSpinIMRTidalLS}{1.57}{PublicationSamples}{1.4}}}
\newcommand{\masstwosourcefourtwofiveplus}[1]{\IfEqCase{#1}{{AlignedSpinInspiralTidalHS}{0.2}{AlignedSpinInspiralTidalLS}{0.09}{AlignedSpinTidalHS}{0.2}{AlignedSpinTidalLS}{0.08}{IMRPhenomDNRTidal-HS}{0.2}{IMRPhenomDNRTidal-LS}{0.08}{IMRPhenomPv2NRTidal-HS}{0.3}{IMRPhenomPv2NRTidal-LS}{0.08}{SEOBNRv4TsurrogateHS}{0.2}{SEOBNRv4TsurrogateLS}{0.08}{SEOBNRv4TsurrogatehighspinRIFT}{0.2}{SEOBNRv4TsurrogatelowspinRIFT}{0.08}{TEOBResumS-HS}{0.2}{TEOBResumS-LS}{0.09}{TaylorF2-HS}{0.2}{TaylorF2-LS}{0.09}{PrecessingSpinIMRTidalHS}{0.3}{PrecessingSpinIMRTidalLS}{0.08}{PublicationSamples}{0.3}}}
\newcommand{\decfourtwofiveminus}[1]{\IfEqCase{#1}{{AlignedSpinInspiralTidalHS}{0.92098}{AlignedSpinInspiralTidalLS}{0.94910}{AlignedSpinTidalHS}{0.88455}{AlignedSpinTidalLS}{0.88762}{IMRPhenomDNRTidal-HS}{0.92222}{IMRPhenomDNRTidal-LS}{0.90469}{IMRPhenomPv2NRTidal-HS}{0.90104}{IMRPhenomPv2NRTidal-LS}{0.97042}{SEOBNRv4TsurrogateHS}{0.87704}{SEOBNRv4TsurrogateLS}{0.89335}{SEOBNRv4TsurrogatehighspinRIFT}{0.89796}{SEOBNRv4TsurrogatelowspinRIFT}{0.90269}{TEOBResumS-HS}{0.88789}{TEOBResumS-LS}{0.89751}{TaylorF2-HS}{0.92183}{TaylorF2-LS}{0.95570}{PrecessingSpinIMRTidalHS}{0.89977}{PrecessingSpinIMRTidalLS}{0.96778}{PublicationSamples}{0.89807}}}
\newcommand{\decfourtwofivemed}[1]{\IfEqCase{#1}{{AlignedSpinInspiralTidalHS}{-0.14949}{AlignedSpinInspiralTidalLS}{-0.10865}{AlignedSpinTidalHS}{-0.15883}{AlignedSpinTidalLS}{-0.13405}{IMRPhenomDNRTidal-HS}{-0.18418}{IMRPhenomDNRTidal-LS}{-0.06597}{IMRPhenomPv2NRTidal-HS}{-0.12984}{IMRPhenomPv2NRTidal-LS}{-0.05685}{SEOBNRv4TsurrogateHS}{-0.14893}{SEOBNRv4TsurrogateLS}{-0.17562}{SEOBNRv4TsurrogatehighspinRIFT}{-0.15053}{SEOBNRv4TsurrogatelowspinRIFT}{-0.12883}{TEOBResumS-HS}{-0.15241}{TEOBResumS-LS}{-0.16207}{TaylorF2-HS}{-0.14824}{TaylorF2-LS}{-0.10463}{PrecessingSpinIMRTidalHS}{-0.13006}{PrecessingSpinIMRTidalLS}{-0.06120}{PublicationSamples}{-0.13133}}}
\newcommand{\decfourtwofiveplus}[1]{\IfEqCase{#1}{{AlignedSpinInspiralTidalHS}{0.99909}{AlignedSpinInspiralTidalLS}{0.93177}{AlignedSpinTidalHS}{0.97810}{AlignedSpinTidalLS}{0.98373}{IMRPhenomDNRTidal-HS}{0.95052}{IMRPhenomDNRTidal-LS}{0.93229}{IMRPhenomPv2NRTidal-HS}{0.96946}{IMRPhenomPv2NRTidal-LS}{0.91574}{SEOBNRv4TsurrogateHS}{1.00436}{SEOBNRv4TsurrogateLS}{1.00055}{SEOBNRv4TsurrogatehighspinRIFT}{0.99086}{SEOBNRv4TsurrogatelowspinRIFT}{0.97101}{TEOBResumS-HS}{0.98792}{TEOBResumS-LS}{1.00135}{TaylorF2-HS}{0.99948}{TaylorF2-LS}{0.92354}{PrecessingSpinIMRTidalHS}{0.96811}{PrecessingSpinIMRTidalLS}{0.91709}{PublicationSamples}{0.96897}}}
\newcommand{\psifourtwofiveminus}[1]{\IfEqCase{#1}{{AlignedSpinInspiralTidalHS}{1.40}{AlignedSpinInspiralTidalLS}{1.41}{AlignedSpinTidalHS}{1.63}{AlignedSpinTidalLS}{1.69}{IMRPhenomDNRTidal-HS}{1.36}{IMRPhenomDNRTidal-LS}{1.41}{IMRPhenomPv2NRTidal-HS}{1.46}{IMRPhenomPv2NRTidal-LS}{1.40}{SEOBNRv4TsurrogateHS}{1.36}{SEOBNRv4TsurrogateLS}{1.43}{SEOBNRv4TsurrogatehighspinRIFT}{2.83}{SEOBNRv4TsurrogatelowspinRIFT}{2.86}{TEOBResumS-HS}{2.86}{TEOBResumS-LS}{2.86}{TaylorF2-HS}{1.41}{TaylorF2-LS}{1.41}{PrecessingSpinIMRTidalHS}{1.46}{PrecessingSpinIMRTidalLS}{1.39}{PublicationSamples}{1.46}}}
\newcommand{\psifourtwofivemed}[1]{\IfEqCase{#1}{{AlignedSpinInspiralTidalHS}{1.55}{AlignedSpinInspiralTidalLS}{1.56}{AlignedSpinTidalHS}{1.80}{AlignedSpinTidalLS}{1.86}{IMRPhenomDNRTidal-HS}{1.50}{IMRPhenomDNRTidal-LS}{1.56}{IMRPhenomPv2NRTidal-HS}{1.61}{IMRPhenomPv2NRTidal-LS}{1.54}{SEOBNRv4TsurrogateHS}{1.50}{SEOBNRv4TsurrogateLS}{1.56}{SEOBNRv4TsurrogatehighspinRIFT}{3.13}{SEOBNRv4TsurrogatelowspinRIFT}{3.14}{TEOBResumS-HS}{3.16}{TEOBResumS-LS}{3.16}{TaylorF2-HS}{1.56}{TaylorF2-LS}{1.55}{PrecessingSpinIMRTidalHS}{1.62}{PrecessingSpinIMRTidalLS}{1.54}{PublicationSamples}{1.61}}}
\newcommand{\psifourtwofiveplus}[1]{\IfEqCase{#1}{{AlignedSpinInspiralTidalHS}{1.42}{AlignedSpinInspiralTidalLS}{1.44}{AlignedSpinTidalHS}{3.54}{AlignedSpinTidalLS}{3.48}{IMRPhenomDNRTidal-HS}{1.50}{IMRPhenomDNRTidal-LS}{1.43}{IMRPhenomPv2NRTidal-HS}{1.38}{IMRPhenomPv2NRTidal-LS}{1.43}{SEOBNRv4TsurrogateHS}{1.47}{SEOBNRv4TsurrogateLS}{1.42}{SEOBNRv4TsurrogatehighspinRIFT}{2.85}{SEOBNRv4TsurrogatelowspinRIFT}{2.83}{TEOBResumS-HS}{2.82}{TEOBResumS-LS}{2.84}{TaylorF2-HS}{1.41}{TaylorF2-LS}{1.44}{PrecessingSpinIMRTidalHS}{1.38}{PrecessingSpinIMRTidalLS}{1.43}{PublicationSamples}{1.38}}}
\newcommand{\networkoptimalsnrfourtwofiveminus}[1]{\IfEqCase{#1}{{AlignedSpinInspiralTidalHS}{1.7}{AlignedSpinInspiralTidalLS}{1.7}{IMRPhenomDNRTidal-HS}{1.7}{IMRPhenomDNRTidal-LS}{1.7}{IMRPhenomPv2NRTidal-HS}{1.7}{IMRPhenomPv2NRTidal-LS}{1.7}{SEOBNRv4TsurrogateHS}{1.7}{SEOBNRv4TsurrogateLS}{1.7}{TaylorF2-HS}{1.7}{TaylorF2-LS}{1.7}{PrecessingSpinIMRTidalHS}{1.7}{PrecessingSpinIMRTidalLS}{1.7}{PublicationSamples}{1.7}}}
\newcommand{\networkoptimalsnrfourtwofivemed}[1]{\IfEqCase{#1}{{AlignedSpinInspiralTidalHS}{12.1}{AlignedSpinInspiralTidalLS}{12.2}{IMRPhenomDNRTidal-HS}{12.0}{IMRPhenomDNRTidal-LS}{12.1}{IMRPhenomPv2NRTidal-HS}{12.0}{IMRPhenomPv2NRTidal-LS}{12.1}{SEOBNRv4TsurrogateHS}{12.0}{SEOBNRv4TsurrogateLS}{12.1}{TaylorF2-HS}{12.1}{TaylorF2-LS}{12.2}{PrecessingSpinIMRTidalHS}{12.0}{PrecessingSpinIMRTidalLS}{12.1}{PublicationSamples}{12.0}}}
\newcommand{\networkoptimalsnrfourtwofiveplus}[1]{\IfEqCase{#1}{{AlignedSpinInspiralTidalHS}{1.7}{AlignedSpinInspiralTidalLS}{1.7}{IMRPhenomDNRTidal-HS}{1.7}{IMRPhenomDNRTidal-LS}{1.7}{IMRPhenomPv2NRTidal-HS}{1.7}{IMRPhenomPv2NRTidal-LS}{1.7}{SEOBNRv4TsurrogateHS}{1.7}{SEOBNRv4TsurrogateLS}{1.7}{TaylorF2-HS}{1.7}{TaylorF2-LS}{1.7}{PrecessingSpinIMRTidalHS}{1.7}{PrecessingSpinIMRTidalLS}{1.7}{PublicationSamples}{1.7}}}
\newcommand{\thetajnfourtwofiveminus}[1]{\IfEqCase{#1}{{AlignedSpinInspiralTidalHS}{0.97}{AlignedSpinInspiralTidalLS}{0.88}{AlignedSpinTidalHS}{0.83}{AlignedSpinTidalLS}{0.83}{IMRPhenomDNRTidal-HS}{0.74}{IMRPhenomDNRTidal-LS}{0.78}{IMRPhenomPv2NRTidal-HS}{0.85}{IMRPhenomPv2NRTidal-LS}{0.84}{SEOBNRv4TsurrogateHS}{0.89}{SEOBNRv4TsurrogateLS}{0.85}{SEOBNRv4TsurrogatehighspinRIFT}{0.86}{SEOBNRv4TsurrogatelowspinRIFT}{0.86}{TEOBResumS-HS}{0.86}{TEOBResumS-LS}{0.88}{TaylorF2-HS}{0.97}{TaylorF2-LS}{0.88}{PrecessingSpinIMRTidalHS}{0.85}{PrecessingSpinIMRTidalLS}{0.84}{PublicationSamples}{0.85}}}
\newcommand{\thetajnfourtwofivemed}[1]{\IfEqCase{#1}{{AlignedSpinInspiralTidalHS}{1.22}{AlignedSpinInspiralTidalLS}{1.11}{AlignedSpinTidalHS}{1.06}{AlignedSpinTidalLS}{1.06}{IMRPhenomDNRTidal-HS}{0.96}{IMRPhenomDNRTidal-LS}{1.02}{IMRPhenomPv2NRTidal-HS}{1.08}{IMRPhenomPv2NRTidal-LS}{1.07}{SEOBNRv4TsurrogateHS}{1.13}{SEOBNRv4TsurrogateLS}{1.09}{SEOBNRv4TsurrogatehighspinRIFT}{1.10}{SEOBNRv4TsurrogatelowspinRIFT}{1.09}{TEOBResumS-HS}{1.09}{TEOBResumS-LS}{1.12}{TaylorF2-HS}{1.22}{TaylorF2-LS}{1.11}{PrecessingSpinIMRTidalHS}{1.08}{PrecessingSpinIMRTidalLS}{1.07}{PublicationSamples}{1.08}}}
\newcommand{\thetajnfourtwofiveplus}[1]{\IfEqCase{#1}{{AlignedSpinInspiralTidalHS}{1.64}{AlignedSpinInspiralTidalLS}{1.75}{AlignedSpinTidalHS}{1.79}{AlignedSpinTidalLS}{1.78}{IMRPhenomDNRTidal-HS}{1.88}{IMRPhenomDNRTidal-LS}{1.83}{IMRPhenomPv2NRTidal-HS}{1.77}{IMRPhenomPv2NRTidal-LS}{1.78}{SEOBNRv4TsurrogateHS}{1.72}{SEOBNRv4TsurrogateLS}{1.76}{SEOBNRv4TsurrogatehighspinRIFT}{1.76}{SEOBNRv4TsurrogatelowspinRIFT}{1.75}{TEOBResumS-HS}{1.77}{TEOBResumS-LS}{1.74}{TaylorF2-HS}{1.63}{TaylorF2-LS}{1.75}{PrecessingSpinIMRTidalHS}{1.77}{PrecessingSpinIMRTidalLS}{1.78}{PublicationSamples}{1.77}}}
\newcommand{\totalmassdetfourtwofiveminus}[1]{\IfEqCase{#1}{{AlignedSpinInspiralTidalHS}{0.06}{AlignedSpinInspiralTidalLS}{0.007}{AlignedSpinTidalHS}{0.04}{AlignedSpinTidalLS}{0.007}{IMRPhenomDNRTidal-HS}{0.06}{IMRPhenomDNRTidal-LS}{0.006}{IMRPhenomPv2NRTidal-HS}{0.08}{IMRPhenomPv2NRTidal-LS}{0.006}{SEOBNRv4TsurrogateHS}{0.03}{SEOBNRv4TsurrogateLS}{0.006}{SEOBNRv4TsurrogatehighspinRIFT}{0.03}{SEOBNRv4TsurrogatelowspinRIFT}{0.007}{TEOBResumS-HS}{0.04}{TEOBResumS-LS}{0.007}{TaylorF2-HS}{0.06}{TaylorF2-LS}{0.007}{PrecessingSpinIMRTidalHS}{0.08}{PrecessingSpinIMRTidalLS}{0.006}{PublicationSamples}{0.08}}}
\newcommand{\totalmassdetfourtwofivemed}[1]{\IfEqCase{#1}{{AlignedSpinInspiralTidalHS}{3.48}{AlignedSpinInspiralTidalLS}{3.42}{AlignedSpinTidalHS}{3.46}{AlignedSpinTidalLS}{3.42}{IMRPhenomDNRTidal-HS}{3.47}{IMRPhenomDNRTidal-LS}{3.42}{IMRPhenomPv2NRTidal-HS}{3.50}{IMRPhenomPv2NRTidal-LS}{3.42}{SEOBNRv4TsurrogateHS}{3.45}{SEOBNRv4TsurrogateLS}{3.42}{SEOBNRv4TsurrogatehighspinRIFT}{3.45}{SEOBNRv4TsurrogatelowspinRIFT}{3.42}{TEOBResumS-HS}{3.46}{TEOBResumS-LS}{3.42}{TaylorF2-HS}{3.48}{TaylorF2-LS}{3.42}{PrecessingSpinIMRTidalHS}{3.50}{PrecessingSpinIMRTidalLS}{3.42}{PublicationSamples}{3.50}}}
\newcommand{\totalmassdetfourtwofiveplus}[1]{\IfEqCase{#1}{{AlignedSpinInspiralTidalHS}{0.3}{AlignedSpinInspiralTidalLS}{0.04}{AlignedSpinTidalHS}{0.3}{AlignedSpinTidalLS}{0.04}{IMRPhenomDNRTidal-HS}{0.4}{IMRPhenomDNRTidal-LS}{0.04}{IMRPhenomPv2NRTidal-HS}{0.3}{IMRPhenomPv2NRTidal-LS}{0.04}{SEOBNRv4TsurrogateHS}{0.2}{SEOBNRv4TsurrogateLS}{0.04}{SEOBNRv4TsurrogatehighspinRIFT}{0.2}{SEOBNRv4TsurrogatelowspinRIFT}{0.04}{TEOBResumS-HS}{0.2}{TEOBResumS-LS}{0.04}{TaylorF2-HS}{0.3}{TaylorF2-LS}{0.04}{PrecessingSpinIMRTidalHS}{0.3}{PrecessingSpinIMRTidalLS}{0.04}{PublicationSamples}{0.3}}}
\newcommand{\redshiftfourtwofiveminus}[1]{\IfEqCase{#1}{{AlignedSpinInspiralTidalHS}{0.02}{AlignedSpinInspiralTidalLS}{0.02}{AlignedSpinTidalHS}{0.02}{AlignedSpinTidalLS}{0.02}{IMRPhenomDNRTidal-HS}{0.02}{IMRPhenomDNRTidal-LS}{0.02}{IMRPhenomPv2NRTidal-HS}{0.02}{IMRPhenomPv2NRTidal-LS}{0.02}{SEOBNRv4TsurrogateHS}{0.02}{SEOBNRv4TsurrogateLS}{0.02}{SEOBNRv4TsurrogatehighspinRIFT}{0.02}{SEOBNRv4TsurrogatelowspinRIFT}{0.02}{TEOBResumS-HS}{0.02}{TEOBResumS-LS}{0.02}{TaylorF2-HS}{0.02}{TaylorF2-LS}{0.02}{PrecessingSpinIMRTidalHS}{0.02}{PrecessingSpinIMRTidalLS}{0.02}{PublicationSamples}{0.02}}}
\newcommand{\redshiftfourtwofivemed}[1]{\IfEqCase{#1}{{AlignedSpinInspiralTidalHS}{0.04}{AlignedSpinInspiralTidalLS}{0.04}{AlignedSpinTidalHS}{0.04}{AlignedSpinTidalLS}{0.03}{IMRPhenomDNRTidal-HS}{0.04}{IMRPhenomDNRTidal-LS}{0.03}{IMRPhenomPv2NRTidal-HS}{0.03}{IMRPhenomPv2NRTidal-LS}{0.03}{SEOBNRv4TsurrogateHS}{0.03}{SEOBNRv4TsurrogateLS}{0.03}{SEOBNRv4TsurrogatehighspinRIFT}{0.04}{SEOBNRv4TsurrogatelowspinRIFT}{0.03}{TEOBResumS-HS}{0.04}{TEOBResumS-LS}{0.03}{TaylorF2-HS}{0.04}{TaylorF2-LS}{0.04}{PrecessingSpinIMRTidalHS}{0.03}{PrecessingSpinIMRTidalLS}{0.03}{PublicationSamples}{0.03}}}
\newcommand{\redshiftfourtwofiveplus}[1]{\IfEqCase{#1}{{AlignedSpinInspiralTidalHS}{0.02}{AlignedSpinInspiralTidalLS}{0.01}{AlignedSpinTidalHS}{0.02}{AlignedSpinTidalLS}{0.01}{IMRPhenomDNRTidal-HS}{0.01}{IMRPhenomDNRTidal-LS}{0.01}{IMRPhenomPv2NRTidal-HS}{0.01}{IMRPhenomPv2NRTidal-LS}{0.01}{SEOBNRv4TsurrogateHS}{0.01}{SEOBNRv4TsurrogateLS}{0.01}{SEOBNRv4TsurrogatehighspinRIFT}{0.02}{SEOBNRv4TsurrogatelowspinRIFT}{0.02}{TEOBResumS-HS}{0.02}{TEOBResumS-LS}{0.01}{TaylorF2-HS}{0.02}{TaylorF2-LS}{0.02}{PrecessingSpinIMRTidalHS}{0.01}{PrecessingSpinIMRTidalLS}{0.01}{PublicationSamples}{0.01}}}
\newcommand{\iotafourtwofiveminus}[1]{\IfEqCase{#1}{{AlignedSpinInspiralTidalHS}{0.97}{AlignedSpinInspiralTidalLS}{0.88}{AlignedSpinTidalHS}{0.83}{AlignedSpinTidalLS}{0.83}{IMRPhenomDNRTidal-HS}{0.74}{IMRPhenomDNRTidal-LS}{0.78}{IMRPhenomPv2NRTidal-HS}{0.85}{IMRPhenomPv2NRTidal-LS}{0.84}{SEOBNRv4TsurrogateHS}{0.89}{SEOBNRv4TsurrogateLS}{0.85}{SEOBNRv4TsurrogatehighspinRIFT}{0.86}{SEOBNRv4TsurrogatelowspinRIFT}{0.86}{TEOBResumS-HS}{0.86}{TEOBResumS-LS}{0.88}{TaylorF2-HS}{0.97}{TaylorF2-LS}{0.88}{PrecessingSpinIMRTidalHS}{0.85}{PrecessingSpinIMRTidalLS}{0.84}{PublicationSamples}{0.85}}}
\newcommand{\iotafourtwofivemed}[1]{\IfEqCase{#1}{{AlignedSpinInspiralTidalHS}{1.22}{AlignedSpinInspiralTidalLS}{1.11}{AlignedSpinTidalHS}{1.06}{AlignedSpinTidalLS}{1.06}{IMRPhenomDNRTidal-HS}{0.96}{IMRPhenomDNRTidal-LS}{1.02}{IMRPhenomPv2NRTidal-HS}{1.09}{IMRPhenomPv2NRTidal-LS}{1.07}{SEOBNRv4TsurrogateHS}{1.13}{SEOBNRv4TsurrogateLS}{1.09}{SEOBNRv4TsurrogatehighspinRIFT}{1.10}{SEOBNRv4TsurrogatelowspinRIFT}{1.09}{TEOBResumS-HS}{1.09}{TEOBResumS-LS}{1.12}{TaylorF2-HS}{1.22}{TaylorF2-LS}{1.11}{PrecessingSpinIMRTidalHS}{1.09}{PrecessingSpinIMRTidalLS}{1.07}{PublicationSamples}{1.09}}}
\newcommand{\iotafourtwofiveplus}[1]{\IfEqCase{#1}{{AlignedSpinInspiralTidalHS}{1.64}{AlignedSpinInspiralTidalLS}{1.75}{AlignedSpinTidalHS}{1.79}{AlignedSpinTidalLS}{1.78}{IMRPhenomDNRTidal-HS}{1.88}{IMRPhenomDNRTidal-LS}{1.83}{IMRPhenomPv2NRTidal-HS}{1.77}{IMRPhenomPv2NRTidal-LS}{1.78}{SEOBNRv4TsurrogateHS}{1.72}{SEOBNRv4TsurrogateLS}{1.76}{SEOBNRv4TsurrogatehighspinRIFT}{1.76}{SEOBNRv4TsurrogatelowspinRIFT}{1.75}{TEOBResumS-HS}{1.77}{TEOBResumS-LS}{1.74}{TaylorF2-HS}{1.63}{TaylorF2-LS}{1.75}{PrecessingSpinIMRTidalHS}{1.77}{PrecessingSpinIMRTidalLS}{1.78}{PublicationSamples}{1.77}}}
\newcommand{\spinonexfourtwofiveminus}[1]{\IfEqCase{#1}{{AlignedSpinInspiralTidalHS}{0.00}{AlignedSpinInspiralTidalLS}{0.00}{AlignedSpinTidalHS}{0.00}{AlignedSpinTidalLS}{0.00}{IMRPhenomDNRTidal-HS}{0.00}{IMRPhenomDNRTidal-LS}{0.00}{IMRPhenomPv2NRTidal-HS}{0.50}{IMRPhenomPv2NRTidal-LS}{0.03}{SEOBNRv4TsurrogateHS}{0.00}{SEOBNRv4TsurrogateLS}{0.00}{SEOBNRv4TsurrogatehighspinRIFT}{0.00}{SEOBNRv4TsurrogatelowspinRIFT}{0.00}{TEOBResumS-HS}{0.00}{TEOBResumS-LS}{0.00}{TaylorF2-HS}{0.00}{TaylorF2-LS}{0.00}{PrecessingSpinIMRTidalHS}{0.50}{PrecessingSpinIMRTidalLS}{0.03}{PublicationSamples}{0.50}}}
\newcommand{\spinonexfourtwofivemed}[1]{\IfEqCase{#1}{{AlignedSpinInspiralTidalHS}{0.00}{AlignedSpinInspiralTidalLS}{0.00}{AlignedSpinTidalHS}{0.00}{AlignedSpinTidalLS}{0.00}{IMRPhenomDNRTidal-HS}{0.00}{IMRPhenomDNRTidal-LS}{0.00}{IMRPhenomPv2NRTidal-HS}{0.00}{IMRPhenomPv2NRTidal-LS}{0.00009}{SEOBNRv4TsurrogateHS}{0.00}{SEOBNRv4TsurrogateLS}{0.00}{SEOBNRv4TsurrogatehighspinRIFT}{0.00}{SEOBNRv4TsurrogatelowspinRIFT}{0.00}{TEOBResumS-HS}{0.00}{TEOBResumS-LS}{0.00}{TaylorF2-HS}{0.00}{TaylorF2-LS}{0.00}{PrecessingSpinIMRTidalHS}{0.00}{PrecessingSpinIMRTidalLS}{0.0001}{PublicationSamples}{0.00}}}
\newcommand{\spinonexfourtwofiveplus}[1]{\IfEqCase{#1}{{AlignedSpinInspiralTidalHS}{0.00}{AlignedSpinInspiralTidalLS}{0.00}{AlignedSpinTidalHS}{0.00}{AlignedSpinTidalLS}{0.00}{IMRPhenomDNRTidal-HS}{0.00}{IMRPhenomDNRTidal-LS}{0.00}{IMRPhenomPv2NRTidal-HS}{0.47}{IMRPhenomPv2NRTidal-LS}{0.03}{SEOBNRv4TsurrogateHS}{0.00}{SEOBNRv4TsurrogateLS}{0.00}{SEOBNRv4TsurrogatehighspinRIFT}{0.00}{SEOBNRv4TsurrogatelowspinRIFT}{0.00}{TEOBResumS-HS}{0.00}{TEOBResumS-LS}{0.00}{TaylorF2-HS}{0.00}{TaylorF2-LS}{0.00}{PrecessingSpinIMRTidalHS}{0.47}{PrecessingSpinIMRTidalLS}{0.03}{PublicationSamples}{0.47}}}
\newcommand{\chirpmassdetfourtwofiveminus}[1]{\IfEqCase{#1}{{AlignedSpinInspiralTidalHS}{0.0005}{AlignedSpinInspiralTidalLS}{0.0003}{AlignedSpinTidalHS}{0.0005}{AlignedSpinTidalLS}{0.0003}{IMRPhenomDNRTidal-HS}{0.0005}{IMRPhenomDNRTidal-LS}{0.0003}{IMRPhenomPv2NRTidal-HS}{0.0006}{IMRPhenomPv2NRTidal-LS}{0.0003}{SEOBNRv4TsurrogateHS}{0.0005}{SEOBNRv4TsurrogateLS}{0.0003}{SEOBNRv4TsurrogatehighspinRIFT}{0.0005}{SEOBNRv4TsurrogatelowspinRIFT}{0.0004}{TEOBResumS-HS}{0.0005}{TEOBResumS-LS}{0.0003}{TaylorF2-HS}{0.0005}{TaylorF2-LS}{0.0003}{PrecessingSpinIMRTidalHS}{0.0006}{PrecessingSpinIMRTidalLS}{0.0003}{PublicationSamples}{0.0006}}}
\newcommand{\chirpmassdetfourtwofivemed}[1]{\IfEqCase{#1}{{AlignedSpinInspiralTidalHS}{1.49}{AlignedSpinInspiralTidalLS}{1.49}{AlignedSpinTidalHS}{1.49}{AlignedSpinTidalLS}{1.49}{IMRPhenomDNRTidal-HS}{1.49}{IMRPhenomDNRTidal-LS}{1.49}{IMRPhenomPv2NRTidal-HS}{1.4873}{IMRPhenomPv2NRTidal-LS}{1.49}{SEOBNRv4TsurrogateHS}{1.49}{SEOBNRv4TsurrogateLS}{1.49}{SEOBNRv4TsurrogatehighspinRIFT}{1.49}{SEOBNRv4TsurrogatelowspinRIFT}{1.49}{TEOBResumS-HS}{1.49}{TEOBResumS-LS}{1.49}{TaylorF2-HS}{1.49}{TaylorF2-LS}{1.49}{PrecessingSpinIMRTidalHS}{1.49}{PrecessingSpinIMRTidalLS}{1.49}{PublicationSamples}{1.49}}}
\newcommand{\chirpmassdetfourtwofiveplus}[1]{\IfEqCase{#1}{{AlignedSpinInspiralTidalHS}{0.0007}{AlignedSpinInspiralTidalLS}{0.0003}{AlignedSpinTidalHS}{0.0007}{AlignedSpinTidalLS}{0.0003}{IMRPhenomDNRTidal-HS}{0.0008}{IMRPhenomDNRTidal-LS}{0.0004}{IMRPhenomPv2NRTidal-HS}{0.0008}{IMRPhenomPv2NRTidal-LS}{0.0003}{SEOBNRv4TsurrogateHS}{0.0006}{SEOBNRv4TsurrogateLS}{0.0003}{SEOBNRv4TsurrogatehighspinRIFT}{0.0006}{SEOBNRv4TsurrogatelowspinRIFT}{0.0003}{TEOBResumS-HS}{0.0005}{TEOBResumS-LS}{0.0003}{TaylorF2-HS}{0.0007}{TaylorF2-LS}{0.0003}{PrecessingSpinIMRTidalHS}{0.0008}{PrecessingSpinIMRTidalLS}{0.0003}{PublicationSamples}{0.0008}}}
\newcommand{\cosiotafourtwofiveminus}[1]{\IfEqCase{#1}{{AlignedSpinInspiralTidalHS}{1.30}{AlignedSpinInspiralTidalLS}{1.40}{AlignedSpinTidalHS}{1.44}{AlignedSpinTidalLS}{1.44}{IMRPhenomDNRTidal-HS}{1.53}{IMRPhenomDNRTidal-LS}{1.48}{IMRPhenomPv2NRTidal-HS}{1.42}{IMRPhenomPv2NRTidal-LS}{1.44}{SEOBNRv4TsurrogateHS}{1.38}{SEOBNRv4TsurrogateLS}{1.42}{SEOBNRv4TsurrogatehighspinRIFT}{1.41}{SEOBNRv4TsurrogatelowspinRIFT}{1.42}{TEOBResumS-HS}{1.42}{TEOBResumS-LS}{1.40}{TaylorF2-HS}{1.30}{TaylorF2-LS}{1.40}{PrecessingSpinIMRTidalHS}{1.42}{PrecessingSpinIMRTidalLS}{1.44}{PublicationSamples}{1.42}}}
\newcommand{\cosiotafourtwofivemed}[1]{\IfEqCase{#1}{{AlignedSpinInspiralTidalHS}{0.34}{AlignedSpinInspiralTidalLS}{0.44}{AlignedSpinTidalHS}{0.49}{AlignedSpinTidalLS}{0.49}{IMRPhenomDNRTidal-HS}{0.58}{IMRPhenomDNRTidal-LS}{0.53}{IMRPhenomPv2NRTidal-HS}{0.46}{IMRPhenomPv2NRTidal-LS}{0.48}{SEOBNRv4TsurrogateHS}{0.43}{SEOBNRv4TsurrogateLS}{0.46}{SEOBNRv4TsurrogatehighspinRIFT}{0.45}{SEOBNRv4TsurrogatelowspinRIFT}{0.46}{TEOBResumS-HS}{0.46}{TEOBResumS-LS}{0.44}{TaylorF2-HS}{0.34}{TaylorF2-LS}{0.44}{PrecessingSpinIMRTidalHS}{0.46}{PrecessingSpinIMRTidalLS}{0.48}{PublicationSamples}{0.46}}}
\newcommand{\cosiotafourtwofiveplus}[1]{\IfEqCase{#1}{{AlignedSpinInspiralTidalHS}{0.62}{AlignedSpinInspiralTidalLS}{0.53}{AlignedSpinTidalHS}{0.49}{AlignedSpinTidalLS}{0.49}{IMRPhenomDNRTidal-HS}{0.40}{IMRPhenomDNRTidal-LS}{0.45}{IMRPhenomPv2NRTidal-HS}{0.51}{IMRPhenomPv2NRTidal-LS}{0.49}{SEOBNRv4TsurrogateHS}{0.54}{SEOBNRv4TsurrogateLS}{0.51}{SEOBNRv4TsurrogatehighspinRIFT}{0.52}{SEOBNRv4TsurrogatelowspinRIFT}{0.51}{TEOBResumS-HS}{0.51}{TEOBResumS-LS}{0.54}{TaylorF2-HS}{0.63}{TaylorF2-LS}{0.53}{PrecessingSpinIMRTidalHS}{0.51}{PrecessingSpinIMRTidalLS}{0.49}{PublicationSamples}{0.51}}}
\newcommand{\comovingdistfourtwofiveminus}[1]{\IfEqCase{#1}{{AlignedSpinInspiralTidalHS}{72}{AlignedSpinInspiralTidalLS}{71}{AlignedSpinTidalHS}{69}{AlignedSpinTidalLS}{70}{IMRPhenomDNRTidal-HS}{69}{IMRPhenomDNRTidal-LS}{69}{IMRPhenomPv2NRTidal-HS}{67}{IMRPhenomPv2NRTidal-LS}{68}{SEOBNRv4TsurrogateHS}{68}{SEOBNRv4TsurrogateLS}{69}{SEOBNRv4TsurrogatehighspinRIFT}{70}{SEOBNRv4TsurrogatelowspinRIFT}{68}{TEOBResumS-HS}{70}{TEOBResumS-LS}{69}{TaylorF2-HS}{72}{TaylorF2-LS}{71}{PrecessingSpinIMRTidalHS}{67}{PrecessingSpinIMRTidalLS}{68}{PublicationSamples}{67}}}
\newcommand{\comovingdistfourtwofivemed}[1]{\IfEqCase{#1}{{AlignedSpinInspiralTidalHS}{156}{AlignedSpinInspiralTidalLS}{157}{AlignedSpinTidalHS}{155}{AlignedSpinTidalLS}{153}{IMRPhenomDNRTidal-HS}{155}{IMRPhenomDNRTidal-LS}{153}{IMRPhenomPv2NRTidal-HS}{151}{IMRPhenomPv2NRTidal-LS}{151}{SEOBNRv4TsurrogateHS}{153}{SEOBNRv4TsurrogateLS}{153}{SEOBNRv4TsurrogatehighspinRIFT}{157}{SEOBNRv4TsurrogatelowspinRIFT}{152}{TEOBResumS-HS}{157}{TEOBResumS-LS}{153}{TaylorF2-HS}{156}{TaylorF2-LS}{157}{PrecessingSpinIMRTidalHS}{151}{PrecessingSpinIMRTidalLS}{151}{PublicationSamples}{151}}}
\newcommand{\comovingdistfourtwofiveplus}[1]{\IfEqCase{#1}{{AlignedSpinInspiralTidalHS}{67}{AlignedSpinInspiralTidalLS}{65}{AlignedSpinTidalHS}{65}{AlignedSpinTidalLS}{63}{IMRPhenomDNRTidal-HS}{63}{IMRPhenomDNRTidal-LS}{64}{IMRPhenomPv2NRTidal-HS}{64}{IMRPhenomPv2NRTidal-LS}{64}{SEOBNRv4TsurrogateHS}{64}{SEOBNRv4TsurrogateLS}{64}{SEOBNRv4TsurrogatehighspinRIFT}{70}{SEOBNRv4TsurrogatelowspinRIFT}{65}{TEOBResumS-HS}{68}{TEOBResumS-LS}{64}{TaylorF2-HS}{67}{TaylorF2-LS}{65}{PrecessingSpinIMRTidalHS}{64}{PrecessingSpinIMRTidalLS}{64}{PublicationSamples}{64}}}
\newcommand{\logpriorfourtwofiveminus}[1]{\IfEqCase{#1}{{AlignedSpinInspiralTidalHS}{8.6}{AlignedSpinInspiralTidalLS}{8.5}{IMRPhenomDNRTidal-HS}{8.6}{IMRPhenomDNRTidal-LS}{8.6}{IMRPhenomPv2NRTidal-HS}{8.6}{IMRPhenomPv2NRTidal-LS}{8.4}{SEOBNRv4TsurrogateHS}{8.4}{SEOBNRv4TsurrogateLS}{8.8}{TaylorF2-HS}{8.6}{TaylorF2-LS}{8.5}{PrecessingSpinIMRTidalHS}{8.6}{PrecessingSpinIMRTidalLS}{8.4}{PublicationSamples}{8.6}}}
\newcommand{\logpriorfourtwofivemed}[1]{\IfEqCase{#1}{{AlignedSpinInspiralTidalHS}{102.5}{AlignedSpinInspiralTidalLS}{106.7}{IMRPhenomDNRTidal-HS}{94.5}{IMRPhenomDNRTidal-LS}{99.2}{IMRPhenomPv2NRTidal-HS}{98.4}{IMRPhenomPv2NRTidal-LS}{97.8}{SEOBNRv4TsurrogateHS}{95.6}{SEOBNRv4TsurrogateLS}{99.0}{TaylorF2-HS}{102.5}{TaylorF2-LS}{106.7}{PrecessingSpinIMRTidalHS}{98.4}{PrecessingSpinIMRTidalLS}{97.8}{PublicationSamples}{98.4}}}
\newcommand{\logpriorfourtwofiveplus}[1]{\IfEqCase{#1}{{AlignedSpinInspiralTidalHS}{6.8}{AlignedSpinInspiralTidalLS}{6.9}{IMRPhenomDNRTidal-HS}{6.8}{IMRPhenomDNRTidal-LS}{6.9}{IMRPhenomPv2NRTidal-HS}{6.7}{IMRPhenomPv2NRTidal-LS}{6.7}{SEOBNRv4TsurrogateHS}{6.9}{SEOBNRv4TsurrogateLS}{6.9}{TaylorF2-HS}{6.8}{TaylorF2-LS}{6.9}{PrecessingSpinIMRTidalHS}{6.7}{PrecessingSpinIMRTidalLS}{6.7}{PublicationSamples}{6.7}}}
\newcommand{\tiltonefourtwofiveminus}[1]{\IfEqCase{#1}{{AlignedSpinInspiralTidalHS}{0.00}{AlignedSpinInspiralTidalLS}{0.00}{AlignedSpinTidalHS}{0.00}{AlignedSpinTidalLS}{0.00}{IMRPhenomDNRTidal-HS}{0.00}{IMRPhenomDNRTidal-LS}{0.00}{IMRPhenomPv2NRTidal-HS}{0.80}{IMRPhenomPv2NRTidal-LS}{0.74}{SEOBNRv4TsurrogateHS}{0.00}{SEOBNRv4TsurrogateLS}{0.00}{SEOBNRv4TsurrogatehighspinRIFT}{0.00}{SEOBNRv4TsurrogatelowspinRIFT}{0.00}{TEOBResumS-HS}{0.00}{TEOBResumS-LS}{0.00}{TaylorF2-HS}{0.00}{TaylorF2-LS}{0.00}{PrecessingSpinIMRTidalHS}{0.80}{PrecessingSpinIMRTidalLS}{0.74}{PublicationSamples}{0.79}}}
\newcommand{\tiltonefourtwofivemed}[1]{\IfEqCase{#1}{{AlignedSpinInspiralTidalHS}{0.00}{AlignedSpinInspiralTidalLS}{0.00}{AlignedSpinTidalHS}{0.00}{AlignedSpinTidalLS}{0.00}{IMRPhenomDNRTidal-HS}{0.00}{IMRPhenomDNRTidal-LS}{0.00}{IMRPhenomPv2NRTidal-HS}{1.31}{IMRPhenomPv2NRTidal-LS}{1.03}{SEOBNRv4TsurrogateHS}{0.00}{SEOBNRv4TsurrogateLS}{0.00}{SEOBNRv4TsurrogatehighspinRIFT}{0.00}{SEOBNRv4TsurrogatelowspinRIFT}{0.00}{TEOBResumS-HS}{0.00}{TEOBResumS-LS}{0.00}{TaylorF2-HS}{0.00}{TaylorF2-LS}{0.00}{PrecessingSpinIMRTidalHS}{1.31}{PrecessingSpinIMRTidalLS}{1.03}{PublicationSamples}{1.31}}}
\newcommand{\tiltonefourtwofiveplus}[1]{\IfEqCase{#1}{{AlignedSpinInspiralTidalHS}{3.14}{AlignedSpinInspiralTidalLS}{3.14}{AlignedSpinTidalHS}{3.14}{AlignedSpinTidalLS}{3.14}{IMRPhenomDNRTidal-HS}{3.14}{IMRPhenomDNRTidal-LS}{3.14}{IMRPhenomPv2NRTidal-HS}{0.66}{IMRPhenomPv2NRTidal-LS}{1.16}{SEOBNRv4TsurrogateHS}{3.14}{SEOBNRv4TsurrogateLS}{3.14}{SEOBNRv4TsurrogatehighspinRIFT}{3.14}{SEOBNRv4TsurrogatelowspinRIFT}{3.14}{TEOBResumS-HS}{3.14}{TEOBResumS-LS}{3.14}{TaylorF2-HS}{3.14}{TaylorF2-LS}{3.14}{PrecessingSpinIMRTidalHS}{0.66}{PrecessingSpinIMRTidalLS}{1.17}{PublicationSamples}{0.66}}}
\newcommand{\spintwoyfourtwofiveminus}[1]{\IfEqCase{#1}{{AlignedSpinInspiralTidalHS}{0.00}{AlignedSpinInspiralTidalLS}{0.00}{AlignedSpinTidalHS}{0.00}{AlignedSpinTidalLS}{0.00}{IMRPhenomDNRTidal-HS}{0.00}{IMRPhenomDNRTidal-LS}{0.00}{IMRPhenomPv2NRTidal-HS}{0.48}{IMRPhenomPv2NRTidal-LS}{0.03}{SEOBNRv4TsurrogateHS}{0.00}{SEOBNRv4TsurrogateLS}{0.00}{SEOBNRv4TsurrogatehighspinRIFT}{0.00}{SEOBNRv4TsurrogatelowspinRIFT}{0.00}{TEOBResumS-HS}{0.00}{TEOBResumS-LS}{0.00}{TaylorF2-HS}{0.00}{TaylorF2-LS}{0.00}{PrecessingSpinIMRTidalHS}{0.48}{PrecessingSpinIMRTidalLS}{0.03}{PublicationSamples}{0.48}}}
\newcommand{\spintwoyfourtwofivemed}[1]{\IfEqCase{#1}{{AlignedSpinInspiralTidalHS}{0.00}{AlignedSpinInspiralTidalLS}{0.00}{AlignedSpinTidalHS}{0.00}{AlignedSpinTidalLS}{0.00}{IMRPhenomDNRTidal-HS}{0.00}{IMRPhenomDNRTidal-LS}{0.00}{IMRPhenomPv2NRTidal-HS}{0.00002}{IMRPhenomPv2NRTidal-LS}{0.00003}{SEOBNRv4TsurrogateHS}{0.00}{SEOBNRv4TsurrogateLS}{0.00}{SEOBNRv4TsurrogatehighspinRIFT}{0.00}{SEOBNRv4TsurrogatelowspinRIFT}{0.00}{TEOBResumS-HS}{0.00}{TEOBResumS-LS}{0.00}{TaylorF2-HS}{0.00}{TaylorF2-LS}{0.00}{PrecessingSpinIMRTidalHS}{0.00}{PrecessingSpinIMRTidalLS}{0.00002}{PublicationSamples}{0.00}}}
\newcommand{\spintwoyfourtwofiveplus}[1]{\IfEqCase{#1}{{AlignedSpinInspiralTidalHS}{0.00}{AlignedSpinInspiralTidalLS}{0.00}{AlignedSpinTidalHS}{0.00}{AlignedSpinTidalLS}{0.00}{IMRPhenomDNRTidal-HS}{0.00}{IMRPhenomDNRTidal-LS}{0.00}{IMRPhenomPv2NRTidal-HS}{0.48}{IMRPhenomPv2NRTidal-LS}{0.03}{SEOBNRv4TsurrogateHS}{0.00}{SEOBNRv4TsurrogateLS}{0.00}{SEOBNRv4TsurrogatehighspinRIFT}{0.00}{SEOBNRv4TsurrogatelowspinRIFT}{0.00}{TEOBResumS-HS}{0.00}{TEOBResumS-LS}{0.00}{TaylorF2-HS}{0.00}{TaylorF2-LS}{0.00}{PrecessingSpinIMRTidalHS}{0.48}{PrecessingSpinIMRTidalLS}{0.03}{PublicationSamples}{0.48}}}
\newcommand{\spintwozfourtwofiveminus}[1]{\IfEqCase{#1}{{AlignedSpinInspiralTidalHS}{0.18}{AlignedSpinInspiralTidalLS}{0.02}{AlignedSpinTidalHS}{0.24}{AlignedSpinTidalLS}{0.02}{IMRPhenomDNRTidal-HS}{0.39}{IMRPhenomDNRTidal-LS}{0.02}{IMRPhenomPv2NRTidal-HS}{0.18}{IMRPhenomPv2NRTidal-LS}{0.02}{SEOBNRv4TsurrogateHS}{0.16}{SEOBNRv4TsurrogateLS}{0.02}{SEOBNRv4TsurrogatehighspinRIFT}{0.18}{SEOBNRv4TsurrogatelowspinRIFT}{0.02}{TEOBResumS-HS}{0.18}{TEOBResumS-LS}{0.02}{TaylorF2-HS}{0.18}{TaylorF2-LS}{0.02}{PrecessingSpinIMRTidalHS}{0.18}{PrecessingSpinIMRTidalLS}{0.02}{PublicationSamples}{0.18}}}
\newcommand{\spintwozfourtwofivemed}[1]{\IfEqCase{#1}{{AlignedSpinInspiralTidalHS}{0.04}{AlignedSpinInspiralTidalLS}{0.008}{AlignedSpinTidalHS}{0.03}{AlignedSpinTidalLS}{0.009}{IMRPhenomDNRTidal-HS}{0.03}{IMRPhenomDNRTidal-LS}{0.009}{IMRPhenomPv2NRTidal-HS}{0.03}{IMRPhenomPv2NRTidal-LS}{0.009}{SEOBNRv4TsurrogateHS}{0.02}{SEOBNRv4TsurrogateLS}{0.009}{SEOBNRv4TsurrogatehighspinRIFT}{0.03}{SEOBNRv4TsurrogatelowspinRIFT}{0.01}{TEOBResumS-HS}{0.03}{TEOBResumS-LS}{0.009}{TaylorF2-HS}{0.04}{TaylorF2-LS}{0.008}{PrecessingSpinIMRTidalHS}{0.03}{PrecessingSpinIMRTidalLS}{0.009}{PublicationSamples}{0.03}}}
\newcommand{\spintwozfourtwofiveplus}[1]{\IfEqCase{#1}{{AlignedSpinInspiralTidalHS}{0.30}{AlignedSpinInspiralTidalLS}{0.03}{AlignedSpinTidalHS}{0.26}{AlignedSpinTidalLS}{0.03}{IMRPhenomDNRTidal-HS}{0.37}{IMRPhenomDNRTidal-LS}{0.03}{IMRPhenomPv2NRTidal-HS}{0.30}{IMRPhenomPv2NRTidal-LS}{0.03}{SEOBNRv4TsurrogateHS}{0.20}{SEOBNRv4TsurrogateLS}{0.03}{SEOBNRv4TsurrogatehighspinRIFT}{0.21}{SEOBNRv4TsurrogatelowspinRIFT}{0.03}{TEOBResumS-HS}{0.21}{TEOBResumS-LS}{0.03}{TaylorF2-HS}{0.30}{TaylorF2-LS}{0.03}{PrecessingSpinIMRTidalHS}{0.30}{PrecessingSpinIMRTidalLS}{0.03}{PublicationSamples}{0.30}}}
\newcommand{\massonesourcefourtwofiveminus}[1]{\IfEqCase{#1}{{AlignedSpinInspiralTidalHS}{0.3}{AlignedSpinInspiralTidalLS}{0.10}{AlignedSpinTidalHS}{0.2}{AlignedSpinTidalLS}{0.09}{IMRPhenomDNRTidal-HS}{0.3}{IMRPhenomDNRTidal-LS}{0.09}{IMRPhenomPv2NRTidal-HS}{0.3}{IMRPhenomPv2NRTidal-LS}{0.09}{SEOBNRv4TsurrogateHS}{0.2}{SEOBNRv4TsurrogateLS}{0.09}{SEOBNRv4TsurrogatehighspinRIFT}{0.2}{SEOBNRv4TsurrogatelowspinRIFT}{0.09}{TEOBResumS-HS}{0.2}{TEOBResumS-LS}{0.09}{TaylorF2-HS}{0.3}{TaylorF2-LS}{0.10}{PrecessingSpinIMRTidalHS}{0.3}{PrecessingSpinIMRTidalLS}{0.09}{PublicationSamples}{0.3}}}
\newcommand{\massonesourcefourtwofivemed}[1]{\IfEqCase{#1}{{AlignedSpinInspiralTidalHS}{2.0}{AlignedSpinInspiralTidalLS}{1.75}{AlignedSpinTidalHS}{1.9}{AlignedSpinTidalLS}{1.75}{IMRPhenomDNRTidal-HS}{2.0}{IMRPhenomDNRTidal-LS}{1.75}{IMRPhenomPv2NRTidal-HS}{2.0}{IMRPhenomPv2NRTidal-LS}{1.74}{SEOBNRv4TsurrogateHS}{1.9}{SEOBNRv4TsurrogateLS}{1.74}{SEOBNRv4TsurrogatehighspinRIFT}{1.9}{SEOBNRv4TsurrogatelowspinRIFT}{1.75}{TEOBResumS-HS}{1.9}{TEOBResumS-LS}{1.75}{TaylorF2-HS}{2.0}{TaylorF2-LS}{1.75}{PrecessingSpinIMRTidalHS}{2.0}{PrecessingSpinIMRTidalLS}{1.74}{PublicationSamples}{2.0}}}
\newcommand{\massonesourcefourtwofiveplus}[1]{\IfEqCase{#1}{{AlignedSpinInspiralTidalHS}{0.5}{AlignedSpinInspiralTidalLS}{0.2}{AlignedSpinTidalHS}{0.6}{AlignedSpinTidalLS}{0.2}{IMRPhenomDNRTidal-HS}{0.7}{IMRPhenomDNRTidal-LS}{0.2}{IMRPhenomPv2NRTidal-HS}{0.6}{IMRPhenomPv2NRTidal-LS}{0.2}{SEOBNRv4TsurrogateHS}{0.5}{SEOBNRv4TsurrogateLS}{0.2}{SEOBNRv4TsurrogatehighspinRIFT}{0.5}{SEOBNRv4TsurrogatelowspinRIFT}{0.2}{TEOBResumS-HS}{0.5}{TEOBResumS-LS}{0.2}{TaylorF2-HS}{0.5}{TaylorF2-LS}{0.2}{PrecessingSpinIMRTidalHS}{0.6}{PrecessingSpinIMRTidalLS}{0.2}{PublicationSamples}{0.6}}}
\newcommand{\geocenttimefourtwofiveminus}[1]{\IfEqCase{#1}{{AlignedSpinInspiralTidalHS}{0.007}{AlignedSpinInspiralTidalLS}{0.008}{AlignedSpinTidalHS}{0.03}{AlignedSpinTidalLS}{0.02}{IMRPhenomDNRTidal-HS}{0.008}{IMRPhenomDNRTidal-LS}{0.01}{IMRPhenomPv2NRTidal-HS}{0.009}{IMRPhenomPv2NRTidal-LS}{0.01}{SEOBNRv4TsurrogateHS}{0.01}{SEOBNRv4TsurrogateLS}{0.008}{SEOBNRv4TsurrogatehighspinRIFT}{0.0}{SEOBNRv4TsurrogatelowspinRIFT}{0.0}{TEOBResumS-HS}{0.0}{TEOBResumS-LS}{0.0}{TaylorF2-HS}{0.007}{TaylorF2-LS}{0.008}{PrecessingSpinIMRTidalHS}{0.009}{PrecessingSpinIMRTidalLS}{0.01}{PublicationSamples}{0.009}}}
\newcommand{\geocenttimefourtwofivemed}[1]{\IfEqCase{#1}{{AlignedSpinInspiralTidalHS}{1240215503.0}{AlignedSpinInspiralTidalLS}{1240215503.0}{AlignedSpinTidalHS}{1240215503.0}{AlignedSpinTidalLS}{1240215503.0}{IMRPhenomDNRTidal-HS}{1240215503.0}{IMRPhenomDNRTidal-LS}{1240215503.0}{IMRPhenomPv2NRTidal-HS}{1240215503.0}{IMRPhenomPv2NRTidal-LS}{1240215503.0}{SEOBNRv4TsurrogateHS}{1240215503.0}{SEOBNRv4TsurrogateLS}{1240215503.0}{SEOBNRv4TsurrogatehighspinRIFT}{1240215503.0}{SEOBNRv4TsurrogatelowspinRIFT}{1240215503.0}{TEOBResumS-HS}{1240215503.0}{TEOBResumS-LS}{1240215503.0}{TaylorF2-HS}{1240215503.0}{TaylorF2-LS}{1240215503.0}{PrecessingSpinIMRTidalHS}{1240215503.0}{PrecessingSpinIMRTidalLS}{1240215503.0}{PublicationSamples}{1240215503.0}}}
\newcommand{\geocenttimefourtwofiveplus}[1]{\IfEqCase{#1}{{AlignedSpinInspiralTidalHS}{0.04}{AlignedSpinInspiralTidalLS}{0.03}{AlignedSpinTidalHS}{0.02}{AlignedSpinTidalLS}{0.02}{IMRPhenomDNRTidal-HS}{0.04}{IMRPhenomDNRTidal-LS}{0.03}{IMRPhenomPv2NRTidal-HS}{0.03}{IMRPhenomPv2NRTidal-LS}{0.03}{SEOBNRv4TsurrogateHS}{0.04}{SEOBNRv4TsurrogateLS}{0.04}{SEOBNRv4TsurrogatehighspinRIFT}{0.0}{SEOBNRv4TsurrogatelowspinRIFT}{0.0}{TEOBResumS-HS}{0.0}{TEOBResumS-LS}{0.0}{TaylorF2-HS}{0.04}{TaylorF2-LS}{0.03}{PrecessingSpinIMRTidalHS}{0.03}{PrecessingSpinIMRTidalLS}{0.03}{PublicationSamples}{0.03}}}
\newcommand{\costilttwofourtwofiveminus}[1]{\IfEqCase{#1}{{AlignedSpinInspiralTidalHS}{2.00}{AlignedSpinInspiralTidalLS}{2.00}{AlignedSpinTidalHS}{2.00}{AlignedSpinTidalLS}{2.00}{IMRPhenomDNRTidal-HS}{2.00}{IMRPhenomDNRTidal-LS}{2.00}{IMRPhenomPv2NRTidal-HS}{0.87}{IMRPhenomPv2NRTidal-LS}{1.13}{SEOBNRv4TsurrogateHS}{2.00}{SEOBNRv4TsurrogateLS}{2.00}{SEOBNRv4TsurrogatehighspinRIFT}{2.00}{SEOBNRv4TsurrogatelowspinRIFT}{2.00}{TEOBResumS-HS}{2.00}{TEOBResumS-LS}{2.00}{TaylorF2-HS}{2.00}{TaylorF2-LS}{2.00}{PrecessingSpinIMRTidalHS}{0.87}{PrecessingSpinIMRTidalLS}{1.12}{PublicationSamples}{0.86}}}
\newcommand{\costilttwofourtwofivemed}[1]{\IfEqCase{#1}{{AlignedSpinInspiralTidalHS}{1.00}{AlignedSpinInspiralTidalLS}{1.00}{AlignedSpinTidalHS}{1.00}{AlignedSpinTidalLS}{1.00}{IMRPhenomDNRTidal-HS}{1.00}{IMRPhenomDNRTidal-LS}{1.00}{IMRPhenomPv2NRTidal-HS}{0.16}{IMRPhenomPv2NRTidal-LS}{0.46}{SEOBNRv4TsurrogateHS}{1.00}{SEOBNRv4TsurrogateLS}{1.00}{SEOBNRv4TsurrogatehighspinRIFT}{1.00}{SEOBNRv4TsurrogatelowspinRIFT}{1.00}{TEOBResumS-HS}{1.00}{TEOBResumS-LS}{1.00}{TaylorF2-HS}{1.00}{TaylorF2-LS}{1.00}{PrecessingSpinIMRTidalHS}{0.16}{PrecessingSpinIMRTidalLS}{0.46}{PublicationSamples}{0.16}}}
\newcommand{\costilttwofourtwofiveplus}[1]{\IfEqCase{#1}{{AlignedSpinInspiralTidalHS}{0.00}{AlignedSpinInspiralTidalLS}{0.00}{AlignedSpinTidalHS}{0.00}{AlignedSpinTidalLS}{0.00}{IMRPhenomDNRTidal-HS}{0.00}{IMRPhenomDNRTidal-LS}{0.00}{IMRPhenomPv2NRTidal-HS}{0.70}{IMRPhenomPv2NRTidal-LS}{0.49}{SEOBNRv4TsurrogateHS}{0.00}{SEOBNRv4TsurrogateLS}{0.00}{SEOBNRv4TsurrogatehighspinRIFT}{0.00}{SEOBNRv4TsurrogatelowspinRIFT}{0.00}{TEOBResumS-HS}{0.00}{TEOBResumS-LS}{0.00}{TaylorF2-HS}{0.00}{TaylorF2-LS}{0.00}{PrecessingSpinIMRTidalHS}{0.70}{PrecessingSpinIMRTidalLS}{0.49}{PublicationSamples}{0.70}}}
\newcommand{\luminositydistancefourtwofiveminus}[1]{\IfEqCase{#1}{{AlignedSpinInspiralTidalHS}{0.08}{AlignedSpinInspiralTidalLS}{0.07}{AlignedSpinTidalHS}{0.07}{AlignedSpinTidalLS}{0.07}{IMRPhenomDNRTidal-HS}{0.07}{IMRPhenomDNRTidal-LS}{0.07}{IMRPhenomPv2NRTidal-HS}{0.07}{IMRPhenomPv2NRTidal-LS}{0.07}{SEOBNRv4TsurrogateHS}{0.07}{SEOBNRv4TsurrogateLS}{0.07}{SEOBNRv4TsurrogatehighspinRIFT}{0.07}{SEOBNRv4TsurrogatelowspinRIFT}{0.07}{TEOBResumS-HS}{0.07}{TEOBResumS-LS}{0.07}{TaylorF2-HS}{0.08}{TaylorF2-LS}{0.08}{PrecessingSpinIMRTidalHS}{0.07}{PrecessingSpinIMRTidalLS}{0.07}{PublicationSamples}{0.07}}}
\newcommand{\luminositydistancefourtwofivemed}[1]{\IfEqCase{#1}{{AlignedSpinInspiralTidalHS}{0.16}{AlignedSpinInspiralTidalLS}{0.16}{AlignedSpinTidalHS}{0.16}{AlignedSpinTidalLS}{0.16}{IMRPhenomDNRTidal-HS}{0.16}{IMRPhenomDNRTidal-LS}{0.16}{IMRPhenomPv2NRTidal-HS}{0.16}{IMRPhenomPv2NRTidal-LS}{0.16}{SEOBNRv4TsurrogateHS}{0.16}{SEOBNRv4TsurrogateLS}{0.16}{SEOBNRv4TsurrogatehighspinRIFT}{0.16}{SEOBNRv4TsurrogatelowspinRIFT}{0.16}{TEOBResumS-HS}{0.16}{TEOBResumS-LS}{0.16}{TaylorF2-HS}{0.16}{TaylorF2-LS}{0.16}{PrecessingSpinIMRTidalHS}{0.16}{PrecessingSpinIMRTidalLS}{0.16}{PublicationSamples}{0.16}}}
\newcommand{\luminositydistancefourtwofiveplus}[1]{\IfEqCase{#1}{{AlignedSpinInspiralTidalHS}{0.07}{AlignedSpinInspiralTidalLS}{0.07}{AlignedSpinTidalHS}{0.07}{AlignedSpinTidalLS}{0.07}{IMRPhenomDNRTidal-HS}{0.07}{IMRPhenomDNRTidal-LS}{0.07}{IMRPhenomPv2NRTidal-HS}{0.07}{IMRPhenomPv2NRTidal-LS}{0.07}{SEOBNRv4TsurrogateHS}{0.07}{SEOBNRv4TsurrogateLS}{0.07}{SEOBNRv4TsurrogatehighspinRIFT}{0.08}{SEOBNRv4TsurrogatelowspinRIFT}{0.07}{TEOBResumS-HS}{0.07}{TEOBResumS-LS}{0.07}{TaylorF2-HS}{0.07}{TaylorF2-LS}{0.07}{PrecessingSpinIMRTidalHS}{0.07}{PrecessingSpinIMRTidalLS}{0.07}{PublicationSamples}{0.07}}}
\newcommand{\spinonezfourtwofiveminus}[1]{\IfEqCase{#1}{{AlignedSpinInspiralTidalHS}{0.14}{AlignedSpinInspiralTidalLS}{0.02}{AlignedSpinTidalHS}{0.15}{AlignedSpinTidalLS}{0.02}{IMRPhenomDNRTidal-HS}{0.22}{IMRPhenomDNRTidal-LS}{0.02}{IMRPhenomPv2NRTidal-HS}{0.12}{IMRPhenomPv2NRTidal-LS}{0.02}{SEOBNRv4TsurrogateHS}{0.11}{SEOBNRv4TsurrogateLS}{0.02}{SEOBNRv4TsurrogatehighspinRIFT}{0.14}{SEOBNRv4TsurrogatelowspinRIFT}{0.02}{TEOBResumS-HS}{0.13}{TEOBResumS-LS}{0.02}{TaylorF2-HS}{0.14}{TaylorF2-LS}{0.02}{PrecessingSpinIMRTidalHS}{0.12}{PrecessingSpinIMRTidalLS}{0.02}{PublicationSamples}{0.12}}}
\newcommand{\spinonezfourtwofivemed}[1]{\IfEqCase{#1}{{AlignedSpinInspiralTidalHS}{0.04}{AlignedSpinInspiralTidalLS}{0.01}{AlignedSpinTidalHS}{0.04}{AlignedSpinTidalLS}{0.01}{IMRPhenomDNRTidal-HS}{0.06}{IMRPhenomDNRTidal-LS}{0.01}{IMRPhenomPv2NRTidal-HS}{0.06}{IMRPhenomPv2NRTidal-LS}{0.01}{SEOBNRv4TsurrogateHS}{0.04}{SEOBNRv4TsurrogateLS}{0.01}{SEOBNRv4TsurrogatehighspinRIFT}{0.04}{SEOBNRv4TsurrogatelowspinRIFT}{0.01}{TEOBResumS-HS}{0.04}{TEOBResumS-LS}{0.01}{TaylorF2-HS}{0.04}{TaylorF2-LS}{0.01}{PrecessingSpinIMRTidalHS}{0.06}{PrecessingSpinIMRTidalLS}{0.01}{PublicationSamples}{0.06}}}
\newcommand{\spinonezfourtwofiveplus}[1]{\IfEqCase{#1}{{AlignedSpinInspiralTidalHS}{0.19}{AlignedSpinInspiralTidalLS}{0.03}{AlignedSpinTidalHS}{0.20}{AlignedSpinTidalLS}{0.03}{IMRPhenomDNRTidal-HS}{0.26}{IMRPhenomDNRTidal-LS}{0.03}{IMRPhenomPv2NRTidal-HS}{0.18}{IMRPhenomPv2NRTidal-LS}{0.03}{SEOBNRv4TsurrogateHS}{0.16}{SEOBNRv4TsurrogateLS}{0.03}{SEOBNRv4TsurrogatehighspinRIFT}{0.16}{SEOBNRv4TsurrogatelowspinRIFT}{0.03}{TEOBResumS-HS}{0.16}{TEOBResumS-LS}{0.03}{TaylorF2-HS}{0.19}{TaylorF2-LS}{0.03}{PrecessingSpinIMRTidalHS}{0.18}{PrecessingSpinIMRTidalLS}{0.03}{PublicationSamples}{0.18}}}
\newcommand{\networkmatchedfiltersnrfourtwofiveminus}[1]{\IfEqCase{#1}{{AlignedSpinInspiralTidalHS}{0.4}{AlignedSpinInspiralTidalLS}{0.4}{IMRPhenomDNRTidal-HS}{0.4}{IMRPhenomDNRTidal-LS}{0.4}{IMRPhenomPv2NRTidal-HS}{0.4}{IMRPhenomPv2NRTidal-LS}{0.4}{SEOBNRv4TsurrogateHS}{0.4}{SEOBNRv4TsurrogateLS}{0.4}{TaylorF2-HS}{0.4}{TaylorF2-LS}{0.4}{PrecessingSpinIMRTidalHS}{0.4}{PrecessingSpinIMRTidalLS}{0.4}{PublicationSamples}{0.4}}}
\newcommand{\networkmatchedfiltersnrfourtwofivemed}[1]{\IfEqCase{#1}{{AlignedSpinInspiralTidalHS}{12.4}{AlignedSpinInspiralTidalLS}{12.5}{IMRPhenomDNRTidal-HS}{12.3}{IMRPhenomDNRTidal-LS}{12.4}{IMRPhenomPv2NRTidal-HS}{12.4}{IMRPhenomPv2NRTidal-LS}{12.5}{SEOBNRv4TsurrogateHS}{12.4}{SEOBNRv4TsurrogateLS}{12.4}{TaylorF2-HS}{12.4}{TaylorF2-LS}{12.5}{PrecessingSpinIMRTidalHS}{12.4}{PrecessingSpinIMRTidalLS}{12.5}{PublicationSamples}{12.4}}}
\newcommand{\networkmatchedfiltersnrfourtwofiveplus}[1]{\IfEqCase{#1}{{AlignedSpinInspiralTidalHS}{0.3}{AlignedSpinInspiralTidalLS}{0.2}{IMRPhenomDNRTidal-HS}{0.3}{IMRPhenomDNRTidal-LS}{0.3}{IMRPhenomPv2NRTidal-HS}{0.3}{IMRPhenomPv2NRTidal-LS}{0.3}{SEOBNRv4TsurrogateHS}{0.3}{SEOBNRv4TsurrogateLS}{0.3}{TaylorF2-HS}{0.3}{TaylorF2-LS}{0.2}{PrecessingSpinIMRTidalHS}{0.3}{PrecessingSpinIMRTidalLS}{0.3}{PublicationSamples}{0.3}}}
\newcommand{\chirpmasssourcefourtwofiveminus}[1]{\IfEqCase{#1}{{AlignedSpinInspiralTidalHS}{0.02}{AlignedSpinInspiralTidalLS}{0.02}{AlignedSpinTidalHS}{0.02}{AlignedSpinTidalLS}{0.02}{IMRPhenomDNRTidal-HS}{0.02}{IMRPhenomDNRTidal-LS}{0.02}{IMRPhenomPv2NRTidal-HS}{0.02}{IMRPhenomPv2NRTidal-LS}{0.02}{SEOBNRv4TsurrogateHS}{0.02}{SEOBNRv4TsurrogateLS}{0.02}{SEOBNRv4TsurrogatehighspinRIFT}{0.02}{SEOBNRv4TsurrogatelowspinRIFT}{0.02}{TEOBResumS-HS}{0.02}{TEOBResumS-LS}{0.02}{TaylorF2-HS}{0.02}{TaylorF2-LS}{0.02}{PrecessingSpinIMRTidalHS}{0.02}{PrecessingSpinIMRTidalLS}{0.02}{PublicationSamples}{0.02}}}
\newcommand{\chirpmasssourcefourtwofivemed}[1]{\IfEqCase{#1}{{AlignedSpinInspiralTidalHS}{1.44}{AlignedSpinInspiralTidalLS}{1.44}{AlignedSpinTidalHS}{1.44}{AlignedSpinTidalLS}{1.44}{IMRPhenomDNRTidal-HS}{1.44}{IMRPhenomDNRTidal-LS}{1.44}{IMRPhenomPv2NRTidal-HS}{1.44}{IMRPhenomPv2NRTidal-LS}{1.44}{SEOBNRv4TsurrogateHS}{1.44}{SEOBNRv4TsurrogateLS}{1.44}{SEOBNRv4TsurrogatehighspinRIFT}{1.44}{SEOBNRv4TsurrogatelowspinRIFT}{1.44}{TEOBResumS-HS}{1.44}{TEOBResumS-LS}{1.44}{TaylorF2-HS}{1.44}{TaylorF2-LS}{1.44}{PrecessingSpinIMRTidalHS}{1.44}{PrecessingSpinIMRTidalLS}{1.44}{PublicationSamples}{1.44}}}
\newcommand{\chirpmasssourcefourtwofiveplus}[1]{\IfEqCase{#1}{{AlignedSpinInspiralTidalHS}{0.02}{AlignedSpinInspiralTidalLS}{0.02}{AlignedSpinTidalHS}{0.02}{AlignedSpinTidalLS}{0.02}{IMRPhenomDNRTidal-HS}{0.02}{IMRPhenomDNRTidal-LS}{0.02}{IMRPhenomPv2NRTidal-HS}{0.02}{IMRPhenomPv2NRTidal-LS}{0.02}{SEOBNRv4TsurrogateHS}{0.02}{SEOBNRv4TsurrogateLS}{0.02}{SEOBNRv4TsurrogatehighspinRIFT}{0.02}{SEOBNRv4TsurrogatelowspinRIFT}{0.02}{TEOBResumS-HS}{0.02}{TEOBResumS-LS}{0.02}{TaylorF2-HS}{0.02}{TaylorF2-LS}{0.02}{PrecessingSpinIMRTidalHS}{0.02}{PrecessingSpinIMRTidalLS}{0.02}{PublicationSamples}{0.02}}}
\newcommand{\phionefourtwofiveminus}[1]{\IfEqCase{#1}{{AlignedSpinInspiralTidalHS}{0.00}{AlignedSpinInspiralTidalLS}{0.00}{AlignedSpinTidalHS}{0.00}{AlignedSpinTidalLS}{0.00}{IMRPhenomDNRTidal-HS}{0.00}{IMRPhenomDNRTidal-LS}{0.00}{IMRPhenomPv2NRTidal-HS}{2.73}{IMRPhenomPv2NRTidal-LS}{2.85}{SEOBNRv4TsurrogateHS}{0.00}{SEOBNRv4TsurrogateLS}{0.00}{SEOBNRv4TsurrogatehighspinRIFT}{0.00}{SEOBNRv4TsurrogatelowspinRIFT}{0.00}{TEOBResumS-HS}{0.00}{TEOBResumS-LS}{0.00}{TaylorF2-HS}{0.00}{TaylorF2-LS}{0.00}{PrecessingSpinIMRTidalHS}{2.73}{PrecessingSpinIMRTidalLS}{2.85}{PublicationSamples}{2.73}}}
\newcommand{\phionefourtwofivemed}[1]{\IfEqCase{#1}{{AlignedSpinInspiralTidalHS}{0.00}{AlignedSpinInspiralTidalLS}{0.00}{AlignedSpinTidalHS}{0.00}{AlignedSpinTidalLS}{0.00}{IMRPhenomDNRTidal-HS}{0.00}{IMRPhenomDNRTidal-LS}{0.00}{IMRPhenomPv2NRTidal-HS}{3.05}{IMRPhenomPv2NRTidal-LS}{3.15}{SEOBNRv4TsurrogateHS}{0.00}{SEOBNRv4TsurrogateLS}{0.00}{SEOBNRv4TsurrogatehighspinRIFT}{0.00}{SEOBNRv4TsurrogatelowspinRIFT}{0.00}{TEOBResumS-HS}{0.00}{TEOBResumS-LS}{0.00}{TaylorF2-HS}{0.00}{TaylorF2-LS}{0.00}{PrecessingSpinIMRTidalHS}{3.05}{PrecessingSpinIMRTidalLS}{3.15}{PublicationSamples}{3.06}}}
\newcommand{\phionefourtwofiveplus}[1]{\IfEqCase{#1}{{AlignedSpinInspiralTidalHS}{0.00}{AlignedSpinInspiralTidalLS}{0.00}{AlignedSpinTidalHS}{0.00}{AlignedSpinTidalLS}{0.00}{IMRPhenomDNRTidal-HS}{0.00}{IMRPhenomDNRTidal-LS}{0.00}{IMRPhenomPv2NRTidal-HS}{2.90}{IMRPhenomPv2NRTidal-LS}{2.83}{SEOBNRv4TsurrogateHS}{0.00}{SEOBNRv4TsurrogateLS}{0.00}{SEOBNRv4TsurrogatehighspinRIFT}{0.00}{SEOBNRv4TsurrogatelowspinRIFT}{0.00}{TEOBResumS-HS}{0.00}{TEOBResumS-LS}{0.00}{TaylorF2-HS}{0.00}{TaylorF2-LS}{0.00}{PrecessingSpinIMRTidalHS}{2.90}{PrecessingSpinIMRTidalLS}{2.83}{PublicationSamples}{2.90}}}
\newcommand{\symmetricmassratiofourtwofiveminus}[1]{\IfEqCase{#1}{{AlignedSpinInspiralTidalHS}{0.03}{AlignedSpinInspiralTidalLS}{0.005}{AlignedSpinTidalHS}{0.03}{AlignedSpinTidalLS}{0.005}{IMRPhenomDNRTidal-HS}{0.04}{IMRPhenomDNRTidal-LS}{0.005}{IMRPhenomPv2NRTidal-HS}{0.03}{IMRPhenomPv2NRTidal-LS}{0.005}{SEOBNRv4TsurrogateHS}{0.03}{SEOBNRv4TsurrogateLS}{0.004}{SEOBNRv4TsurrogatehighspinRIFT}{0.02}{SEOBNRv4TsurrogatelowspinRIFT}{0.005}{TEOBResumS-HS}{0.03}{TEOBResumS-LS}{0.005}{TaylorF2-HS}{0.03}{TaylorF2-LS}{0.005}{PrecessingSpinIMRTidalHS}{0.03}{PrecessingSpinIMRTidalLS}{0.005}{PublicationSamples}{0.03}}}
\newcommand{\symmetricmassratiofourtwofivemed}[1]{\IfEqCase{#1}{{AlignedSpinInspiralTidalHS}{0.242}{AlignedSpinInspiralTidalLS}{0.249}{AlignedSpinTidalHS}{0.245}{AlignedSpinTidalLS}{0.249}{IMRPhenomDNRTidal-HS}{0.243}{IMRPhenomDNRTidal-LS}{0.249}{IMRPhenomPv2NRTidal-HS}{0.240}{IMRPhenomPv2NRTidal-LS}{0.249}{SEOBNRv4TsurrogateHS}{0.246}{SEOBNRv4TsurrogateLS}{0.249}{SEOBNRv4TsurrogatehighspinRIFT}{0.246}{SEOBNRv4TsurrogatelowspinRIFT}{0.249}{TEOBResumS-HS}{0.245}{TEOBResumS-LS}{0.249}{TaylorF2-HS}{0.242}{TaylorF2-LS}{0.249}{PrecessingSpinIMRTidalHS}{0.240}{PrecessingSpinIMRTidalLS}{0.249}{PublicationSamples}{0.240}}}
\newcommand{\symmetricmassratiofourtwofiveplus}[1]{\IfEqCase{#1}{{AlignedSpinInspiralTidalHS}{0.007}{AlignedSpinInspiralTidalLS}{0.0008}{AlignedSpinTidalHS}{0.005}{AlignedSpinTidalLS}{0.0008}{IMRPhenomDNRTidal-HS}{0.007}{IMRPhenomDNRTidal-LS}{0.0008}{IMRPhenomPv2NRTidal-HS}{0.010}{IMRPhenomPv2NRTidal-LS}{0.0007}{SEOBNRv4TsurrogateHS}{0.004}{SEOBNRv4TsurrogateLS}{0.0007}{SEOBNRv4TsurrogatehighspinRIFT}{0.004}{SEOBNRv4TsurrogatelowspinRIFT}{0.0008}{TEOBResumS-HS}{0.005}{TEOBResumS-LS}{0.0009}{TaylorF2-HS}{0.007}{TaylorF2-LS}{0.0008}{PrecessingSpinIMRTidalHS}{0.010}{PrecessingSpinIMRTidalLS}{0.0007}{PublicationSamples}{0.010}}}

\acrodef{MDC}[MDC]{mock data challenge}
\acrodef{MECO}[MECO]{minimum energy circular orbit}
\acrodef{QNM}[QNM]{quasi-normal mode}

\begin{document}

\newcommand{\TGRIIINUMTESTS}{\reviewed{seven}\xspace}
\newcommand{\TGRIIINUMEVENTS}{\reviewed{42}\xspace}

\DeclareRobustCommand{\pyRingHierarchicalMuDeltaFMedian}{\reviewed{0.11}}
\DeclareRobustCommand{\pyRingHierarchicalMuDeltaFPlus}{\reviewed{0.13}}
\DeclareRobustCommand{\pyRingHierarchicalMuDeltaFMinus}{\reviewed{0.10}}
\DeclareRobustCommand{\pyRingHierarchicalSigmaDeltaFMedian}{\reviewed{0.08}}
\DeclareRobustCommand{\pyRingHierarchicalSigmaDeltaFPlus}{\reviewed{0.13}}
\DeclareRobustCommand{\pyRingHierarchicalSigmaDeltaFMinus}{\reviewed{0.07}}
\DeclareRobustCommand{\pyRingHierarchicalSigmaDeltaFBound}{\reviewed{0.18}}
\DeclareRobustCommand{\pyRingHierarchicalMuDeltaTauMedian}{\reviewed{0.18}}
\DeclareRobustCommand{\pyRingHierarchicalMuDeltaTauPlus}{\reviewed{0.17}}
\DeclareRobustCommand{\pyRingHierarchicalMuDeltaTauMinus}{\reviewed{0.17}}
\DeclareRobustCommand{\pyRingHierarchicalSigmaDeltaTauMedian}{\reviewed{0.09}}
\DeclareRobustCommand{\pyRingHierarchicalSigmaDeltaTauPlus}{\reviewed{0.16}}
\DeclareRobustCommand{\pyRingHierarchicalSigmaDeltaTauMinus}{\reviewed{0.08}}
\DeclareRobustCommand{\pyRingHierarchicalSigmaDeltaTauBound}{\reviewed{0.22}}
\DeclareRobustCommand{\pyRingHierarchicalRhoMedian}{\reviewed{-0.16}}
\DeclareRobustCommand{\pyRingHierarchicalRhoPlus}{\reviewed{0.83}}
\DeclareRobustCommand{\pyRingHierarchicalRhoMinus}{\reviewed{0.65}}

\DeclareRobustCommand{\pyRingHierarchicalQuantileOneDMuDeltaTauValue}{\reviewed{92}}
\DeclareRobustCommand{\pyRingHierarchicalQuantileOneDMuDeltaTauPlus}{\reviewed{8}}
\DeclareRobustCommand{\pyRingHierarchicalQuantileOneDMuDeltaTauMinus}{\reviewed{70}}
\DeclareRobustCommand{\pyRingHierarchicalQuantileOneDMuDeltaTauPluslnBcut}{\reviewed{18}}
\DeclareRobustCommand{\pyRingHierarchicalQuantileOneDMuDeltaTauMinuslnBcut}{\reviewed{61}}
\DeclareRobustCommand{\pyRingHierarchicalQuantileFourDPlus}{\reviewed{5.3}}
\DeclareRobustCommand{\pyRingHierarchicalQuantileFourDMinus}{\reviewed{17.9}}
\DeclareRobustCommand{\pyRingHierarchicalQuantileFourDValuelnBcut}{\reviewed{58}}
\DeclareRobustCommand{\pyRingHierarchicalQuantileFourDPluslnBcut}{\reviewed{36}}
\DeclareRobustCommand{\pyRingHierarchicalQuantileFourDMinuslnBcut}{\reviewed{30}}

\DeclareRobustCommand{\pseobJointDeltaFMedian}{\reviewed{0.00}}
\DeclareRobustCommand{\pseobJointDeltaFPlus}{\reviewed{0.03}}
\DeclareRobustCommand{\pseobJointDeltaFMinus}{\reviewed{0.02}}
\DeclareRobustCommand{\pseobJointDeltaTauMedian}{\reviewed{0.17}}
\DeclareRobustCommand{\pseobJointDeltaTauPlus}{\reviewed{0.11}}
\DeclareRobustCommand{\pseobJointDeltaTauMinus}{\reviewed{0.11}}

\DeclareRobustCommand{\pseobHierarchicalMuDeltaFMedian}{\reviewed{0.00}}
\DeclareRobustCommand{\pseobHierarchicalMuDeltaFPlus}{\reviewed{0.03}}
\DeclareRobustCommand{\pseobHierarchicalMuDeltaFMinus}{\reviewed{0.03}}
\DeclareRobustCommand{\pseobHierarchicalSigmaDeltaFBound}{\reviewed{0.06}}
\DeclareRobustCommand{\pseobHierarchicalMuDeltaTauMedian}{\reviewed{0.16}}
\DeclareRobustCommand{\pseobHierarchicalMuDeltaTauPlus}{\reviewed{0.11}}
\DeclareRobustCommand{\pseobHierarchicalMuDeltaTauMinus}{\reviewed{0.10}}
\DeclareRobustCommand{\pseobHierarchicalSigmaDeltaTauBound}{\reviewed{0.15}}
\DeclareRobustCommand{\pseobHierarchicalRhoMedian}{\reviewed{-0.02}}
\DeclareRobustCommand{\pseobHierarchicalRhoPlus}{\reviewed{0.74}}
\DeclareRobustCommand{\pseobHierarchicalRhoMinus}{\reviewed{0.72}}

\DeclareRobustCommand{\pseobDeltaFMedian}[1]{\IfEqCase{#1}{{GW150914}{\reviewed{0.02}} {GW170104}{\reviewed{-0.02}} {GW190519_153544}{\reviewed{-0.15}} {GW190521_074359}{\reviewed{0.06}} {GW190630_185205}{\reviewed{-0.06}} {GW190828_063405}{\reviewed{0.10}} {GW190910_112807}{\reviewed{0.01}} {GW191109_010717}{\reviewed{1.06}} {GW200129_065458}{\reviewed{-0.01}} {GW200208_130117}{\reviewed{0.17}} {GW200224_222234}{\reviewed{0.01}} {GW200311_115853}{\reviewed{0.02}} {GW230628_231200}{\reviewed{0.08}} {GW230914_111401}{\reviewed{-0.12}} {GW230927_153832}{\reviewed{-0.01}} {GW231028_153006}{\reviewed{-0.23}} {GW231102_071736}{\reviewed{0.25}} {GW231206_233901}{\reviewed{-0.03}} {GW231226_101520}{\reviewed{0.02}}}}
\DeclareRobustCommand{\pseobDeltaFPlus}[1]{\IfEqCase{#1}{{GW150914}{\reviewed{0.10}} {GW170104}{\reviewed{0.16}} {GW190519_153544}{\reviewed{0.19}} {GW190521_074359}{\reviewed{0.19}} {GW190630_185205}{\reviewed{0.14}} {GW190828_063405}{\reviewed{0.13}} {GW190910_112807}{\reviewed{0.13}} {GW191109_010717}{\reviewed{1.36}} {GW200129_065458}{\reviewed{0.06}} {GW200208_130117}{\reviewed{0.98}} {GW200224_222234}{\reviewed{0.15}} {GW200311_115853}{\reviewed{0.17}} {GW230628_231200}{\reviewed{0.17}} {GW230914_111401}{\reviewed{0.15}} {GW230927_153832}{\reviewed{0.09}} {GW231028_153006}{\reviewed{0.19}} {GW231102_071736}{\reviewed{0.32}} {GW231206_233901}{\reviewed{0.12}} {GW231226_101520}{\reviewed{0.14}}}}
\DeclareRobustCommand{\pseobDeltaFMinus}[1]{\IfEqCase{#1}{{GW150914}{\reviewed{0.07}} {GW170104}{\reviewed{0.14}} {GW190519_153544}{\reviewed{0.13}} {GW190521_074359}{\reviewed{0.10}} {GW190630_185205}{\reviewed{0.19}} {GW190828_063405}{\reviewed{0.12}} {GW190910_112807}{\reviewed{0.10}} {GW191109_010717}{\reviewed{0.45}} {GW200129_065458}{\reviewed{0.08}} {GW200208_130117}{\reviewed{0.30}} {GW200224_222234}{\reviewed{0.12}} {GW200311_115853}{\reviewed{0.09}} {GW230628_231200}{\reviewed{0.11}} {GW230914_111401}{\reviewed{0.11}} {GW230927_153832}{\reviewed{0.07}} {GW231028_153006}{\reviewed{0.11}} {GW231102_071736}{\reviewed{0.27}} {GW231206_233901}{\reviewed{0.09}} {GW231226_101520}{\reviewed{0.07}}}}
\DeclareRobustCommand{\pseobDeltaTauMedian}[1]{\IfEqCase{#1}{{GW150914}{\reviewed{0.10}} {GW170104}{\reviewed{0.43}} {GW190519_153544}{\reviewed{0.20}} {GW190521_074359}{\reviewed{-0.04}} {GW190630_185205}{\reviewed{-0.04}} {GW190828_063405}{\reviewed{0.17}} {GW190910_112807}{\reviewed{0.61}} {GW191109_010717}{\reviewed{-0.11}} {GW200129_065458}{\reviewed{0.16}} {GW200208_130117}{\reviewed{-0.11}} {GW200224_222234}{\reviewed{0.22}} {GW200311_115853}{\reviewed{0.15}} {GW230628_231200}{\reviewed{0.31}} {GW230914_111401}{\reviewed{0.26}} {GW230927_153832}{\reviewed{0.35}} {GW231028_153006}{\reviewed{0.17}} {GW231102_071736}{\reviewed{0.02}} {GW231206_233901}{\reviewed{0.01}} {GW231226_101520}{\reviewed{0.04}}}}
\DeclareRobustCommand{\pseobDeltaTauPlus}[1]{\IfEqCase{#1}{{GW150914}{\reviewed{0.35}} {GW170104}{\reviewed{0.94}} {GW190519_153544}{\reviewed{0.56}} {GW190521_074359}{\reviewed{0.37}} {GW190630_185205}{\reviewed{0.62}} {GW190828_063405}{\reviewed{0.56}} {GW190910_112807}{\reviewed{0.63}} {GW191109_010717}{\reviewed{0.41}} {GW200129_065458}{\reviewed{0.38}} {GW200208_130117}{\reviewed{0.68}} {GW200224_222234}{\reviewed{0.48}} {GW200311_115853}{\reviewed{1.51}} {GW230628_231200}{\reviewed{0.47}} {GW230914_111401}{\reviewed{0.47}} {GW230927_153832}{\reviewed{0.45}} {GW231028_153006}{\reviewed{0.32}} {GW231102_071736}{\reviewed{0.44}} {GW231206_233901}{\reviewed{0.36}} {GW231226_101520}{\reviewed{0.25}}}}
\DeclareRobustCommand{\pseobDeltaTauMinus}[1]{\IfEqCase{#1}{{GW150914}{\reviewed{0.28}} {GW170104}{\reviewed{0.67}} {GW190519_153544}{\reviewed{0.39}} {GW190521_074359}{\reviewed{0.28}} {GW190630_185205}{\reviewed{0.46}} {GW190828_063405}{\reviewed{0.48}} {GW190910_112807}{\reviewed{0.50}} {GW191109_010717}{\reviewed{0.30}} {GW200129_065458}{\reviewed{0.28}} {GW200208_130117}{\reviewed{0.43}} {GW200224_222234}{\reviewed{0.34}} {GW200311_115853}{\reviewed{0.45}} {GW230628_231200}{\reviewed{0.38}} {GW230914_111401}{\reviewed{0.40}} {GW230927_153832}{\reviewed{0.43}} {GW231028_153006}{\reviewed{0.28}} {GW231102_071736}{\reviewed{0.35}} {GW231206_233901}{\reviewed{0.28}} {GW231226_101520}{\reviewed{0.20}}}}
\DeclareRobustCommand{\pseobFMedian}[1]{\IfEqCase{#1}{{GW150914}{\reviewed{253.3}} {GW170104}{\reviewed{284.8}} {GW190519_153544}{\reviewed{121.7}} {GW190521_074359}{\reviewed{204.0}} {GW190630_185205}{\reviewed{247.6}} {GW190828_063405}{\reviewed{252.4}} {GW190910_112807}{\reviewed{174.5}} {GW191109_010717}{\reviewed{124.1}} {GW200129_065458}{\reviewed{247.6}} {GW200208_130117}{\reviewed{213.0}} {GW200224_222234}{\reviewed{195.8}} {GW200311_115853}{\reviewed{239.8}} {GW230628_231200}{\reviewed{224.2}} {GW230914_111401}{\reviewed{122.1}} {GW230927_153832}{\reviewed{375.6}} {GW231028_153006}{\reviewed{78.5}} {GW231102_071736}{\reviewed{108.1}} {GW231206_233901}{\reviewed{203.5}} {GW231226_101520}{\reviewed{192.3}}}}
\DeclareRobustCommand{\pseobFPlus}[1]{\IfEqCase{#1}{{GW150914}{\reviewed{17.9}} {GW170104}{\reviewed{25.3}} {GW190519_153544}{\reviewed{11.8}} {GW190521_074359}{\reviewed{24.4}} {GW190630_185205}{\reviewed{32.5}} {GW190828_063405}{\reviewed{20.3}} {GW190910_112807}{\reviewed{12.2}} {GW191109_010717}{\reviewed{14.3}} {GW200129_065458}{\reviewed{12.5}} {GW200208_130117}{\reviewed{187.9}} {GW200224_222234}{\reviewed{11.0}} {GW200311_115853}{\reviewed{23.1}} {GW230628_231200}{\reviewed{12.8}} {GW230914_111401}{\reviewed{8.2}} {GW230927_153832}{\reviewed{26.5}} {GW231028_153006}{\reviewed{3.8}} {GW231102_071736}{\reviewed{11.5}} {GW231206_233901}{\reviewed{17.9}} {GW231226_101520}{\reviewed{8.2}}}}
\DeclareRobustCommand{\pseobFMinus}[1]{\IfEqCase{#1}{{GW150914}{\reviewed{12.9}} {GW170104}{\reviewed{36.8}} {GW190519_153544}{\reviewed{14.3}} {GW190521_074359}{\reviewed{12.7}} {GW190630_185205}{\reviewed{49.4}} {GW190828_063405}{\reviewed{18.2}} {GW190910_112807}{\reviewed{8.4}} {GW191109_010717}{\reviewed{8.8}} {GW200129_065458}{\reviewed{16.8}} {GW200208_130117}{\reviewed{47.6}} {GW200224_222234}{\reviewed{10.7}} {GW200311_115853}{\reviewed{18.9}} {GW230628_231200}{\reviewed{13.2}} {GW230914_111401}{\reviewed{7.9}} {GW230927_153832}{\reviewed{21.9}} {GW231028_153006}{\reviewed{4.5}} {GW231102_071736}{\reviewed{6.4}} {GW231206_233901}{\reviewed{13.4}} {GW231226_101520}{\reviewed{7.1}}}}
\DeclareRobustCommand{\pseobTauMedian}[1]{\IfEqCase{#1}{{GW150914}{\reviewed{4.51}} {GW170104}{\reviewed{4.91}} {GW190519_153544}{\reviewed{8.55}} {GW190521_074359}{\reviewed{5.49}} {GW190630_185205}{\reviewed{3.81}} {GW190828_063405}{\reviewed{6.41}} {GW190910_112807}{\reviewed{9.49}} {GW191109_010717}{\reviewed{14.98}} {GW200129_065458}{\reviewed{5.09}} {GW200208_130117}{\reviewed{5.03}} {GW200224_222234}{\reviewed{6.89}} {GW200311_115853}{\reviewed{5.24}} {GW230628_231200}{\reviewed{6.81}} {GW230914_111401}{\reviewed{9.14}} {GW230927_153832}{\reviewed{3.66}} {GW231028_153006}{\reviewed{13.90}} {GW231102_071736}{\reviewed{13.18}} {GW231206_233901}{\reviewed{4.80}} {GW231226_101520}{\reviewed{5.66}}}}
\DeclareRobustCommand{\pseobTauPlus}[1]{\IfEqCase{#1}{{GW150914}{\reviewed{1.41}} {GW170104}{\reviewed{3.58}} {GW190519_153544}{\reviewed{4.75}} {GW190521_074359}{\reviewed{1.90}} {GW190630_185205}{\reviewed{2.49}} {GW190828_063405}{\reviewed{2.84}} {GW190910_112807}{\reviewed{3.46}} {GW191109_010717}{\reviewed{3.71}} {GW200129_065458}{\reviewed{1.72}} {GW200208_130117}{\reviewed{4.45}} {GW200224_222234}{\reviewed{2.57}} {GW200311_115853}{\reviewed{6.17}} {GW230628_231200}{\reviewed{2.14}} {GW230914_111401}{\reviewed{3.78}} {GW230927_153832}{\reviewed{1.24}} {GW231028_153006}{\reviewed{5.25}} {GW231102_071736}{\reviewed{5.67}} {GW231206_233901}{\reviewed{1.87}} {GW231226_101520}{\reviewed{1.26}}}}
\DeclareRobustCommand{\pseobTauMinus}[1]{\IfEqCase{#1}{{GW150914}{\reviewed{1.11}} {GW170104}{\reviewed{2.29}} {GW190519_153544}{\reviewed{2.90}} {GW190521_074359}{\reviewed{1.53}} {GW190630_185205}{\reviewed{1.79}} {GW190828_063405}{\reviewed{2.70}} {GW190910_112807}{\reviewed{2.82}} {GW191109_010717}{\reviewed{3.17}} {GW200129_065458}{\reviewed{1.32}} {GW200208_130117}{\reviewed{2.37}} {GW200224_222234}{\reviewed{1.99}} {GW200311_115853}{\reviewed{2.07}} {GW230628_231200}{\reviewed{1.95}} {GW230914_111401}{\reviewed{2.97}} {GW230927_153832}{\reviewed{1.16}} {GW231028_153006}{\reviewed{3.73}} {GW231102_071736}{\reviewed{5.32}} {GW231206_233901}{\reviewed{1.34}} {GW231226_101520}{\reviewed{1.06}}}}
\DeclareRobustCommand{\pseobRemnantMassMedian}[1]{\IfEqCase{#1}{{GW150914}{\reviewed{71.7}} {GW170104}{\reviewed{71.6}} {GW190519_153544}{\reviewed{142.4}} {GW190521_074359}{\reviewed{87.8}} {GW190630_185205}{\reviewed{69.1}} {GW190828_063405}{\reviewed{84.6}} {GW190910_112807}{\reviewed{123.2}} {GW191109_010717}{\reviewed{180.1}} {GW200129_065458}{\reviewed{77.3}} {GW200208_130117}{\reviewed{78.3}} {GW200224_222234}{\reviewed{100.9}} {GW200311_115853}{\reviewed{80.0}} {GW230628_231200}{\reviewed{93.1}} {GW230914_111401}{\reviewed{147.6}} {GW230927_153832}{\reviewed{53.1}} {GW231028_153006}{\reviewed{226.2}} {GW231102_071736}{\reviewed{187.7}} {GW231206_233901}{\reviewed{81.5}} {GW231226_101520}{\reviewed{92.0}}}}
\DeclareRobustCommand{\pseobRemnantMassPlus}[1]{\IfEqCase{#1}{{GW150914}{\reviewed{10.9}} {GW170104}{\reviewed{13.6}} {GW190519_153544}{\reviewed{36.5}} {GW190521_074359}{\reviewed{14.8}} {GW190630_185205}{\reviewed{16.9}} {GW190828_063405}{\reviewed{12.1}} {GW190910_112807}{\reviewed{16.4}} {GW191109_010717}{\reviewed{21.1}} {GW200129_065458}{\reviewed{10.4}} {GW200208_130117}{\reviewed{33.5}} {GW200224_222234}{\reviewed{14.4}} {GW200311_115853}{\reviewed{26.5}} {GW230628_231200}{\reviewed{10.6}} {GW230914_111401}{\reviewed{26.7}} {GW230927_153832}{\reviewed{7.2}} {GW231028_153006}{\reviewed{37.6}} {GW231102_071736}{\reviewed{29.7}} {GW231206_233901}{\reviewed{17.3}} {GW231226_101520}{\reviewed{10.7}}}}
\DeclareRobustCommand{\pseobRemnantMassMinus}[1]{\IfEqCase{#1}{{GW150914}{\reviewed{12.6}} {GW170104}{\reviewed{22.0}} {GW190519_153544}{\reviewed{34.3}} {GW190521_074359}{\reviewed{17.8}} {GW190630_185205}{\reviewed{18.1}} {GW190828_063405}{\reviewed{19.7}} {GW190910_112807}{\reviewed{19.6}} {GW191109_010717}{\reviewed{21.0}} {GW200129_065458}{\reviewed{12.3}} {GW200208_130117}{\reviewed{25.2}} {GW200224_222234}{\reviewed{17.5}} {GW200311_115853}{\reviewed{21.8}} {GW230628_231200}{\reviewed{14.4}} {GW230914_111401}{\reviewed{33.0}} {GW230927_153832}{\reviewed{10.3}} {GW231028_153006}{\reviewed{42.6}} {GW231102_071736}{\reviewed{49.5}} {GW231206_233901}{\reviewed{17.7}} {GW231226_101520}{\reviewed{12.2}}}}
\DeclareRobustCommand{\pseobRemnantSpinMedian}[1]{\IfEqCase{#1}{{GW150914}{\reviewed{0.76}} {GW170104}{\reviewed{0.86}} {GW190519_153544}{\reviewed{0.70}} {GW190521_074359}{\reviewed{0.75}} {GW190630_185205}{\reviewed{0.69}} {GW190828_063405}{\reviewed{0.90}} {GW190910_112807}{\reviewed{0.90}} {GW191109_010717}{\reviewed{0.93}} {GW200129_065458}{\reviewed{0.81}} {GW200208_130117}{\reviewed{0.77}} {GW200224_222234}{\reviewed{0.84}} {GW200311_115853}{\reviewed{0.81}} {GW230628_231200}{\reviewed{0.88}} {GW230914_111401}{\reviewed{0.75}} {GW230927_153832}{\reviewed{0.85}} {GW231028_153006}{\reviewed{0.73}} {GW231102_071736}{\reviewed{0.86}} {GW231206_233901}{\reviewed{0.65}} {GW231226_101520}{\reviewed{0.73}}}}
\DeclareRobustCommand{\pseobRemnantSpinPlus}[1]{\IfEqCase{#1}{{GW150914}{\reviewed{0.11}} {GW170104}{\reviewed{0.11}} {GW190519_153544}{\reviewed{0.20}} {GW190521_074359}{\reviewed{0.13}} {GW190630_185205}{\reviewed{0.20}} {GW190828_063405}{\reviewed{0.06}} {GW190910_112807}{\reviewed{0.05}} {GW191109_010717}{\reviewed{0.03}} {GW200129_065458}{\reviewed{0.10}} {GW200208_130117}{\reviewed{0.20}} {GW200224_222234}{\reviewed{0.08}} {GW200311_115853}{\reviewed{0.15}} {GW230628_231200}{\reviewed{0.06}} {GW230914_111401}{\reviewed{0.15}} {GW230927_153832}{\reviewed{0.07}} {GW231028_153006}{\reviewed{0.15}} {GW231102_071736}{\reviewed{0.08}} {GW231206_233901}{\reviewed{0.19}} {GW231226_101520}{\reviewed{0.10}}}}
\DeclareRobustCommand{\pseobRemnantSpinMinus}[1]{\IfEqCase{#1}{{GW150914}{\reviewed{0.23}} {GW170104}{\reviewed{0.43}} {GW190519_153544}{\reviewed{0.45}} {GW190521_074359}{\reviewed{0.30}} {GW190630_185205}{\reviewed{0.53}} {GW190828_063405}{\reviewed{0.26}} {GW190910_112807}{\reviewed{0.12}} {GW191109_010717}{\reviewed{0.05}} {GW200129_065458}{\reviewed{0.25}} {GW200208_130117}{\reviewed{0.55}} {GW200224_222234}{\reviewed{0.23}} {GW200311_115853}{\reviewed{0.43}} {GW230628_231200}{\reviewed{0.15}} {GW230914_111401}{\reviewed{0.38}} {GW230927_153832}{\reviewed{0.24}} {GW231028_153006}{\reviewed{0.34}} {GW231102_071736}{\reviewed{0.30}} {GW231206_233901}{\reviewed{0.40}} {GW231226_101520}{\reviewed{0.18}}}}

\DeclareRobustCommand{\IMRFMedian}[1]{\IfEqCase{#1}{{GW150914}{\reviewed{251.3}} {GW170104}{\reviewed{292.4}} {GW190519_153544}{\reviewed{127.9}} {GW190521_074359}{\reviewed{197.7}} {GW190630_185205}{\reviewed{260.7}} {GW190828_063405}{\reviewed{240.2}} {GW190910_112807}{\reviewed{177.1}} {GW191109_010717}{\reviewed{119.3}} {GW200129_065458}{\reviewed{250.4}} {GW200208_130117}{\reviewed{190.4}} {GW200224_222234}{\reviewed{195.5}} {GW200311_115853}{\reviewed{235.5}} {GW230628_231200}{\reviewed{216.8}} {GW230914_111401}{\reviewed{128.4}} {GW230927_153832}{\reviewed{384.1}} {GW231028_153006}{\reviewed{81.8}} {GW231102_071736}{\reviewed{108.0}} {GW231206_233901}{\reviewed{209.8}} {GW231226_101520}{\reviewed{191.5}}}}
\DeclareRobustCommand{\IMRFPlus}[1]{\IfEqCase{#1}{{GW150914}{\reviewed{8.1}} {GW170104}{\reviewed{10.7}} {GW190519_153544}{\reviewed{9.3}} {GW190521_074359}{\reviewed{7.2}} {GW190630_185205}{\reviewed{11.1}} {GW190828_063405}{\reviewed{9.9}} {GW190910_112807}{\reviewed{8.3}} {GW191109_010717}{\reviewed{7.6}} {GW200129_065458}{\reviewed{6.9}} {GW200208_130117}{\reviewed{14.1}} {GW200224_222234}{\reviewed{9.6}} {GW200311_115853}{\reviewed{9.3}} {GW230628_231200}{\reviewed{10.0}} {GW230914_111401}{\reviewed{6.5}} {GW230927_153832}{\reviewed{9.7}} {GW231028_153006}{\reviewed{3.1}} {GW231102_071736}{\reviewed{7.8}} {GW231206_233901}{\reviewed{7.4}} {GW231226_101520}{\reviewed{4.3}}}}
\DeclareRobustCommand{\IMRFMinus}[1]{\IfEqCase{#1}{{GW150914}{\reviewed{7.4}} {GW170104}{\reviewed{22.2}} {GW190519_153544}{\reviewed{8.6}} {GW190521_074359}{\reviewed{7.3}} {GW190630_185205}{\reviewed{19.1}} {GW190828_063405}{\reviewed{10.8}} {GW190910_112807}{\reviewed{8.2}} {GW191109_010717}{\reviewed{6.3}} {GW200129_065458}{\reviewed{8.2}} {GW200208_130117}{\reviewed{16.1}} {GW200224_222234}{\reviewed{8.9}} {GW200311_115853}{\reviewed{11.5}} {GW230628_231200}{\reviewed{9.4}} {GW230914_111401}{\reviewed{6.4}} {GW230927_153832}{\reviewed{19.0}} {GW231028_153006}{\reviewed{3.1}} {GW231102_071736}{\reviewed{6.7}} {GW231206_233901}{\reviewed{11.5}} {GW231226_101520}{\reviewed{4.4}}}}
\DeclareRobustCommand{\IMRTauMedian}[1]{\IfEqCase{#1}{{GW150914}{\reviewed{4.10}} {GW170104}{\reviewed{3.47}} {GW190519_153544}{\reviewed{9.49}} {GW190521_074359}{\reviewed{5.38}} {GW190630_185205}{\reviewed{4.05}} {GW190828_063405}{\reviewed{4.65}} {GW190910_112807}{\reviewed{5.83}} {GW191109_010717}{\reviewed{7.96}} {GW200129_065458}{\reviewed{4.39}} {GW200208_130117}{\reviewed{5.24}} {GW200224_222234}{\reviewed{5.58}} {GW200311_115853}{\reviewed{4.38}} {GW230628_231200}{\reviewed{4.80}} {GW230914_111401}{\reviewed{8.26}} {GW230927_153832}{\reviewed{2.72}} {GW231028_153006}{\reviewed{16.46}} {GW231102_071736}{\reviewed{9.76}} {GW231206_233901}{\reviewed{4.87}} {GW231226_101520}{\reviewed{5.29}}}}
\DeclareRobustCommand{\IMRTauPlus}[1]{\IfEqCase{#1}{{GW150914}{\reviewed{0.31}} {GW170104}{\reviewed{0.31}} {GW190519_153544}{\reviewed{1.84}} {GW190521_074359}{\reviewed{0.59}} {GW190630_185205}{\reviewed{0.39}} {GW190828_063405}{\reviewed{0.56}} {GW190910_112807}{\reviewed{0.75}} {GW191109_010717}{\reviewed{1.92}} {GW200129_065458}{\reviewed{0.44}} {GW200208_130117}{\reviewed{0.82}} {GW200224_222234}{\reviewed{0.71}} {GW200311_115853}{\reviewed{0.49}} {GW230628_231200}{\reviewed{0.55}} {GW230914_111401}{\reviewed{1.35}} {GW230927_153832}{\reviewed{0.12}} {GW231028_153006}{\reviewed{2.80}} {GW231102_071736}{\reviewed{1.46}} {GW231206_233901}{\reviewed{0.41}} {GW231226_101520}{\reviewed{0.29}}}}
\DeclareRobustCommand{\IMRTauMinus}[1]{\IfEqCase{#1}{{GW150914}{\reviewed{0.25}} {GW170104}{\reviewed{0.28}} {GW190519_153544}{\reviewed{1.54}} {GW190521_074359}{\reviewed{0.37}} {GW190630_185205}{\reviewed{0.25}} {GW190828_063405}{\reviewed{0.41}} {GW190910_112807}{\reviewed{0.55}} {GW191109_010717}{\reviewed{1.13}} {GW200129_065458}{\reviewed{0.30}} {GW200208_130117}{\reviewed{0.67}} {GW200224_222234}{\reviewed{0.53}} {GW200311_115853}{\reviewed{0.40}} {GW230628_231200}{\reviewed{0.40}} {GW230914_111401}{\reviewed{1.09}} {GW230927_153832}{\reviewed{0.08}} {GW231028_153006}{\reviewed{2.42}} {GW231102_071736}{\reviewed{1.16}} {GW231206_233901}{\reviewed{0.32}} {GW231226_101520}{\reviewed{0.25}}}}
\DeclareRobustCommand{\IMRRemnantMassMedian}[1]{\IfEqCase{#1}{{GW150914}{\reviewed{67.6}} {GW170104}{\reviewed{57.7}} {GW190519_153544}{\reviewed{146.5}} {GW190521_074359}{\reviewed{87.8}} {GW190630_185205}{\reviewed{66.4}} {GW190828_063405}{\reviewed{74.5}} {GW190910_112807}{\reviewed{96.1}} {GW191109_010717}{\reviewed{134.9}} {GW200129_065458}{\reviewed{70.9}} {GW200208_130117}{\reviewed{87.5}} {GW200224_222234}{\reviewed{90.2}} {GW200311_115853}{\reviewed{72.4}} {GW230628_231200}{\reviewed{79.0}} {GW230914_111401}{\reviewed{134.9}} {GW230927_153832}{\reviewed{44.7}} {GW231028_153006}{\reviewed{242.2}} {GW231102_071736}{\reviewed{159.9}} {GW231206_233901}{\reviewed{80.7}} {GW231226_101520}{\reviewed{87.8}}}}
\DeclareRobustCommand{\IMRRemnantMassPlus}[1]{\IfEqCase{#1}{{GW150914}{\reviewed{3.6}} {GW170104}{\reviewed{4.0}} {GW190519_153544}{\reviewed{16.3}} {GW190521_074359}{\reviewed{6.2}} {GW190630_185205}{\reviewed{4.6}} {GW190828_063405}{\reviewed{5.4}} {GW190910_112807}{\reviewed{8.5}} {GW191109_010717}{\reviewed{19.1}} {GW200129_065458}{\reviewed{4.2}} {GW200208_130117}{\reviewed{10.3}} {GW200224_222234}{\reviewed{7.5}} {GW200311_115853}{\reviewed{5.6}} {GW230628_231200}{\reviewed{6.0}} {GW230914_111401}{\reviewed{13.4}} {GW230927_153832}{\reviewed{1.8}} {GW231028_153006}{\reviewed{18.6}} {GW231102_071736}{\reviewed{15.0}} {GW231206_233901}{\reviewed{4.6}} {GW231226_101520}{\reviewed{3.4}}}}
\DeclareRobustCommand{\IMRRemnantMassMinus}[1]{\IfEqCase{#1}{{GW150914}{\reviewed{3.2}} {GW170104}{\reviewed{3.6}} {GW190519_153544}{\reviewed{15.6}} {GW190521_074359}{\reviewed{4.5}} {GW190630_185205}{\reviewed{3.4}} {GW190828_063405}{\reviewed{4.7}} {GW190910_112807}{\reviewed{7.1}} {GW191109_010717}{\reviewed{15.0}} {GW200129_065458}{\reviewed{3.4}} {GW200208_130117}{\reviewed{9.1}} {GW200224_222234}{\reviewed{6.4}} {GW200311_115853}{\reviewed{5.1}} {GW230628_231200}{\reviewed{5.2}} {GW230914_111401}{\reviewed{12.8}} {GW230927_153832}{\reviewed{1.0}} {GW231028_153006}{\reviewed{19.6}} {GW231102_071736}{\reviewed{14.8}} {GW231206_233901}{\reviewed{4.1}} {GW231226_101520}{\reviewed{3.2}}}}
\DeclareRobustCommand{\IMRRemnantSpinMedian}[1]{\IfEqCase{#1}{{GW150914}{\reviewed{0.68}} {GW170104}{\reviewed{0.67}} {GW190519_153544}{\reviewed{0.79}} {GW190521_074359}{\reviewed{0.71}} {GW190630_185205}{\reviewed{0.70}} {GW190828_063405}{\reviewed{0.74}} {GW190910_112807}{\reviewed{0.69}} {GW191109_010717}{\reviewed{0.61}} {GW200129_065458}{\reviewed{0.73}} {GW200208_130117}{\reviewed{0.66}} {GW200224_222234}{\reviewed{0.73}} {GW200311_115853}{\reviewed{0.69}} {GW230628_231200}{\reviewed{0.69}} {GW230914_111401}{\reviewed{0.71}} {GW230927_153832}{\reviewed{0.69}} {GW231028_153006}{\reviewed{0.84}} {GW231102_071736}{\reviewed{0.70}} {GW231206_233901}{\reviewed{0.67}} {GW231226_101520}{\reviewed{0.67}}}}
\DeclareRobustCommand{\IMRRemnantSpinPlus}[1]{\IfEqCase{#1}{{GW150914}{\reviewed{0.05}} {GW170104}{\reviewed{0.06}} {GW190519_153544}{\reviewed{0.07}} {GW190521_074359}{\reviewed{0.06}} {GW190630_185205}{\reviewed{0.06}} {GW190828_063405}{\reviewed{0.07}} {GW190910_112807}{\reviewed{0.07}} {GW191109_010717}{\reviewed{0.18}} {GW200129_065458}{\reviewed{0.06}} {GW200208_130117}{\reviewed{0.09}} {GW200224_222234}{\reviewed{0.07}} {GW200311_115853}{\reviewed{0.07}} {GW230628_231200}{\reviewed{0.07}} {GW230914_111401}{\reviewed{0.09}} {GW230927_153832}{\reviewed{0.04}} {GW231028_153006}{\reviewed{0.05}} {GW231102_071736}{\reviewed{0.09}} {GW231206_233901}{\reviewed{0.06}} {GW231226_101520}{\reviewed{0.04}}}}
\DeclareRobustCommand{\IMRRemnantSpinMinus}[1]{\IfEqCase{#1}{{GW150914}{\reviewed{0.05}} {GW170104}{\reviewed{0.08}} {GW190519_153544}{\reviewed{0.12}} {GW190521_074359}{\reviewed{0.06}} {GW190630_185205}{\reviewed{0.07}} {GW190828_063405}{\reviewed{0.07}} {GW190910_112807}{\reviewed{0.08}} {GW191109_010717}{\reviewed{0.19}} {GW200129_065458}{\reviewed{0.05}} {GW200208_130117}{\reviewed{0.13}} {GW200224_222234}{\reviewed{0.07}} {GW200311_115853}{\reviewed{0.08}} {GW230628_231200}{\reviewed{0.06}} {GW230914_111401}{\reviewed{0.13}} {GW230927_153832}{\reviewed{0.03}} {GW231028_153006}{\reviewed{0.08}} {GW231102_071736}{\reviewed{0.10}} {GW231206_233901}{\reviewed{0.07}} {GW231226_101520}{\reviewed{0.04}}}}

\DeclareRobustCommand{\QuantileJointGWTCThree}{\reviewed{93.8}\%}
\DeclareRobustCommand{\QuantileHierGWTCThree}{\reviewed{66.1}\%}
\DeclareRobustCommand{\QuantileHierGWTCFour}{\reviewed{85.1}\%}
\DeclareRobustCommand{\QuantileHierGWTCThreeTauTwoD}{\reviewed{83.5}\%}
\DeclareRobustCommand{\QuantileHierGWTCThreeTauOneD}{\reviewed{94.9}\%}
\DeclareRobustCommand{\QuantileHierGWTCFourTauTwoD}{\reviewed{95.9}\%}

\DeclareRobustCommand{\FractionOfCatalogsAboveJointQuantile}{\reviewed{0.7\%}}
\DeclareRobustCommand{\FractionOfCatalogsAboveHierQuantile}{\reviewed{3.5\%}}
\DeclareRobustCommand{\FractionOfCatalogsAboveHierQuantileTauTwoD}{\reviewed{1.5\%}}
\DeclareRobustCommand{\FractionOfCatalogsAboveHierQuantileTauOneD}{\reviewed{0.2\%}}

\DeclareRobustCommand{\QuantileHierGWTCFourUncertainty}{\reviewed{85.1^{+14.9}_{-19.7}}}
\DeclareRobustCommand{\QuantileJointGWTCFourUncertainty}{\reviewed{98.6^{+1.4}_{-9.4}}}
\DeclareRobustCommand{\QuantileHierGWTCThreeUncertainty}{\reviewed{66.1^{+31.9}_{-34.6}}}
\DeclareRobustCommand{\QuantileJointGWTCThreeUncertainty}{\reviewed{93.8^{+6.1}_{-20.0}}}
\DeclareRobustCommand{\QuantileHierGWTCFourUncertaintyTauTwoD}{\reviewed{95.8^{+4.1}_{-13.0}}}
\DeclareRobustCommand{\QuantileHierGWTCFourUncertaintyTauOneD}{\reviewed{99.3^{+0.7}_{-4.5}}}
\DeclareRobustCommand{\QuantileHierGWTCThreeUncertaintyTauTwoD}{\reviewed{83.5^{+14.7}_{-35.9}}}
\DeclareRobustCommand{\QuantileHierGWTCThreeUncertaintyTauOneD}{\reviewed{94.9^{+4.4}_{-18.2}}}

\DeclareRobustCommand{\QuantileHierGWTCFourPlusTwoFiveZeroOneOneFour}{\reviewed{73.0}\%}
\DeclareRobustCommand{\QuantileHierGWTCFourTauTwoDPlusTwoFiveZeroOneOneFour}{\reviewed{85.6}\%}

\DeclareRobustCommand{\QuantileJointGWTCFourMinusOneNineZeroNineOneZero}{\reviewed{96.0}\%}
\DeclareRobustCommand{\QuantileHierGWTCFourMinusOneNineZeroNineOneZero}{\reviewed{78.1}\%}
\DeclareRobustCommand{\QuantileHierGWTCFourTauTwoDMinusOneNineZeroNineOneZero}{\reviewed{93.2}\%}
\DeclareRobustCommand{\QuantileHierGWTCFourTauOneDMinusOneNineZeroNineOneZero}{\reviewed{98.3}\%}

\DeclareRobustCommand{\TGRQNMRFSnrSelection}{9}
\DeclareRobustCommand{\TGRQNMRFSegmentLength}{0.2~\text{s}}

\DeclareRobustCommand{\Mandchi}{\{M,\chi\}}

\DeclareRobustCommand{\BFADA}[1]{\IfEqCase{#1}{
{GW230601\_224134}{\reviewed{$-0.2$}}
{GW230605\_065343}{\reviewed{$0.0$}}
{GW230606\_004305}{\reviewed{$0.0$}}
{GW230609\_064958}{\reviewed{$-0.1$}}
{GW230624\_113103}{\reviewed{$-1.1$}}
{GW230627\_015337}{\reviewed{$-2.3$}}
{GW230628\_231200}{\reviewed{$-0.1$}}
{GW230630\_234532}{\reviewed{$0.0$}}
{GW230702\_185453}{\reviewed{$-0.1$}}
{GW230731\_215307}{\reviewed{$-0.3$}}
{GW230811\_032116}{\reviewed{$-2.2$}}
{GW230814\_061920}{\reviewed{$0.0$}}
{GW230824\_033047}{\reviewed{$-0.6$}}
{GW230904\_051013}{\reviewed{$1.1$}}
{GW230914\_111401}{\reviewed{$-0.1$}}
{GW230919\_215712}{\reviewed{$-0.2$}}
{GW230920\_071124}{\reviewed{$-0.3$}}
{GW230922\_020344}{\reviewed{$-0.2$}}
{GW230922\_040658}{\reviewed{$0.6$}}
{GW230924\_124453}{\reviewed{$-0.5$}}
{GW230927\_043729}{\reviewed{$-0.2$}}
{GW230927\_153832}{\reviewed{$-0.2$}}
{GW230928\_215827}{\reviewed{$-0.2$}}
{GW231001\_140220}{\reviewed{$-0.1$}}
{GW231020\_142947}{\reviewed{$-1.5$}}
{GW231028\_153006}{\reviewed{$-0.2$}}
{GW231102\_071736}{\reviewed{$-0.2$}}
{GW231104\_133418}{\reviewed{$-0.2$}}
{GW231108\_125142}{\reviewed{$-0.4$}}
{GW231110\_040320}{\reviewed{$-0.6$}}
{GW231113\_200417}{\reviewed{$-0.9$}}
{GW231114\_043211}{\reviewed{$-0.2$}}
{GW231118\_005626}{\reviewed{$-2.4$}}
{GW231118\_090602}{\reviewed{$-0.7$}}
{GW231123\_135430}{\reviewed{$-2.5$}}
{GW231206\_233134}{\reviewed{$-0.4$}}
{GW231206\_233901}{\reviewed{$-0.4$}}
{GW231213\_111417}{\reviewed{$-0.1$}}
{GW231223\_032836}{\reviewed{$-0.2$}}
{GW231224\_024321}{\reviewed{$-0.7$}}
{GW231226\_101520}{\reviewed{$-1.3$}}
}}

\DeclareRobustCommand{\BFADABestVal}{\reviewed{$1.1$}}
\DeclareRobustCommand{\BFADABestEvent}{GW230904\_051013}

\DeclareRobustCommand{\BFBHP}[1]{\IfEqCase{#1}{
{GW230605\_065343}{\reviewed{$ 0.0 $}}
{GW230606\_004305}{\reviewed{$ -0.1 $}}
{GW230628\_231200}{\reviewed{$ -0.4$}}
{GW230630\_234532}{\reviewed{$0.0$}}
{GW230702\_185453}{\reviewed{$ -0.1 $}}
{GW230731\_215307}{\reviewed{$ 0.2  $}}
{GW230811\_032116}{\reviewed{$ -0.4 $}}
{GW230914\_111401}{\reviewed{$ 0.0$}}
{GW230919\_215712}{\reviewed{$ -0.2$}}
{GW230920\_071124}{\reviewed{$-0.1  $}}
{GW230922\_020344}{\reviewed{$ 0.1$}}
{GW230924\_124453}{\reviewed{$ 0.1  $}}
{GW230927\_043729}{\reviewed{$0.1 $}}
{GW230927\_153832}{\reviewed{$ -0.1 $}}
{GW231001\_140220}{\reviewed{$-0.2  $}}
{GW231020\_142947}{\reviewed{$ -1.2$}}
{GW231102\_071736}{\reviewed{$ -0.5$}}
{GW231108\_125142}{\reviewed{$-0.5$}}
{GW231110\_040320}{\reviewed{$0.0$}}
{GW231123\_135430}{\reviewed{$ -5.9$}}
{GW231206\_233134}{\reviewed{$ -0.1 $}}
{GW231206\_233901}{\reviewed{$  -0.7$}}
{GW231213\_111417}{\reviewed{$ 0.1 $}}
{GW231223\_032836}{\reviewed{$ -0.2 $}}
{GW231226\_101520}{\reviewed{$ -1.9 $}}
}}

\DeclareRobustCommand{\BFBHPBestVal}{\reviewed{$0.2$}}
\DeclareRobustCommand{\BFBHPBestEvent}{GW230731\_215307}

\DeclareRobustCommand{\BFBWBestVal}{\reviewed{$-1.8$}}
\DeclareRobustCommand{\BFBWBestEvent}{GW230630\_234532}

\DeclareRobustCommand{\cWBpValue}[1]{\IfEqCase{#1}{
{GW230601\_224134}{\reviewed{$ 0.516^{+0.010}_{-0.010} $}}
{GW230606\_004305}{\reviewed{$ 0.063^{+0.006}_{-0.005} $}}
{GW230609\_064958}{\reviewed{$ 0.431^{+0.012}_{-0.012} $}}
{GW230624\_113103}{\reviewed{$ 0.646^{+0.014}_{-0.014} $}}
{GW230627\_015337}{\reviewed{$ 0.760^{+0.007}_{-0.007} $}}
{GW230628\_231200}{\reviewed{$ 0.530^{+0.009}_{-0.009} $}}
{GW230702\_185453}{\reviewed{$ 1.0000^{+0}_{-0.0003} $}}
{GW230731\_215307}{\reviewed{$ 0.463^{+0.015}_{-0.015} $}}
{GW230811\_032116}{\reviewed{$ 0.769^{+0.006}_{-0.006} $}}
{GW230814\_061920}{\reviewed{$ 0.223^{+0.006}_{-0.006} $}}
{GW230824\_033047}{\reviewed{$ 0.570^{+0.007}_{-0.007} $}}
{GW230914\_111401}{\reviewed{$ 0.075^{+0.004}_{-0.004} $}}
{GW230919\_215712}{\reviewed{$ 0.312^{+0.007}_{-0.007} $}}
{GW230920\_071124}{\reviewed{$ 0.114^{+0.006}_{-0.006} $}}
{GW230922\_020344}{\reviewed{$ 0.714^{+0.008}_{-0.008} $}}
{GW230922\_040658}{\reviewed{$ 0.258^{+0.007}_{-0.007} $}}
{GW230924\_124453}{\reviewed{$ 0.350^{+0.007}_{-0.007} $}}
{GW230927\_153832}{\reviewed{$ 0.711^{+0.007}_{-0.007} $}}
{GW230927\_043729}{\reviewed{$ 0.574^{+0.008}_{-0.008} $}}
{GW230928\_215827}{\reviewed{$ 1.0000^{+0}_{-0.0002} $}}
{GW231001\_140220}{\reviewed{$ 0.050^{+0.004}_{-0.004} $}}
{GW231028\_153006}{\reviewed{$ 0.485^{+0.009}_{-0.009} $}}
{GW231102\_071736}{\reviewed{$ 0.186^{+0.006}_{-0.006} $}}
{GW231108\_125142}{\reviewed{$ 0.736^{+0.008}_{-0.008} $}}
{GW231113\_200417}{\reviewed{$ 0.617^{+0.031}_{-0.030} $}}
{GW231123\_135430}{\reviewed{$ 0.861^{+0.006}_{-0.006} $}}
{GW231206\_233134}{\reviewed{$ 0.646^{+0.009}_{-0.009} $}}
{GW231206\_233901}{\reviewed{$ 0.684^{+0.009}_{-0.009} $}}
{GW231213\_111417}{\reviewed{$ 0.676^{+0.010}_{-0.010} $}}
{GW231223\_032836}{\reviewed{$ 0.510^{+0.012}_{-0.012} $}}
{GW231226\_101520}{\reviewed{$ 0.581^{+0.010}_{-0.010} $}}
}}

\DeclareRobustCommand{\cWBSNRon}[1]{\IfEqCase{#1}{
{GW230601\_224134}{\reviewed{$<0.1 $}}
{GW230606\_004305}{\reviewed{$ 0.3 $}}
{GW230609\_064958}{\reviewed{$<0.1 $}}
{GW230624\_113103}{\reviewed{$<0.1 $}}
{GW230627\_015337}{\reviewed{$1.2 $}}
{GW230628\_231200}{\reviewed{$<0.1 $}}
{GW230702\_185453}{\reviewed{$<0.1 $}}
{GW230731\_215307}{\reviewed{$<0.1 $}}
{GW230811\_032116}{\reviewed{$<0.1 $}}
{GW230814\_061920}{\reviewed{$<0.1 $}}
{GW230824\_033047}{\reviewed{$<0.1 $}}
{GW230914\_111401}{\reviewed{$1.1$}}
{GW230919\_215712}{\reviewed{$0.2 $}}
{GW230920\_071124}{\reviewed{$<0.1 $}}
{GW230922\_020344}{\reviewed{$<0.1 $}}
{GW230922\_040658}{\reviewed{$<0.1 $}}
{GW230924\_124453}{\reviewed{$0.1$}}
{GW230927\_153832}{\reviewed{$0.25$}}
{GW230927\_043729}{\reviewed{$<0.1 $}}
{GW230928\_215827}{\reviewed{$<0.1 $}}
{GW231001\_140220}{\reviewed{$2.9$}}
{GW231028\_153006}{\reviewed{$<0.1 $}}
{GW231102\_071736}{\reviewed{$<0.1 $}}
{GW231108\_125142}{\reviewed{$<0.1  $}}
{GW231113\_200417}{\reviewed{$<0.1 $}}
{GW231123\_135430}{\reviewed{$<0.1 $}}
{GW231206\_233134}{\reviewed{$<0.1 $}}
{GW231206\_233901}{\reviewed{$0.2 $}}
{GW231213\_111417}{\reviewed{$<0.1 $}}
{GW231223\_032836}{\reviewed{$<0.1  $}}
{GW231226\_101520}{\reviewed{$0.3 $}}
}}

\DeclareRobustCommand{\SNRcWBBestVal}{\reviewed{$2.9$}}
\DeclareRobustCommand{\SNRcWBBBestEvent}{GW231001\_140220}

\title{\paperscommonname III. Tests of the Remnants}

\iftoggle{endauthorlist}{
 \let\mymaketitle\maketitle
 \let\myauthor\author
 \let\myaffiliation\affiliation
 \author{\LVKCollabAuthors}
 {\def\thefootnote{}\footnotetext{\LVKCorrespondence}}
 \email{~~~~~~lvc.publications@ligo.org}
}{
 \iftoggle{fullauthorlist}{


\author[0000-0003-4786-2698]{A.~G.~Abac}
\affiliation{Max Planck Institute for Gravitational Physics (Albert Einstein Institute), D-14476 Potsdam, Germany}
\author{I.~Abouelfettouh}
\affiliation{LIGO Hanford Observatory, Richland, WA 99352, USA}
\author{F.~Acernese}
\affiliation{Dipartimento di Farmacia, Universit\`a di Salerno, I-84084 Fisciano, Salerno, Italy}
\affiliation{INFN, Sezione di Napoli, I-80126 Napoli, Italy}
\author[0000-0002-8648-0767]{K.~Ackley}
\affiliation{University of Warwick, Coventry CV4 7AL, United Kingdom}
\author[0000-0001-5525-6255]{C.~Adamcewicz}
\affiliation{OzGrav, School of Physics \& Astronomy, Monash University, Clayton 3800, Victoria, Australia}
\author[0009-0004-2101-5428]{S.~Adhicary}
\affiliation{The Pennsylvania State University, University Park, PA 16802, USA}
\author{D.~Adhikari}
\affiliation{Max Planck Institute for Gravitational Physics (Albert Einstein Institute), D-30167 Hannover, Germany}
\affiliation{Leibniz Universit\"{a}t Hannover, D-30167 Hannover, Germany}
\author[0000-0002-4559-8427]{N.~Adhikari}
\affiliation{University of Wisconsin-Milwaukee, Milwaukee, WI 53201, USA}
\author[0000-0002-5731-5076]{R.~X.~Adhikari}
\affiliation{LIGO Laboratory, California Institute of Technology, Pasadena, CA 91125, USA}
\author{V.~K.~Adkins}
\affiliation{Louisiana State University, Baton Rouge, LA 70803, USA}
% ...

 }{
 \author{\LVKCollabAuthors}
}
}

\date[\relax]{Compiled: \today}

\begin{abstract}

This is the third paper of the set recording the results of the suite of tests of \GR
    performed on the signals from the fourth Gravitational-Wave Transient Catalog (GWTC-4.0),
    where we focus on the remnants of the binary mergers.
We examine for the first time \TGRNUMEVENTS events from the first part of the fourth observing run
    of the LIGO--Virgo--KAGRA detectors,
    alongside events from the previous observation runs,
    restricting our analysis to the confident signals,
    which were measured in at least two detectors
    and that have false alarm rates \TGRFARTHRESH.
This paper focuses on \TGRIIINUMTESTS tests of the coalescence remnants.
Three of these are tests of the ringdown
    and its consistency with the expected quasi-normal mode (QNM) spectrum of a Kerr black hole.
Specifically, two tests analyze just the ringdown in the time domain,
    and the
    third test analyzes the entire signal in the frequency domain.
Four tests allow for the existence of possible echoes arriving after the end of the ringdown.
As such echoes are not expected in \GR,
    we consider two families of proposed waveform templates,
    and
    two independent searches for general excess
    coherent
    power after the
    merger.
We find overall consistency of the remnants with \GR,
    and the tightest single-event constraint on the damping time of the dominant $(2, 2, 0)$ QNM of all the GWTC-4.0 events
    is found for \FULLNAME{GW231226_101520} by the frequency domain ringdown analysis.
When combining events by multiplying likelihoods (hierarchically),
    that analysis finds that the GR prediction lies at the boundary of the
    $\QuantileJointGWTCFourUncertainty \%$ ($\QuantileHierGWTCFourUncertaintyTauOneD \%$) credible region,
    an increase from $\QuantileJointGWTCThreeUncertainty \%$ ($\QuantileHierGWTCThreeUncertaintyTauOneD \%$) for GWTC-3.0. Here the ranges of values comes from bootstrapping to
    account for the finite number of events analyzed and suggest that some of the apparently significant deviation could be attributed to variance due to the finite catalog. Since the significance
    also decreases to $\QuantileJointGWTCFourPlusTwoFiveZeroOneOneFour$ ($\QuantileHierGWTCFourTauOneDPlusTwoFiveZeroOneOneFour$) when including the more recent very loud event GW250114, there is no strong evidence for a GR deviation.
We find no evidence for post-merger echoes in the events that were analyzed.

\end{abstract}

\subpapersection{Overview}\label{sec:paper III intro}

In this paper, we examine whether the \acp{GW} emitted by remnants of \acp{CBC} behave as predicted by \acf{GR}.
The preceding two papers presented tests for general consistency with \GR \citep[Paper I;][]{GWTC:TGR-I}
    and parameterized tests \citep[Paper II;][]{GWTC:TGR-II}.
This third testing \GR paper specifically summarizes the results of the ringdown-based tests for \ac{BH} remnants and echo searches.
The expected remnant of a \ac{BBH} merger is an isolated Kerr \ac{BH}, a simple object whose perturbations are
    well studied mathematically \citep{1983mtbh.book.....C,2022hgwa.bookE..38P},
    making it an excellent candidate for clean tests of strong-field gravity.
In particular, a linearly perturbed Kerr \ac{BH} is expected to shed its perturbations by emitting radiation described by \acp{QNM}, which have complex frequencies determined by the \ac{BH}'s mass and spin \citep{Vishveshwara:1970zz,Berti:2009kk}.
    Thus, the \ac{QNM} signal decays exponentially and in practice only a handful of cycles are detectable even for strong signals at current sensitivities, such as the event GW250114~\citep{GW250114} from the \ac{O4b}.
Analyzing this ringdown signal in the linear regime allows one to make tests of the consistency of multiple \acp{QNM} with the \ac{GR} predictions.
No further signals are expected in \ac{GR} after the ringdown of the system, although later signals, termed echoes,
    have been predicted in some alternative theories.
These ringdowns and echoes are the subject of the tests in this paper.

The tests were performed on the events reported by the \ac{LVK} in the fourth \ac{GW} transient catalog \citep[GWTC-4.0;][]{GWTC:Introduction, GWTC:Results},
    which were observed with at least two detectors and have a false-alarm rate of \TGRFARTHRESH.
Table~\ref{tab:selectionIII} details which events were examined for each test.
These include events from the \ac{O4a}, which are new \citep{GWTC:Results},
    as well as some events from previous runs,
    O1 \citep{LIGOScientific:2016dsl},
    O2 \citep{GWTC1,LIGOScientific:2019fpa},
    O3a \citep{GWTC2p1,LIGOScientific:2020tif}, and O3b \citep{GWTC3,LIGOScientific:2021sio},
    which have been used for a subset of the tests.
Of the \ac{O4a} events tested here, \FULLNAME{GW230518_125908} has masses consistent with a neutron star--\ac{BH} binary, while the rest all have
masses consistent with \acp{BBH}. The loud event \FULLNAME{GW230814_230901}
\clearpage

\startlongtable
\begin{deluxetable*}{c l cccc ccc c cc c cc}
\tabcolsep=0.06cm
\centerwidetable
\tablecaption{Event selection table for the analyses in this paper, from \ac{O4a}
       and the previous observing runs O1--O3}
\tablehead{
\colhead{} & \colhead{} & \colhead{} & \colhead{} & \colhead{} & \colhead{} &
             \multicolumn{3}{c}{Ringdown} &
             \colhead{} &
             \multicolumn{5}{c}{Echoes}
            \\
\cline{7-9} \cline{11-15}
\colhead{} & \colhead{} & \colhead{} & \colhead{} & \colhead{} & \colhead{} &
             \multicolumn{3}{c}{} &
             \colhead{} &
             \multicolumn{2}{c}{Waveforms} &
             \colhead{} &
             \multicolumn{2}{c}{Min. Mod.} \\
\cline{11-12}\cline{14-15}
\colhead{Run} & 
 \colhead{Event Name} & 
 \colhead{SNR} & 
 \colhead{$(1\!\!+\!\!z)M_\mathrm{f}/\Msun$} & 
 \colhead{$\chi_{\mathrm{f}}$} & 
 \colhead{$q$} & 
 \colhead{~pyRing~} & 
 \colhead{~pSEOBNR~} & 
 \colhead{~QNMRF~} & 
 \colhead{} & 
 \colhead{ADA} & 
 \colhead{BHP} & 
 \colhead{} & 
 \colhead{\BW} &
 \colhead{\CWB}
}
\startdata
 O1 &
 \cellcolor{lightgray}{GW150914} & 
 \cellcolor{lightgray}{$\networkmatchedfiltersnrgwtctwopointoneuncert{GW150914_095045}$} & 
 \cellcolor{lightgray}{$\finalmassdetgwtctwopointoneuncert{GW150914_095045}$} & 
 \cellcolor{lightgray}{$\finalspingwtctwopointoneuncert{GW150914_095045}$} & 
 \cellcolor{lightgray}{$\massratiogwtctwopointoneuncert{GW150914_095045}$} & 
 \cellcolor{lightgray}{\checkprevlvk} & 
 \cellcolor{lightgray}{\checkmark} & 
 \cellcolor{lightgray}{\checkprevext} & 
 \cellcolor{lightgray}{ } & 
 \cellcolor{lightgray}{\checkprevlvk} & 
 \cellcolor{lightgray}{\checkprevext} & 
 \cellcolor{lightgray}{ } & 
 \cellcolor{lightgray}{\checkprevext} & 
 \cellcolor{lightgray}{\checkprevext} \\
   & 
 GW151012* & 
$\networkmatchedfiltersnrgwtctwopointoneuncert{GW151012_095443}$ & 
$\finalmassdetgwtctwopointoneuncert{GW151012_095443}$ & 
$\finalspingwtctwopointoneuncert{GW151012_095443}$ & 
$\massratiogwtctwopointoneuncert{GW151012_095443}$ & 
 \checknodata & 
 \checknodata & 
 \checknodata & 
   & 
 \checkprevext  & 
  \checkprevext & 
   & 
 \checkprevext & 
 \checkprevext \\
   & 
 \cellcolor{lightgray}{GW151226} & 
 \cellcolor{lightgray}{$\networkmatchedfiltersnrgwtctwopointoneuncert{GW151226_033853}$} & 
 \cellcolor{lightgray}{$\finalmassdetgwtctwopointoneuncert{GW151226_033853}$} & 
 \cellcolor{lightgray}{$\finalspingwtctwopointoneuncert{GW151226_033853}$} & 
 \cellcolor{lightgray}{$\massratiogwtctwopointoneuncert{GW151226_033853}$} & 
 \cellcolor{lightgray}{\checknodata} & 
 \cellcolor{lightgray}{\checknodata} & 
 \cellcolor{lightgray}{\checknodata} & 
 \cellcolor{lightgray}{ } & 
 \cellcolor{lightgray}{\checkprevlvk} & 
 \cellcolor{lightgray}{\checkprevext} & 
 \cellcolor{lightgray}{ } & 
 \cellcolor{lightgray}{\checkprevext} & 
 \cellcolor{lightgray}{\checkprevext} \\
\enddata
\tablecomments{Event selection table for the analyses in this paper, from \ac{O4a} \citep{GWTC:Results}
       and the previous observing runs O1 \citep{LIGOScientific:2016dsl},
       O2 \citep{GWTC1},
       O3a \citep{GWTC2,GWTC2p1},
       and O3b \citep{GWTC3},
       with binary parameters from each observing run's catalog paper.
       The starred events GW151012 and GW170729 had previously been used for tests of GR
        in \citet{LIGOScientific:2016dsl,LIGOScientific:2019fpa},
        but do not meet the current selection criteria, though we list analyses performed on them in external work.
        A \checkmark{} indicates an event meeting our selection criteria for an analysis, and thus included in our results.
    A \checkbar{} indicates an event meeting the criteria, but not included yet due to runtime constraints (see Section \ref{sec:modeled})
    A \checkprevext{} indicates existing results from previous runs in external works,
    while a \checkprevlvk{} indicates previous results by the LVK exist, although are not used for the new bounds here.
    A \checknodata indicates events not meeting selection criteria.
        }

\label{tab:selectionIII}

\end{deluxetable*}

\vspace{-22pt}
\noindent\citep[shortened to \COMMONNAME{GW230814single};][]{GW230814} is not covered by this paper, as it was a single-detector event.

In Section~\ref{sec:ringdown}, we discuss the three tests performed on the ringdown of remnants of \ac{BBH} mergers,
    which are expected to behave like vacuum Kerr \acp{BH}.
These check if the observed ringdown is consistent with the predicted spectrum of \acp{QNM} of a Kerr \BH.
Of these, \soft{pyRing} (Section~\ref{sec:pyring}) performs various analyses of just the post-inspiral signal in the time domain,
    pSEOBNR (Section~\ref{sec:pseob}) analyzes the entire signal in the frequency domain, and the QNM rational filter analysis (QNMRF; Section~\ref{subsubsec:QNMRF})
    again just considers the post-inspiral signal in the time domain, applying a filter to determine the \acp{QNM} present in the signal.
We summarize the ringdown results in Section~\ref{subsubsec:conc-RD}.
Section~\ref{sec:echoes} describes searches for echoes, i.e.,
    post-ringdown signals on longer timescales than expected for a pure GR ringdown.
These include both searches with waveform models (Echoes WFM), both ADA and BHP, as
    described in Section~\ref{sec:modeled},
    and minimally modeled searches (Echoes MM),
    similar to searches for bursts of \acp{GW} coherent between the detectors,
    namely \BAYESWAVE (\BW) and \CWB, as described in Sections~\ref{sec:BW} and~\ref{sec:cWB}, respectively.
We summarize the echoes results in Section~\ref{sec:echoDiscussion}.
We give the overall conclusion in Section~\ref{sec:paper III conclusions},
    and additional details about the \soft{pyRing} analysis in the \hyperref[app:pyring]{Appendix}.

All masses used in this paper are the redshifted masses $(1+z)m$
    (sometimes denoted with a ``det'' superscript, for the detector frame),
    unless otherwise specified (Section \ref{sec:cWB}),
    with $m$ either the total original mass $M$ or the final remnant mass $M_{\mathrm{f}}$.
When the mass is of interest, we use it in units of the solar mass $M_\odot$, namely $(1+z)m/M_\odot$,
    while when interested in the time scale derived from the mass, we use the conversion $t_m=G(1+z)m/c^3$, in seconds.

\subpapersection{Ringdown Tests}\label{sec:ringdown}

In Section 4.2 of Paper I, we examined the overall consistency of the early to late parts of the \ac{IMR} signals.
Here, we focus on the post-merger signal, consisting of GWs emitted from the remnant as it relaxes to a equilibrium state.
In binary BH mergers, this relaxation is highly dynamic. However, at sufficiently late times, it can be modeled using BH perturbation theory as a linear combination of \acp{QNM} with fixed amplitudes (i.e., just having the expected exponential decay).
Here, we present the results obtained by three ringdown-based tests of GR: (1) a time-domain analysis that examines only the post-merger signal (\soft{pyRing}), (2) a frequency-domain analysis that considers the entire signal (pSEOBNR), and (3) a time-domain analysis that filters out specific \acp{QNM} from the post-merger signal (QNMRF).

Traditional perturbation theory-based analyses, such as the \texttt{Kerr} analysis within \soft{pyRing} and the QNMRF analysis, concentrate solely on this late-time regime, more commonly known as the ringdown phase.
In contrast, the \texttt{KerrPostmerger} \soft{pyRing} analysis and pSEOBNR incorporate the entire post-merger signal.
Additionally, pSEOBNR includes the pre-merger portion as well, assuming it adheres to GR, whereas the other methods exclude this region from the stretch of data analyzed.

Detecting multiple QNMs and using them to perform \BH spectroscopy \citep{Detweiler:1980gk,Dreyer:2003bv,Berti:2005ys,Berti:2025hly} is a goal of many ringdown analyses. BH spectroscopy aims to test GR by using the observation of multiple QNMs and checking their consistency with the spectrum predicted for a Kerr BH. Similarly, when a subdominant mode is identified in addition to the dominant 220 QNM, 
verifying that the detector-frame remnant mass $(1+z)M_\mathrm{f}$ and spin $\chi_\mathrm{f}$ inferred from ringdown analysis agree with those from the full IMR analysis 
    constitutes a self-consistency test of GR. However, \BH spectroscopy is complicated by the fact that we have an incomplete understanding of the relaxation dynamics in the early post-merger phase of the binary's evolution. Using a ringdown model based on a superposition of QNMs can thus introduce systematic uncertainties if the spacetime has not relaxed enough to admit a stationary Kerr perturbative description.
The early post-merger phase is influenced by transient effects driven by (i) initial conditions \citep{Berti:2006wq, Albanesi:2023bgi, Lagos:2022otp, Chavda:2024awq, DeAmicis:2025xuh}; (ii) nonlinearities, which have been investigated using both numerical-relativity \citep[NR;][]{London:2014cma, Bhagwat:2017tkm, Baibhav:2023clw, Cheung:2022rbm, Mitman:2022qdl, Bourg:2024jme} and perturbative approaches \citep{Gleiser:1996yc, Sberna:2021eui, Bucciotti:2023ets, Lagos:2022otp, Bucciotti:2024zyp, Ma:2024qcv, Perrone:2023jzq, Redondo-Yuste:2023seq}; and (iii) variations in the remnant BH’s mass and spin \citep{Sberna:2021eui, Capuano:2024qhv, May:2024rrg, Redondo-Yuste:2023ipg, Zhu:2024dyl}.
Additionally, the mode amplitudes are traditionally assumed to be constant, with no time variations after the exponential decay has been factored out, for simplicity. However, amplitude growth has recently been computed in toy models \citep{Lagos:2022otp, Chavda:2024awq} and perturbative binary settings \citep{DeAmicis:2025xuh}. Finally, higher harmonics peak significantly later than the fundamental mode, requiring a QNM description starting at later times \citep{Nagar:2019wds}.

None of the aforementioned effects that exist in early post-merger are accounted for in the QNM-based analyses that use fixed amplitudes, which could potentially introduce bias if we start our analysis before the post-merger admits this simplistic description. This should be kept in mind as an important caveat.
Currently, only phenomenological descriptions exist for the earlier dynamical QNM regime in comparable mass systems \citep{Baker:2008mj, Damour:2014yha, Estelles:2020twz, Pompili:2023tna}.
Thus, the accuracy of pure QNM superpositions crucially depends on the validity of the stationary QNM description for the ringdown, determined by the analysis start time, $t_{\rm start}$. This is set by the model's assumptions, as detailed below, and is a key ingredient for accurate detection of multiple QNMs.

Finally, since astrophysical BHs are expected to be uncharged \citep{Wald:1974np, Gibbons:1975kk, Blandford:1977ds}, we
disregard electric or magnetic charges in all these analyses. Studies have indicated that the impact of a remnant U(1) charge
in ringdown measurements should be negligible at current sensitivities \citep{Carullo:2021oxn, Gu:2023eaa}.

For the \soft{pyRing} \texttt{Kerr} and QNMRF analyses, the ringdown signal is modeled as a superposition of QNMs.
For a given $(\ell, |m|, n)$ mode, the waveform can be written using the
    spin-weighted spheroidal harmonics $S^{\ell, m, n}_s$ \citep{Teukolsky:1973ha},
    which extend the spin-weighted spherical harmonics $Y^{\ell, m}_s$ \citep{Gelfand1958,Newman:1966ub,Creighton:2011zz}.
    Here we need only the harmonics for spin $s=-2$ and have evaluated the spin-weighted spheroidal harmonics at the QNM frequency,
    hence the $n$ overtone index. We thus have
\begin{widetext}
\begin{equation}
\begin{split}
\label{eq:qnm_wf}
h_+ - i h_\times &= A_{\ell, +m, n}\exp{\left[i\left(\frac{2\pi f_{\ell |m| n} t}{1+z} + \phi_{\ell, +m, n}\right)\right]}\exp{\left[-\frac{t}{(1+z)\tau_{\ell |m| n}}\right]} S^{\ell, +m, n}_{-2}(\iota, \varphi, \chi_\mathrm{f})\\
                      & \quad + A_{\ell, -m, n}\exp{\left[i\left(-\frac{2\pi f_{\ell |m| n} t}{1+z} + \phi_{\ell, -m, n}\right)\right]}\exp{\left[-\frac{t}{(1+z)\tau_{\ell |m| n}}\right]}S^{\ell, -m, n}_{-2}(\iota, \varphi, \chi_\mathrm{f})\, ,
\end{split}
\end{equation}
\end{widetext}
where $\iota$ is the inclination angle, $f_{\ell |m| n}, \tau_{\ell |m| n}$ are the QNM frequency and damping times,
    $A_{\ell, +m, n}$ and $A_{\ell, -m, n}$ are the amplitudes for the left- and right-handed polarizations,
    and $\phi_{\ell, +m, n}$ and $\phi_{\ell, -m, n}$ are the corresponding phases.
In GR, $f_{\ell |m| n}, \tau_{\ell |m| n}$ are uniquely determined by the remnant mass and spin through the Kerr spectrum.
The \soft{pyRing} analysis approximates the spheroidal harmonics by the spherical harmonics. The mixing between spherical harmonic modes
created by this approximation is expected to be negligible for the systems being considered, particularly for the dominant 220 QNM \citep{Isi:2021iql}.

For non-precessing binaries, modes with opposite azimuthal index are related by reflection symmetry,
\begin{equation}
X_{\ell,-m} = (-1)^{\ell} X_{\ell,m}^{*},
\end{equation}
where $X_{\ell m}$ denotes either (frequency domain) waveform multipoles $\tilde{h}_{\ell m}$ or QNM amplitudes $A_{\ell, +m, n}$.
This symmetry also holds approximately for systems with generic spins.

For the \soft{pyRing} \texttt{KerrPostmerger} and pSEOBNR analyses, which allow for deviations from GR, we quantify the consistency with the null hypothesis by the GR quantile, which corresponds to the fraction of the posterior enclosed by the isoprobability contour that passes through the GR value~\citep{Ghosh:2017gfp}, and is defined such that $0\%$ ($100\%$) indicates full consistency (full inconsistency) with the null hypothesis. For these analyses, we obtain combined results on many events hierarchically, and denote the GR quantile in the full four-dimensional space of hyperparameters for the hierarchical analysis by $Q_{\mathrm{GR}}^{\mathrm{4D}}$. For pSEOBNR, we also compute the joint posterior of the two deviation parameters when combining events, and denote the GR quantile in that case by $Q_{\mathrm{GR}}^{\mathrm{2D}}$.

\subpapersubsection{The \soft{pyRing} analysis}\label{sec:pyring}

\begin{table*}[ht]
\caption{\label{tab:remnant_params}
Results from the \soft{pyRing} analysis
}
\begin{center}
\resizebox{\textwidth}{!}{%
\small 
\setlength{\tabcolsep}{3pt} 
\begin{tabular}{@{}l @{\hspace{6pt}}c@{\hspace{6pt}}c @{\hspace{6pt}}c@{\hspace{6pt}} *{3}{c}@{\hspace{6pt}}c@{\hspace{6pt}}*{3}{c}@{\hspace{6pt}}c@{\hspace{6pt}}*{2}{c}@{}}
\toprule
Event & & $\log_{10} \mathcal{B}_\mathrm{Noise}^{22}$ & &\multicolumn{3}{c}{$(1+z)M_\mathrm{f}/M_{\odot}$} & & \multicolumn{3}{c}{$\chi_\mathrm{f}$} & & \multicolumn{2}{c}{$\log_{10} \mathcal{B}^{\rm HM}_{{\rm H}_0}$} \\[0.075cm]
\cmidrule(lr){5-7}
\cmidrule(lr){9-11}
\cmidrule(lr){13-14}
& & & & IMR & \texttt{Kerr} & \texttt{KerrPostmerger} & & IMR & \texttt{Kerr} & \texttt{KerrPostmerger} & & \texttt{Kerr} & \texttt{KerrPostmerger}   \\
\midrule

GW230601\_224134 & & $10.81$ & &$\finalmassdetuncert{GW230601_224134}$ & $200_{-140}^{+230}$ & $180_{-22}^{+13}$ & & $\finalspinuncert{GW230601_224134}$ & $0.83_{-0.72}^{+0.15}$ & $0.84_{-0.25}^{+0.08}$ & & $-0.71$ & $0.16$ \\
GW230609\_064958 & & $2.60$ & &$\finalmassdetuncert{GW230609_064958}$ & $82_{-70}^{+390}$ & $94_{-13}^{+12}$ & & $\finalspinuncert{GW230609_064958}$ & $0.41_{-0.36}^{+0.46}$ & $0.75_{-0.26}^{+0.15}$ & & $-0.68$ & $0.11$ \\
GW230628\_231200 & & $6.79$ & &$\finalmassdetuncert{GW230628_231200}$ & $88_{-25}^{+21}$ & $87.3_{-8.8}^{+6.9}$ & & $\finalspinuncert{GW230628_231200}$ & $0.80_{-0.48}^{+0.14}$ & $0.85_{-0.14}^{+0.07}$ & & $-0.42$ & $0.20$ \\
GW230811\_032116 & & $3.19$ & &$\finalmassdetuncert{GW230811_032116}$ & $74_{-38}^{+120}$ & $75_{-11}^{+11}$ & & $\finalspinuncert{GW230811_032116}$ & $0.56_{-0.49}^{+0.38}$ & $0.73_{-0.23}^{+0.15}$ & & $-1.22$ & $0.12$ \\
GW230814\_061920 & & $9.19$ & &$\finalmassdetuncert{GW230814_061920}$ & $177_{-42}^{+64}$ & $193_{-20}^{+17}$ & & $\finalspinuncert{GW230814_061920}$ & $0.61_{-0.52}^{+0.30}$ & $0.79_{-0.17}^{+0.09}$ & & $-1.29$ & $0.46$ \\
GW230824\_033047 & & $7.23$ & &$\finalmassdetuncert{GW230824_033047}$ & $166_{-50}^{+51}$ & $161_{-19}^{+13}$ & & $\finalspinuncert{GW230824_033047}$ & $0.78_{-0.60}^{+0.17}$ & $0.80_{-0.24}^{+0.10}$ & & $-0.98$ & $-0.04$ \\
GW230914\_111401 & & $18.91$ & &$\finalmassdetuncert{GW230914_111401}$ & $144_{-34}^{+34}$ & $146_{-13}^{+11}$ & & $\finalspinuncert{GW230914_111401}$ & $0.66_{-0.52}^{+0.22}$ & $0.78_{-0.15}^{+0.09}$ & & $-1.15$ & $0.23$ \\
GW230922\_020344 & & $3.37$ & &$\finalmassdetuncert{GW230922_020344}$ & $75_{-14}^{+26}$ & $86.2_{-9.9}^{+8.7}$ & & $\finalspinuncert{GW230922_020344}$ & $0.37_{-0.33}^{+0.41}$ & $0.78_{-0.21}^{+0.11}$ & & $-0.91$ & $0.17$ \\
GW230922\_040658 & & $14.26$ & &$\finalmassdetuncert{GW230922_040658}$ & $227_{-49}^{+65}$ & $240_{-20}^{+16}$ & & $\finalspinuncert{GW230922_040658}$ & $0.62_{-0.51}^{+0.28}$ & $0.77_{-0.16}^{+0.10}$ & & $-0.96$ & $0.17$ \\
GW230924\_124453 & & $4.47$ & &$\finalmassdetuncert{GW230924_124453}$ & $79_{-25}^{+28}$ & $72.8_{-9.5}^{+7.9}$ & & $\finalspinuncert{GW230924_124453}$ & $0.76_{-0.60}^{+0.19}$ & $0.79_{-0.21}^{+0.11}$ & & $-1.35$ & $0.00$ \\
GW230927\_043729 & & $1.95$ & &$\finalmassdetuncert{GW230927_043729}$ & $220_{-150}^{+190}$ & $92_{-13}^{+11}$ & & $\finalspinuncert{GW230927_043729}$ & $0.43_{-0.39}^{+0.45}$ & $0.77_{-0.25}^{+0.13}$ & & $-0.74$ & $0.00$ \\
GW230927\_153832 & & $9.76$ & &$\finalmassdetuncert{GW230927_153832}$ & $44_{-11}^{+14}$ & $48.0_{-5.2}^{+3.7}$ & & $\finalspinuncert{GW230927_153832}$ & $0.65_{-0.54}^{+0.27}$ & $0.80_{-0.19}^{+0.10}$ & & $-0.77$ & $0.14$ \\
GW230928\_215827 & & $2.05$ & &$\finalmassdetuncert{GW230928_215827}$ & $150_{-130}^{+260}$ & $134_{-20}^{+16}$ & & $\finalspinuncert{GW230928_215827}$ & $0.64_{-0.56}^{+0.33}$ & $0.80_{-0.27}^{+0.11}$ & & $-0.91$ & $0.06$ \\
GW231001\_140220 & & $12.89$ & &$\finalmassdetuncert{GW231001_140220}$ & $172_{-24}^{+42}$ & $215_{-19}^{+16}$ & & $\finalspinuncert{GW231001_140220}$ & $0.24_{-0.22}^{+0.41}$ & $0.77_{-0.17}^{+0.10}$ & & $-0.49$ & $0.55$ \\
GW231028\_153006 & & $60.05$ & &$\finalmassdetuncert{GW231028_153006}$ & $251_{-32}^{+26}$ & $227.8_{-10.0}^{+9.9}$ & & $\finalspinuncert{GW231028_153006}$ & $0.81_{-0.17}^{+0.08}$ & $0.78_{-0.07}^{+0.06}$ & & $0.01$ & $-0.07$ \\
GW231102\_071736 & & $16.41$ & &$\finalmassdetuncert{GW231102_071736}$ & $152_{-43}^{+49}$ & $166_{-14}^{+12}$ & & $\finalspinuncert{GW231102_071736}$ & $0.76_{-0.49}^{+0.16}$ & $0.77_{-0.17}^{+0.09}$ & & $-0.70$ & $-0.01$ \\
GW231108\_125142 & & $2.27$ & &$\finalmassdetuncert{GW231108_125142}$ & $45_{-12}^{+98}$ & $50.5_{-6.5}^{+5.6}$ & & $\finalspinuncert{GW231108_125142}$ & $0.40_{-0.36}^{+0.45}$ & $0.80_{-0.23}^{+0.11}$ & & $-0.82$ & $-0.02$ \\
GW231206\_233134 & & $8.07$ & &$\finalmassdetuncert{GW231206_233134}$ & $127_{-35}^{+26}$ & $99.4_{-10.0}^{+8.6}$ & & $\finalspinuncert{GW231206_233134}$ & $0.90_{-0.32}^{+0.07}$ & $0.81_{-0.17}^{+0.10}$ & & $-1.17$ & $0.10$ \\
GW231206\_233901 & & $19.85$ & &$\finalmassdetuncert{GW231206_233901}$ & $86_{-20}^{+28}$ & $81.5_{-7.8}^{+6.8}$ & & $\finalspinuncert{GW231206_233901}$ & $0.59_{-0.49}^{+0.29}$ & $0.67_{-0.19}^{+0.13}$ & & $-1.00$ & $0.04$ \\
GW231213\_111417 & & $4.93$ & &$\finalmassdetuncert{GW231213_111417}$ & $87_{-23}^{+34}$ & $105_{-12}^{+11}$ & & $\finalspinuncert{GW231213_111417}$ & $0.55_{-0.48}^{+0.33}$ & $0.77_{-0.24}^{+0.13}$ & & $-1.23$ & $0.01$ \\
GW231223\_032836 & & $4.12$ & &$\finalmassdetuncert{GW231223_032836}$ & $118_{-72}^{+79}$ & $129_{-19}^{+18}$ & & $\finalspinuncert{GW231223_032836}$ & $0.47_{-0.42}^{+0.44}$ & $0.73_{-0.26}^{+0.16}$ & & $-0.91$ & $0.10$ \\
GW231226\_101520 & & $75.67$ & &$\finalmassdetuncert{GW231226_101520}$ & $97_{-20}^{+17}$ & $87.0_{-5.6}^{+4.9}$ & & $\finalspinuncert{GW231226_101520}$ & $0.72_{-0.39}^{+0.16}$ & $0.67_{-0.12}^{+0.09}$ & & $-1.15$ & $0.02$ \\

\bottomrule
\end{tabular}%
}
\end{center}
\tablecomments{The median and symmetric 90\% credible intervals of the redshifted final mass and final spin, inferred from the full IMR analysis \citep{GWTC:Results} 
    and the \soft{pyRing} analysis with two waveform models (\texttt{Kerr} and \texttt{KerrPostmerger}) at their nominal validity time $t_{\mathrm{nom}}$ 
    (see the \hyperref[app:pyring]{Appendix}). 
    The selection criteria for an event to be included in the \soft{pyRing} analysis is $\log_{10} \mathcal{B}_\mathrm{Noise}^{22}\gtrsim 1$. 
    For values \ensuremath{\log_{10} \mathcal{B}^{\rm HM}_{{\rm H}_0} > 1} the data would provide support for higher modes (HMs) over the single mode null hypothesis ($\mathrm{H}_0$). 
    The error on each Bayes factor from the nested sampling stopping criterion is $\sim 0.09$.}
\end{table*}

The \soft{pyRing} \citep{Carullo:2019flw} analysis is designed to isolate and analyze the post-merger phase of 
binary BH mergers by employing a purely time-domain likelihood formulation \citep{DelPozzo:2016kmd, Carullo:2019flw, Isi:2021iql}.
This analysis uses a Bayesian framework, allowing for independent ringdown-only estimation of the remnant BH’s mass, spin, and QNM amplitudes, 
as well as the measurement of QNM frequencies and damping times for performing BH spectroscopy \citep{Berti:2025hly}.
It also realizes parameterized tests for non-GR features by introducing agnostic deviations in Kerr QNM frequencies and damping times.

The \soft{pyRing} analysis uses a hierarchical modeling strategy, using three template families that progressively incorporate more information.
The most agnostic model, known as \texttt{DampedSinusoids} (\texttt{DS}), uses a linear superposition of fixed-amplitude damped sinusoids with free frequencies and damping times.
The next model, the \texttt{Kerr} template, constrains the frequency and damping time spectrum to match the QNM spectrum of a Kerr BH. Finally, the \texttt{KerrPostmerger} model incorporates progenitor information and uses amplitude models calibrated to NR simulations up to the signal peak.
\texttt{KerrPostmerger}  is the most comprehensive model, as it phenomenologically accounts for time-dependent amplitudes and extracts the most information from the data. 

To decrease computational cost, we select only systems with a sufficient observable ringdown signal.
Specifically, events with a $\log_{10} \mathcal{B}_\mathrm{Noise}^{22} \gtrsim 1$ are selected, where $\mathcal{B}_\mathrm{Noise}^{22}$ is the signal-to-noise Bayes factor for \texttt{KerrPostmerger}
(i.e., the ratio of the evidence for the presence of a \texttt{KerrPostmerger} signal with only the $22$-mode contribution to that for Gaussian noise).
This threshold is low enough to include all events with a ringdown signal that can be confidently distinguished from noise, while excluding those with negligible ringdown \SNR.
Estimates of the remnant parameters from IMR analyses in GR, for O4a events that pass the above criterion, are reported in Table~\ref{tab:remnant_params}. Due to a technical issue only discovered at a late
stage in the preparation of this paper, $\mathcal{B}_\mathrm{Noise}^{22}$ was not computed for GW230606\_004305 and GW231118\_005626. Thus, those events are currently excluded a priori from
the \soft{pyRing} analysis.

\begin{figure*}[t]
    \centering
    \includegraphics[width=\TGRFigureWidthPage]{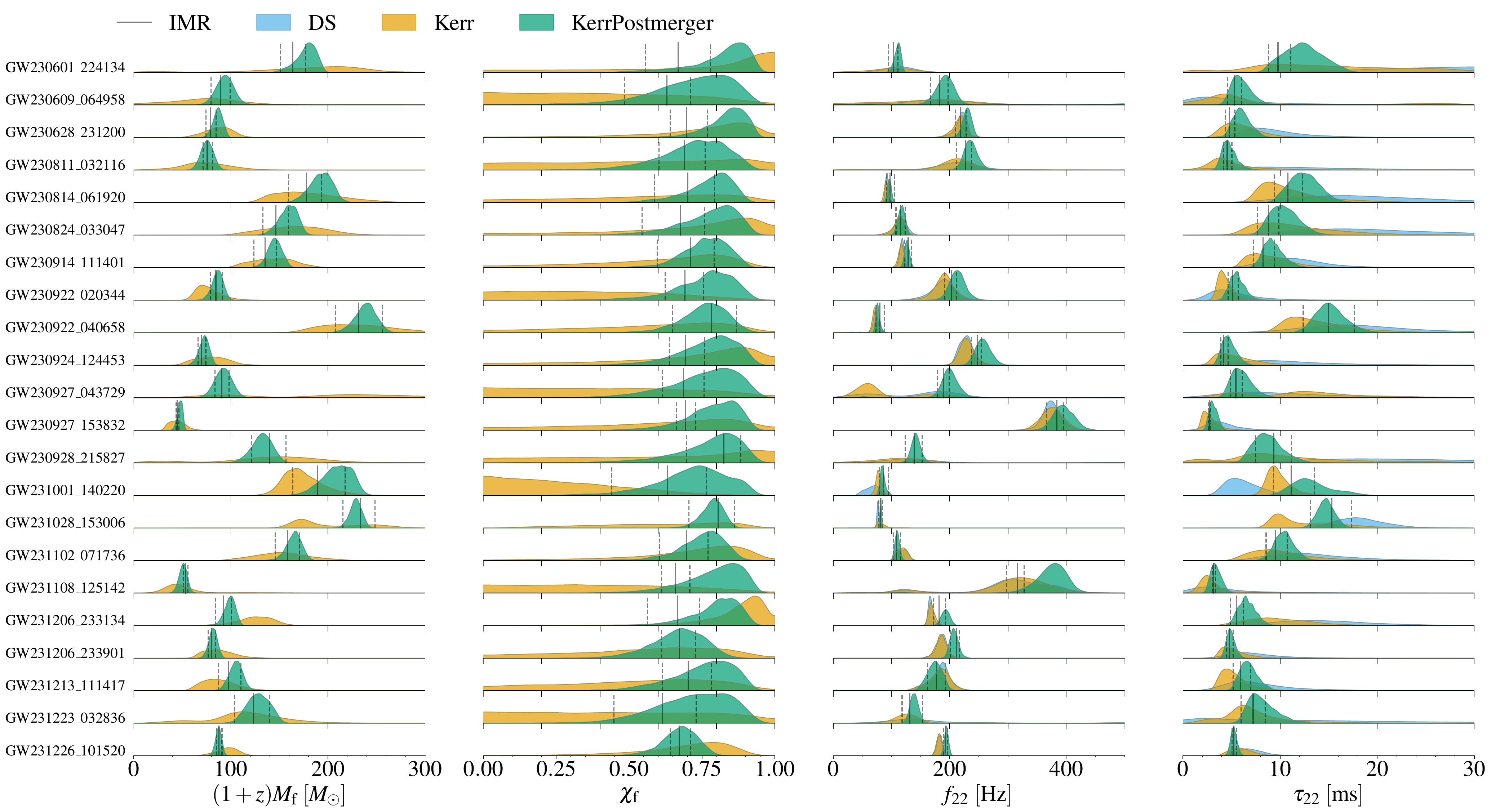}
    \caption{Comparison of final mass, final spin, fundamental mode ringdown frequency and damping time at their nominal validity time $t_\mathrm{nom}$ (see the \hyperref[app:pyring]{Appendix}) for all events analyzed by \soft{pyRing}.
    Different posterior colors represent the templates used in the analysis: \texttt{DS} (blue) and \texttt{Kerr} (yellow) each with the highest evidence mode combination at the nominal time and \texttt{KerrPostmerger} (green) using all available
    HMs. The \texttt{DS} analysis just provides results for $f_{22}$ and $\tau_{22}$.
    IMR PE median values (solid vertical black lines) with 90\% credible intervals (dashed vertical black lines) from \citet{GWTC:Results} are shown alongside the corresponding ringdown estimates, assessing consistency with GR expectations.
    \label{fig:event_comparison}
    }
\end{figure*}

The analysis start time, specified below for the different models, is defined relative to a reference time, $t_0$. 
chosen as the median time of the peak strain $h_+^2 + h_\times^2$ from the IMR GR analysis \citep{GWTC:Results} with the \SURSEVENDQFOUR \citep{Varma:2019csw} waveform when available for the given event. By convention, we use the strain in the Hanford detector to estimate the position $t_0$ of the peak. To set the start time in other detectors we use the appropriate time shifts given the source's inferred sky location. 
For events where results with \SURSEVENDQFOUR are not available, we instead determine $t_0$ using the \IMRPhenomXPHMST waveform \citep[henceforth \IMRPhenomXPHM for brevity;][]{Pratten:2020ceb,Colleoni:2024knd}.
For NR-calibrated ringdown templates like \texttt{KerrPostmerger}, the reference time $t_0^{22}$ is computed as the median time of the peak of the $h_{22}$ mode, consistent with its NR calibration.

We repeat the analysis for multiple starting times to confirm consistency with GR predictions and to check for potential anomalies,
    but for simplicity in Figure~\ref{fig:event_comparison} and Table~\ref{tab:remnant_params} we report results at a single characteristic time, $t_{\rm nom}$,
    set by the model's regime of validity.
The choice of $t_{\rm nom}$, and further discussion on how multiple start times
    are incorporated in the detection of HMs, can be found in the  \hyperref[app:pyring]{Appendix}.
Due to the current truncated time-segment formulation used in the analysis \citep{Isi:2021iql},
    in which the ringdown analysis starts at the signal peak,
    the sky location is fixed to the maximum-likelihood value obtained from the full IMR analysis \citep{GWTC:Results}.
For certain exceptional events \citep[e.g., in][]{GW231123}, we have explicitly verified that changing the sky location within the $90\%$ credible region of the IMR posterior did not significantly affect the ringdown results. 
However, this analysis is expensive enough that we reserve it for exceptional cases.

\subpapersubsubsection{Results}

\begin{figure*}[htbp]
    \centering
    \includegraphics[width=\TGRFigureWidthPage]{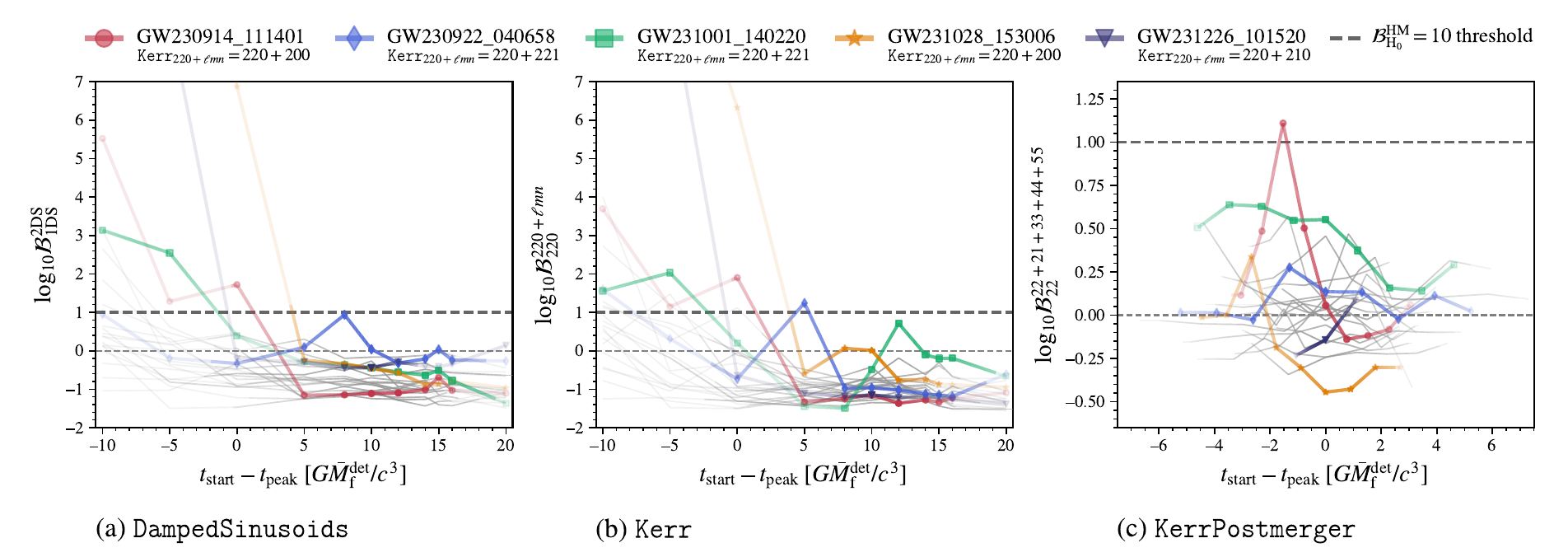}
    \caption{
        Logarithmic Bayes factor vs.\ start time of the analysis across different template families, used to evaluate the evidence for HMs (here extended to also include the two-mode hypothesis for \texttt{DampedSinusoids}) over the single mode null hypothesis ($\mathrm{H}_0$).
        The dashed horizontal lines mark the $\log_{10} \mathcal{B}^{\mathrm{HM}}_{\mathrm{H}_0} = 0$ border and the $\log_{10}\mathcal{B}^{\mathrm{HM}}_{\mathrm{H}_0} = 1$ detection threshold. 
        The colored shading represents the $90\%$ credible interval of the $t_\mathrm{start}$ posterior, while regions outside that interval have a constant faint shading.
        We highlight five events in color, while the bulk of events is depicted in gray.
        The $\texttt{DampedSinusoids}$ model Bayes factors, depicted in panel (a), compares the agnostic two mode vs.\ the single mode hypothesis.
        For the $\texttt{Kerr}$ template, the most favored HM combination is reported in panel (b). 
        The second line of the legend states the most favored $\texttt{Kerr}$ HM combination for the highlighted events.
        The $\texttt{KerrPostmerger}$ analysis, depicted in panel (c), includes the $22$, $21$, $33$, $44$, and $55$ modes in the HM template.
        No statistically significant presence of HMs was found for all times in the $90\%$ credible region of the $t_\mathrm{start}$ posterior.}
    \label{fig:bayes_factors_summary}
\end{figure*}

\textit{\texttt{DampedSinusoids} --} The \texttt{DS} template serves as a minimally modeled test of the ringdown emission's consistency with GR,
since  no specific assumptions are made about the underlying spacetime metric nor the nature of the emitting object. 
In this approach, the frequency, damping time, and constant complex amplitude of each mode are treated as free parameters, assuming left-circular polarization. 
The analysis employs uniform priors for the frequency, damping time, logarithm of the amplitude, and phase.
The evidence for an additional frequency component in the data is quantified by the logarithm of the Bayes factor between  
the two-mode (\texttt{2DS}) and one-mode damped sinusoid (\texttt{1DS}) models ($\log_{10} \mathcal{B}^{\rm \texttt{2DS}}_{\rm\texttt{1DS}}$) shown in the \hyperref[app:pyring]{Appendix}.
Bayes factors for a wide range of starting times are shown in Figure~\ref{fig:bayes_factors_summary}(a).
Within the validity regime of the model (see the \hyperref[app:pyring]{Appendix}), no statistically significant presence of multiple modes was found.
Parameter-estimation results using the favored model at its nominal validity starting time are shown in Figure~\ref{fig:event_comparison} and reported in Table~\ref{tab:remnant_params}. The favored model is \texttt{1DS} for all events except \FULLNAME{GW230922_040658}, for which \texttt{2DS} is marginally preferred; see Figure~\ref{fig:bayes_factors_summary}(a).
For all events, 90\% credible level (CL) overlap is found with GR results, signalling consistency with the GR prediction.
Since we do not find any evidence for a second mode, we do not extend the search beyond two damped sinusoids.

\textit{\texttt{Kerr} --} A further semi-agnostic test to quantify the agreement with QNMs predictions from GR involves assuming that 
the frequencies depend on the asymptotically late value of the remnant Kerr BH mass and spin, as predicted by perturbation theory. 
This assumption characterizes all the models considered below. We adopt the QNM waveform model given in Equation~\eqref{eq:qnm_wf}.
In this model, $\varphi$ is set to $0$ given its degeneracy with the modes' phases.
This model ignores counter-rotating modes as their contribution is expected to be suppressed for the parameter space considered here \citep{Cheung:2023vki}.

Compared to \texttt{DS}, the only additional assumption of the \texttt{Kerr} model is that the mode frequencies and damping times are functions of the mass and spin of the remnant BH, which are treated as free parameters, together with the amplitudes and phases of the modes.

For the \texttt{Kerr}  analysis, we start with the fundamental (2,2,0) mode (expected to dominate the signal for most comparable-mass binary configurations) and 
then systematically add one additional mode at a time in the form of $220+\ell mn$, where $\ell mn$ is any of the modes from the set 
$\Lambda^{\text{Kerr}} = \{221, 210, 200, 330, 320, 440\}$. These are the most excited modes in NR simulations of binary BH~\citep{Berti:2007fi, Buonanno:2006ui}.
To reduce computational cost, we impose that the \texttt{Kerr} amplitudes obey the reflection symmetry stated in Section~\ref{sec:ringdown}.
Since the difference between the $\pm m$ modes is expected to be an order of magnitude smaller than the overall mode amplitudes \citep{Nobili:2025ydt}, we do not expect this assumption to affect the analysis at current sensitivity. 
The contribution of higher modes (HMs) is quantified by the Bayes factor between the model with an additional mode,
$220+\ell mn$ and the null hypothesis $\mathrm{H}_0$ of detecting only the fundamental mode in the data ($\log_{10} \mathcal{B}^{220+\ell mn}_{220}$).

Bayes factors for a wide range of starting times are shown in Figure~\ref{fig:bayes_factors_summary}(b).
Within the validity regime of the model (see the \hyperref[app:pyring]{Appendix}), no statistically significant presence of multiple modes was found.
Parameter-estimation results using the most favored model at its nominal validity starting time $t_{\rm nom}$ are shown in Figure~\ref{fig:event_comparison} and reported in Table~\ref{tab:remnant_params}. The favored model includes only the fundamental mode for all models except \FULLNAME{GW231028_153006}, where there is a very slight preference for the model also containing the $200$ mode; see Figure~\ref{fig:bayes_factors_summary}(b).
For all events, 90\% CL overlap is found with GR results, signalling consistency with the GR prediction.
Since we do not find any evidence for a second mode, we do not extend the search beyond two modes.

For the event \FULLNAME{GW231123_135430} (henceforth shortened to \COMMONNAME{GW231123}), where we present results in \citet{GW231123}, we find evidence for additional modes until a ringdown start time of $32.3$~ms ($ 21.6 G\bar{M}_\mathrm{f}^\mathrm{det}/c^3$) past the peak of the signal.
It is unclear if the ringdown start times, for which such evidence is recovered, lie within the regime of validity of the stationary ringdown templates deployed. This is due to the complex signal morphology of \COMMONNAME{GW231123}, as discussed in \citet{GW231123}.
Among the multi-mode ringdown templates with positive evidence, the ones yielding consistency with IMR results contain mode contributions that are not predicted to be significantly excited in the IMR template.
This discrepancy may be due to waveform modeling uncertainties, as discussed in \citet{GW231123}.
However, the single-mode results are consistent with IMR estimates for the final mass and spin, and all mode combinations favor a massive remnant, in agreement with IMR estimates.

\textit{\texttt{KerrPostmerger} --} The \texttt{KerrPostmerger} analysis employs an NR-calibrated template for spin-aligned binaries that uses a phenomenological prescription to model the time-dependent amplitudes and phases, effectively capturing nonlinearities, overtone excitations, and transient effects in the early post-merger.
The \texttt{KerrPostmerger} model uses the \texttt{TEOBPM} ansatz, developed within the TEOBResumS family of waveforms \citep{Damour:2014yha, DelPozzo:2016kmd, Nagar:2018zoe, Nagar:2019wds, Nagar:2020pcj} and also used in other waveforms \citep{Bohe:2016gbl, Cotesta:2018fcv, Estelles:2020osj, Estelles:2021gvs}.
\texttt{KerrPostmerger} is defined from the peak of the mode $(\ell,m) = (2,2)$ in the full IMR waveform, $t_0 = t_{22}^{\text{peak}} \equiv 0M$, 
and includes the dominant spherical multipole moments with $\ell \leq 5$, specifically $(2,1)$, $(3,3)$, $(4,4)$, and $(5,5)$.
The excitation amplitudes of the different modes are expressed as a function of the progenitor parameters, here the binary's two (redshifted) component masses, and two aligned spin components,
all sampled with uniform priors.
In the current implementation \citep{Gennari:2023gmx}, each mode included in the analysis brings one additional free initial phase $\phi_{\ell m}$.

\begin{figure}[htbp]
     \centering
     \includegraphics[width=\TGRFigureWidth]{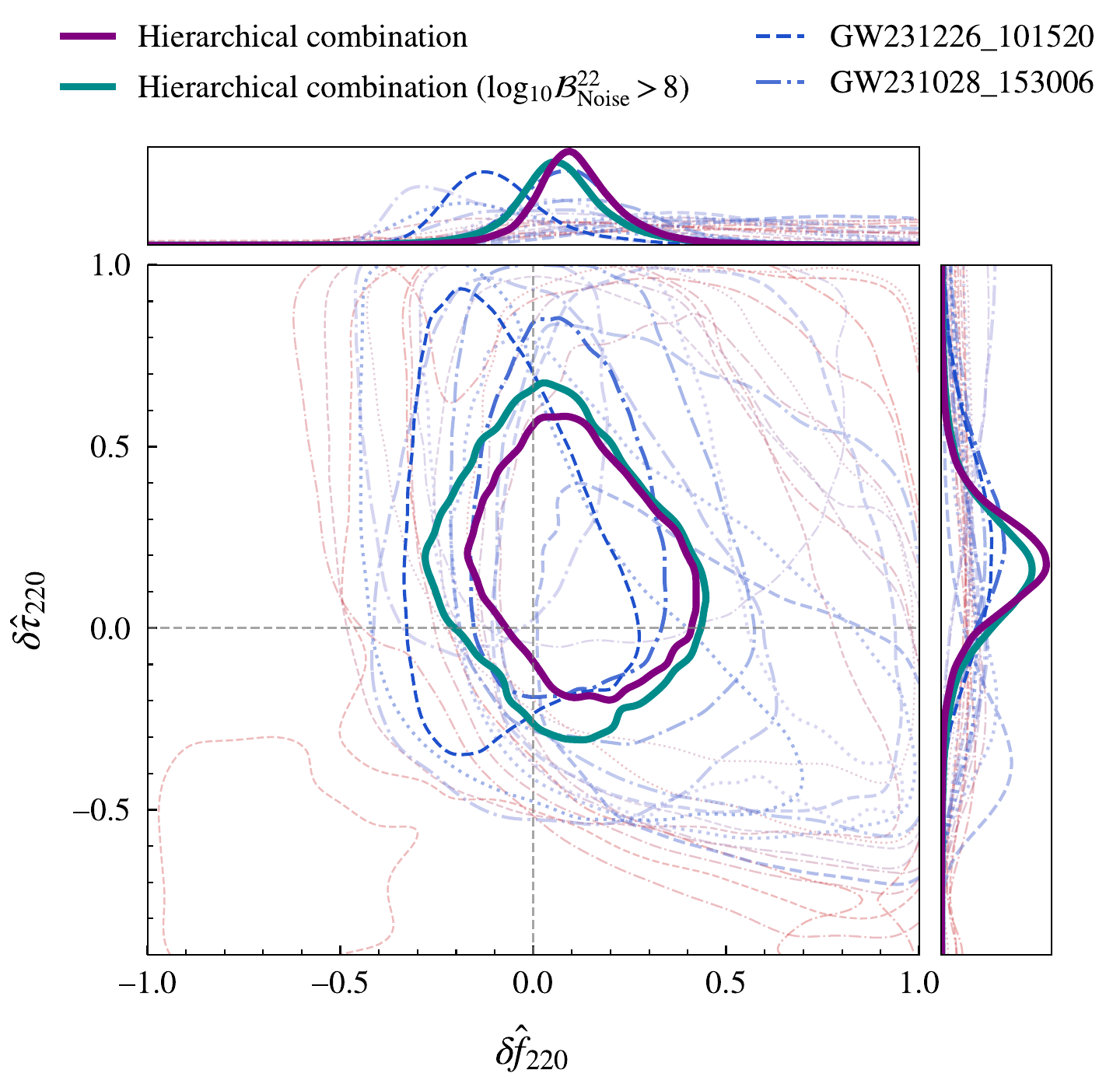}
     \caption{$90\%$ contours for the posterior probability distribution of frequency deviation $\delta \hat{f}_{220}$ and damping time $\delta \hat{\tau}_{220}$ for the analysis with a \texttt{KerrPostmerger} template including all HMs and fractional deviations in the $(\ell,m,n)=(2,2,0)$ mode (light contours, with opacity and color determined by $\mathrm{log}_{10} \mathcal{B}_\mathrm{Noise}^{22}$), along with the hierarchically combined results (heavy contours), including with the $\mathrm{log}_{10} \mathcal{B}_\mathrm{Noise}^{22} > 8$ constraint. The hierarchical combination is applied to the O4a events listed in Table~\ref{tab:remnant_params}, which are also the events plotted; the contours of the two individual events with the largest $\mathrm{log}_{10} \mathcal{B}_\mathrm{Noise}^{22}$ value are explicitly marked in the legend.
     }
     \label{fig:combined}
\end{figure}

Incorporating the largest amount of information, \texttt{KerrPostmerger} is the most accurate and sensitive model in \soft{pyRing} \citep{Gennari:2023gmx}, ideal for searching for HMs 
and small GR deviations, at the cost of being less flexible and agnostic about unmodeled physics.
The model does not account for precession or eccentricity and does not include mode mixing. 
For the mode detection analysis, we use a template that includes only the $(2,2)$ mode for the null hypothesis $\mathrm{H}_0$ and
an HM template which incorporates the $(2,1)$, $(3,3)$, $(4,4)$, and $(5,5)$ modes.
To avoid missing a mode detection, we also search separately for the presence of either the $(2,1)$ mode or the $(3,3)$ mode in addition to the $(2,2)$ mode, since those are the two HMs expected to have the largest contribution for comparable-mass spin-aligned quasi-circular systems \citep{Kamaretsos:2011um, Kamaretsos:2012bs, London:2014cma, Bhagwat:2019bwv, Bhagwat:2019dtm, JimenezForteza:2020cve, London:2018gaq, Ota:2021ypb, Gennari:2023gmx}.
Bayes factors for the full set of HMs for a wide range of starting times are shown in Figure~\ref{fig:bayes_factors_summary}(c).
No statistically significant presence of multiple modes was found. 
Parameter-estimation results using the HM model are shown in Figure~\ref{fig:event_comparison} and reported in Table~\ref{tab:remnant_params}.
For all events, we find 90\% CL overlap with GR results, signaling consistency with the GR prediction.

\textit{\texttt{Parametrized tests} --} Finally, to explore potential deviations in the ringdown spectrum of the remnant BH, 
we allow for deviations in the frequency and damping time of the GR QNMs.
The parameter-estimation is carried out using the same set of parameters as in the GR template, with the addition of the fractional frequency deviation $\delta\hat{f}_{220}$ constrained to the range $[-1,1]$ and the fractional damping time deviation $\delta\hat{\tau}_{220}$ constrained to $[-0.9,1]$ \citep[][]{LIGOScientific:2021sio}.
We use uniform priors for the deviation parameters.
If GR accurately describes the ringdown emission, the posterior distributions of the deviation parameters should encompass zero, with Bayesian evidence excluding the non-GR hypothesis. 

We search for such deviations in the mode $(\ell,m) = (2,2)$ with the \texttt{KerrPostmerger} model, using the most accurate GR baseline that includes all HMs.
All modes are included to avoid false deviations induced by the presence of additional unmodeled modes in the data.
Although the model is only valid for spin-aligned binaries, we have analyzed precessing, comparable-mass signals simulated within GR and found that at current sensitivity deviation parameters recover values consistent with GR.
Having excitation amplitudes informed by NR allows us to constrain GR deviations even if only one detectable mode is present in the signal \citep{LIGOScientific:2020tif, Ghosh:2021mrv, Gennari:2023gmx, LIGOScientific:2021sio}.
This allows us to constrain parameterized deviations even when no significant evidence for HMs is found, as in the current dataset.

We hierarchically combine the deviation measurements from all O4a events analyzed, listed in Table~\ref{tab:remnant_params}. While the pre-O4 events were analyzed with \texttt{KerrPostmerger} in \citet{Gennari:2023gmx}, that analysis used an approximation for the peak time that has since been found to lead to nonnegligible shifts in the results in some cases. Thus, a reanalysis of the pre-O4 events will be added later. 
We combine the $(\delta\hat{f}_{220}, \delta \hat{\tau}_{220})$ 2D posteriors from the \texttt{KerrPostmerger} template with all HMs starting at the peak, and adopt a multivariate Gaussian distribution for the population of deviations \citep{Zhong:2024pwb}.
We obtain a hierarchical constraint on the deviation in frequency and damping time equal to
\begin{equation}
    \delta \hat{f}_{220}^\mathrm{hier} = {\pyRingHierarchicalDeltaFMedian}_{-\pyRingHierarchicalDeltaFMinus}^{+\pyRingHierarchicalDeltaFPlus};
    \quad\quad
    \delta \hat{\tau}_{220}^\mathrm{hier} = {\pyRingHierarchicalDeltaTauMedian}_{-\pyRingHierarchicalDeltaTauMinus}^{+\pyRingHierarchicalDeltaTauPlus},
\end{equation}
and show its posterior probability distribution in Figure~\ref{fig:combined}. 
The corresponding hyperparameters inferred, i.e., the means and standard deviations of the multivariate Gaussian used to model the population of deviations, as well as the correlation $\rho$, are
\begin{equation}
        \begin{aligned}
     \mu_{\delta \hat{f}_{220}} = {\pyRingHierarchicalMuDeltaFMedian}_{-\pyRingHierarchicalMuDeltaFMinus}^{+\pyRingHierarchicalMuDeltaFPlus} ;&\quad\quad \sigma_{\delta \hat{f}_{220}}<\pyRingHierarchicalSigmaDeltaFBound \\
     \mu_{\delta \hat{\tau}_{220}} = {\pyRingHierarchicalMuDeltaTauMedian}_{-\pyRingHierarchicalMuDeltaTauMinus}^{+\pyRingHierarchicalMuDeltaTauPlus} ;&\quad\quad \sigma_{\delta \hat{\tau}_{220}}<\pyRingHierarchicalSigmaDeltaTauBound \\
                \rho_{\delta \hat{f}_{220} \delta \hat{\tau}_{220}} &= {\pyRingHierarchicalRhoMedian}_{-\pyRingHierarchicalRhoMinus}^{+\pyRingHierarchicalRhoPlus}.
        \end{aligned}
\end{equation}
The distributions of $\delta \hat{f}_{220}$ and $\delta \hat{\tau}_{220}$ both tend towards a positive deviation, but are consistent with GR within their $90\%$ credible interval. In terms of the hyperparameters, GR is located slightly outside the $90\%$ credible interval of the inferred one-dimensional deviation distribution for the mean of $\hat{f}_{220}$, but lies within the $95\%$ credible interval.

We observe a systematic tendency for events with relatively low SNR to yield more samples with positive deviations in at least one of the deviation parameters, even though their posteriors are mostly uninformative. 
We suspect that this effect arises due to an unbroken degeneracy between the total mass of the system and the deviation parameters at low SNR \citep{Ghosh:2021mrv}.
To assess the influence of the less informative events on the combined analysis, we repeat the hierarchical inference using only the $11$ events with $\log_{10} \mathcal{B}_\mathrm{Noise}^{22} > 8$.
As shown in Figure~\ref{fig:combined}, this selection results in a broadening of the posterior, but also reduces the positive bias in the distribution's median,
suggesting that weakly informative events may affect the analysis. Further work is needed to better characterize this behavior, including investigating noise modeling and selection effects in the hierarchical inference. Neglecting precession and eccentricity can also introduce systematic biases \citep{Gupta:2024gun}, leading to apparent deviations from GR, although such effects are expected to impact higher-SNR events more strongly, contrary to what we observe.

We quantify consistency with GR using the GR quantile defined in Section~\ref{sec:ringdown}, obtaining
\begin{equation}
    Q_\mathrm{GR}^\mathrm{4D} = {\pyRingHierarchicalQuantileFourDValue}_{-\pyRingHierarchicalQuantileFourDMinus}^{+\pyRingHierarchicalQuantileFourDPlus}\%;
    \quad\quad
    Q_\mathrm{GR}^{\mathrm{4D},\, \mathcal{B}_\mathrm{Noise}^{22}>10^8} = {\pyRingHierarchicalQuantileFourDValuelnBcut}_{-\pyRingHierarchicalQuantileFourDMinuslnBcut}^{+\pyRingHierarchicalQuantileFourDPluslnBcut}\%.
\end{equation}
The uncertainty of $Q_\mathrm{GR}^\mathrm{4D}$ quantifies the variance due to a finite catalog size and is estimated via bootstrapping, using 1000 synthetic catalogs created by resampling the original event set with replacement~\citep{Pacilio:2023uef}. We see that changing the selection criterion for the consistency test increases the agreement with GR significantly, at the cost of a higher variance of the quantile, a direct consequence of the reduced set of events available for the test. In fact, while the four-dimensional GR quantile is larger than the one-dimensional GR quantiles for the individual hyperparameters for the full set of O4a events, this is no longer the case with the $\mathrm{log}_{10} \mathcal{B}_\mathrm{Noise}^{22} > 8$ constraint, where the largest one-dimensional GR quantile is ${\pyRingHierarchicalQuantileOneDMuDeltaTauValuelnBcut}_{-\pyRingHierarchicalQuantileOneDMuDeltaTauMinuslnBcut}^{+\pyRingHierarchicalQuantileOneDMuDeltaTauPluslnBcut}\%$ for $\mu_{\delta \hat{\tau}_{220}}$, though this is still less than than the corresponding value of ${\pyRingHierarchicalQuantileOneDMuDeltaTauValue}_{-\pyRingHierarchicalQuantileOneDMuDeltaTauMinus}^{+\pyRingHierarchicalQuantileOneDMuDeltaTauPlus}\%$ with the full set of O4a events.
We conclude overall agreement with GR, although when combining all events, GR is only found at the very boundary of the $90\%$ credible region of the posteriors when including the bootstrapping estimate of catalog variance.

\subpapersubsection{The pSEOBNR analysis}\label{sec:pseob}

\begin{table*}[t]
        \caption{\label{tab:pSEOBNR_table}
        Results from the pSEOBNR analysis
        }
	\begin{center}
\begin{tabular}{lllllll}
    \toprule
    Event & $\delta \hat{f}_{220}$ & $\delta \hat{\tau}_{220}$ & $f_{220}$ [Hz] & $\tau_{220}$ [ms] & $(1+z)M_{\mathrm{f}}/M_{\odot}$ & $\chi_{\mathrm{f}}$ \\[0.075cm]
    \midrule
    GW150914 \!\!\! & \!\!\! $\pseobDeltaFMedian{GW150914}^{+\pseobDeltaFPlus{GW150914}}_{-\pseobDeltaFMinus{GW150914}}$ \!\!\! & \!\!\! $\pseobDeltaTauMedian{GW150914}^{+\pseobDeltaTauPlus{GW150914}}_{-\pseobDeltaTauMinus{GW150914}}$ \!\!\! & \!\!\! $\pseobFMedian{GW150914}^{+\pseobFPlus{GW150914}}_{-\pseobFMinus{GW150914}}$  {\tiny ($\IMRFMedian{GW150914}^{+\IMRFPlus{GW150914}}_{-\IMRFMinus{GW150914}}$)} \!\!\! & \!\!\! $\pseobTauMedian{GW150914}^{+\pseobTauPlus{GW150914}}_{-\pseobTauMinus{GW150914}}$  {\tiny ($\IMRTauMedian{GW150914}^{+\IMRTauPlus{GW150914}}_{-\IMRTauMinus{GW150914}}$)} \!\!\! & \!\!\! $\pseobRemnantMassMedian{GW150914}^{+\pseobRemnantMassPlus{GW150914}}_{-\pseobRemnantMassMinus{GW150914}}$  {\tiny ($\IMRRemnantMassMedian{GW150914}^{+\IMRRemnantMassPlus{GW150914}}_{-\IMRRemnantMassMinus{GW150914}}$)} \!\!\! & \!\!\! $\pseobRemnantSpinMedian{GW150914}^{+\pseobRemnantSpinPlus{GW150914}}_{-\pseobRemnantSpinMinus{GW150914}}$  {\tiny ($\IMRRemnantSpinMedian{GW150914}^{+\IMRRemnantSpinPlus{GW150914}}_{-\IMRRemnantSpinMinus{GW150914}}$)}\\
\bottomrule
\end{tabular}

        \end{center}
	\tablecomments{Median values and symmetric 90\% credible intervals for the one-dimensional marginalised posteriors of the fractional deviations in the frequency and damping time of the $(2, 2,0)$ QNM,
					$(\delta \hat{f}_{220},\delta \hat{\tau}_{220})$;
				the reconstructed frequency and damping time of the $(2,2,0)$ QNM;
				and the mass and spin of the remnant object, from different events that are analyzed using the \pSEOBNRFIVEPHM method.
			 For the last four quantities, estimates from IMR analyses assuming GR \citep{GWTC3, GWTC2p1, GWTC:Results} are shown in a small font in parentheses for comparison. Events marked with an asterisk are excluded from the combined results due to indications of noise-related systematics.}
\end{table*}

\begin{figure*}
	\centering
	\begin{minipage}[t]{0.63\textwidth}
		\centering
		\includegraphics[height=10.0cm, keepaspectratio]{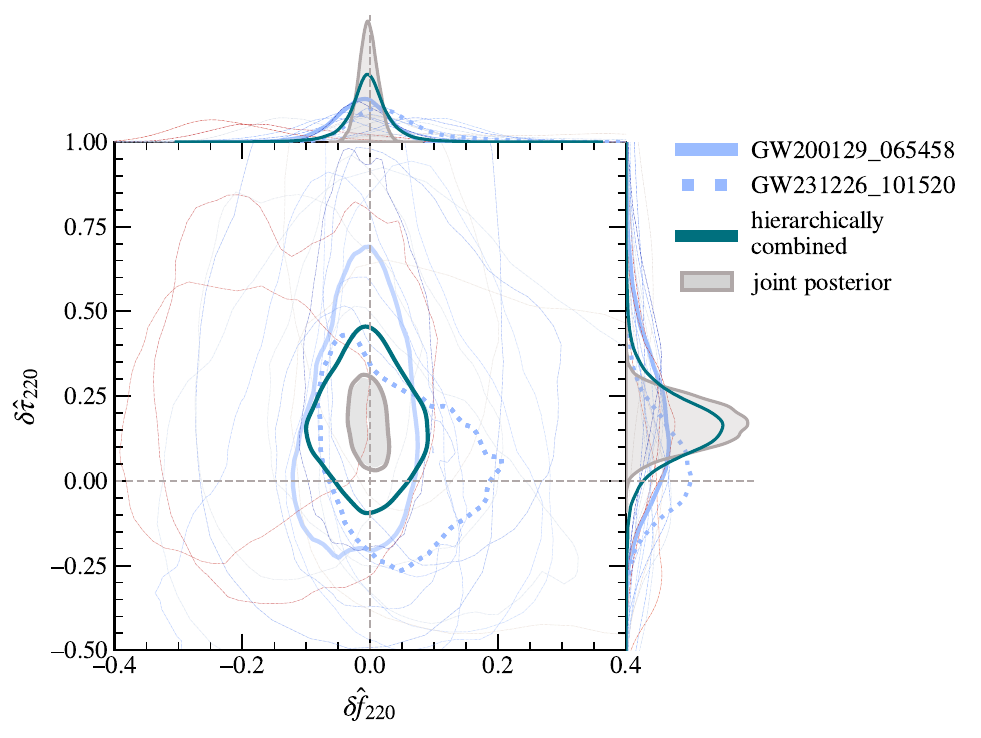}
	\end{minipage}
	\hfill
	\begin{minipage}[t]{0.27\textwidth}
		\centering
		\includegraphics[height=8.0cm, keepaspectratio]{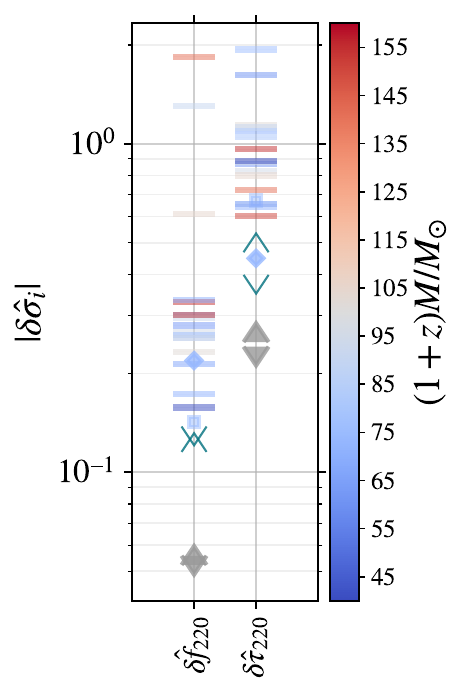}
	\end{minipage}
	\caption{%
        \emph{Left panel}: The 90\% credible regions of the posterior probability distribution of the fractional deviations in the frequency and damping time of the $(2,2,0)$ QNM,
				$(\delta \hat{f}_{220},\delta \hat{\tau}_{220})$,
				and their corresponding one-dimensional marginalized posterior distributions,
				for events passing a SNR threshold of $8$ in both the inspiral and post-inspiral signal.
				We highlight the posteriors for GW200129\_065458 and GW231226\_101520.
				The joint constraints on $(\delta \hat{f}_{220},\delta \hat{\tau}_{220})$
				obtained multiplying the posteriors (given a flat prior) from individual events are given by the filled grey contours,
				while the hierarchical method of combination yields the teal contours.
				The apparent deviation from GR in the joint posterior is not as significant as it appears,
				since including the uncertainty inferred by bootstrapping yields a GR quantile of $\QuantileJointGWTCFourUncertainty\%$.
		\emph{Right panel}: Width of the 90\% credible interval for the one-dimensional marginalised posteriors of
				$(\delta \hat{f}_{220},\delta \hat{\tau}_{220})$, colored by the median redshifted total mass $(1+z)M/M_\odot$,
				inferred assuming GR. Filled gray (unfilled teal) downward V-shaped markers indicate the constraints obtained when all the events are combined by multiplying posteriors (hierarchically). For comparison, we mark bounds from GWTC-3.0 results, using the \pSEOBNRFIVEPHM model~\citep{Pompili:2025cdc}, with filled/unfilled upward V-shaped markers. The bounds from GW200129\_065458 (square) and GW231226\_101520 (diamond) are indicated with separate markers.}
	\label{fig:pSEOBNR_220_combined}
\end{figure*}

The \pSEOBNRFIVEPHM analysis~\citep{Brito:2018rfr, Ghosh:2021mrv, Maggio:2022hre, Toubiana:2023cwr, Pompili:2025cdc} introduces fractional deviations $(\delta \hat{f}_{\ell m 0}$, $\delta \hat{\tau}_{\ell m 0})$ to the frequency and decay time of the fundamental QNMs in the ringdown description of the underlying \SEOBNRFIVEPHM waveform model~\citep{Khalil:2023kep,Pompili:2023tna,Ramos-Buades:2023ehm,vandeMeent:2023ols} as:
\begin{eqnarray}
	f_{\ell m 0} &=& f_{\ell m 0}^{\text{GR}}\, (1 + \delta \hat{f}_{\ell m 0})\,,\label{eq:nongr_freqs_a} \\
	\tau _{\ell m 0} &=& \tau _{\ell m 0}^{\text{GR}}\, (1 + \delta \hat{\tau}_{\ell m 0})\,. \label{eq:nongr_freqs_b}
\end{eqnarray}
The final mass and spin of the remnant BH, computed from the masses and spins of the components using NR fits~\citep{Hofmann:2016yih,Jimenez-Forteza:2016oae}, are used to predict the GR values of the frequency and damping time of the $(\ell, m, 0)$ QNM~\citep{Berti:2009kk}.

The \SEOBNRFIVEPHM waveform model describes BBHs with spin precession and includes the subdominant modes $(\ell, |m|)=(2,1)$, $(3,3)$, $(3,2)$, $(4,4)$, and $(4,3)$, in addition to the dominant $(2,2)$ mode, in the coprecessing frame. While the $(5,5)$ mode is also modeled, it is not included by default for computational reasons and is not used in the analyses presented here, as its contribution is typically negligible.
We denote the spherical-harmonic modes in the coprecessing frame as $\tilde{h}_{\ell m}$.
Negative-$m$ modes are derived from the positive-$m$ ones using the reflection symmetry stated in Section~\ref{sec:ringdown}.
While this is exact for aligned-spin binaries, it is not so for precessing-spin binaries~\citep{Boyle:2014ioa}, even in the coprecessing frame.
However, mode asymmetries in the co-precessing frame are a subdominant effect and are not currently included in \SEOBNRFIVEPHM.
In the following, we restrict the discussion to $(\ell, m)$ modes with $m > 0$.
The \SEOBNRFIVEPHM waveform is constructed by attaching the merger--ringdown waveform, $\tilde{h}_{\ell m}^{\text{merger-RD}}(t)$, to the inspiral--plunge waveform, $\tilde{h}_{\ell m}^{\text{insp--plunge}}(t)$, in the coprecessing frame at a matching time, $t_{\text{match}}$, corresponding to the peak amplitude of the $(2,2)$ harmonic.
The merger--ringdown for each mode can be written as
\begin{equation}
	\tilde{h}_{\ell m}^{\text{merger--RD}} \!=\! \eta \tilde{A}_{\ell m}(t) \exp{\left[i \tilde{\phi}_{\ell m}(t)\right]} \exp{\left[-i \tilde \sigma_{\ell m 0} \left(t\!-\!t_{\text{match}}\right)\right]} \,,
\end{equation}
where $\eta$ is the symmetric mass ratio of the binary and $\tilde \sigma_{\ell m 0}$ are the complex frequencies of the least-damped QNM of the remnant BH, in the coprecessing frame. The functions $\tilde{A}_{\ell m}(t)$ and $\tilde{\phi}_{\ell m}(t)$ are constrained by the requirement that the amplitude and phase of $\tilde{h}_{\ell m}(t)$ are continuously differentiable at the matching time and are calibrated to a large set of NR simulations of BBHs with aligned spins~\citep{Pompili:2023tna}.

In BH perturbation theory, the mode decomposition used to define QNMs assumes a frame where the $z$-axis is aligned with the final angular momentum of the system ($\boldsymbol{J}_{\mathrm{f}}$-frame). The QNM frequencies in this frame, $\sigma_{\ell m 0}$, can be mapped to effective QNM frequencies in the coprecessing frame, $\tilde \sigma_{\ell m 0}$~\citep{Hamilton:2023znn}. The complex frequencies are related to the QNM oscillation frequency and damping time as:
\begin{equation} 
	f_{\ell m 0} = \frac{1}{2 \pi} \operatorname{Re}\left(\sigma_{\ell m 0}\right); \quad
	\tau_{\ell m 0} = -\frac{1}{\operatorname{Im}\left(\sigma_{\ell m 0}\right)}.
\end{equation}

Because the pSEOBNR analysis directly modifies parameters in an IMR waveform model, it takes full advantage of the signal length and its SNR, and avoids ambiguities associated with selecting the start time of ringdown. 
This analysis requires $\text{SNR} \geq 8$ in both the inspiral and post-inspiral parts of the signal, since a reasonable inspiral SNR is needed to constrain the remnant properties expected in the GR prediction of the QNMs, and break the degeneracy between the fundamental ringdown frequency deviation parameter and the remnant mass~\citep{Ghosh:2021mrv}.
More specifically, for events from O4a that have been analyzed with the \SEOBNRFIVEPHM model in \citet{GWTC:Results}, we use the median SNR from the GR analyses performed with this waveform. For events up to O3b, where results with \SEOBNRFIVEPHM are not available, we follow~\cite{LIGOScientific:2021sio} and use the maximum a posteriori estimate from the IMR analyses performed with \IMRPhenomXPHMMSA~\citep{Pratten:2020ceb}.

The current version of the analysis focuses exclusively on constraining the fractional deviations of the dominant (least-damped) QNM by sampling over $\delta \hat{f}_{220}$ and $\delta \hat{\tau}_{220}$ in addition to the GR parameters of the waveform model. The prior range for the fractional deviations is uniform in the ranges $\delta \hat{f}_{220} \in [-0.8,2]$ and $\delta \hat{\tau}_{220} \in [-0.8,2]$. The events GW170104, GW191109\_010717, and GW200208\_130117 exhibit railing (i.e., significant posterior probability density right up to at least one prior boundary), so we extend the prior range to $[-0.8,4]$.

Table~\ref{tab:pSEOBNR_table} presents the median values and symmetric 90\% credible intervals on the one-dimensional marginalised posteriors of the fractional deviations in the frequency and damping time of the $(2,2,0)$ QNM, $(\delta \hat{f}_{220},\delta \hat{\tau}_{220})$. Additionally, the table reports the reconstructed frequency and damping time of the $(2,2,0)$ QNM, derived using Equations~\eqref{eq:nongr_freqs_a} and~\eqref{eq:nongr_freqs_b}, and the mass and spin of the remnant BH, estimated from the complex QNM frequencies by inverting the fitting formula from~\cite{Berti:2005ys}. For all events analyzed, the two-dimensional posteriors for the reconstructed frequency and damping time of the $(2,2,0)$ QNM, as well as the inferred remnant mass and spin, are consistent at the 90\% CL with the estimates from IMR analyses \citep{GWTC3, GWTC2p1, GWTC:Results}.
For GW190910\_112807, however, the one-dimensional posterior for the $(2,2,0)$ damping time is shifted toward larger values, with the respective 90\% credible intervals from the \pSEOBNRFIVEPHM and IMR analyses being marginally incompatible.
For GW191109\_010717, the posterior for the frequency deviation $\delta \hat{f}_{220}$ is also shifted away from zero, and the corresponding reconstructed remnant mass and spin show tension with the IMR estimates at the 90\% CL. This behavior is attributed to possible noise-related systematics, as indicated by follow-up investigations using synthetic signals in neighboring data segments carried out in~\cite{LIGOScientific:2021sio}. Consistent conclusions were obtained in subsequent analyses using the \pSEOBNRFIVEPHM model~\citep{Pompili:2025cdc}. For this reason, GW191109\_010717 (as well as GW200208\_130117) is excluded from the combined results, consistent with the treatment adopted in GWTC-3.0~\citep{LIGOScientific:2021sio}.

The results of the analysis are summarised in Figure~\ref{fig:pSEOBNR_220_combined}. The left panel of the figure shows the two-dimensional posteriors (along with the corresponding marginalized one-dimensional posteriors) for the fractional deviations in the frequency and damping time of the $(2,2,0)$ QNM, for all events listed in Table~\ref{tab:pSEOBNR_table}. The contours are colored according to the median detector-frame total mass $(1 + z)M$ of the corresponding binary. We specifically highlight the posteriors from two events, GW200129\_065458 and GW231226\_101520~\citep{GWTC3, GWTC:Results}, which are among the loudest observed so far and provide the strongest single-event bounds on the frequency and damping time deviations, respectively. Combined constraints are shown both by the joint posterior, obtained by multiplying individual posteriors and depicted as the filled grey contours, and by hierarchically combining events, represented by the teal contours.
For the hierarchical analysis, we model the population-level fractional deviations $\delta \hat{f}_{220}$ and $\delta \hat{\tau}_{220}$ as a two-dimensional Gaussian distribution, with means $(\mu_{\delta \hat{f}_{220}}, \mu_{\delta \hat{\tau}_{220}})$, standard deviations $(\sigma_{\delta \hat{f}_{220}}, \sigma_{\delta \hat{\tau}_{220}})$, and a correlation coefficient $\rho_{\delta \hat{f}_{220} \delta \hat{\tau}_{220}}$~\citep{Zhong:2024pwb}. 
The right panel of Figure~\ref{fig:pSEOBNR_220_combined} summarizes the 90\% credible intervals on the one-dimensional marginalized posteriors, color-coded by the median detector-frame mass of each binary.

The events up to O3 were previously analyzed in the corresponding testing GR paper~\citep{LIGOScientific:2021sio}, using an earlier version of the pSEOBNR model for aligned-spin binaries (\soft{pSEOBNRv4HM}; \citealt{Ghosh:2021mrv}). These events have now been reanalyzed with the updated \pSEOBNRFIVEPHM model~\citep{Pompili:2025cdc}, which includes spin-precession effects, giving broadly consistent results. The most noticeable differences arise for the events GW200129\_065458 and GW200311\_115853, due to correlations between QNM deviations, distance, inclination, and spin precession~\citep{Pompili:2025cdc}.
From the event GW190910\_112807, we infer a value of $\delta \hat \tau_{220}$ shifted toward positive values, with the GR prediction lying at the edge of the 90\% credible region. The results for this event are consistent with those reported in GWTC-3.0~\citep{LIGOScientific:2021sio}. We have verified that the inclusion of this event does not significantly affect the combined results, and we keep it in the analysis as in previous works.

Among the O4a events, GW231226\_101520 is the loudest currently analyzed, with a median SNR of $\networkmatchedfiltersnrmed{GW231226_101520}$, and yields the tightest single-event constraints on the $(2,2,0)$ damping in GWTC-4.0.
The posterior shows a slight tail toward positive values of $\delta \hat f_{220}$, correlated with support for unequal mass ratios, which is not present in the corresponding GR run. The maximum-likelihood parameters lie in this region, despite the bulk of the posterior being centered around $\delta \hat f_{220} = 0$ and equal masses. The posterior structure remains stable under different sampler settings, supporting the robustness of this feature.

The event GW231028\_153006 places the GR prediction at the edge of the 90\% credible region. This is a loud event with median SNR of $\networkmatchedfiltersnrmed{GW231028_153006}$, occurring in a region of parameter space (with support for unequal masses, high total mass, and strong spin precession) where waveform systematics are expected to be significant~\citep{Dhani:2024jja}. This was further investigated using  synthetic signals simulated in zero noise. 
An analysis of a signal simulated using \SURSEVENDQFOUR, with maximum-likelihood parameters from a corresponding GR run \citep{GWTC:Results}, shows a deviation in $\delta \hat f_{220}$ similar to that observed for the real event, with the GR prediction at the edge of the 90\% credible region. However, waveform systematics alone do not fully explain the observed behavior. A zero-noise synthetic signal using \SEOBNRFIVEPHM and its maximum-likelihood parameters also shows a qualitatively similar shift, although with reduced magnitude, such that GR is now found within the 90\% credible region. These deviations appear to correlate with the mass ratio and effective inspiral spin of the binary. Simulated signals with unequal masses and positive effective inspiral spin tend to be recovered with more comparable masses and lower spin magnitudes. The recovered maximum-likelihood parameters lie at the tail of the posterior, close to the simulated parameters and showing no GR deviations, suggesting a potential influence of non-uniform priors, particularly on the spins. This behavior mirrors that observed in the actual event, where the maximum-likelihood parameters, characterized by $\delta \hat f_{220} \simeq 0$, unequal masses, and non-zero spins, lie at the tail of the posterior distribution. Similar effects of priors for this signal are seen for other analyses in Appendices~A and~B of Paper~II.

The event GW231102\_071736 shows a bimodal posterior in total mass and $\delta \hat f_{220}$, with one mode near the GR prediction and another at positive $\delta \hat f_{220}$ and higher total mass, strongly correlated with each other. The persistence of this degeneracy suggests an insufficient inspiral SNR~\citep{Ghosh:2021mrv}, which is indeed the lowest (9.4) among O4a events.

When considering the joint posterior results, a shift toward positive values in the damping time deviation, previously noted in the GWTC-3.0 analysis~\citep{LIGOScientific:2021sio, Pompili:2025cdc}, is now recovered with increased probability, with the GR prediction $(0,0)$ lying slightly outside the 90\% credible region. The distribution for the frequency deviation is more sharply peaked around zero compared to GWTC-3.0, while the damping time deviation shows a more pronounced shift towards positive values. A similar trend is observed in the hierarchical distribution for $\delta \hat \tau_{220}$, which also peaks at positive values but features larger statistical uncertainties compared to the corresponding joint posterior.

The joint bounds read
\begin{equation}
	\label{eq:pseob-joint-gwtc4}
	\delta \hat{f}_{220}^\mathrm{joint}=\pseobJointDeltaFMedian_{-\pseobJointDeltaFMinus}^{+\pseobJointDeltaFPlus} ; \quad
	\delta \hat{\tau}_{220}^\mathrm{joint}=\pseobJointDeltaTauMedian_{-\pseobJointDeltaTauMinus}^{+\pseobJointDeltaTauPlus}
\end{equation}
by multiplying the posteriors and
\begin{equation}
	\begin{aligned}
		\delta \hat{f}_{220}^\mathrm{hier}=\pseobHierarchicalDeltaFMedian_{-\pseobHierarchicalDeltaFMinus}^{+\pseobHierarchicalDeltaFPlus} \quad &
     \left[\mu_{\delta \hat{f}_{220}}=\pseobHierarchicalMuDeltaFMedian_{-\pseobHierarchicalMuDeltaFMinus}^{+\pseobHierarchicalMuDeltaFPlus} ; \sigma_{\delta \hat{f}_{220}}<\pseobHierarchicalSigmaDeltaFBound \right] ; \\
		\delta \hat{\tau}_{220}^\mathrm{hier}=\pseobHierarchicalDeltaTauMedian_{-\pseobHierarchicalDeltaTauMinus}^{+\pseobHierarchicalDeltaTauPlus} \quad &
     \left[\mu_{\delta \hat{\tau}_{220}} = \pseobHierarchicalMuDeltaTauMedian_{-\pseobHierarchicalMuDeltaTauMinus}^{+\pseobHierarchicalMuDeltaTauPlus} ; \sigma_{\delta \hat{\tau}_{220}}<\pseobHierarchicalSigmaDeltaTauBound \right] ; \\
		\rho_{\delta \hat{f}_{220} \delta \hat{\tau}_{220}} & = \pseobHierarchicalRhoMedian_{-\pseobHierarchicalRhoMinus}^{+\pseobHierarchicalRhoPlus}
	\end{aligned}
\end{equation}
by combining hierarchically. The values in square brackets correspond to the inferred hyperparameters.

\begin{figure}[t]
	\includegraphics[width=\TGRFigureWidth]{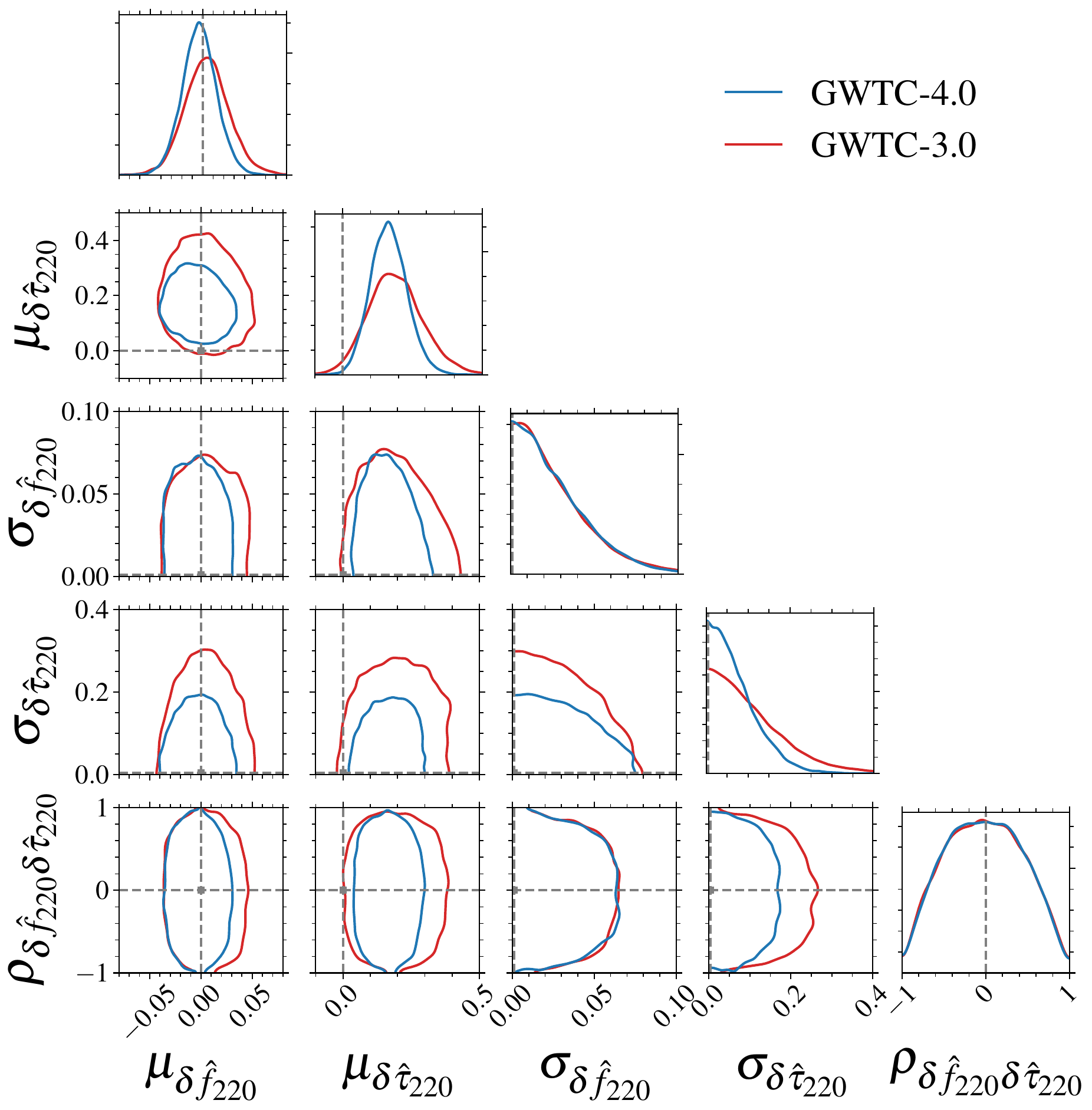}
	\caption{
        Posterior distribution for the hyperparameters of the fractional deviations in the $(2,2,0)$ QNM frequency, $\delta \hat{f}_{220}$, and damping time, $\delta \hat{\tau}_{220}$, obtained from the pSEOBNR analysis. Blue lines show the GWTC-4.0 constraints, while red lines correspond to GWTC-3.0 results \citep[based on][]{Pompili:2025cdc}, both using the \pSEOBNRFIVEPHM model. Contours mark 90\% credible regions. The GR prediction corresponds to $\mu_{\delta \hat{f}_{220}}=\mu_{\delta \hat{\tau}_{220}}=\sigma_{\delta \hat{f}_{220}}=\sigma_{\delta \hat{\tau}_{220}}= 0$, irrespective of $\rho_{\delta \hat{f}_{220} \delta \hat{\tau}_{220}}$. The apparent deviation from GR in the GWTC-4.0 $\mu_{\delta \hat{\tau}_{220}}$ results is not as significant as it appears, since including the uncertainty inferred by bootstrapping yields a GR quantile of $\QuantileHierGWTCFourUncertaintyTauOneD\%$, and including the \ac{O4b} event GW250114~\citep{GW250114, GW250114_TGR} reduces the GR quantile to $\QuantileHierGWTCFourTauOneDPlusTwoFiveZeroOneOneFour$.}
	\label{fig:pSEOB_hyper}
\end{figure}

Figure~\ref{fig:pSEOB_hyper} shows the posterior distribution for the hyperparameters, with contours indicating 90\% credible regions. The point $\mu_{\delta \hat{f}_{220}}=\mu_{\delta \hat{\tau}_{220}}=\sigma_{\delta \hat{f}_{220}}=\sigma_{\delta \hat{\tau}_{220}}= 0$ corresponds to the GR prediction, irrespective of $\rho_{\delta \hat{f}_{220} \delta \hat{\tau}_{220}}$. The inclusion of additional events leads to tighter constraints on the hyperparameters. The distribution for the frequency deviation becomes more sharply peaked around $(\mu_{\delta \hat{f}_{220}}, \sigma_{\delta \hat{f}_{220}}) = (0, 0)$, indicating increased consistency with the GR prediction. In contrast, the damping time deviation shows reduced consistency with GR: while $\sigma_{\delta \hat{\tau}_{220}}$ remains consistent with zero, the mean shifts toward positive values, and the point $\mu_{\delta \hat{\tau}_{220}}=0$ is excluded from the 90\% credible region, consistent with Figure~\ref{fig:pSEOBNR_220_combined}.

We quantify consistency with GR using the GR quantile defined in Section~\ref{sec:ringdown}, obtaining the results summarized in Table~\ref{tab:gr_quantiles}.

\begin{table*}
    \caption{\label{tab:gr_quantiles}
    GR quantiles for combined pSEOBNR results}
    \begin{center}
    \renewcommand{\arraystretch}{1.2}
    \setlength{\tabcolsep}{6pt}
    \begin{tabular}{lccc}
        \hline
        Catalog & Joint \(Q_{\mathrm{GR}}^{\mathrm{2D}}\) & Hierarchical \(Q_{\mathrm{GR}}^{\mathrm{1D}, \mu_{\delta \hat{\tau}_{220}}}\) & Hierarchical \(Q_{\mathrm{GR}}^{\mathrm{4D}}\)\\
        \hline
        GWTC-3.0 & $\QuantileJointGWTCThreeUncertainty \%$ & $\QuantileHierGWTCThreeUncertaintyTauOneD \%$ & $\QuantileHierGWTCThreeUncertainty \%$ \\
		GWTC-4.0 & $\QuantileJointGWTCFourUncertainty \%$ & $\QuantileHierGWTCFourUncertaintyTauOneD \%$ & $\QuantileHierGWTCFourUncertainty \%$\\
        \hline
    \end{tabular}
    \end{center}
    \tablecomments{Comparison of GR quantiles for joint (\(Q_{\mathrm{GR}}^{\mathrm{2D}}\)) and hierarchical (\(Q_{\mathrm{GR}}^{\mathrm{1D}, \mu_{\delta \hat{\tau}_{220}}}\) and \(Q_{\mathrm{GR}}^{\mathrm{4D}}\)) results of the \pSEOBNRFIVEPHM analysis, where the GWTC-3.0 results are based on \citet{Pompili:2025cdc}. Smaller GR quantiles indicate better consistency with general relativity. The reported values are computed from the actual catalog, with uncertainties estimated via bootstrapping over the event set to quantify variance from the finite catalog size.}
\end{table*}

The analysis of GWTC-3.0 events using the \pSEOBNRFIVEPHM model results in $Q_{\mathrm{GR}}^{\mathrm{2D}}=\QuantileJointGWTCThree$ when multiplying the posteriors, while the hierarchical combination yields $Q_{\mathrm{GR}}^{\mathrm{1D}, \mu_{\delta \hat{\tau}_{220}}} = \QuantileHierGWTCThreeTauOneD$ when just considering the mean of the deviation in the damping time which exhibits the largest deviation and $Q_{\mathrm{GR}}^{\mathrm{4D}} = \QuantileHierGWTCThree$ when considering all four parameters (two means and two standard deviations).
Adding the O4a events results in a shift toward higher GR quantiles, indicating reduced consistency with GR. We find $Q_{\mathrm{GR}}^{\mathrm{2D}}=\QuantileJointGWTCFour$ when multiplying the posteriors, while the hierarchical combination yields $Q_{\mathrm{GR}}^{\mathrm{1D}, \mu_{\delta \hat{\tau}_{220}}} = \QuantileHierGWTCFourTauOneD$ and $Q_{\mathrm{GR}}^{\mathrm{4D}} = \QuantileHierGWTCFour$.
Excluding GW190910\_112807, the event with $\delta \hat{\tau}_{220}$ most shifted toward positive values, reduces the GWTC-4.0 GR quantiles to $Q_{\mathrm{GR}}^{\mathrm{2D}}=\QuantileJointGWTCFourMinusOneNineZeroNineOneZero$, $Q_{\mathrm{GR}}^{\mathrm{1D}, \mu_{\delta \hat{\tau}_{220}}} = \QuantileHierGWTCFourTauOneDMinusOneNineZeroNineOneZero$, and $Q_{\mathrm{GR}}^{\mathrm{4D}}=\QuantileHierGWTCFourMinusOneNineZeroNineOneZero$.

To quantify the variance due to the finite number of events in the catalog, we estimate uncertainties by bootstrapping over the event set~\citep{Pacilio:2023uef}. Specifically, we build 1000 synthetic catalogs by resampling events with replacement and recompute the combined quantiles for each realization. The central 90\% interval of the resulting bootstrap distribution is reported as the uncertainty on $Q_{\mathrm{GR}}^{\mathrm{2D}}$, $Q_{\mathrm{GR}}^{\mathrm{1D}, \mu_{\delta \hat{\tau}_{220}}}$, and $Q_{\mathrm{GR}}^{\mathrm{4D}}$ in Table~\ref{tab:gr_quantiles}.
In particular, the joint posterior quantile is $\QuantileJointGWTCFourUncertainty \%$, showing that the apparent deviation is sensitive to the limited number of events in the catalog, and that the degree of tension with GR may be less severe than the nominal value suggests.

\begin{figure*}
	\includegraphics[width=3.5507in]{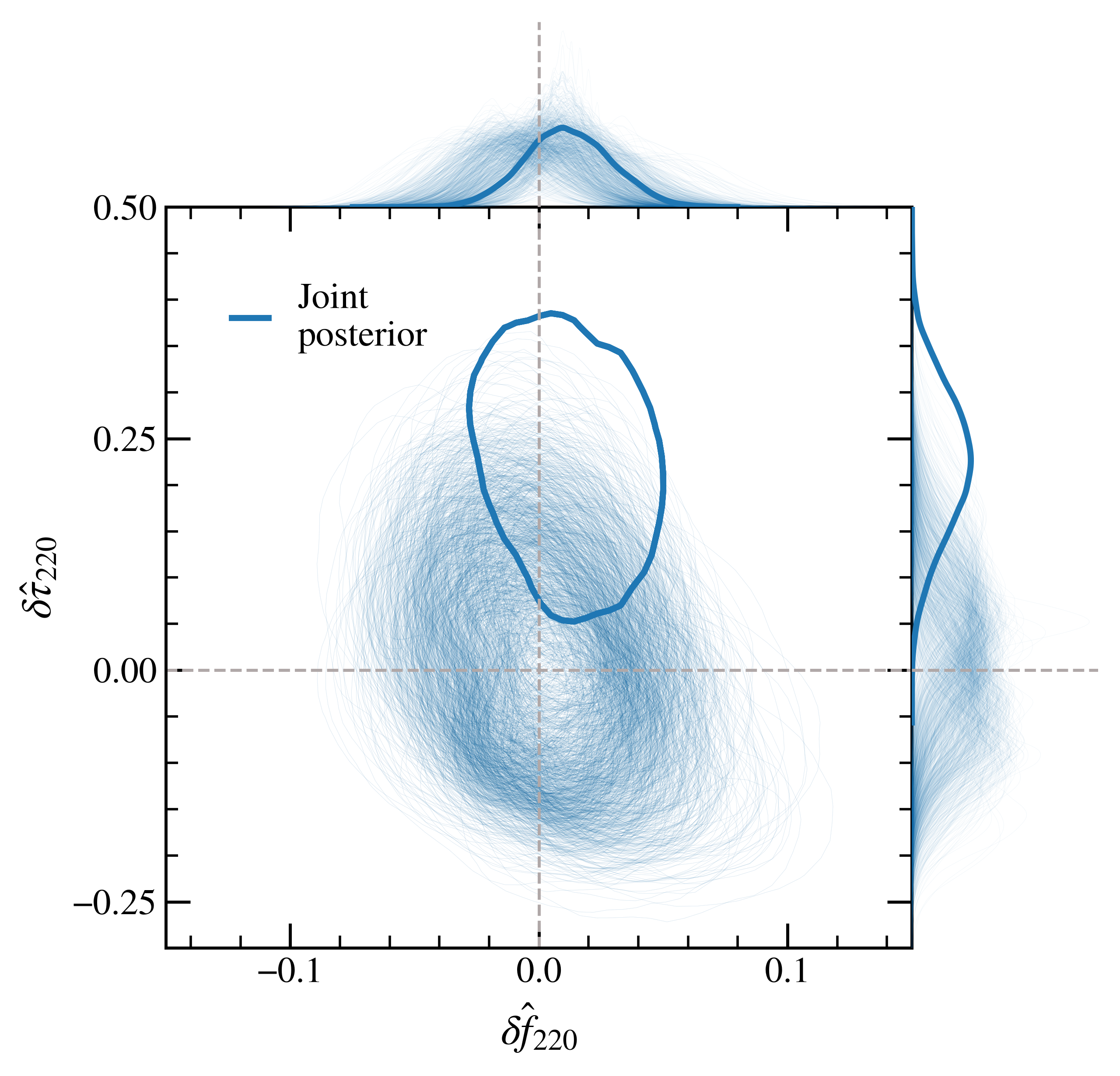}
	\includegraphics[width=3.5507in]{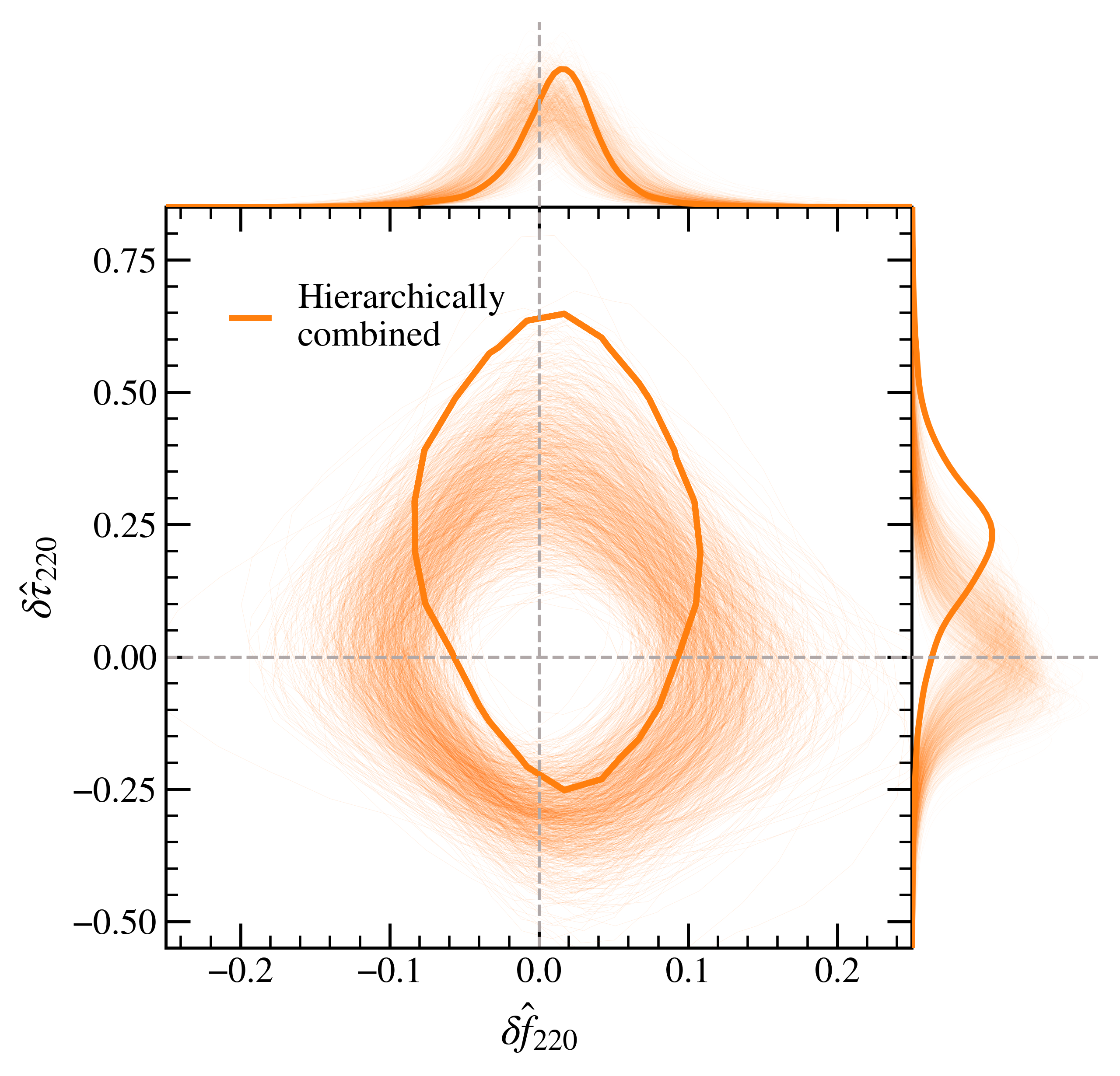}
	\caption{
		The 90\% credible regions of the posterior probability distribution of the fractional deviations in the frequency and damping time of the $(2,2,0)$ QNM, $(\delta \hat{f}_{220},\delta \hat{\tau}_{220})$, and their corresponding one-dimensional marginalized posterior distributions, for 1000 mock catalogs. 
		Each catalog, shown as thin lines, contains 17 signals simulated in Gaussian noise with the \SURSEVENDQFOUR model and recovered with \pSEOBNRFIVEPHM. The left panel shows the joint constraints obtained by multiplying individual posteriors, while the right panel shows hierarchically combined results. The thick lines highlight the catalog that yields the largest apparent deviation from GR in the joint analysis. This illustrates that apparent deviations from GR, qualitatively similar to those seen in real data, can appear in a small subset of simulated catalogs generated assuming GR.
        }
	\label{fig:pSEOB_mock_cat}
\end{figure*}

This discrepancy with GR could arise from a variety of factors, including non-Gaussian or non-stationary noise~\citep{LIGOScientific:2020tif, Ghosh:2021mrv}, parameter correlations~\citep{LIGOScientific:2021sio}, systematic errors from waveform modeling, intrinsic variance due to the limited number of events in the catalog~\citep{Pacilio:2023uef}, or selection effects. Incorporating additional events from future observing runs could help clarify this behavior.
To quantify these effects, we perform a large number of synthetic signal simulations assuming GR, using binaries drawn from a distribution consistent with the GWTC-3.0 population~\citep{GWTC3pop}.
Signals are simulated in both zero-noise and Gaussian noise using a three-detector configuration (LIGO Hanford, LIGO Livingston, and Virgo). We use the noise curves \texttt{aLIGO\_O4\_high} for the LIGO detectors and \texttt{AdV} for Virgo~\citep{Aasi:2013wya, psds_for_simulations}, as named in \BILBY \citep{Ashton:2018jfp,Romero-Shaw:2020owr}.
Signals are generated with both the \SURSEVENDQFOUR and \SEOBNRFIVEPHM models, and recovered with \pSEOBNRFIVEPHM. To assess the impact on combined results, we bootstrap 1000 mock catalogs~\citep{Pacilio:2023uef}, each containing 17 simulated signals, corresponding to the number of events included in the combined analysis of the real catalog.

Comparing \SEOBNRFIVEPHM and \SURSEVENDQFOUR signals in zero noise, we find that waveform modeling uncertainties have a small impact at current detector sensitivities, both for individual events and for the combined results. This is expected, as most events lie in regions of parameter space where different waveform models agree well, and the typical SNRs are moderate~\citep{Dhani:2024jja}. When simulating \SURSEVENDQFOUR signals in zero noise, we are unable to reproduce GR quantiles as large as those observed in the actual data in simulated catalogs of comparable size.

By comparing the previous results to \SURSEVENDQFOUR signals in Gaussian noise, we find that Gaussian noise can have a larger effect than waveform uncertainties. Given the limited number of events, it is possible for statistical fluctuations to shift multiple measurements in the same direction, creating a bias in the combined analysis.
Furthermore, while our noise model generally accounts for shifts induced by Gaussian noise, selection effects and differences between the assumed priors and the true source population can interact non-trivially with noise fluctuations and potentially lead to biases.
Nonetheless, we observe large GR quantiles less frequently than would be expected under a uniform distribution. 
Indeed, GR quantiles are expected to be uniformly distributed only when the simulated source population matches the prior assumed in the analysis, while in our case, we simulate only GR-consistent signals~\citep{Ghosh:2017gfp, Chua:2020oxn, Pacilio:2023uef}. 

Still, we find that GR quantiles as large as those observed in the actual data occur in a small fraction of the simulated catalogs, approximately \FractionOfCatalogsAboveJointQuantile{} for the joint posterior analysis and \FractionOfCatalogsAboveHierQuantileTauOneD{}, \FractionOfCatalogsAboveHierQuantile{} for the one-dimensional $\mu_{\delta \hat{\tau}_{220}}$, four-dimensional hierarchically combined results, respectively. These mock datasets can qualitatively reproduce features seen in the real analysis, such as frequency deviations consistent with GR and damping time shifted to positive values.
This is illustrated in Figure~\ref{fig:pSEOB_mock_cat}, which shows the 90\% credible regions for $(\delta \hat{f}_{220},\delta \hat{\tau}_{220})$ from the joint and hierarchically combined analyses of the 1000 mock catalogs simulated with \SURSEVENDQFOUR in Gaussian noise.
Therefore, we cannot exclude the possibility that the observed deviation is caused by statistical fluctuations due to Gaussian noise, potentially amplified by unaccounted selection effects and correlations among parameters. 

Real detector noise may produce even larger deviations than Gaussian noise, increasing the likelihood of observing spurious GR violations. 
For example, GR-consistent signals simulated in both Gaussian and real detector noise performed for \COMMONNAME{GW230814single}~\citep{GW230814},
the loud single-detector event not included in our main analysis, show that real noise leads to more frequent and larger-magnitude deviations than Gaussian noise, when recovered with \pSEOBNRFIVEPHM.
Among 10 signals simulated with \SEOBNRFIVEPHM in Gaussian noise, only one yielded a GR quantile above $95 \%$, whereas 6 out of 20 exceeded this threshold when simulated in real noise. The impact of real noise on combined results and in events observed by multiple detectors remains to be investigated.

Although we currently observe large GR quantiles less frequently than would be expected under a uniform distribution, this may not be true as catalog size increases, and statistical errors shrink, unless analysis assumptions remain valid. 
As previously noted, several events exhibit non-trivial correlations between the deviations $\delta \hat{f}_{220}$ and $\delta \hat{\tau}_{220}$ and the binary's mass and spin parameters, which are themselves influenced by prior choices that may not reflect the true source population~\citep{Payne:2023kwj}. These correlations should be accounted for by jointly modeling the GR deviations along with the relevant astrophysical parameters.
Additionally, selection effects are not accounted for in the current hierarchical inference. 
For example, the pSEOBNR selection criteria may favor events with positive damping time deviations, as these tend to have a larger ringdown SNR, potentially contributing to the observed deviation.
We therefore expect that properly accounting for selection effects would shift $\delta \hat{\tau}_{220}$ toward lower values and reduce the apparent level of inconsistency with GR.
These limitations should be addressed in future analyses to avoid potential biases.

At the same time, there remains significant statistical uncertainty due to the relatively small number of events currently available. The apparent deviation is sensitive to the size of the catalog, and adding more events will likely have a substantial impact. For example, GW250114~\citep{GW250114}, the exceptionally loud SNR $\sim 80$ event observed in \ac{O4b}, provides a single-event constraint tighter than the combined GWTC-4.0 results, while remaining in excellent agreement with GR~\citep{GW250114_TGR}. When combined with the GWTC-4.0 events, GW250114 shifts the results toward improved consistency with GR, with the joint-posterior GR quantile reduced to $Q_{\mathrm{GR}}^{\mathrm{2D}} = \QuantileJointGWTCFourPlusTwoFiveZeroOneOneFour$, and the GR quantiles for the hierarchically combined results reduced to $Q_{\mathrm{GR}}^{\mathrm{1D}, \mu_{\delta \hat{\tau}_{220}}} = \QuantileHierGWTCFourTauOneDPlusTwoFiveZeroOneOneFour$, $Q_{\mathrm{GR}}^{\mathrm{4D}} = \QuantileHierGWTCFourPlusTwoFiveZeroOneOneFour$. This highlights that conclusions drawn from the present catalog should be interpreted with caution, as the inclusion of a few additional high-SNR events can significantly alter the inferred level of agreement with GR.

\subpapersubsection{The QNM rational filter analysis}
\label{subsubsec:QNMRF}
The QNM rational filter \citep[QNMRF;][]{Ma:2022wpv,Ma:2023vvr,Ma:2023cwe,Lu:2025mwp} analyzes post-merger signals from binary BH systems to identify the QNMs present and,
    if more than one mode is present, perform BH spectroscopy. 
In this section we denote the mass and spin pair as $\vartheta_\mathrm{f} = \{(1+z)M_\mathrm{f}, \chi_\mathrm{f}\}$.
The QNMRF applies filters in the frequency domain to eliminate specific complex-valued QNMs from the ringdown signal,
    and then compares the residual with colored Gaussian noise.
For a ringdown model with a chosen set of QNMs, a rational filter is constructed for a given pair $\vartheta_\mathrm{f}$ of remnant BH mass and spin.
By applying filters to the frequency-domain strain data and transforming the filtered data back to the time domain,
    the method removes the QNMs associated with each $\vartheta_\mathrm{f}$ pair.
Since the filters remove QNMs without requiring prior knowledge of their amplitudes or phases,
    the likelihood function $\mathcal{L}(\vartheta_\mathrm{f})$ remains two-dimensional regardless of the number of modes considered.

To determine the set QNMs present in a ringdown signal, we compare the Bayesian evidences for different hypotheses, each of which contains a different set of QNMs and has a different analysis start time. We compute the Bayesian evidence $\mathcal{Z}(d|\mathcal{H})$ for the hypothesis
 $\mathcal{H}$ by integrating the likelihood over the remnant mass and spin of the BH for the given set of QNMs and analysis start time. We then define a detection statistic $\mathcal{D}$ which quantifies how much the data supports a hypothesis $\mathcal{H}$
 over an alternative hypothesis $\mathcal{H}'$:
\begin{equation}
\mathcal{D}(\mathcal{H}:\mathcal{H}') = \log_{10}\frac{\mathcal{Z}(d | \mathcal{H})}{\mathcal{Z}(d | \mathcal{H'})} \, .\label{eq:BF}
\end{equation}
Here $\mathcal{D}$ is analogous to a log Bayes factor but formally differs from the Bayes factors computed by other time-domain ringdown analyses.
Specifically, the QNMRF likelihood is closely connected to the usual time-domain likelihood when using the maximum likelihood estimation for mode amplitudes under the assumption of white noise \citep[Appendix A]{Lu:2025mwp}.

For each hypothesis, we also compute the joint posterior quantile of the remnant mass and spin inferred from the full IMR analysis:
\begin{align}
    p_\mathrm{IMR} = \frac{\sum_{\mathcal{L}(d|\vartheta_\mathrm{f}) > \mathcal{L}(d | \vartheta_\mathrm{f}^\mathrm{IMR})}
                                                    \mathcal{L}(d|\vartheta_\mathrm{f})}
                                                {\sum_{\vartheta_\mathrm{f}}\mathcal{L}(d|\vartheta_\mathrm{f})}
                                         \, , \label{eq:posterior_quantile}
\end{align}
where summation over the grid on which the likelihoods are computed is used instead of integration.
A lower $p_\mathrm{IMR}$ value indicates a better match between the IMR analysis and a specific QNM hypothesis.
We set $\vartheta_\mathrm{f}^\mathrm{IMR}$ and other IMR-inferred quantities to the values with the maximum likelihood in the IMR parameter estimation, using the combined samples from the different waveform approximants from \citet{GWTC:Results}.
In this analysis, we focus on the detection of subdominant ringdown modes by comparing a \{220\}-only QNM model to models with a different secondary mode.
Specifically, we test for the presence of the 221, 210, 200, 330, and 440 modes.

\begin{figure}[htb]
\includegraphics[width=\TGRFigureWidth,clip=true]{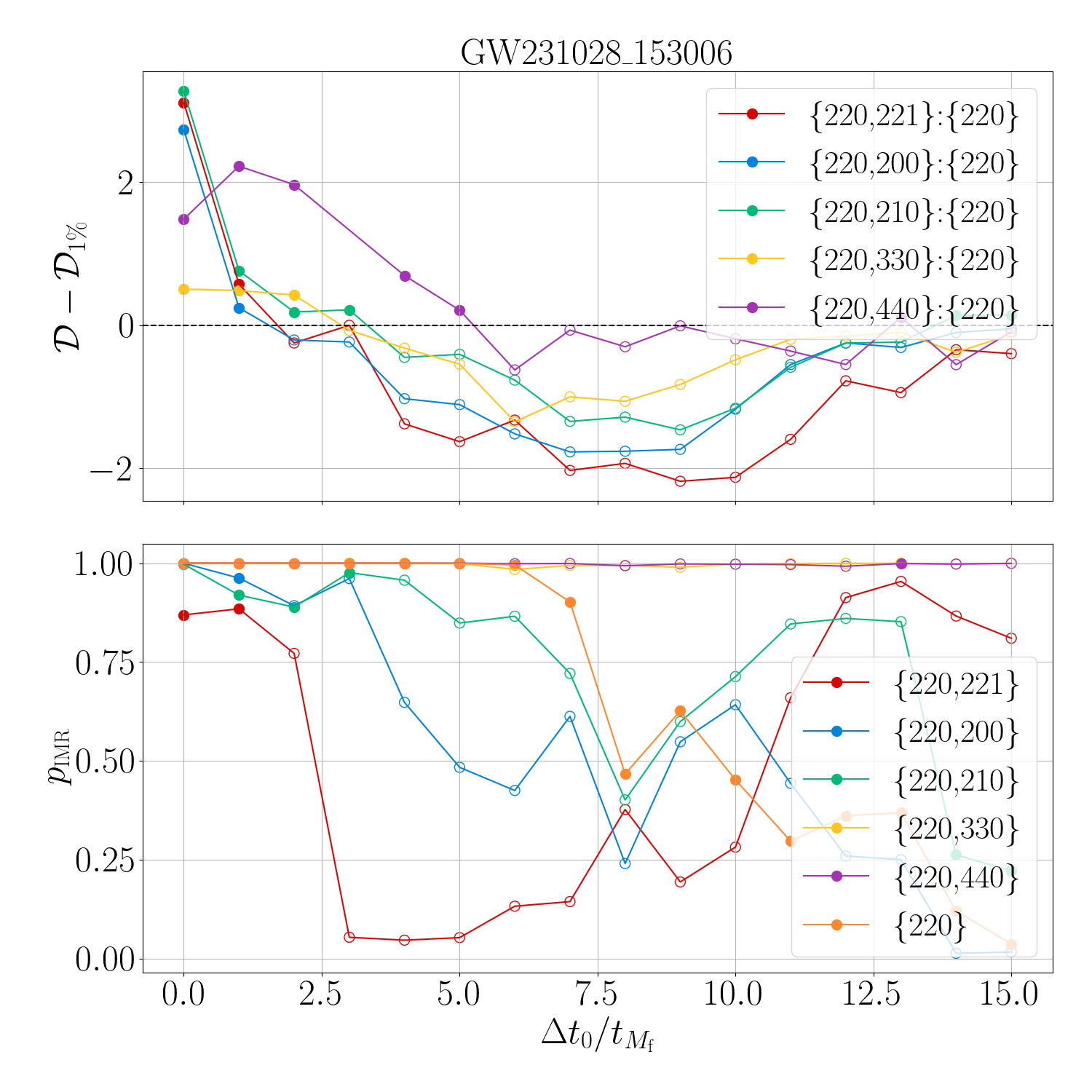}
 \caption{\label{fig:QNMRF_S231028bg} QNM rational filter results for GW231028\_153006. The top panel shows $\mathcal{D}-\mathcal{D}_{1\%}$ when comparing different candidates for the secondary ringdown mode against the \{220\}-only hypothesis at different starting times. Values above zero are unlikely to be caused by the noise background and are denoted by filled markers, while values below zero are denoted by unfilled markers.
 The bottom panel shows $p_\mathrm{IMR}$ for the different mode hypotheses.
 The \{220, 221\} mode combination yields a $\mathcal{D}-\mathcal{D}_{1\%}>0$  for a range of times and improves $p_\mathrm{IMR}$ the most compared to all other modes. }
\end{figure}

\begin{figure}[htb]
\includegraphics[width=\TGRFigureWidth,clip=true]{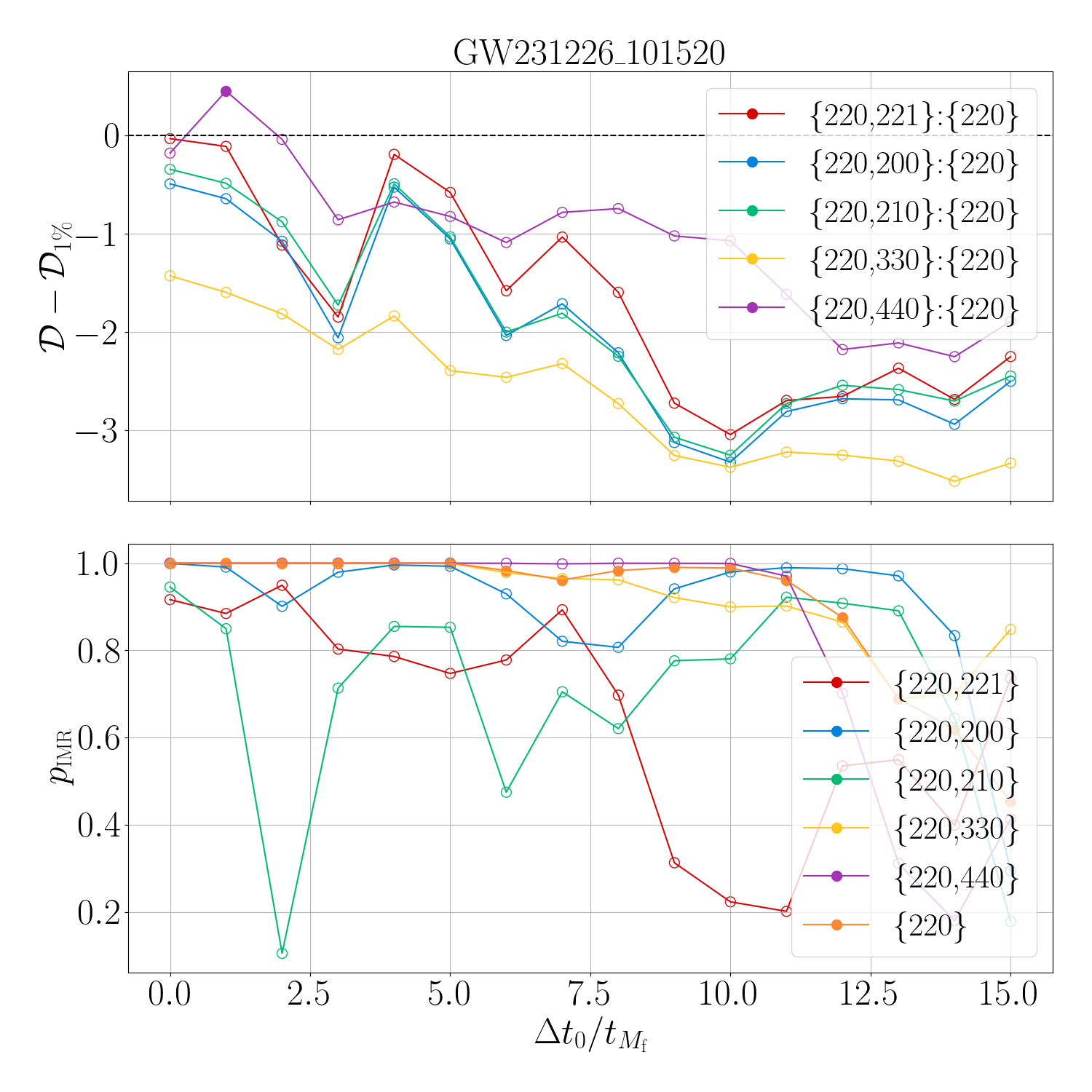}
 \caption{\label{fig:QNMRF_S231226av} (Similar to Figure~\ref{fig:QNMRF_S231028bg}.) QNM rational filter results for GW231226\_101520.
 The \{220,440\} mode combination has a $\mathcal{D}-\mathcal{D}_{1\%}>0$ at $\Delta t_0=1t_{M_\mathrm{f}}$ but does not improve  $p_\mathrm{IMR}$ compared to the \{220\}-only model and thus does not indicate the presence of a 440 QNM. }
\end{figure}

To evaluate a detection statistic preferring a specific two-mode hypothesis to the single-mode hypothesis, we compute a threshold on $\mathcal{D}$, which corresponds to a statistical significance with a 1\% false-alarm probability (FAP).
We do this by injecting a single QNM with randomized properties into the detector noise around the event (i.e., analyzing a simulated observation of a QNM added to the detector noise).
We then compare the detection statistic between an overfiltered two-mode hypothesis (for a specific secondary mode) and the correct single-mode hypothesis. By performing 300 injections with the detector noise around each event we find a distribution
 of false-alarm detection statistics and then take the $\mathcal{D}$ value that corresponds to a 1\% FAP as a threshold $\mathcal{D}_{1\%}$.
 Observing a $\mathcal{D} -\mathcal{D}_{1\%}>0$ indicates that the preference for a secondary mode is unlikely to be caused by the noise background. However, it is common to find $\mathcal{D} -\mathcal{D}_{1\%}>0$ for multiple two-mode hypotheses because the power of a specific subdominant QNM can be partially recovered when searching for other QNMs.
 In this case, the preferred secondary QNM is identified by selecting the hypothesis that reduces $p_\mathrm{IMR}$ the most \citep{Lu:2025mwp}.
 A subdominant mode is, therefore, statistically significant if \emph{both} $\mathcal{D} -\mathcal{D}_{1\%}>0$ \emph{and} it leads to the largest decrease in $p_\mathrm{IMR}$.

Since the onset of the stationary QNM regime is uncertain, we evaluate multiple reference times $\Delta t_0 > 0$, where $\Delta t_0 = 0$ marks the merger time as determined by the IMR analysis.
The likelihood is computed with the data within $[\Delta t_0, \Delta t_0 + \TGRQNMRFSegmentLength]$,
 with the sky position inferred from the IMR analysis.
To determine the events to which we apply this analysis, we considered the four events with the highest total SNRs and then selected the two with the largest redshifted final masses, excluding \COMMONNAME{GW231123} due to waveform uncertainties \citep{GW231123}.
The two selected events are GW231028\_153006 and GW231226\_101520.
 We do not report results for any events from previous observing runs, though GW150914 was analyzed in \citet{Ma:2023vvr,Ma:2023cwe,Lu:2025mwp}.
 
The results for the analysis of GW231028\_153006 are shown in Figure~\ref{fig:QNMRF_S231028bg}.
We observe that for the 221 QNM at $\Delta t_0<2t_{M_\mathrm{f}}$,
     where $t_{M_\mathrm{f}}$ is defined from $M_\mathrm{f}^\textrm{IMR}$,
    the detection statistic is above the threshold and the observation has a FAP $<$ 1\% compared to the noise background, suggesting that these statistics favor the presence of the 221 mode.
While other subdominant modes also have $\mathcal{D}-\mathcal{D}_{1\%}>0$, the 221 QNM improves $p_\mathrm{IMR}$ the most at times when $\mathcal{D}  -\mathcal{D}_{1\%}>0$. 
However, the caveats discussed in Section~\ref{sec:ringdown} still apply: (i) at such an early time, systematic uncertainties due to time-dependent effects, unaccounted for in QNM superpositions, may affect the results; (ii) there is a statistical uncertainty in the merger time itself ($t_\text{geo}{}^{+6.7t_{M_\mathrm{f}}}_{-6.0t_{M_\mathrm{f}}}$).
Therefore, while a 221 mode is found with $\mathcal{D}$ above threshold, its interpretation remains uncertain due to potential early-time systematic uncertainties and merger time uncertainty. These results should not, by themselves, be interpreted as evidence for a constant-amplitude 221 QNM in the signal. A recent study \citep{Wang:2025baj} reported evidence for the 221 mode for GW231028\_153006 but finds that the overtone is present at $\Delta t_0=10t_{M_\mathrm{f}}$, a time at which the QNMRF does not find statistical preference for the presence of an overtone. However, \citet{Wang:2025baj} does not perform the background study that QNMRF does, which may explain the difference in results.

The results for the analysis of GW231226\_101520 are shown in Figure~\ref{fig:QNMRF_S231226av}.
No compelling evidence for a subdominant QNM is found. While the \{220,440\} mode combination yields a $\mathcal{D} >\mathcal{D}^\mathrm{thr}$ at $\Delta t_0=1t_{M_\mathrm{f}}$, it does not improve $p_\mathrm{IMR}$ compared to the \{220\}-only model so no indicative evidence is found for the presence of a 440 mode.

\subpapersubsection{Summary of ringdown tests}
\label{subsubsec:conc-RD}

All the analyses conducted using \soft{pyRing} show no statistically significant evidence for the presence of multiple modes. The lack of detection of multiple modes means that it is not possible to perform \BH spectroscopy with these signals. Furthermore, for all events, there were 90\% CL overlaps with the results of the \ac{IMR} analysis, indicating overall consistency with the GR. However, when combining results from all events, we obtain shifts away from GR. The \soft{pyRing} \texttt{KerrPostmerger} analysis finds a shifts towards larger frequencies and damping times, with a (four-dimensional) hierarchical GR quantile for O4a events of ${\pyRingHierarchicalQuantileFourDValue}_{-\pyRingHierarchicalQuantileFourDMinus}^{+\pyRingHierarchicalQuantileFourDPlus}\%$, where the uncertainty comes from a bootstrapping analysis. However, restricting to the events with significant ringdowns reduces the four-dimensional GR quantile to ${\pyRingHierarchicalQuantileFourDValuelnBcut}_{-\pyRingHierarchicalQuantileFourDMinuslnBcut}^{+\pyRingHierarchicalQuantileFourDPluslnBcut}\%$, and the one-dimensional GR quantile of ${\pyRingHierarchicalQuantileOneDMuDeltaTauValuelnBcut}_{-\pyRingHierarchicalQuantileOneDMuDeltaTauMinuslnBcut}^{+\pyRingHierarchicalQuantileOneDMuDeltaTauPluslnBcut}\%$ is then the largest. Thus, there is no evidence for a significant GR deviation.

The pSEOBNR analysis finds a shift toward higher GR quantiles for the damping time when O4a events are added to the analysis; $Q_{\mathrm{GR}}^{\mathrm{2D}}=\QuantileJointGWTCFourUncertainty\%$ is obtained from the joint posterior, while the hierarchical combination yields $Q_{\mathrm{GR}}^{\mathrm{4D}}=\QuantileHierGWTCFourUncertainty\%$ for all four parameters and $Q_{\mathrm{GR}}^{\mathrm{1D}, \mu_{\delta \hat{\tau}_{220}}} = \QuantileHierGWTCFourUncertaintyTauOneD\%$ for the mean of the damping time alone. However, the bootstrapping uncertainties indicate that these results are influenced by the small size of the event catalog, and this conclusion is bolstered by the reduction in these GR quantiles to $\QuantileJointGWTCFourPlusTwoFiveZeroOneOneFour$ (joint) and $\QuantileHierGWTCFourPlusTwoFiveZeroOneOneFour$, $\QuantileHierGWTCFourTauOneDPlusTwoFiveZeroOneOneFour$ (hierarchical) with the inclusion of the loud \ac{O4b} event GW250114~\citep{GW250114, GW250114_TGR}.

Finally, the QNM filter analysis finds that for GW231028\_153006 at $\Delta t_0<2t_{M_\textrm{f}}$ the detection statistic favors the presence of the 221 mode. However, interpreting this result remains delicate and uncertain, as possible systematic effects cannot be ruled out, and overtone analyses are particularly sensitive to such systematic uncertainties.

Additionally, there are results from the ringdown analyses of the loud event \COMMONNAME{GW230814single} that show support for apparent deviations from GR, excluded from this paper since that event was only observed by a single detector.
We discuss why these apparent deviations do not constitute evidence for a violation of GR in \cite{GW230814}. To summarize, the results for \COMMONNAME{GW230814single} can likely be explained by a combination of effects of detector noise (amplified by only having data from a single detector) and inaccuracies in waveform modeling.

\subpapersection{Echo Tests}\label{sec:echoes}
There are a variety of compact objects proposed as alternatives to \acp{BH}, such as boson stars \citep{Kaup:1968zz,Ruffini:1969qy,Liebling:2012fv}, gravastars \citep{Mazur:2004fk}, fuzzballs \citep{Mathur:2005zp}, and firewalls \citep{Almheiri:2012rt}.
Some of them are compact enough to possess a light ring and have a surface instead of an event horizon.
In such case, the ingoing merger--ringdown signal may be reflected at the surface and at the potential barrier iteratively.
This iterative reflection of the signal can also happen between two potential barriers for a traversable wormhole \citep{Morris:1988tu}.
As a result, we may observe additional signals after the merger of compact binaries \citep{Cardoso:2016rao,Cardoso:2016oxy,Cardoso:2019rvt,Siemonsen:2024snb}.
These signals are called \ac{GW}  echoes.
Some specific quantum BHs lead to echoes as well, since they only absorb signals with specific discrete frequencies \citep{Cardoso:2019apo, Wang:2019rcf,Oshita:2019sat,Agullo:2020hxe,Chakraborty:2022zlq}.  
Therefore, the detection of echoes can be evidence of a modification in the vicinity of the classical event horizon.

Here we employ both template-based and model agnostic searches for echoes.
A template-based search is the most effective method to detect signals if we can model the signals accurately.
So far, various studies have attempted to model echoes
    \citep{Ashton:2016xff,Abedi:2016hgu,Mark:2017dnq,Maselli:2017tfq,Nakano:2017fvh,Testa:2018bzd,Maggio:2019zyv,Wang:2019szm,Sago:2020avw,Conklin:2021cbc,Xin:2021zir,Wu:2023wfv,Zimmerman:2023hua}.
However, since we do not know the exact physics of the echo mechanism,
    we in principle need a large number of models to detect all possible signals,
    which is infeasible in practice.
Thus, we here select only two representative models, one a phenomenological frequency-independent model (ADA; \citealt{Abedi:2016hgu}),
    and the other a model with a physically motivated frequency dependence from \ac{BH} perturbation theory (BHP; \citealt{Nakano:2017fvh}).

On the other hand, the model agnostic search can cover a wider range of possible echo morphologies.
We use two methods for the model agnostic search: the \BW analysis \citep{Cornish:2014kda,Cornish:2020dwh,bayeswave} and the \CWB analysis \citep{Klimenko:2015ypf,Drago:2020kic}.
These analyses are able to detect GW signals without a detailed model for the signal.
The \BW echo analysis models the echoes as a sum of sine--Gaussians and computes the evidence for echoes via a signal-to-noise Bayes factor.
The \CWB echo analysis considers the coherent energy excess among the detectors, which is completely model independent, and computes a \textit{p}-value for the presence of echoes.
Both analyses have less dependence on the IMR signal compared to the template-based analysis.

Various studies have searched for echoes from O1 to O3 events so far \citep{Ashton:2016xff,Abedi:2016hgu,Westerweck:2017hus,Abedi:2018npz,Lo:2018sep,Nielsen:2018lkf,Uchikata:2019frs,Tsang:2019zra,Wang:2020ayy,LIGOScientific:2020tif,Ren:2021xbe,Abedi:2021tti,Miani:2023,Uchikata:2023zcu,LIGOScientific:2021sio,Abedi:2022bph}.
No strong evidence of echoes from BBH has been reported. 
\citet{Wu:2025enn} also analyzes two O4 events with a different method than those used in this paper, similarly reporting no detection of echo signals. 
Weak-to-moderate evidence of echoes from \COMMONNAME{GW231123} has been reported by \citet{Lai:2026yvm}.
However, the echoes they consider overlap with the merger--ringdown, so their analysis does not allow for the post-merger echoes we consider.
Thus, their results are not comparable to ours.

\subpapersubsection{Waveform template based analysis}\label{sec:modeled}

For the waveform template based echoes analysis, we use waveform models that consist of a BBH IMR waveform plus echoes.
If we assume the echoes are a consequence of repeated reflection of the merger--ringdown signal between the surface and the barrier, we can construct the echo waveform based on the merger-ringdown part of the preceding IMR waveform.
The series of echoes is then characterized by a decay rate $\gamma$ and delay time $\Delta t_{\rm echo}$.

\begin{table}[b]
  \caption{\label{tab:modeled_echo_prior}
    Prior ranges for echo parameters for the modeled analyses.
    }
\begin{center}
 \begin{tabular}{ l c c  }
\hline \hline
Echo parameters &  ADA & BHP  \\
\hline 
$\log_{10} A$ & $[-2, 0]$ & $[-3, 0]$ \\      
$\log_{10} \gamma$&  $[-2, 0]$ & \nodata \\
 $ t_0  $&$[-100, 10]\, t_{\Mtot} $ &  $[-100, 10]\, t_{\Mtot} $ \\
$t_{\mathrm{echo}} $  & $[10,10^3]\, t_{\Mtot}$ & $[10,10^3]\,t_{\Mf} $\\
$ \Delta t_{\mathrm{echo}} $ & $[10,10^3]\,t_{\Mtot}$ & \nodata\\
 $\phi$ (rad)& \nodata &  $[0,2\pi]$ \\
\hline 
 \end{tabular}
 \end{center}
 \tablecomments{We sample the parameters uniformly over these ranges.
  Here, $t_{\Mtot}$ and $t_{\Mf}$ are defined respectively from $M^{\mathrm{maxL}}$ and $\Mf^{\mathrm{maxL}}$,
    where ${\mathrm{maxL}}$ denotes the maximum-likelihood value.}
\end{table}

We consider two approaches to determine $\gamma$ and $\Delta t_{\rm echo}$.
One approach, the ADA model, models the signal phenomenologically, treating $\gamma$ and $\Delta t_{\rm echo}$ as free parameters \citep{Abedi:2016hgu,Lo:2018sep,LIGOScientific:2020tif}.
The phase shift at each echo is fixed to $\pi$ assuming the phase is inverted at each iteration.
We searched for such modeled echoes in previous catalogs (up to GWTC-2.0), as have other groups \citep{Ashton:2016xff,Abedi:2016hgu,Westerweck:2017hus,Lo:2018sep,Nielsen:2018lkf,Uchikata:2019frs,Uchikata:2023zcu},
  but the analysis here has been rewritten and upgraded to use \BILBY.
In the other approach, the BHP model, these parameters are numerically computed using a physical model
  of BH perturbation theory \citep{Nakano:2017fvh}.
We compute the reflection rate at the potential barrier, which is related to the decay rate,
  and as a result, the reflection rate becomes frequency dependent.
While the phase shift $\phi$ is treated as a free parameter,
  the time delay of each echo is obtained from the remnant mass and spin \citep{Uchikata:2019frs,Uchikata:2023zcu},
  themselves calculated from the binary components masses and spins using \soft{NRSur7dq4Remnant} \citep{Varma:2019csw}.

In both models, we employ \IMRPhenomXPHM for the IMR waveform.
The same waveform model is used to create the echo waveform by removing the inspiral part of the model.
The point at which we choose the inspiral to end is parametrized by $t_0$, which we treat as a free parameter.
In addition to the parameters described above, we also vary the relative amplitude of the echoes $A$ (compared to the truncated IMR waveform) and the start time of the first echo $t_{\rm echo}$.

We assess the evidence for echoes using the above two models by performing Bayesian parameter estimation as described in Paper~I, evaluating the statistical evidence for echoes using the Bayes factor for IMR plus echoes models to IMR only, $\mathcal{B}^{\mathrm{IMRE}}_{\mathrm{IMR}}$.
We vary the extrinsic parameters and echo parameters described above but fix the intrinsic parameters to their maximum-likelihood values from parameter-estimation results obtained using \IMRPhenomXPHM \citep{GWTC:Results} to reduce the computational cost.
We have confirmed using injection studies that fixing the intrinsic parameters will not affect the detectability of echoes.
We set a  threshold for the Bayes factor  $\log_{10} \mathcal{B}^{\mathrm{IMRE}}_{\mathrm{IMR}} \sim 2.1$, 
above which we would follow up with an analysis that varies all the IMR parameters.
The threshold corresponds to a $3.3$~$\sigma$ detection in O1 data \citep{Lo:2018sep}.
The priors for the echo parameters are summarized in Table~\ref{tab:modeled_echo_prior}.
We sample the echo parameters uniformly over the ranges shown in the table.
For the ADA model, the number of echoes is fixed to five, while for the BHP model, it is determined by the duration of the post-merger data and $\Delta t_{\mathrm{echo}}$.

\begin{figure}[tb]
 \includegraphics[width=\TGRFigureWidth,clip=true]{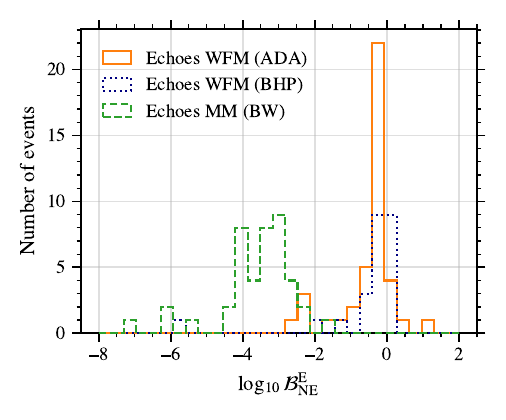}
  \caption{Histogram of $\log_{10}$ Bayes factors for echoes (E) versus no echoes (NE) for the ADA and BHP modeled analyses and minimally modeled BW analysis.
          Here $\log_{10}\mathcal{B}^\mathrm{E}_\mathrm{NE} $ refers to
            $\log_{10}\mathcal{B}^\mathrm{IMRE}_\mathrm{IMR}$ for the ADA and BHP analyses and $ \log_{10}\mathcal{B}^\mathrm{signal}_\mathrm{noise}  $ for the BW analysis.
          For the modeled analyses, the Bayes factors compare IMR + echoes to IMR,
            so $\log_{10}\mathcal{B}^\mathrm{E}_\mathrm{NE} \lesssim 0$ in the absence of echoes.
          The minimally modeled analysis using \BW provides a signal-to-noise Bayes factor for echoes,
            so $\log_{10}\mathcal{B}^\mathrm{E}_\mathrm{NE} < 0$ in the absence of echoes.
            Thus, all the results are consistent with a lack of echoes.}
  \label{fig:echoes_histogram}
\end{figure}

For the ADA model, we analyze events with both component masses larger than $3 M_{\odot}$, so they are likely BBHs, and which have a mass ratio larger than $0.1$.
We restrict to these mass ratios since they are the ones for which we have confirmed that the analysis gives accurate results in injection studies, though we expect that the model can be extended to more unequal mass ratios, since the underlying IMR waveform model
has a larger domain of validity.
All 41 O4a BBH events listed in Table~\ref{tab:selectionIII} have mass ratio larger than $0.1$ and are thus analyzed with the ADA model.
For the BHP model, which focuses on echoes in a narrow frequency band, we analyze events whose maximum-likelihood $220$ mode QNM frequency is less than $1000$~Hz, since the detectors are less sensitive to higher frequencies \citep{GWTC:Introduction}.
Since the reflection rate is calibrated for final spins $0.6 < \chi_\mathrm{f} < 0.8$, we also exclude events whose final spin lies outside the range.
Furthermore, we exclude events whose mass ratio is smaller than $1/6$, since \soft{NRSur7dq4Remnant} is not reliable for such mass ratios.
We apply all these restrictions using the maximum-likelihood results from \citet{GWTC:Results}.
Based on an earlier version of the results given in \citet{GWTC:Results}, 31 events in O4a pass the BHP model's selection criteria.

\begin{figure}[th]
\begin{center}
\includegraphics[width=\columnwidth]{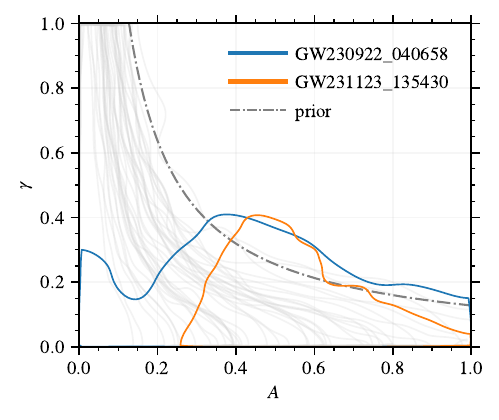}
\caption{\label{fig:echoes_ADA_A_gamma_2d}Distribution of the amplitude $A$ and the decay rate $\gamma$ for the ADA modeled echoes analysis. The contours are the $90\%$ credible regions, and the posteriors that deviate from the prior are highlighted in color. The prior distribution is shown in the dash-dotted curve. }
\end{center}
\end{figure}

\begin{figure}[hb]
\begin{center}
\includegraphics[width=\columnwidth]{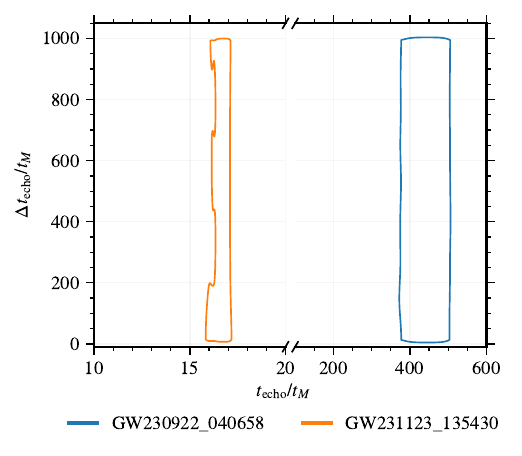}
\caption{\label{fig:echoes_ADA_t_echo_delta_t_echo_2d}Distribution of the time of the first echo relative to the merger $t_{\rm echo}$ and the time delay between echoes $\Delta t_{\rm echo}$, as inferred by the ADA modeled echoes analysis for the events that are highlighted in Figure~\ref{fig:echoes_ADA_A_gamma_2d}.}
\end{center}
\end{figure}

We summarize the values of $\log_{10} \mathcal{B}^{\mathrm{IMRE}}_{\mathrm{IMR}} $ in Table~\ref{tab:modeled_echo}.
For both models, all $\log_{10} \mathcal{B}^{\mathrm{IMRE}}_{\mathrm{IMR}} $ values are below the threshold.
In the absence of echoes, we expect $\log_{10} \mathcal{B}^{\mathrm{IMRE}}_{\mathrm{IMR}}$ values around or below zero, which we indeed see in Figure~\ref{fig:echoes_histogram}.
Therefore, we conclude that we have not found any significant echoes that can be modeled by the ADA and BHP models in the O4a events.

\vfill \break
\begin{table}[H]
  \caption{\label{tab:modeled_echo}
    Values of $\log_{10} \mathcal{B}^{\mathrm{IMRE}}_{\mathrm{IMR}} $ for the template-based echoes analyses.
    }
 \begin{center}
 \begin{tabular}{ l c c  }
\hline \hline
    Event & $\log_{10} \mathcal{B}^{\mathrm{IMRE(ADA)}}_{\mathrm{IMR}}  $  & $\log_{10} \mathcal{B}^{\mathrm{IMRE(BHP)}}_{\mathrm{IMR}}  $  \\[0.075cm]
    \hline
    GW230601\_224134 & \BFADA{GW230601\_224134} & \nodata \\
    GW230605\_065343 & \BFADA{GW230605\_065343}&   \BFBHP{GW230605\_065343}  \\
    GW230606\_004305 & \BFADA{GW230606\_004305}& \BFBHP{GW230606\_004305}   \\
    GW230609\_064958 & \BFADA{GW230609\_064958}& \nodata    \\
    GW230624\_113103 & \BFADA{GW230624\_113103}&  \nodata \\
    GW230627\_015337 & \BFADA{GW230627\_015337} & \nodata  \\
    GW230628\_231200 & \BFADA{GW230628\_231200} &  \BFBHP{GW230628\_231200} \\
    GW230630\_234532 & \BFADA{GW230630\_234532} &  \BFBHP{GW230630\_234532} \\
    GW230702\_185453 & \BFADA{GW230702\_185453} &\BFBHP{GW230702\_185453}    \\
    GW230731\_215307 & \BFADA{GW230731\_215307} &  \BFBHP{GW230731\_215307} \\
    GW230811\_032116 & \BFADA{GW230811\_032116} &  \BFBHP{GW230811\_032116}   \\
    GW230814\_061920 & \BFADA{GW230814\_061920} & {\checkbar} \\
    GW230824\_033047 & \BFADA{GW230824\_033047}&  \nodata  \\
    GW230904\_051013 & \BFADA{GW230904\_051013} & {\checkbar}  \\
    GW230914\_111401 & \BFADA{GW230914\_111401} & \BFBHP{GW230914\_111401}   \\
    GW230919\_215712 & \BFADA{GW230919\_215712} &  \BFBHP{GW230919\_215712}  \\
    GW230920\_071124 & \BFADA{GW230920\_071124} &  \BFBHP{GW230920\_071124} \\
    GW230922\_020344 & \BFADA{GW230922\_020344} &\BFBHP{GW230922\_020344}    \\
    GW230922\_040658 & \BFADA{GW230922\_040658} & \nodata \\
    GW230924\_124453 & \BFADA{GW230924\_124453} & \BFBHP{GW230924\_124453}   \\
    GW230927\_043729 & \BFADA{GW230927\_043729} &  \BFBHP{GW230927\_043729} \\
    GW230927\_153832 & \BFADA{GW230927\_153832} &   \BFBHP{GW230927\_153832} \\
    GW230928\_215827 & \BFADA{GW230928\_215827} &  \nodata \\
    GW231001\_140220 & \BFADA{GW231001\_140220} &  \BFBHP{GW231001\_140220} \\
    GW231020\_142947 & \BFADA{GW231020\_142947} &  \BFBHP{GW231020\_142947} \\
    GW231028\_153006 & \BFADA{GW231028\_153006} &  \nodata \\
    GW231102\_071736 & \BFADA{GW231102\_071736} &   \BFBHP{GW231102\_071736}  \\
    GW231104\_133418 & \BFADA{GW231104\_133418} &  {\checkbar}  \\
    GW231108\_125142 & \BFADA{GW231108\_125142} &  \BFBHP{GW231108\_125142}  \\
    GW231110\_040320 & \BFADA{GW231110\_040320} &     \BFBHP{GW231110\_040320}   \\
    GW231113\_200417 & \BFADA{GW231113\_200417}&  {\checkbar}\\
    GW231114\_043211 & \BFADA{GW231114\_043211} &   {\checkbar} \\
    GW231118\_005626 & \BFADA{GW231118\_005626} &  \nodata \\
    GW231118\_090602 & \BFADA{GW231118\_090602}&  \nodata  \\
    GW231123\_135430 & \BFADA{GW231123\_135430}&  \BFBHP{GW231123\_135430}   \\
    GW231206\_233134 & \BFADA{GW231206\_233134}&   \BFBHP{GW231206\_233134} \\
    GW231206\_233901 & \BFADA{GW231206\_233901}&   \BFBHP{GW231206\_233901}  \\
    GW231213\_111417 & \BFADA{GW231213\_111417} & \BFBHP{GW231213\_111417}   \\
    GW231223\_032836 & \BFADA{GW231223\_032836}&  \BFBHP{GW231223\_032836} \\
    GW231224\_024321 & \BFADA{GW231224\_024321} & {\checkbar} \\
    GW231226\_101520 & \BFADA{GW231226\_101520}&  \BFBHP{GW231226\_101520}   \\
    \hline
  \end{tabular}
 \end{center}
 \tablecomments{Ellipses indicate events that do not satisfy the event selection criteria for a given model
                and {\checkbar}'s indicate events that satisfy the criteria but are not yet included, because their analysis would take infeasibly long to complete.}
\end{table}

We show the joint posteriors of the amplitude $A$ and the decay rate $\gamma$ for the ADA model in Figure~\ref{fig:echoes_ADA_A_gamma_2d}, with two events that show some support for $A > 0$ highlighted.
While $A>0 $ is favored for these two events, $\gamma$ is constrained to be less than $0.5$, which is a stronger constraint compared to its prior.
For the other events, both parameters are uninformative or $A=0$ is supported more strongly than by the prior.

Furthermore, we show the joint posterior distributions of $t_{\rm echo}$ and $\Delta t_{\rm echo}$ for the above two events in Figure~\ref{fig:echoes_ADA_t_echo_delta_t_echo_2d}.
For GW230922\_040658, we visually inspected the data quality around the event and found that the echo inference results may be associated with some excess power after the merger in both detectors.
For \COMMONNAME{GW231123}, the $t_{\rm echo}$ posterior distributions are constrained to $\sim 17 t_M$, which means that the analysis is latching onto features in the early post-merger data. 
For additional discussion of possible interpretations and effects in \COMMONNAME{GW231123}, see \citet{GW231123, GWTC:Lensing}.
For both events, the $\Delta t_{\rm echo}$ posterior distributions are uninformative.
To summarize, for these two events, the post-ringdown data are fit well with only one echo, which is inconsistent with the model assuming multiple echoes used in this analysis.

\subpapersubsection{Minimally modeled analysis with \BAYESWAVE}\label{sec:BW}

We perform a minimally modeled search for the echoes using \BW~\citep{Tsang:2018uie}.
We use a train of sine--Gaussians as basis functions to describe a potential echoes signal.
The individual sine--Gaussians are parameterized by an amplitude, a damping time, a reference frequency, a reference phase, and a central time.
A train of sine--Gaussians includes four additional parameters, namely a time separation, a relative phase shift, a damping factor, and a widening factor between successive sine--Gaussians.
While we cannot expect that a potential echoes signal exactly matches a train of sine--Gaussians, it has been demonstrated that a wide range of echoes signals can be represented by the superposition of such basis functions~\citep{Tsang:2018uie}.
This aspect makes the search minimally modeled.

In particular, we analyze 4~s of data starting at $t_\mathrm{event} + 3\tau_{220}$ for each signal, where $t_\mathrm{event}$ is the merger time of the observed GW \citep[][]{GWTC:Results} and $\tau_{220}$ is the decay time of the 220 QNM,
    estimated as a function of the final object's mass and spin through the fit presented in \citet{Berti:2005ys}.
We use a conservative estimate for $\tau_{220}$ obtained from the upper limit of the $90\%$ credible interval for the $\tau_{220}$ posterior, thus ensuring that the analyzed data are not contaminated by the ringdown signal,
    which decays exponentially.
We use \BW to compute the \ac{PSD} of the echoes analysis segment itself, rather than using the \ac{PSD} used for the analysis of the \ac{CBC} signal in \cite{GWTC:Results}.

The end product of the analysis is an echoes signal-to-noise Bayes factor $\mathcal{B}^{\mathrm{signal}}_{\mathrm{noise}}$ which serves as the detection statistic.
We analyze all 42 \ac{O4a} events listed in Table~\ref{tab:selectionIII} and present the distribution of $\log_{10} \mathcal{B}^{\mathrm{signal}}_{\mathrm{noise}}$ for those events in Figure~\ref{fig:echoes_histogram}.
We find that $\log_{10} \mathcal{B}^{\mathrm{signal}}_{\mathrm{noise}} \le 0$, meaning that no evidence for the presence of an echoes signal is found.

\vfill \break

\subpapersubsection{Minimally modeled analysis with coherent WaveBurst}\label{sec:cWB}
The \CWB search for echo signals is a minimally modeled search method \citep{Miani:2023}, meaning it does not rely on prior assumptions about the waveform morphology. Instead, it identifies coherent energy excesses in the data collected by the detector network and extracts these as candidate signals. Specifically, the analysis focuses on the coherent energy content within a time window that follows the CBC signal under study, in the frequency band [16, 1024]~Hz.

The time window that is analyzed starts at $t_{\text{echo}}^{(1)} - 0.05$~s after the coalescence time, where $t_{\text{echo}}^{(1)}$ is the predicted arrival time of the first echo pulse according to Equation (2) of \citet{Abedi:2016hgu}, using the maximum-likelihood values of the source-frame remnant mass and remnant spin
from \citet{GWTC:Results}.
The end time of the analyzed time window is set at $4 t_{\text{echo}}^{(1)} \ +0.05$~s.
We empirically checked that this time window ensures that contributions from the primary CBC signal are excluded. Given the uncertainties in the echo model and distance estimates, we fixed the time window for each event without applying a cosmological redshift correction, hence underestimating $t_{\text{echo}}^{(1)}$. The resulting earlier starting time of the window is conservative, in that it allows for an earlier arrival of the first echo pulse than predicted by the expression from \citet{Abedi:2016hgu} used for $t_{\text{echo}}^{(1)}$.
Assuming the model in \citet{Abedi:2016hgu} and the mean redshift estimates, the first three echo pulses would occur inside the analyzed time window for more than half of the events reported in Table~\ref{tab:unmodeled_echo}.

\begin{table}[ht]
 \caption{\label{tab:unmodeled_echo}
 Results from the minimally modeled \CWB echo analysis}
\begin{center}
 \begin{tabular}{ c c c }
   \hline \hline
     \multirow{2}{*}{Event} & \multirow{2}{*}{\textit{p}-value} $ $  & Post-merger \\
     & & SNR  \\
     \hline
    GW230601\_224134  & \cWBpValue{GW230601\_224134} & \cWBSNRon{GW230601\_224134}  \\
    GW230606\_004305  & \cWBpValue{GW230606\_004305}& \cWBSNRon{GW230606\_004305}  \\
    GW230609\_064958  & \cWBpValue{GW230609\_064958}& \cWBSNRon{GW230609\_064958}    \\
    GW230624\_113103  & \cWBpValue{GW230624\_113103} &  \cWBSNRon{GW230624\_113103}  \\
    GW230627\_015337  & \cWBpValue{GW230627\_015337}& \cWBSNRon{GW230627\_015337}  \\
    GW230628\_231200  & \cWBpValue{GW230628\_231200} & \cWBSNRon{GW230628\_231200} \\
    GW230702\_185453  & \cWBpValue{GW230702\_185453}& \cWBSNRon{GW230702\_185453}   \\
    GW230731\_215307  & \cWBpValue{GW230731\_215307} &  \cWBSNRon{GW230731\_215307}  \\
    GW230811\_032116  & \cWBpValue{GW230811\_032116} &  \cWBSNRon{GW230811\_032116} \\
    GW230814\_061920  & \cWBpValue{GW230814\_061920} & \cWBSNRon{GW230814\_061920} \\
    GW230824\_033047  & \cWBpValue{GW230824\_033047} &  \cWBSNRon{GW230824\_033047} \\
    GW230914\_111401  & \cWBpValue{GW230914\_111401}& \cWBSNRon{GW230914\_111401} \\
    GW230919\_215712  & \cWBpValue{GW230919\_215712} & \cWBSNRon{GW230919\_215712} \\
    GW230920\_071124  & \cWBpValue{GW230920\_071124} &  \cWBSNRon{GW230920\_071124} \\
    GW230922\_020344  & \cWBpValue{GW230922\_020344} &  \cWBSNRon{GW230922\_020344} \\
    GW230922\_040658  & \cWBpValue{GW230922\_040658} & \cWBSNRon{GW230922\_040658}  \\
    GW230924\_124453  & \cWBpValue{GW230924\_124453} & \cWBSNRon{GW230924\_124453} \\
    GW230927\_043729  & \cWBpValue{GW230927\_043729}  &  \cWBSNRon{GW230927\_043729} \\
    GW230927\_153832  & \cWBpValue{GW230927\_153832} & \cWBSNRon{GW230927\_153832}   \\
    GW230928\_215827  & \cWBpValue{GW230928\_215827}  & \cWBSNRon{GW230928\_215827} \\
    GW231001\_140220  & \cWBpValue{GW231001\_140220} &  \cWBSNRon{GW231001\_140220} \\
    GW231028\_153006  & \cWBpValue{GW231028\_153006} & \cWBSNRon{GW231028\_153006} \\
    GW231102\_071736  & \cWBpValue{GW231102\_071736}  &  \cWBSNRon{GW231102\_071736} \\
    GW231108\_125142  & \cWBpValue{GW231108\_125142} & \cWBSNRon{GW231108\_125142}  \\
    GW231113\_200417  & \cWBpValue{GW231113\_200417}& \cWBSNRon{GW231113\_200417} \\
    GW231123\_135430 & \cWBpValue{GW231123\_135430} & \cWBSNRon{GW231123\_135430} \\
    GW231206\_233134  & \cWBpValue{GW231206\_233134} & \cWBSNRon{GW231206\_233134} \\
    GW231206\_233901  & \cWBpValue{GW231206\_233901} &  \cWBSNRon{GW231206\_233901} \\
    GW231213\_111417  & \cWBpValue{GW231213\_111417} & \cWBSNRon{GW231213\_111417}  \\
    GW231223\_032836  & \cWBpValue{GW231223\_032836} & \cWBSNRon{GW231223\_032836} \\
    GW231226\_101520  & \cWBpValue{GW231226\_101520} &  \cWBSNRon{GW231226\_101520} \\
     \hline \hline
  \end{tabular}
  \end{center}
  \tablecomments{We give the \textit{p}-value for the null hypothesis and related 90\% confidence interval along with the network SNR statistic, as reconstructed by \CWB in post-merger.}
\end{table}

Of the O4a events considered in this paper, we excluded those with a network SNR $\lesssim 7$ as reconstructed by \CWB. This threshold ensures a negligible false dismissal probability and avoids selection biases in the analysis of CBC posterior waveform samples.
While the median matched-filter network SNRs of all O4a events considered in this paper are $\geq \networkmatchedfiltersnrmedminTGRevents$, the SNR reconstructed by \CWB can be lower than the matched filter SNR, as in the case of events with low chirp masses, $\lesssim 6$--$7\ M_{\odot}$, or long duration, $> 5$ s above $16$~Hz.
Hence, this analysis reports results for $31$ O4a events out of $\TGRNUMEVENTS$.

The coherent signal energy (SNR) within the analyzed time window after each CBC event (\textit{on-source} result) is reported in Table~\ref{tab:unmodeled_echo}.
To determine the statistical significance of the on-source result,
    we compare it to the empirical distribution due to noise, obtained by performing the same analysis over $\sim 10^4$ injections of signals generated using random CBC posterior samples,
    without echoes, in real LIGO noise.
This off-source dataset typically covers five weeks of data centered on the event, allowing us to marginalize noise effects over a sufficiently long observing time.
The \textit{p}-value of the null hypothesis is estimated by the fraction of reconstructed off-source injections with an SNR greater than the on-source SNR.
Our analysis is agnostic with respect to the waveform model of the CBC posterior samples used for the off-source injections;
    for practical reasons related to the availability of CBC parameter-estimation results at the time of our analysis,
    we adopted \IMRPhenomXPHM for the events through \COMMONNAME{GW231123} and \SURSEVENDQFOUR for the remaining five events.

\begin{figure}[htb]
 \includegraphics[width=\TGRFigureWidth,clip=true]{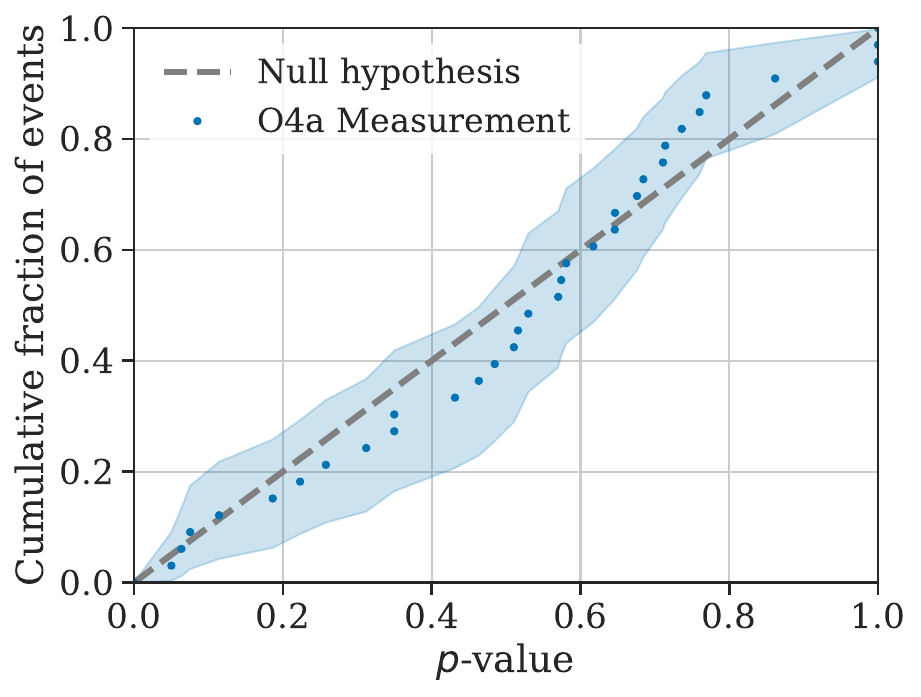}
  \caption{Probability--probability plot for the 31 events in O4a using the \CWB minimally modeled echoes analysis.}
 \label{fig:echo_pp}
\end{figure}

The results of this analysis for all analyzed O4a events are summarized in Table~\ref{tab:unmodeled_echo} and in Figure~\ref{fig:echo_pp}. The distribution of \textit{p}-values is in agreement with the null hypothesis, and in particular the lowest $p$-value found, $0.05$ for \FULLNAME{GW231001_140220}, cannot be considered evidence for the presence of echoes, given the $31$ independent trials.
\FULLNAME{GW231001_140220} also has the largest post-merger SNR $\simeq 3$, which is the largest value for O4a events, but still in the range of plausible noise outliers, and smaller than that for two pre-O4 events \citep{Miani:2023}. This analysis also does not find any evidence for echoes for pre-O4 events \citep{Miani:2023}.%

\subpapersubsection{Discussion}\label{sec:echoDiscussion}

We have performed four analyses allowing for the presence of possible echo signals after the ringdown signal of the CBCs considered in this paper: two analyses that use a waveform model (one phenomenological and one physical) and two using minimally modeled approaches.
The results of all of these analyses are consistent with the absence of echoes.
The waveform-template modeled analyses and the \BW minimally modeled echoes analysis produce Bayes factors, which we summarize in Figure~\ref{fig:echoes_histogram}.
The \CWB minimally modeled analysis produces \textit{p}-values, which are shown in Figure~\ref{fig:echo_pp}.

If we assume that any echoes originate from the merger--ringdown signal or are affected by the merger, we would expect that the echo amplitude is proportional to the amplitude of the signal at merger.
However, we do not see any significant echoes result from our analyses for either the most massive \ac{BBH} detected to date, \COMMONNAME{GW231123} \citep{GW231123}, with a redshifted final mass of $\finalmassdetuncert{GW231123_135430}M_\odot$, where the merger--ringdown is the dominant signal, or the highest SNR event analyzed in this paper, GW231226\_101520, with a median matched-filter SNR of $\networkmatchedfiltersnrmed{GW231226_101520}$.

Our results are also consistent with those of \citet{GW230814},
    which found no echo signals after \COMMONNAME{GW230814single},
    the loudest event in O4a, with a median matched-filter SNR of $\networkmatchedfiltersnrmed{GW230814_230901}$;
    \COMMONNAME{GW230814single} is not included in this paper as it was only observed by a single detector.

\subpapersection{Conclusions}\label{sec:paper III conclusions}

We presented \TGRIIINUMTESTS tests of \GR, \reviewed{four} of which are new,
    focusing on the post-merger stages, i.e., the ringdown and possible echoes.
Overall, our tests find that the individual signals we analyze are in agreement with our expectations from GR.
Moreover, in the pSEOBNR analysis, the high-\SNR event \FULLNAME{GW231226_101520} gives the tightest single-event constraint
    on the damping time of the dominant $(2, 2, 0)$ QNM of all the GWTC-4.0 events,
    though the very loud \ac{O4b} event GW250114 gives even better constraints \citep{GW250114,GW250114_TGR}.
The QNM rational filter analysis of \FULLNAME{GW231028_153006} found marginal Bayesian support for a secondary mode,
    albeit at times close to merger where the validity of
    modeling the signal as a superposition of QNMs is in doubt.

For the echo analyses, the minimally modeled \CWB analysis finds
    that the \textit{p}-values are $\geq \reviewed{0.05}$, a result consistent with noise when considering the number of events tested.
The three Bayesian analyses for post-merger echoes find that all
    the $\log_{10}$ Bayes factors for echoes are either at most \BFBWBestVal{} (for the minimally modeled \BW analysis), or at most \BFADABestVal{} for waveform-modeled echoes.
The larger values for the waveform-modeled analyses are expected,
    since these models compare waveforms composed of the full \ac{IMR} signal plus echoes versus \ac{IMR}-only waveforms, like the minimally modeled analysis,
    and the maximum value is still less than the threshold of $\sim 2.1$ to trigger follow-up analyses.
Thus, the results of all the echo analyses are consistent with a lack of GW echoes,
    and rather with finding statistical noise fluctuations.

Of the \TGRIIINUMEVENTS O4a events covered in this paper, only one event, \COMMONNAME{GW231123},
    showed deviations from the GR expectations (for ECH-WFM-ADA).
These deviations could be due to inaccuracies in waveform modeling, wave-optics lensing, or other features,
    as described in our dedicated paper on that event \citep{GW231123}.
Additionally, in \citet{GW230814}, we applied ringdown tests to the loud event \COMMONNAME{GW230814single} and found apparent deviations from \GR. That event does not appear in this paper since it was only observed by a single detector.
We investigated these apparent deviations carefully and found that they can likely be attributed to a combination of detector noise and waveform modeling inaccuracies.

Also, we do find the \GR value in the tails of the distribution in some cases when combining together multiple events.
Specifically, the \soft{pyRing} \texttt{KerrPostmerger} analysis finds a \ac{GR} quantile of ${\pyRingHierarchicalQuantileFourDValue}_{-\pyRingHierarchicalQuantileFourDMinus}^{+\pyRingHierarchicalQuantileFourDPlus}\%$,
    though this reduces to ${\pyRingHierarchicalQuantileOneDMuDeltaTauValuelnBcut}_{-\pyRingHierarchicalQuantileOneDMuDeltaTauMinuslnBcut}^{+\pyRingHierarchicalQuantileOneDMuDeltaTauPluslnBcut}\%$ when only
    considering events with a significant ringdown, so there is no significant evidence for a \ac{GR} deviation. Here the error bars come from a bootstrapping analysis and indicate that some of the apparent
    significance could also be due to catalog variance \citep[cf.][]{Pacilio:2023uef}.
The pSEOBNR parameterized ringdown analysis finds the \GR value at $\QuantileJointGWTCFourUncertainty\%$ credibility in the joint posterior analysis and $\QuantileHierGWTCFourUncertaintyTauOneD\%$
in the hierarchical analysis, while the analysis of GWTC-3.0 events only finds the \GR value at $\QuantileJointGWTCThreeUncertainty\%$ and $\QuantileHierGWTCThreeUncertaintyTauOneD\%$ credibility,
respectively. The error bars from the bootstrapping analysis again indicate that some of the apparent significance could be due to catalog variance, which is supported by the significance decreasing to
$\QuantileJointGWTCFourPlusTwoFiveZeroOneOneFour$ and $\QuantileHierGWTCFourTauOneDPlusTwoFiveZeroOneOneFour$
when including the loud \ac{O4b} event GW250114 \citep{GW250114, GW250114_TGR}. Investigations of simulated observations
indicate that the pSEOBNR result is not likely to be due to waveform-modeling uncertainties.
Analyses of future high-SNR signals as well as injections into real detector noise could help clarify these combined results.

We will perform further tests of GR using detections from the remainder of the fourth observing run and future runs \citep{Aasi:2013wya}.\footnote{LVK observing run plans \url{https://observing.docs.ligo.org/plan}} Applying the pSEOBNR analysis to additional detections will show if the combined results end up disfavoring GR more strongly with more events, or if further studies of the noise and other systematics determine that the current tension is just a statistical fluctuation or systematic effect.%
Regardless, improvements in detector sensitivity along with advances in analysis and modeling techniques will let us place ever more stringent constraints on potential deviations from GR in the ringdown.

All strain data analyzed in this paper are available from the Gravitational Wave Open Science Center \citep{OpenData}. The data and scripts used to prepare the figures and tables are available at \citet{PaperIII_DCC_release}.

\section*{Acknowledgements}

This material is based upon work supported by NSF's LIGO Laboratory, which is a
major facility fully funded by the National Science Foundation.
The authors also gratefully acknowledge the support of
the Science and Technology Facilities Council (STFC) of the
United Kingdom, the Max-Planck-Society (MPS), and the State of
Niedersachsen/Germany for support of the construction of Advanced LIGO 
and construction and operation of the GEO\,600 detector. 
Additional support for Advanced LIGO was provided by the Australian Research Council.
The authors gratefully acknowledge the Italian Istituto Nazionale di Fisica Nucleare (INFN),  
the French Centre National de la Recherche Scientifique (CNRS) and
the Netherlands Organization for Scientific Research (NWO)
for the construction and operation of the Virgo detector
and the creation and support  of the EGO consortium. 
The authors also gratefully acknowledge research support from these agencies as well as by 
the Council of Scientific and Industrial Research of India, 
the Department of Science and Technology, India,
the Science \& Engineering Research Board (SERB), India,
the Ministry of Human Resource Development, India,
the Spanish Agencia Estatal de Investigaci\'on (AEI),
the Spanish Ministerio de Ciencia, Innovaci\'on y Universidades,
the European Union NextGenerationEU/PRTR (PRTR-C17.I1),
the ICSC - CentroNazionale di Ricerca in High Performance Computing, Big Data
and Quantum Computing, funded by the European Union NextGenerationEU,
the Comunitat Auton\`oma de les Illes Balears through the Conselleria d'Educaci\'o i Universitats,
the Conselleria d'Innovaci\'o, Universitats, Ci\`encia i Societat Digital de la Generalitat Valenciana and
the CERCA Programme Generalitat de Catalunya, Spain,
the Polish National Agency for Academic Exchange,
the National Science Centre of Poland and the European Union - European Regional
Development Fund;
the Foundation for Polish Science (FNP),
the Polish Ministry of Science and Higher Education,
the Swiss National Science Foundation (SNSF),
the Russian Science Foundation,
the European Commission,
the European Social Funds (ESF),
the European Regional Development Funds (ERDF),
the Royal Society, 
the Scottish Funding Council, 
the Scottish Universities Physics Alliance, 
the Hungarian Scientific Research Fund (OTKA),
the French Lyon Institute of Origins (LIO),
the Belgian Fonds de la Recherche Scientifique (FRS-FNRS), 
Actions de Recherche Concert\'ees (ARC) and
Fonds Wetenschappelijk Onderzoek - Vlaanderen (FWO), Belgium,
the Paris \^{I}le-de-France Region, 
the National Research, Development and Innovation Office of Hungary (NKFIH), 
the National Research Foundation of Korea,
the Natural Sciences and Engineering Research Council of Canada (NSERC),
the Canadian Foundation for Innovation (CFI),
the Brazilian Ministry of Science, Technology, and Innovations,
the International Center for Theoretical Physics South American Institute for Fundamental Research (ICTP-SAIFR), 
the Research Grants Council of Hong Kong,
the National Natural Science Foundation of China (NSFC),
the Israel Science Foundation (ISF),
the US-Israel Binational Science Fund (BSF),
the Leverhulme Trust, 
the Research Corporation,
the National Science and Technology Council (NSTC), Taiwan,
the United States Department of Energy,
and
the Kavli Foundation.
The authors gratefully acknowledge the support of the NSF, STFC, INFN and CNRS for provision of computational resources.

This work was supported by MEXT,
the JSPS Leading-edge Research Infrastructure Program,
JSPS Grant-in-Aid for Specially Promoted Research 26000005,
JSPS Grant-in-Aid for Scientific Research on Innovative Areas 2402: 24103006,
24103005, and 2905: JP17H06358, JP17H06361 and JP17H06364,
JSPS Core-to-Core Program A.\ Advanced Research Networks,
JSPS Grants-in-Aid for Scientific Research (S) 17H06133 and 20H05639,
JSPS Grant-in-Aid for Transformative Research Areas (A) 20A203: JP20H05854,
the joint research program of the Institute for Cosmic Ray Research,
University of Tokyo,
the National Research Foundation (NRF),
the Computing Infrastructure Project of the Global Science experimental Data hub
Center (GSDC) at KISTI,
the Korea Astronomy and Space Science Institute (KASI),
the Ministry of Science and ICT (MSIT) in Korea,
Academia Sinica (AS),
the AS Grid Center (ASGC) and the National Science and Technology Council (NSTC)
in Taiwan under grants including the Science Vanguard Research Program,
the Advanced Technology Center (ATC) of NAOJ,
and the Mechanical Engineering Center of KEK.

Additional acknowledgements for support of individual authors may be found in the following document: \\
\texttt{https://dcc.ligo.org/LIGO-M2300033/public}.

For the purpose of open access, the authors have applied a Creative Commons Attribution (CC BY)
license to any Author Accepted Manuscript version arising.
We request that citations to this article use `A. G. Abac {\it et al.} (LIGO-Virgo-KAGRA Collaboration), ...' or similar phrasing, depending on journal convention.

\emph{The following open-source software has been used:}

Calibration of the \ac{LIGO} strain data was performed with a \GSTLAL{}-based
    calibration software pipeline~\citep{Viets:2017yvy}.
    Calibration of the Virgo strain data is performed with C-based software~\citep{VIRGO:2021kfv}.
Data-quality products and event-validation results were computed using the
    \soft{DMT}{}~\citep{DMTdocumentation}, \soft{DQR}{}~\citep{DQRdocumentation},
    \soft{DQSEGDB}{}~\citep{Fisher:2020pnr}, \soft{gwdetchar}{}~\citep{gwdetchar-software},
    \soft{hveto}{}~\citep{Smith:2011an}, \soft{iDQ}{}~\citep{Essick:2020qpo},
    \soft{Omicron}{}~\citep{Robinet:2020lbf} and
    \soft{PythonVirgoTools}{}~\citep{pythonvirgotools} software packages and contributing
    software tools.
Analyses in this catalog relied upon the \LALSUITE{} software library~\citep{lalsuite-software,Wette:2020air}.
The detection of the signals and subsequent significance evaluations in this catalog were performed with the
    \GSTLAL{}-based inspiral software pipeline~\citep{Messick:2016aqy,Sachdev:2019vvd,Hanna:2019ezx,Cannon:2020qnf},
    with the \MBTA{} pipeline~\citep{Adams:2015ulm,Aubin:2020goo}, and with the
    \PYCBC{}~\citep{Usman:2015kfa,Nitz:2017svb,Davies:2020tsx} and the
    \CWB{}~\citep{Klimenko:2004qh,Klimenko:2011hz,Klimenko:2015ypf} packages.
Estimates of the noise spectra and glitch models were obtained using
    \BAYESWAVE{}~\citep{Cornish:2014kda,Littenberg:2015kpb,Cornish:2020dwh,Gupta:2023jrn}.
Noise subtraction for one candidate was also performed with \soft{gwsubtract}{}~\citep{Davis:2022ird}.
Source-parameter estimation was performed with the \BILBY{}
    and \PBILBY{} libraries~\citep{Ashton:2018jfp,Romero-Shaw:2020owr,Smith:2019ucc} using the
    \DYNESTY{} nested sampling package~\citep{Speagle:2020spe}.

\SEOBNRFIVEPHM waveforms used in parameter estimation were generated using \soft{pySEOBNR}~\citep{Mihaylov:2023bkc}.
\pSEOBNRFIVEPHM waveforms used for testing \GR were generated using \BILBYTGR~\citep{ashton_2025_15676285}.
Echoes M.\ waveforms used for constraining echoes were generated using \soft{echoes\_waveform\_models}~\citep{echowfm}.
\soft{cpnest}~\citep{cpnest} and \soft{pyRing}~\citep{pyRing} were used to perform ringdown analyses.
Quasinormal mode frequencies were computed using \soft{QNM}~\citep{Stein:2019mop}.
The QNMRF analysis used \citet{qnmrf-soft}. The multi-dimensional hierarchical analysis results were produced using \soft{hierfit}~\citep{hierfit}.
\PESUMMARY{} was used to postprocess and collate parameter-estimation
results~\citep{Hoy:2020vys}.  The various stages of the parameter-estimation
analysis were managed with the \ASIMOV{} library~\citep{Williams:2022pgn} together with \soft{CBCFlow}~\citep{cbcflow}.
Plots were prepared with \soft{Matplotlib}~\citep{Hunter:2007ouj},
\SEABORN{}~\citep{Waskom:2021psk}, and \GWPY{}~\citep{gwpy-software}.
\NUMPY~\citep{Harris:2020xlr} and \SCIPY~\citep{Virtanen:2019joe} were used for analyses in the manuscript.

\phantomsection

\appendix

 \label{app:pyring} 

The selection of the analysis start time is a crucial step in BH spectroscopy, since it determines the extent to which the signal can be reliably modeled as a superposition of QNMs.
If the analysis begins too early, residual dynamical effects may bias the results, while starting too late reduces the available SNR.
The discussion below explains the procedure adopted in this work for the \soft{pyRing} analysis, and explains how potential systematics are controlled.

We initially vary the starting time over a broad interval to ensure that the estimated quantities evolve toward their GR values as expected, 
and to search for any unexpected anomaly. Subsequently, we refine the analysis by running over a restricted interval around $t_{\rm start}$
to verify that the linear model aligns with GR within the expected time range, mitigating systematics and checking for stability of the results. The systematic variation of start times provides crucial validation of our results.

For simplicity, results for each model are reported only at the characteristic start time $t_{\rm nom}$ when a QNM superposition model becomes valid, defined relative to each model peak time as described in Section~\ref{sec:pyring}.
As shown in Table~\ref{tab:remnant_params} for the \texttt{Kerr} and \texttt{KerrPostmerger} models and in Table~\ref{tab:ds_bayes_factors} for the \texttt{DS} model, there is no significant evidence for additional modes at $t_{\rm nom}$.
Since the IMR peak time has a non-negligible uncertainty \citep{Carullo:2019flw, Finch:2021qph, Cotesta:2022pci, Crisostomi:2023tle}, we have also verified that this result remains robust across the full range of plausible start times within
the 90\% credible interval of the IMR peak time measurement, centered on $t_{\rm nom}$.

For the \texttt{DS} and \texttt{Kerr} models, $t_{\rm nom} = 10t_{M^\mathrm{IMR}_\mathrm{f}}$,
    i.e., ten times the time-scaled median IMR value of the (redshifted) remnant mass from \citet{GWTC:Results}.
This time is expected to be sufficiently late for the QNM description to be valid given current sensitivity \citep{Bhagwat:2017tkm, Carullo:2018sfu},
    as the model mismatch is $O(10^{-3})$ for parameters compatible with equal-mass low-spin binary progenitors.
The analysis start time $t_{\rm nom}$ for \texttt{KerrPostmerger} is 0.

\clearpage

\begin{table}[ht]
\caption{\label{tab:ds_bayes_factors}
    Bayes factors between two-mode and one-mode \texttt{DS} models from the \soft{pyRing} analysis
    }
\begin{center}
\begin{tabular}{lclc}
\hline
Events & \ensuremath{\log_{10} \mathcal{B}^{\rm \texttt{2DS}}_{\rm \texttt{1DS}}} &
Events & \ensuremath{\log_{10} \mathcal{B}^{\rm \texttt{2DS}}_{\rm \texttt{1DS}}} \\
\hline
GW230601\_224134 \!\! & \!\! $-0.166$ \!\! & \!\! GW230927\_153832 \!\! & \!\! $-0.433$ \\
GW230609\_064958 \!\! & \!\! $-0.870$ \!\! & \!\! GW230928\_215827 \!\! & \!\! $-0.654$\\
GW230628\_231200 \!\! & \!\! $-0.891$ \!\! & \!\! GW231001\_140220 \!\! & \!\! $-0.449$ \\
GW230811\_032116 \!\! & \!\! $-0.705$ \!\! & \!\! GW231028\_153006 \!\! & \!\! $-0.459$ \\
GW230814\_061920 \!\! & \!\! $-0.839$ \!\! & \!\! GW231102\_071736 \!\! & \!\! $-1.050$ \\
GW230824\_033047 \!\! & \!\! $-1.132$ \!\! & \!\! GW231108\_125142 \!\! & \!\! $-1.062$ \\
GW230914\_111401 \!\! & \!\! $-1.114$ \!\! & \!\! GW231206\_233134 \!\! & \!\! $-1.029$ \\
GW230919\_215712 \!\! & \!\! $-0.746$ \!\! & \!\! GW231206\_233901 \!\! & \!\! $-0.843$ \\
GW230922\_020344 \!\! & \!\! $-0.404$ \!\! & \!\! GW231213\_111417 \!\! & \!\! $-1.246$ \\
GW230922\_040658 \!\! & \!\! $0.038$  \!\! & \!\! GW231223\_032836 \!\! & \!\! $-0.266$ \\
GW230924\_124453 \!\! & \!\! $-1.071$ \!\! & \!\! GW231226\_101520 \!\! & \!\! $-0.443$ \\
GW230927\_043729 \!\! & \!\! $-0.810$ \\
\hline
\end{tabular}
\end{center}
\tablecomments{The Bayes factors are computed starting at $t_{\rm nom}$.
         A value of \ensuremath{\log_{10} \mathcal{B}^{\rm \texttt{2DS}}_{\rm \texttt{1DS}} > 1} indicates support for HMs in the data.
         The error on each Bayes factor from the nested-sampling stopping criterion is $\sim 0.09$.}
\end{table}

\bibliography{}

\clearpage

\iftoggle{endauthorlist}{
 \let\author\myauthor
 \let\affiliation\myaffiliation
 \let\maketitle\mymaketitle
 \title{Authors}
 \pacs{}
 
 \newpage
 \maketitle
}

\end{document}